%% file: main.tex
\documentclass[sigconf,dvipsnames]{acmart}
\AtBeginDocument{%
  \providecommand\BibTeX{{%
    \normalfont B\kern-0.5em{\scshape i\kern-0.25em b}\kern-0.8em\TeX}}}

\setcopyright{acmcopyright}
\copyrightyear{2024}
\acmYear{2024}
\setcopyright{acmlicensed}
\acmConference[CHI '24]{Proceedings of the CHI Conference on Human Factors in Computing Systems}{May 11--16, 2024}{Honolulu, HI, USA}
\acmBooktitle{Proceedings of the CHI Conference on Human Factors in Computing Systems (CHI '24), May 11--16, 2024, Honolulu, HI, USA}
\acmDOI{10.1145/3613904.3642697}
\acmISBN{979-8-4007-0330-0/24/05}

\usepackage{xspace}
\usepackage{xcolor}
\usepackage{soul}

\usepackage{graphicx}
\usepackage{multirow}   

\usepackage{bbding}
\usepackage{hyperref}

\usepackage{colortbl}

\makeatletter
\@ifpackageloaded{arydshln}{
   \newcommand{\dashedline}[1]{
    \cdashline{#1}[.4pt/1pt]\noalign{\vskip 1pt}
  }
}{
  \newcommand{\dashedline}[1]{
    \cmidrule[.0025mm]{#1}
  }
}
\makeatother

\begin{document}

\title{A Design Space for Intelligent and Interactive Writing Assistants}


\author{Mina Lee\textsuperscript{1}, Katy Ilonka Gero\textsuperscript{1}, John Joon Young Chung\textsuperscript{1},
Simon Buckingham Shum\textsuperscript{2}, Vipul Raheja\textsuperscript{2}}
\author{Hua Shen\textsuperscript{2}, Subhashini Venugopalan\textsuperscript{2}, Thiemo Wambsganss\textsuperscript{2}, David Zhou\textsuperscript{2},
Emad A. Alghamdi\textsuperscript{3}}
\author{Tal August\textsuperscript{3}, Avinash Bhat\textsuperscript{3}, Madiha Zahrah Choksi\textsuperscript{3}, Senjuti Dutta\textsuperscript{3}, Jin L.C. Guo\textsuperscript{3}, Md Naimul Hoque\textsuperscript{3}, Yewon Kim\textsuperscript{3}, Simon Knight\textsuperscript{3}, Seyed Parsa Neshaei\textsuperscript{3}, Agnia Sergeyuk\textsuperscript{3}, Antonette Shibani\textsuperscript{3}, Disha Shrivastava\textsuperscript{3}, Lila Shroff\textsuperscript{3}, Jessi Stark\textsuperscript{3}, Sarah Sterman\textsuperscript{3}, Sitong Wang\textsuperscript{3},
Antoine Bosselut\textsuperscript{4}} 
\author{Daniel Buschek\textsuperscript{4}, Joseph Chee Chang\textsuperscript{4}, Sherol Chen\textsuperscript{4}, Max Kreminski\textsuperscript{4}, Joonsuk Park\textsuperscript{4}}
\author{Roy Pea\textsuperscript{4}, Eugenia H. Rho\textsuperscript{4}, Shannon Zejiang Shen\textsuperscript{4}, Pao Siangliulue\textsuperscript{4}}

					
						


\authornote{
Corresponding author: Mina Lee at the University of Chicago (\url{mnlee@uchicago.edu}) and Microsoft Research (\url{minal@microsoft.com}).
We denote each author's self-assigned role with the following superscripts:
1 for project leads,
2 for team leads (alphabetical),
3 for team members (alphabetical), and
4 for advisors (alphabetical).
Please see Appendix~\ref{app:author} for the full author list with their roles, affiliations, and contributions.
}

\renewcommand{\shortauthors}{Lee et al.}

\input{macro}

\begin{abstract}
\input{sections/abstract}
\end{abstract}

\maketitle

\section{Introduction}
\label{sec:introduction}
\input{sections/introduction}

\input{figures/designspace}

\vspace{-0.1cm}
\section{Background}
\label{sec:background}
\input{sections/background}

\vspace{-0.2cm}
\section{Approach}
\label{sec:methodology}
\input{sections/methodology}

\section{Design Space}
\label{sec:designspace}
\input{sections/designspace}

\section{Discussion}
\label{sec:discussion}
\input{sections/discussion}

\section{Conclusion}
\input{sections/conclusion}

\begin{acks}
We thank
CHI 2024 ACs and reviewers, 
Carly Schnitzler, 
Daniel Jiang, 
Rishi Bommasani, 
Advait Bhat, 
Martin Zinkevich, 
Tania Bedrax-Weiss, 
and
Minsuk Chang
for their valuable feedback on the manuscript.
We disclose the use of various intelligent and interactive writing assistants in the process of writing this manuscript. 
However, we note that the use was primarily limited to editing the authors' own text and the authors checked for plagiarism, misrepresentation, fabrication, and falsification of content.
Among our authors,
Simon Buckingham Shum is supported by University of Technology Sydney Learning \& Teaching Grant: AcaWriter.
Jin L.C. Guo and Avinash Bhat are supported by Natural Sciences and Engineering Research Council of Canada (NSERC) and Fonds de recherche du Québec (FRQNT).
Yewon Kim is supported by Institute of Information \& Communications Technology Planning \& Evaluation (IITP) grant funded by the Korea government (MSIT) (No.2022-0-00495, On-Device Voice Phishing Call Detection).	
Daniel Buschek is supported by the Bavarian State Ministry of Science and the Arts in a project coordinated by the Bavarian Research Institute for Digital Transformation (BIDT). 
Daniel Buschek is also supported by a Google Research Scholar Award.
Antoine Bosselut is supported by Swiss National Science Foundation (No. 215390), Innosuisse (PFFS-21-29), Sony Group Corporation, and Allen Institute for AI.
Hua Shen is supported by the Carnegie Fund from the University of Michigan and Carnegie Foundation. 
Roy Pea is supported as a Faculty Affiliate at the Institute for Human-Centered Artificial Intelligence (HAI) at Stanford University. 
\end{acks}

\bibliographystyle{ACM-Reference-Format}
\bibliography{main,all,technology}

\newpage

\appendix

\input{sections/appendix}

\end{document}

%% file: macro.tex
\newcommand{\note}[1]{{\color{red}\noindent\textbf{NOTE:} #1}\xspace}
\newcommand{\todo}[1]{{\color{red}\emph{TODO: #1}}\xspace}
\newcommand{\todocite}[0]{{\color{red}\textbf{[CITE]}}\xspace}
\newcommand{\placeholder}[1]{{\color{red}#1}\xspace}

\newenvironment{change}{}{\xspace}

\newenvironment{googlechange}{}{\xspace}

\newcommand{\teamauthors}[1]{
    \vspace{-0.1cm} \noindent {\fontfamily{lmss}\selectfont\small{\hyperref[app:author]{Team}: #1}}
}

\newcommand\eg{e.g.,~}
\newcommand\ie{i.e.,~}
\newcommand\dash{---}

\newcommand\lm{LM\xspace}
\newcommand\lms{LMs\xspace}

\newcommand{\numpaper}{{\change{115}}\xspace}
\newcommand{\numhcipaper}{{\change{60}}\xspace}
\newcommand{\numnlppaper}{{\change{55}}\xspace}
\newcommand{\numdim}{{\change{35}}\xspace}
\newcommand{\numcode}{{\change{143}}\xspace}
\newcommand{\numplaceholder}{{\color{red}n}\xspace}

\definecolor{mina-skyblue}{rgb}{0.1098, 0.5686, 0.7411}
\definecolor{mina-yellow}{rgb}{0.9450, 0.6117, 0}
\definecolor{mina-blue}{rgb}{0.3333, 0.3764, 0.6627}
\definecolor{mina-orange}{rgb}{0.8392, 0.3921, 0.2666}
\definecolor{mina-green}{rgb}{0.0274, 0.4274, 0.2352}

\definecolor{url-blue}{rgb}{0, 0, 0.5429}
\newcommand{\website}{\textcolor{url-blue}{\url{https://writing-assistant.github.io}}\xspace}

\newcommand{\textft}[1]{#1}
\newcommand{\customul}[2][black]{\setulcolor{#1}\ul{#2}\setulcolor{black}}

\newcommand{\taskdimension}[1]{\textbf{\textcolor{mina-skyblue}{#1:}}}
\newcommand{\userdimension}[1]{\textbf{\textcolor{mina-yellow}{#1:}}}
\newcommand{\technologydimension}[1]{\textbf{\textcolor{mina-blue}{#1:}}}
\newcommand{\interactiondimension}[1]{\textbf{\textcolor{mina-orange}{#1:}}}
\newcommand{\ecosystemdimension}[1]{\textbf{\textcolor{mina-green}{#1:}}}

\newcommand{\taskdimensiont}[1]{\emph{\textcolor{mina-skyblue}{#1}}}
\newcommand{\userdimensiont}[1]{\emph{\textcolor{mina-yellow}{#1}}}
\newcommand{\technologydimensiont}[1]{\emph{\textcolor{mina-blue}{#1}}}
\newcommand{\interactiondimensiont}[1]{\emph{\textcolor{mina-orange}{#1}}}
\newcommand{\ecosystemdimensiont}[1]{\emph{\textcolor{mina-green}{#1}}}

\newcommand{\taskcode}[1]{{\textft{\customul[mina-skyblue]{#1}}}}
\newcommand{\usercode}[1]{{\textft{\customul[mina-yellow]{#1}}}}
\newcommand{\technologycode}[1]{{\textft{\customul[mina-blue]{#1}}}}
\newcommand{\interactioncode}[1]{{\textft{\customul[mina-orange]{#1}}}}
\newcommand{\ecosystemcode}[1]{{\textft{\customul[mina-green]{#1}}}}

\newcommand{\question}[1]{\textbf{#1}}

\newcommand{\systemname}[1]{\textsc{#1}}

\newcommand{\mina}[1]{{\color{blue}\emph{\noindent \textbf{Mina}: #1}}\xspace}

\newcommand{\intframe}{\textsl{\texttt{Interaction Framework}}\xspace}

\definecolor{interaction}{HTML}{2B8C40}
\newcommand{\daniel}[1]{{\small\textcolor{orange}{\bf [#1 --Daniel]}}}
\newcommand{\sarah}[1]{{\small\textcolor{orange}{\bf [#1 --Sarah]}}}
\newcommand{\joseph}[1]{{\small\textcolor{WildStrawberry}{\bf [#1 --Joseph]}}}
\newcommand{\pao}[1]{{\small\textcolor{orange}{\bf [#1 --Pao]}}}
\newcommand{\shannon}[1]{{\small\textcolor{orange}{\bf [#1 --Shannon]}}}
\newcommand{\md}[1]{{\small\textcolor{interaction}{\bf [#1 --Md Naimul]}}}
\newcommand{\maxk}[1]{{\small\textcolor{interaction}{\bf [#1 --MaxK]}}}
\newcommand{\jin}[1]{{\small\textcolor{interaction}{\bf [#1 --Jin]}}}
\newcommand{\avinash}[1]{{\small\textcolor{interaction}{\bf [#1 --Avinash]}}}
\newcommand{\tal}[1]{{\small\textcolor{interaction}{\bf [#1 --Tal]}}}
\newcommand{\thiemo}[1]{{\small\textcolor{interaction}{\bf [#1 --Thiemo]}}}
\newcommand{\hua}[1]{{\small\textcolor{interaction}{\bf [#1 --Hua]}}}
\newcommand{\david}[1]{{\color{purple}\bf{[#1 --David]}\normalfont}}
\newcommand{\senjuti}[1]{{\color{blue}\bf{[#1 --Senjuti]}\normalfont}}
\newcommand{\vipul}[1]{{\small\textcolor{mina-blue}{\bf [#1]}}}
\newcommand{\parsa}[1]{{\color{purple}{#1}}}

%% file: sections/abstract.tex
In our era of rapid technological advancement, the research landscape for writing assistants has become increasingly fragmented across various research communities.
We seek to address this challenge by proposing a \emph{design space} as a structured way to examine and explore the multidimensional space of intelligent and interactive writing assistants. 
Through a large community collaboration, we explore five aspects of writing assistants: task, user, technology, interaction, and ecosystem. 
Within each aspect, we define dimensions (\ie fundamental components of an aspect) and codes (\ie potential options for each dimension) by systematically reviewing \numpaper papers. 
Our design space aims to offer researchers and designers a practical tool to navigate, comprehend, and compare the various possibilities of writing assistants, and aid in the envisioning and design of new writing assistants.

%% file: sections/introduction.tex
A \emph{writing assistant} is a computational system that assists users with improving the quality and effectiveness of their writing, from grammar and spelling checks to idea generation, text restructuring, and stylistic improvement.
In our current era of rapid technological advancement, however, the research landscape for writing assistants is becoming increasingly fragmented across various communities.
While numerous writing assistants have emerged in recent years, quite disparate research communities like Natural Language Processing (NLP), Human-Computer Interaction (HCI), and Computational Social Science (CSS) study writing assistants with different emphases, such as model performance, user interaction, and social phenomena.
There are even more specific areas like creativity support, second language acquisition, and disability studies, each of which may struggle to stay up to date with work happening across other communities.
This splintering poses a significant challenge for researchers and designers seeking a holistic view, making it essential to bridge these gaps for effectively navigating the complexities of sociotechnical systems~\citep{leavitt1965applied,trist1981evolution}.

In this paper, we contribute a \emph{design space} to provide a structured way to explore the multidimensional space of intelligent and interactive writing assistants.
Design spaces are taxonomies that present critical aspects of a design \citep{card1990thedesignspace,maclean1991questions,card1991morphological,romer2004thedesignspace,morris2023design} with three main uses.
The first is to establish a shared vocabulary that can help streamline communication and collaboration between researchers, designers, and other stakeholders.
The second use case is to provide support for understanding existing designs. 
They allow one to reflect on why certain design choices were made, and why a given design may succeed in certain ways and fail in others. 
The third is to support envisioning new designs. 
By thinking about regions of a design space that may be ``empty'' (\ie given a set of dimensions, no design considers all of them), we can think about what a design in that space would look like, and whether it might be worth pursuing.

Through a large community collaboration, we create a design space based on five key \emph{aspects} of writing assistants---task, user, technology, interaction, and ecosystem (Figure~\ref{fig:designspace})---based on the sociotechnical systems perspective.
Within each aspect, we identify \emph{dimensions} (\ie fundamental components of an aspect) and \emph{codes} (\ie potential options for each dimension) by systematically reviewing \numpaper papers and employing an iterative coding process.
As a result, our design space contains \numdim dimensions and \numcode codes (Table~\ref{tab:task-designspace}--\ref{tab:ecosystem-designspace}), which includes writing contexts (\eg academic, creative, and journalistic) and users' relationships to a system (\eg agency, ownership, trust, and privacy) that are closely related to interaction metaphors (\eg agent and tool).
It also includes diverse learning problems for technology (\eg classification, regression, and generation) as well as wider ecosystem considerations, such as digital infrastructure (\eg usability consistency and technical interoperability), norms and rules (\eg laws and conventions), and change over time (\eg written artifacts, information environment).
\googlechange{We provide two illustrative scenarios that demonstrate how our design space can be utilized for a range of stakeholders (\eg researchers and policymakers) in Section~\ref{sec:usecases}.}

Our main contribution is the development of the design space by systematically reviewing existing work, identifying important components within each aspect, and connecting those elements under one framework. 
While we recognize that this design space may not be exhaustive or permanent, we believe it provides researchers and designers with a practical means for exploring, understanding, and comparing the diverse potential of writing assistants beyond their immediate fields. 
We anticipate that this work will foster dialogue and research in the realm of writing assistants, ultimately aiding in the creation of innovative, ethical writing assistant designs.
We publicly release our annotated papers as a living artifact at \website to promote community involvement in refining and extending the design space.

%% file: figures/designspace.tex
\begin{figure*}
    \centering
    \includegraphics[width=1\textwidth,angle=0,origin=c]{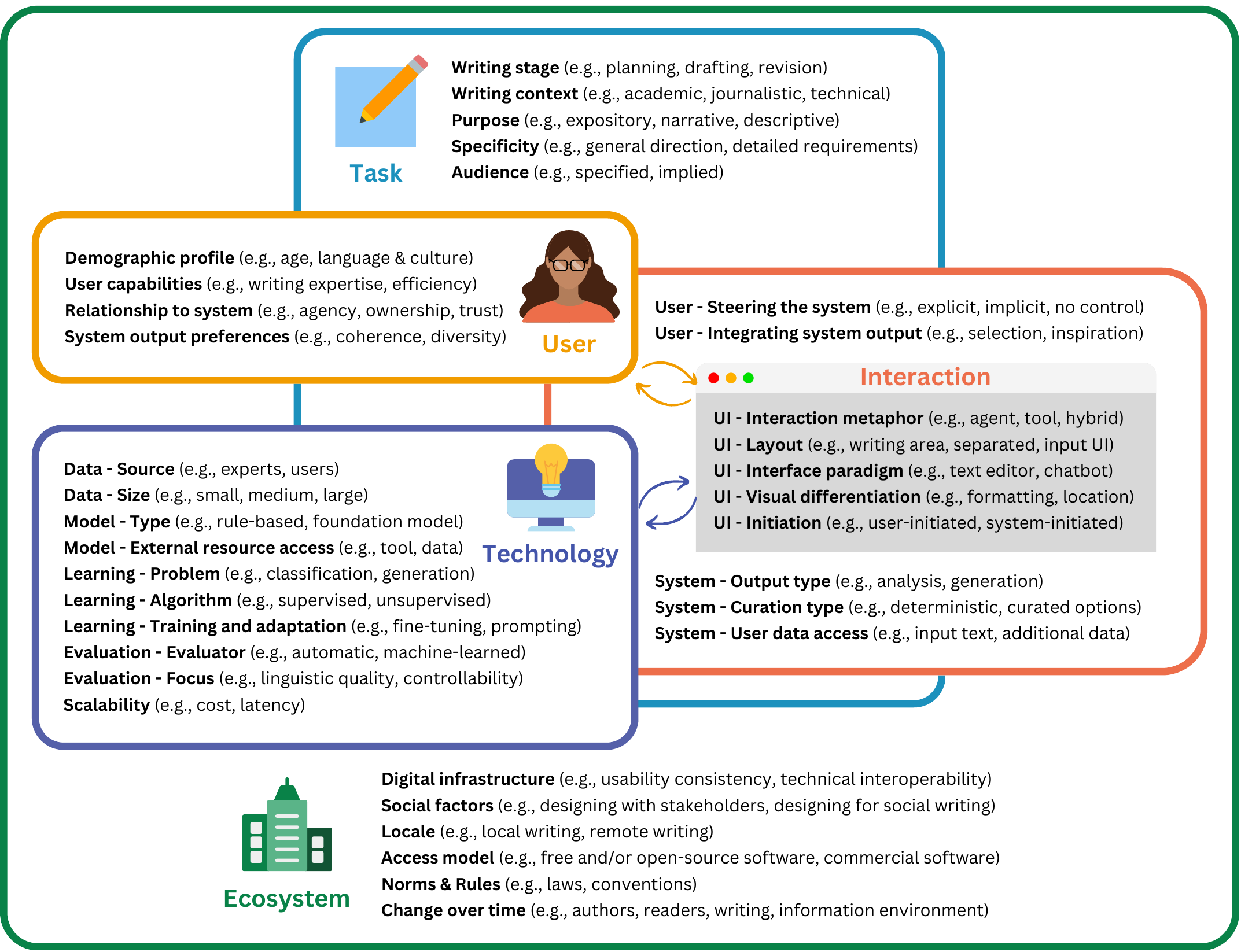}
    \caption[]{Our design space for intelligent and interactive writing assistants consists of five key \emph{aspects}---task, user, technology, interaction, and ecosystem---that are interconnected and interdependent. Within each aspect, we define \emph{dimensions} (bold texts) that represent fundamental components of the aspect and \emph{codes} (examples associated with bold texts) that represent possible options for each dimension.
    \change{When necessary, we group semantically relevant dimensions together within each aspect and use a prefix to denote the group name; the interaction dimensions are grouped by user, user interface (UI), and system; likewise, the technology dimensions are grouped by data, model, learning, and evaluation.}
    }
    \label{fig:designspace}
    \Description{The figure shows the integrated design space for writing assistants. It is comprised of five boxes, Ecosystem, Task, User, Interaction, and Technology. 
    Ecosystem is the biggest box that surrounds the rest, and has the following items: Digital infrastructure (e.g., usability consistency, technical interoperability), Social factors (e.g., designing with stakeholders, desigining for social writing), Locale (e.g., local writing, remote writing), Access model (e.g., free and/or open-source software, commercial software), Norms and Rules (e.g., laws, conventions), and Change over time (e.g., authors, readers, writing, information environment). 
    The top part inside Ecosystem box is occupied by Task box, which overlaps with User, Interaction, and Technology boxes. Task box has the following items: Writing stage (e.g., planning, drafting, revision), Writing context (e.g., academic, journalistic, technical), Purpose (e.g., expository, narrative, descriptive), Specificity (e.g., general direction, detailed requirements), and Audience (e.g., specified, implied).
    Interaction box is positioned in the middle of the User and Technology boxes, with arrows connecting to and from these boxes. Interaction has the following points: User - Steering the system (e.g., explicit, implicit, no control), User - Integrating system output (e.g., selection,  inspiration), UI - Interface paradigm (e.g., text editor, chatbot),  UI - Layout (e.g., writing area, separated, input UI), UI - Visual differentiation (e.g., formatting, location), UI - Interaction metaphor (e.g., agent, tool, hybrid), UI - Initiative (e.g., user-initiated, system-initiated), Technology - Output type (e.g., analysis, generation), Technology - Curation type (e.g., deterministic, curated options), and Technology - User Data access (e.g., input text, additional data). 
    User box is on the left of the Interaction box, with the following points: Demographic profile (e.g., age, language and culture), User capabilities (e.g., writing expertise, efficiency), Relationship to system (e.g., agency, ownership, trust), and System output preferences (e.g., coherence, diversity). 
    Technology box is on the left of Interaction box and below the User box, with the following points: Data - Source (e.g., experts, users), Data - Size (e.g., small, medium, large), Model - Type (e.g., rule-based, foundation models), Model - External resource access (e.g., tool, data), Learning - Problem (e.g., classification, generation), Learning - Algorithm (e.g., supervised, unsupervised), Learning - Training and adaptation (e.g., fine-tuning, prompting), Evaluation - Evaluator (e.g., automatic, machine-learned), Evaluation - Focus (e.g., linguistic quality, controllability), and Scalability (e.g., cost, latency).}
\end{figure*}

%% file: sections/background.tex
Technology has had a significant impact on the way we write.  
Some technologies long precede the appearance of computers, such as clay tablets known as cuneiform writing~\cite{olson2002writing, schmandt1992before}, and in the 19th and 20th century, typewriter technologies~\cite{pea1987cognitive}, each of which dramatically changed the way we produce written artifacts.
Technical advances introduced computer-powered text-entering systems, which further developed into word processors. 
Word processors allowed users to flexibly edit texts with functions like deletion, copy-paste, and find-and-replace~\cite{pea1987cognitive}. 
These systems further evolved to support the writer's \emph{cognitive process}, or the distinctive thought processes the writers go through when composing writings~\cite{flower1981cognitive,hayes1986writing}. 
Examples of such support include providing text analysis~\cite{frase1983unix, macdonald1983unix} and text-planning support~\cite{collins1988thecomputer} (see Appendix~\ref{app:background} for more examples).

While supporting the cognitive process of writing can help people improve their writing proficiency and efficiency, how to do so consistently and persistently remains a challenge. 
For example, what kinds of support should we provide to the human writing process and with which interactions?  
How can we technically enable such support? 
And when should we ``fade'' this support as a temporary ``scaffolding''~\cite{pea2018social} to encourage independent writing?
One thread of work is understanding the human writing process, such as the cognitive process theory of writing~\cite{flower1981cognitive,hayes2012modeling}. 
Another thread has been on extending the ``intelligence'' of writing assistants, so that these assistants can provide a wider range of support, such as those that require ``understanding'' of long and complicated texts. 
For example, for a writing assistant that provides suggestions on filling in between already written story events~\cite{donahue2020infilling,gero2023social}, the system would need to grasp what happened in the preceding and subsequent events in order to coherently connect them.
Fortunately, relevant technologies have advanced rapidly within the fields of NLP, where most recently, language models (\lms)~\cite{brown2020language,openai2022chatgpt,openai2023gpt4} and their prompt- and example-based usage showed impressive capabilities in generating coherent text.

Integrating scientific knowledge of human writing and technical advances, researchers designed and built many intelligent and interactive writing assistants in recent years. 
Some of these aimed to support existing human writing tasks and domains, such as help request writing in professional settings~\cite{hui2018introassist}, instant message writing in affectionate relationships~\cite {kim2019love}, or tweetorial writing for scientific communication~\cite{Gero2022sparks}. 
Researchers also studied how writers interact with these new writing assistants and how they influence human writing. 
For example, \citet{Buschek2021emails} studied how generative writing suggestions can influence email writing behaviors. 
Similarly, \citet{lee2022coauthor} collected a dataset of human-LM interactions during creative and argumentative writing to analyze how people interact with and get support from these technologies. 
Some work studied how writers leverage different prompting strategies for  \lms~\cite{dang2023choice,zamfirescu2023johnny} and introduced novel interaction paradigms to steer LM outputs, such as visual sketching~\cite{Chung2022talebrush}.

\googlechange{As new writing assistants are being introduced at an increasingly fast pace,
there are growing concerns that we lack a comprehensive understanding of 
when and in what ways it is desirable to use these writing assistants in order to avoid potential unforeseen consequences or ethical issues~\cite{mieczkowski2023thesis,jakesch2023cowriting}.
For instance, in the academic community, there are concerns about students' use of AI during writing and homework~\cite{meyer2023chatgpt,dagostino2023chatgpt,bowman2022chatbot,ku2023ai}.}
Moreover, while there are many design considerations for these assistants (\eg related to task, user, interaction, ecosystem), often only a few are considered (\eg technology), \googlechange{which can lead to a model-centric approach to building systems~\cite{kim2022interaction}.
}
Reflecting upon these concerns, some researchers strive to deepen our understanding about intelligent and interactive writing assistants as a whole. 
For example, \citet{wan2022user} studied interaction patterns of recent human-AI writing assistants. 
Similarly, \citet{gero2022design} analyzed existing writing assistants along the dimensions of writing goals and technologies. 
Another effort involved understanding user perspectives in getting support for their writing processes~\cite{gero2023social}. 
However, we are still far from having a comprehensive picture of the landscape. 
Outside of writing assistants, researchers studied various aspects of creativity support tools with the goal of comprehensive understanding of the space~\cite{chung2021intersection, frich2019mapping}. 
However, such a study has not yet been conducted specifically on writing assistants. 


%% file: sections/methodology.tex
To develop a design space for intelligent and interactive writing assistants,
we first decide on the scope (Section~\ref{sec:scope}) and core aspects of writing assistants (Section~\ref{sec:categories}).
Then, we perform a systematic literature review while employing an iterative coding process and create the design space (Section~\ref{sec:litreview}, Figure~\ref{fig:process}).
We highlight that this process was a collaborative effort consisting of a large team of researchers from a variety of disciplines, including HCI, NLP, Information Systems (IS), and Education.

\vspace{-0.2cm}
\subsection{Scope}
\label{sec:scope}

We define the scope of our work by specifying what we consider as \emph{intelligent} and \emph{interactive} \emph{writing} assistants.
Note that these are working definitions that align with the objectives of our study, and are not intended to be imposed as a universally accepted definition.
\begin{itemize}
    \item \textbf{Intelligent}: We consider systems to be intelligent if they are capable of autonomous decision-making and/or text generation, with a special focus on modern AI, such as language models (\lms).
    \item \textbf{Interactive}: We consider systems to be interactive if they reflect human input and/or output (as opposed to generating text without human involvement) and facilitate an iterative process to produce a written artifact.
    \item \textbf{Writing}: We consider systems to be relevant to writing if at some point human users translate their thoughts into written language, perhaps via system support, and produce text \googlechange{in a natural language (\eg English)} as a final artifact.
\end{itemize}
To have a stronger and tighter focus on intelligent and interactive writing assistants, we exclude the following types of work: 
programming tools (to focus on writing as a means to communicate with humans as opposed to machines), 
collaborative tools (to focus on the dynamics between a user and a system as opposed to multiple users), 
text-entry tools (to focus on tools that facilitate the cognitive process of writing, rather than input methods, such as \googlechange{gestural, hand-writing, or speech-to-text}).
Throughout the paper, we interchangeably use ``interactive and intelligent writing assistants'' and ``writing assistants'' to refer to the writing assistants that fall within our scope (see Appendix~\ref{app:terminology} for our definitions of similar terms like ``technology,'' ``system,'' and ``model'').

\subsection{Five Aspects of Writing Assistants}
\label{sec:categories}

\change{Writing assistants are \emph{sociotechnical systems} that involve interaction between both technical (\eg AI and hardware infrastructure) and social (\eg user behaviors and societal norms) components.
The sociotechnical systems perspective is an interdisciplinary approach~\citep{leavitt1965applied,bostrom1977mis,sawyer2014sociotechnical}, 
widely used in various research fields like HCI~\citep{knittel2019bitcoin, lewicki2023service} and IS~\citep{weber2021pedagogical,chen2008aperspective} to analyze complex interactions involving technology, workflow design, and organizational changes.

A core assumption of the sociotechnical systems perspective is that technology does not exist in a vacuum and is always interconnected with people and other systems.
The perspective considers ``task,'' ``user,'' ``technology,'' and ``structure'' as integral and interdependent parts of a complex system~\citep{leavitt1965applied}.
We adopt this point of view with the following modifications:
We first split ``technology'' into two---``technology'' and ``interaction''---to be more aligned with current research in NLP and HCI, respectively, and rename ``structure'' as ``ecosystem'' to account for the broader context in which writing assistants operate.
As a result,}
we designate these five key \emph{aspects} of writing assistants.

\begin{itemize}
    \item \textbf{\textcolor{mina-skyblue}{Task}}: Writing stages, contexts, and purposes that writing assistants aim to support. This involves understanding the purpose of writing, the constraints imposed on writing, and the intended audience for the written content.
    \item \textbf{\textcolor{mina-yellow}{User}}: Characteristics and preferences of users of writing assistants, providing insights into how different user groups may prioritize different attributes in their interactions with these systems.
    \item \textbf{\textcolor{mina-blue}{Technology}}: Building blocks of developing underlying models that power writing assistants, involving the quality and quantity of training data, modeling problems and techniques, and evaluation methods.
    \item \textbf{\textcolor{mina-orange}{Interaction}}: Diverse interaction paradigms and essential user interface components of writing assistants, that contribute to the dynamic interplay between the user, the interface, and technology powering the interface.
    \item \textbf{\textcolor{mina-green}{Ecosystem}}: Issues stemming from the broader context in which writing assistants operate. This involves economic, social, and regulatory considerations that impact how these systems are developed, used, and evolved.
\end{itemize}

\subsection{Systematic Literature Review}
\label{sec:litreview}

\input{figures/process}

\change{Figure~\ref{fig:process} shows the overall process of our systematic literature review and the design space creation.}

\subsubsection{\change{Paper Selection}}
\label{sec:sampling}
To understand the current research literature on writing assistants, we performed a systematic literature review.
We first identified fields closely related to intelligent and interactive writing assistants.
We selected the fields of HCI and NLP as our core fields\footnote{
While there are other relevant fields (\eg Machine Learning, Cognitive Science, Writing, and Education), we excluded them as they usually do not consider all of the three elements (\ie intelligent, interactive, and writing) that define our scope (Section~\ref{sec:scope}).
\change{Nevertheless, we note that many of the selected papers tend to be interdisciplinary and cover various subcommunities.}
}
and their relevant associated venues listed in Appendix~\ref{sec:venues}.
From the \href{https://dl.acm.org/}{ACM Digital Library} and \href{https://aclanthology.org/}{ACL Anthology}, we retrieved $419$ candidate papers (as of August 2023) that included the participles of the verb ``write'' \change{and its variations} either in the title or keywords, if provided by authors \change{(see Appendix~\ref{sec:litreview} for details)}.
Then, three of the authors filtered candidates encompassed by the scope of the project (Section~\ref{sec:scope}).
These authors first independently annotated the relevance of candidate papers, discussed disagreements, and then refined the scope. 
\change{As a result, we selected \numpaper out of $419$ papers that fulfilled the inclusion criteria for our review.}

Note that the above sampling method targeted a specific set of \emph{existing} writing assistants, as opposed to all possible options for \emph{future} writing assistants.
Concretely, papers under our review explicitly included a specific technology, a human user, and interaction between the user and the technology through an interface in the context of writing.
If a paper did not present a strong connection to all of these components, we excluded it from the review.
For instance, we excluded a technical paper that rewrites content in a target style even though it could be used in writing assistants (\ie it did not explicitly consider user and interaction aspects, but studied style transfer in isolation).
In the following section, we describe how we complement the strict exclusion of these works by leveraging the expertise of our paper's authors by adding flexibility to the coding process.

\subsubsection{\change{Code Development}}
\label{sec:coding}
\change{With the \numpaper papers, the authors were split into five teams based on the five aspects (Section~\ref{sec:categories}) to develop codes for each aspect.
Then, every team sub-selected the papers that were relevant to their aspect.
For instance, some papers might allude to potential users but lack descriptions of exact target users; 
the team focusing on the user aspect (henceforth ``the user team'') excluded those papers from their consideration.
At least two authors from each team read each paper to decide the relevance while resolving any disagreement through discussions.}

To go beyond the understanding of \emph{existing} writing assistants and think about possibilities for \emph{future} writing assistants, 
\change{we intentionally allowed flexibility in the code development process in the following ways.
First, while some teams (task and user) derived the initial codes mostly from their selected papers (\ie inductive coding), other teams
(technology, interaction, and ecosystem) created the initial codes based on external knowledge (\ie deductive coding).}
\change{Second, to provide insights on recent advancements in technology that are not yet leveraged for writing assistants, the technology team brought in $25$ additional papers (outside of the \numpaper papers) based on relevance and importance (see Appendix~\ref{app:external} for more details) and referenced them while iteratively refining the codes.}
Third, most teams considered relevant works that are not necessarily about writing assistants (\eg papers solely about \lms) to inform their codes. 
\change{Note that we used these external references to aid our code development but did not code them as part of our review process.}

With the subset of \numpaper papers selected by each team, the authors in the team iteratively coded papers while refining the codes.
The iterative coding process began with the distribution of papers to members of each team. 
We distributed papers so that each paper was read by at least two members,
while maximizing the number of combinations of readers to facilitate discussion among team members.
With distributed papers, each team member first independently read and coded papers with the up-to-date version of the codes. 
Then, team members had a discussion session to share and agree on how they coded each paper.
If team members decided that a specific paper could not be adequately coded with the existing code structure, they updated the code structure either by revising, merging, splitting, adding, or removing codes. 
As codes were updated, team members returned to already-reviewed papers to ensure that papers were coded with the up-to-date code structure. 
The process of coding papers repeated until the team reviewed all papers under the final set of codes.

\subsubsection{\change{Final Design Space and Coding}}
\change{After each team developed their codes, all teams gathered to create the final design space by combining the dimensions and codes from all teams.
During the process, we removed overlapping dimensions and codes, improved their consistency across the aspects (\eg inclusion criteria and granularity), and revised their names and definitions for clarity.
Once we finalized the design space, we repeated the iterative coding process, where each paper was read by two authors (same as the initial iterative coding in Section~\ref{sec:coding}). 
This time, all \numpaper papers were coded with the full set of \numdim dimensions and \numcode codes from the five aspects. 
We briefly analyze trends and gaps based on this final coding of the papers (Section~\ref{sec:analysis}) and report two metrics for inter-coder reliability: 
Percent agreement (mean: 0.93, std: 0.06) and Krippendorff's alpha (mean: 0.69, std, 0.17).\footnote{\change{Krippendorff's alpha is applicable when multiple coders see only partial instances, which is our case as many coders contributed to coding all the papers. However, \citet{krippendorff2011agreement} suggest that this metric is only reliable for binary classification (in our case, codes either occur or do not) when certain conditions (\eg minimum code occurrence) are met; for this reason we excluded codes that occur less than 50 times and considered 72 codes for calculating Krippendorff's alpha.}}
We release our coded papers as a living artifact at \website and encourage others to contribute by adding papers beyond those we covered in this work to track future developments in this space.}

%% file: figures/process.tex
\begin{figure*}
    \centering
    \includegraphics[width=0.8\textwidth]{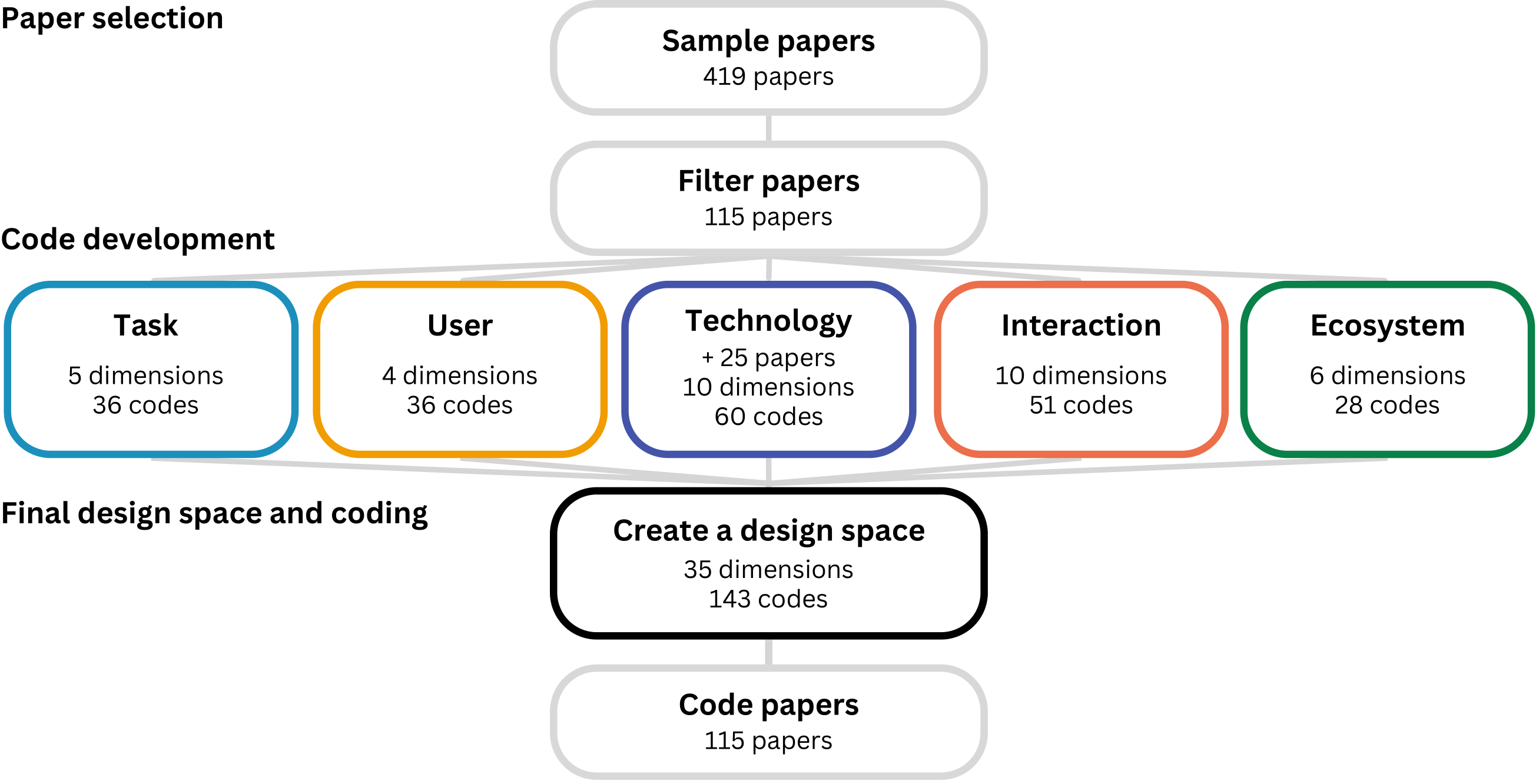}
    \caption[]{\change{Our systematic literature review has three stages.
    First, we sampled and filtered papers relevant to writing assistants from HCI and NLP venues.
    Second, the authors split into five teams based on the five key aspects of writing assistants and developed codes by reviewing papers and employing an iterative coding process.
    In the final stage, all teams gathered to create the final design space by combining all dimensions and codes and coded all papers.
    }}
    \label{fig:process}
    \Description{The figure shows a flow chart for the overall process of our systematic literature review in three stages: paper selection, code development, and final design space and coding. The first stage (paper selection) is comprised of two boxes. On the top, there is a box labeled "sample papers" with the downward arrow pointing to the other box labeled "filter papers", located below. Then, there are five outward arrows pointing to the five boxes in the next stage (code development), which corresponds to the five aspects of writing assistants. These arrows represent the use of sampled and filtered papers in all five aspects. The five boxes are labeled as "task", "user", "technology", "interaction", and "ecosystem". Then, in the third stage (final design space and coding), there are five downward arrows from the five boxes in the previous stage to the "create a design space" box to represent the use of the combined set of dimensions and codes from all five aspects. Finally, this box has a downward arrow to the "code papers" box to represent the use of all dimensions and codes to code a full set of papers.
    }
\end{figure*}

%% file: sections/designspace.tex
In this section, we present a \emph{design space} as a structured way to examine and explore the multidimensional space of writing assistants (Figure~\ref{fig:designspace}).
Our design space encompasses five key \emph{aspects}---task, user, technology, interaction, and ecosystem (Section~\ref{sec:categories})---which are interconnected and play vital roles in the realm of writing assistants.
Within each aspect, we define \emph{dimensions} that represent the fundamental components of that aspect and \emph{codes} that denote potential options for each dimension, based on our systematic literature review (Section~\ref{sec:litreview}).
In the following sections, we provide detailed explanations regarding the dimensions and codes specific to each aspect along with concrete examples of research papers associated with each code. 
When doing so, we formulate our dimensions as questions and present codes as possible answers to the questions, similar to questions and options in \citet{maclean1991questions}.

\subsection{Task}
\input{sections/task}

\subsection{User}
\input{sections/user}

\subsection{Technology}
\label{sec:technology}
\input{sections/technology}

\subsection{Interaction}
\label{sec:interaction}
\input{sections/interaction}

\subsection{Ecosystem}
\input{sections/ecosystem}

%% file: sections/task.tex
\input{tables/task-designspace}

In the design space of writing assistants, we define a \emph{task} as a rhetorical purpose of a written artifact identified by a user (\eg to persuade) at a specific stage of writing (\eg revision) and in a particular context (\eg academic). 
This task can be articulated to a writing assistant with varying degrees of requirements (from no explicit guidelines to detailed instructions) and a particular audience. 
With this definition, the task aspect embodies user-driven objectives, facilitating a nuanced, user-centered approach.

\subsubsection{Dimensions and Codes}
We describe dimensions and codes for the task aspect. Figure~\ref{fig:designspace} (``Task'') shows task dimensions in a broad context, while Table~\ref{tab:task-designspace} lists all dimensions, codes, and definitions.

\paragraph{\taskdimension{Writing Stage}}
\looseness-1 \question{At what point in the writing process is the task taking place?} 
In the intricate and iterative process of writing, 
a writing assistant can be designed to support specific writing stages\footnote{
Note that our writing stages differ from the cognitive processes of writing proposed by \citet{flower1981cognitive}, despite the similarity in terminology   
(they use ``planning,'' ``translating,'' and ``reviewing'' to describe writing subprocesses).
Rather, our writing stages resemble (yet are not the same as) the stage models of writing, such as  \citet{rohman1965pre,britton1975development}, that model the growth of the written artifact rather than the inner workings of the writer.
We choose this approach because we find it more intuitive to design writing assistants to support a writing stage as a high-level cluster of relevant cognitive processes, compared to designing one writing assistant for each cognitive process.
To illustrate the distinction, consider writing an outline of a paper as an example of a task in the \taskcode{planning} stage.
There are multiple cognitive processes involved in the task, such as figuring out what to write (``planning''), jotting these ideas down as bullet points (``translating''), and reorganizing these points to enhance the overall flow (``reviewing'').
Therefore, there is a natural one-to-many relationship between our writing stages and the cognitive processes of writing.
} to provide more targeted and relevant support. 
For example, during the \taskcode{idea generation} stage, writers brainstorm and develop concepts.  
In this stage, writing assistants can offer ideas for potential topics and content~\cite{Gero2022sparks, schmitt2021characterchat}.
In the \taskcode{planning} phase, writers focus on organizing the structure and content outline. 
Writing assistants can play a crucial role in aiding writers during this phase, for instance, by reducing planning time \cite{kaur2018using, babaian2002awriters}. 
During the \taskcode{drafting} stage, writers may require assistance with thought facilitation or initial drafts \cite{natalie2023supporting, park2021iwrote}.
The \taskcode{revision} stage involves identifying and rectifying mistakes, making necessary corrections, and enhancing the content for its purpose(s). 
Writing assistants at this stage might be used to provide feedback on various aspects of the written text
\cite{afrin2021effective, wambsganss2022adaptive}.

\paragraph{\taskdimension{Writing \change{Context}}}
\question{\change{What combination of stylistic norms, audience expectations, and domain-specific conventions characterize the approach to the task?}}
This dimension delves into the diverse contexts of writing that shape the task's approach based on community and domain-specific norms.
The \taskcode{academic} context is characterized by rigorous research, thorough analysis, and the formal presentation of knowledge. Within this context, writers may benefit from assistance that upholds formal practices in genres such as scientific writing \cite{august2020writing} and theses \cite{rapp2015thesis}. 
In the \taskcode{creative} context, the focus lies on imagination, artistic expression, and original storytelling. An intelligent writing assistant within this context can provide support for diverse creative endeavors, including writing lyrics~\cite{watanabe2017lyrisys}, crafting metaphors~\cite{Gero2019metaphoria}, and offering inspiration and assistance for storytelling~\cite{osone2021buncho, schmitt2021characterchat}.
The \taskcode{journalistic} context centers around factual reporting, news coverage, and effective communication of information to the public. 
Writing assistants can play supportive roles in this process by assisting with science writing~\cite{kim2023metaphorian} and generating suggestions based on keywords~\cite{clark2018creative}. 
In the \taskcode{technical} context, authors convey complex information, provide instructions, and explain specialized topics, such as by writing figure captions ~\cite{padmakumar2022machine}. 
In the \taskcode{professional} context, precise and formal writing is essential. Writing assistants in this context can be tailored to aid in producing reports and documents related to companies, trade, and professional work~\cite{wambsganss2022modeling, hui2023lettersmith}.
In the \taskcode{personal} context, individuals engage in private or public reflections on their thoughts, experiences, and emotions. Writing assistants tailored for this context can provide support for tasks that involve emotional expression in writing~\cite{wang2018mirroru, park2021iwrote}, as well as fostering empathetic relations with readers~\cite{peng2020exploring}.

\paragraph{\taskdimension{Purpose}}
\question{What is the purpose of the written artifact?} 
A writing task may be motivated by a broad range of writing goals that guide the writing process.
One fundamental goal is to explain; \taskcode{expository} applies to writing tasks whose purpose is to convey relevant facts and knowledge \citep{kinnunen2012swan, rapp2015thesis}. 
Another common purpose is telling a story; \taskcode{narrative} applies to tasks whose goal is to convey an account of real \citep{gonzales2010motivating} and imagined experiences \citep{goldfarb2019plan, roemmele2018automated, hsu2019on}. 
Some texts are written to be \taskcode{descriptive} in order to convey emotion \citep{wang2018mirroru} and provide expressive details, particularly with evocative language \citep{Gero2019metaphoria, gero2019how, liu2019neuralbased}. 
Meanwhile, \taskcode{persuasive} writing aims to convince or influence the audience through text \citep{jakesch2023cowriting,afrin2021effective}, requiring high readability \citep{karolus2023your} and conciseness \citep{hui2018introassist}. 
Sometimes, the purpose of the written artifact can be \taskcode{educational}
(\eg readers learning about a topic by reading the written artifact). 
In this context, writing assistance could
assist writers in conveying educational ideas in an informative and clear manner \citep{babaian2002awriters, Gero2022sparks}.
With \taskcode{entertainment} as its purpose, a writing task can aim to engage and amuse readers through text,  
requiring coherent storytelling \citep{kim2023metaphorian, mirowski2023cowriting} to engage the audience and may use the structure of the writing itself to convey surprise \citep{gabriel2015inkwell} and pleasure \citep{watanabe2017lyrisys, lee2019icomposer}. 
\taskcode{Analytical} writing provides in-depth examination, which benefits from reflection and iteration \citep{Dang2022beyond, huang2018feedback}.
\taskcode{Accessibility} describes tasks that aim to support inclusivity; for example helping neurodivergent individuals explain and write about common social situations \citep{kim2008common}. 
Likewise, second language writing can be difficult due to unfamiliar spelling and grammar rules \citep{liu2000pens} as well as vocabulary \citep{huang2012transahead}, and writing assistants can help by aiding in  \taskcode{translation} to ease the difficulty of writing in an unfamiliar language \citep{zomer2021beyond, liu2000pens, huang2012transahead, huang2012transahead}. 
Lastly, \taskcode{feedback} describes tasks that provide evaluative comments, such as customer reviews \citep{dong2012first, bhat2023interacting} and constructive responses \citep{schmitt2021characterchat}.

\paragraph{\taskdimension{Specificity}} 
\question{How detailed are the task requirements?} 
To answer this question, we examine design choices that add specification to the writing task.
\taskcode{Nonspecific} is applied to tasks that do not have specific objectives. 
Systems supporting these tasks often include generic drafting environments that bundle many functionalities \citep{howe2009rita, lee2019icomposer} or general-purpose systems that do not specify a particular objective \citep{peng2020exploring, roemmele2018automated}. 
Tasks that provide a broad indication of the desired outcome without specifying the precise steps needed to achieve it are considered to have \taskcode{general direction} as their specificity.
These tasks may suggest the direction of an outcome through writing conventions \citep{park2008is}, writing reflection variations \citep{gabriel2015inkwell, kim2008common}, or writing format \citep{clark2018creative, dong2012towards}, but will not define how the desired outcome should be realized.
On the other hand, some writing tasks have \taskcode{specific objectives} that directly connect to the overall goal.
For example, while a general direction might be a target writing format, specific objectives may suggest the inclusion of certain subsections \citep{dacunha2017artext, hui2018introassist} or information \citep{padmakumar2022machine} that contribute to that overall goal.
Finally, tasks with \taskcode{detailed requirements} come with detailed and measurable instructions for the task. 
\citet{weber2023structured}, for example, outline a task for supporting legal writing based on the case solution's major claim, definition, subsumption, and conclusion, as well as elements and relations in the subsumption, which are highly specific requirements about what should be done in the task.

\paragraph{\taskdimension{Audience}} 
\question{Who is the intended recipient of the written artifact?}
This dimension describes the audiences for whom the output of the task is intended. 
\taskcode{Specified} audiences are specifically mentioned in the paper, such as the academic community \citep{Gero2022sparks}, people on the autistic spectrum \citep{kim2008common}, people with limited vision \citep{natalie2023supporting}, and writers themselves \citep{belakova2021sonami,sadauskas2015mining}. 
An \taskcode{implied} audience is inferred by the system design or characterization of the textual artifact, for example online shoppers for a tool designed to help writers review products \citep{dong2012towards}, business stakeholders for a system that aids with writing introductory help requests \citep{hui2018introassist}, and writers themselves for a tool that assists with personal reflection \citep{wang2018mirroru}.

%% file: tables/task-designspace.tex
\begin{table*}
    \resizebox{1.\linewidth}{!}{
        \renewcommand{\arraystretch}{1.45}
        \setlength{\tabcolsep}{4pt}
        \begin{tabular}{p{0.035\textwidth}p{0.25\textwidth}p{0.8\textwidth}}
            \toprule
            & \textbf{Code}             & \textbf{Definition}                                                                            \\
            \midrule
            \multirow{5}{*}{\rotatebox{90}{\textbf{\taskdimensiont{Writing Stage}}}} &
            \multicolumn{2}{l}{\textit{At what point in the writing process is the task taking place?}} \\
            \dashedline{2-3}
            & \taskcode{Idea generation}           & Brainstorming and developing concepts or content themes                                        \\
            & \taskcode{Planning}                  & Organizing the structure and content outline                                                   \\
            & \taskcode{Drafting}                  & Composing the written content                                                                  \\
            & \taskcode{Revision}                  & Reviewing and refining the written material                                                    \\
          
            \midrule
            \multirow{7}{*}[-.75em]{\rotatebox{90}{\textbf{\taskdimensiont{Writing Context}}}} &
            \multicolumn{2}{p{.95\linewidth}}{\textit{What combination of stylistic norms, audience expectations, and domain-specific conventions characterize the approach to the task?}} \\
            \dashedline{2-3}
            &
            \taskcode{Academic}                    &
            Focuses on research, analysis, and formal presentation of knowledge within educational contexts \\
            
            & \taskcode{Creative}                  & Focuses on imagination, narrative, artistic elements, and original storytelling \\  
            & \taskcode{Journalistic}              & Focuses on factual reporting, news coverage, and conveying information to the public \\
            & \taskcode{Technical}                 & Focuses on complex information, instructions, or explanations in specialized fields \\
            &
            \taskcode{Professional}                &
            Focuses on precise and formal writing like reports and documents \\
            
            & \taskcode{Personal}                  & Focuses on individual thoughts, experiences, and emotions; can be private or communicative \\                              
            
            \midrule
            \multirow{12}{*}{\rotatebox{90}{\textbf{\taskdimensiont{Purpose}}}} &
            \multicolumn{2}{l}{\textit{What is the purpose of the written artifact?}} \\
            \dashedline{2-3}
            & \taskcode{Expository}                &Intending to convey factual and informative content to the audience         \\
            & \taskcode{Narrative}                 & Intending to convey a story                                                                    \\
            & \taskcode{Descriptive}               & Intending to provide expressive details                                                        \\
            & \taskcode{Persuasive}                & Intending to convince or influence opinions or actions                          \\
            & \taskcode{Educational}               & Intending to teach and help people learn                                                       \\
            & \taskcode{Entertainment}             & Intending to engage and amuse  for leisure or enjoyment                             \\
            & \taskcode{Analytical}                & Intending to provide in-depth examination, analysis, or evaluation                                                  \\
            & \taskcode{Accessibility}             & Intending to support individuals with  health conditions or impairments \\
            & \taskcode{Translation}               & Intending to convert content from one language to another                                             \\
            & \taskcode{Feedback}                  & Intending to provide evaluative comments or responses to content                                          \\
            \midrule
            \multirow{5}{*}{\rotatebox{90}{\textbf{\taskdimensiont{Specificity}}}} &
            \multicolumn{2}{l}{\textit{How detailed are the task requirements?}} \\
            \dashedline{2-3}
            & \taskcode{Nonspecific}               & Having no explicit guidelines, instructions, or objectives                                      \\
            & \taskcode{General direction}         & Having a broad indication of the desired outcome without specifying precise steps           \\
            & \taskcode{Specific objectives}       & Having some specific objectives contributing to the task                                            \\
            & \taskcode{Detailed requirements}     & Having detailed and measurable instructions                                                  \\
            \midrule
            \multirow{4}{*}[0.5em]{\rotatebox{90}{\textbf{\taskdimensiont{Audience}}}} &
            \multicolumn{2}{l}{\textit{Who is the intended recipient of the written artifact?}} \\
            \dashedline{2-3}

            & \taskcode{Specified}   & Audience is clearly identified or explicitly stated                          \\
            & \taskcode{Implied}      & Audience is assumed or inferred without explicit mention                                                                 \\
            \bottomrule
        \end{tabular}
    }
\caption{Task dimensions, codes, and definitions.}
\label{tab:task-designspace}
\end{table*}

%% file: sections/user.tex
\input{tables/user-designspace}

In this section,
we ask who stands to benefit from these assistants and illuminate the varied needs and preferences of users that might influence design considerations for their systems. 
These distinctions remind us that the effectiveness of a writing assistant often lies in its attunement to a user's unique requirements.

\subsubsection{Dimensions and Codes}
Figure~\ref{fig:designspace} (``User'') shows user dimensions in a broad context, while \googlechange{Table~\ref{tab:user-designspace-1}} lists all dimensions, codes, and definitions.

\paragraph{\userdimension{Demographic Profile}}
\question{What are the demographic details of the users considered?}
The demographic profile of users reflects a broad spectrum of their characteristics, highlighting the diversity inherent across various groups of users. 
Some systems ensure equal representation of minority and marginalized groups by incorporating \usercode{gender} and \usercode{race} into their design~\cite{hoque2022dramatvis}. 
Others assist the writing process of users with diverse \usercode{socioeconomic status}, as it influences the availability and usage of technology, with consequences for user experiences and needs~\cite{goncalves2015you}.
\usercode{Language and culture}, reflecting users' fluent languages and cultural backgrounds, uniquely shape their writing experiences and outcomes. A number of studies focus on non-native English writers~\cite{chang2015writeahead2, Buschek2021emails, zomer2021beyond} in English writing contexts to address challenges in non-native language writing and to enhance language learning experiences.
Understanding \usercode{age} is important when tailoring systems to accommodate the developmental and cognitive characteristics of users across a wide age spectrum, ranging from young, pre-literate children~\cite{zarei2020investigating} to adolescents~\cite{goncalves2015you, sadauskas2015mining}.
Similarly,  \usercode{education} level and background of users, whether they are high school students ~\cite{sadauskas2015mining} or university graduates ~\cite{wambsganss2022supporting, afrin2021effective}, indicates varying cognitive and learning competencies, which in turn may influence how users interact with and critically engage in systems. 
Finally, some studies emphasize tailoring system designs to a specific \usercode{profession}, such as professional creative writers~\cite{gabriel2015inkwell,mirowski2023cowriting}, ensuring that the systems meet the unique writing needs of different professions.

\paragraph{\userdimension{User Capabilities}}
\question{What user attributes associated with the writing process can be influenced by writing assistants?}
User capabilities represent a distinct category from demographic attributes, which remain largely unchanged while interacting with writing assistants; user capabilities are often targeted for improvement by writing assistants.
\usercode{Writing expertise} captures the user's writing proficiency or expertise (\eg amateur vs. professional writer) in terms of writing quality~\cite{osone2021buncho, padmakumar2022machine} or genre specializations (\eg science writing vs. email writing)~\cite{Gero2022sparks, hui2018introassist}.
\usercode{Efficiency}, or how efficiently the user can complete specific writing tasks, was also considered in many papers. Several studies have utilized metrics such as word count, time, and effort expended by the user to quantify improvements in the user's writing efficiency before, during, and after using a writing assistant~\cite{arnold2020predictive, Buschek2021emails}. 
\usercode{Technical proficiency} relates to the extent to which a user is knowledgeable about the underlying technology of a writing assistant. 
Understanding how a \lm functions, for instance, has been shown to influence how effectively the user engages with a writing assistant~\cite{osone2021buncho}. 
Several papers in the literature focus on enhancing user capabilities related to emotional and cognitive aspects. 
For example, several studies capture the user's \usercode{confidence} in both the writing process and the final product, possibly expressed as self-efficacy or perceived skill level~\cite{hui2018introassist, wu2019design}. 
\usercode{Creativity} examines how the writing assistant can foster the user's creative exploration or curiosity when performing a writing task, for instance, by supporting idea generation~\cite{Gero2022sparks, yuan2022wordcraft, Chung2022talebrush}. 
\usercode{Emotion} refers to the user's emotional state before, during, and after interacting with the writing assistant~\cite{bixler2013detecting, wang2018mirroru, park2021iwrote}. 
\usercode{Empathy} focuses on the user's ability to emotionally and cognitively empathize with others in the writing process.
This empathetic focus was observed within educational contexts, where students are instructed to write more empathetic peer reviews~\cite{wambsganss2022adaptive, wambsganss2022supporting}. 
\usercode{Cognition} looks at cognitive aspects like the user's focus, sense of immersion, and cognitive load, as systems can increase cognitive engagement by tackling phenomena like writer's block~\cite{Chung2022talebrush,belakova2021sonami, schmitt2021characterchat}.
\change{Finally, \usercode{neurodiversity} encompasses considerations for users with diverse neurological profiles, such as aphasia~\cite{neate2019empowering} and dyslexia~\cite{fan2019character}.}

\paragraph{\userdimension{Relationship to System}} 
\question{What aspects characterize users' perceptions and their relationship with the system?}
As users engage with a system, they gradually develop a mental model of its functioning, which subsequently shapes their interaction and engagement with the system.
First, \usercode{agency} refers to users' sense of control over the system or the writing process. 
It is typically facilitated by providing users with options to steer the model outputs, either through adjusting model parameters~\cite{goldfarb2019plan, clark2018creative} or by using customized prompts~\cite{yuan2022wordcraft,dang2023choice}. 
The \usercode{ownership} or authenticity of a final product can be influenced by system design. 
A writer's sense of ownership may diminish as the proportion of system-generated text increases~\cite{lee2022coauthor}, yet this issue could be mitigated by personalizing writing assistants to mimic a writer's unique style~\cite{gero2023social}, or by designing assistants with greater agency~\cite{neate2019empowering}.
Similarly, maintaining a sense of \usercode{integrity} is an important factor when assisted by AI.
This encompasses worries about unintentional plagiarism and the moral implications of using writing assistants~\cite{biermann2022fromtool, Gero2022sparks}. 
\usercode{Trust} is another critical facet of users' perceptions, referring to their perception of the system's capabilities and their sentiments toward the technology itself. 
The level of trust users hold towards AI could influence the human-AI collaborative writing experience~\cite{biermann2022fromtool, liu2022will}. 
A system's \usercode{availability} emerged in the context of comparing human support to computer support, where human writers (\eg friends) are not always readily available, while computer programs are typically perceived as constantly accessible~\cite{gero2023social}. 
\usercode{Privacy} highlights users' concerns regarding how their data is handled by a system, including a sense of surveillance over their writing process~\cite{park2021iwrote, wu2019design, belakova2021sonami}.
Lastly, users are concerned with the \usercode{transparency} \citep{liao2023ai} of writing assistants 
as they seek clarity on how systems operate~\cite{bhat2023interacting, peng2020exploring}, how data is used~\cite{peng2020exploring}, and AI's role in these systems~\cite{park2021iwrote, carrera2022watch, bhat2023interacting}.

\paragraph{\userdimension{System Output Preferences}}
\question{What influences users' perceptions of system outputs?}
Understanding how writers evaluate system outputs, such as writing suggestions, is crucial as it can influence their interaction and engagement with the system.
One common consideration is \usercode{textual coherence}, which underlines the need for grammatically and contextually coherent outputs~\cite{Gero2022sparks, lee2022coauthor}.
Another significant dimension is \usercode{textual diversity}, which emphasizes the importance of offering varied system outputs to foster creativity in writing~\cite{Gero2019metaphoria, Gero2022sparks, singh2022where}.
The \usercode{explainability} of the system can also influence users' perceptions of its outputs. Providing additional information to explain the rationale behind system outputs may enhance user understanding and engagement~\cite{park2008is}.
Additionally, system-generated content may exhibit various forms of \usercode{bias}, ranging from skewed perspectives on topics~\cite{poddar2023aiwriting, jakesch2023cowriting} to societal stereotypes~\cite{hoque2022dramatvis}.
Lastly, the \usercode{personalization} of system outputs, which involves adapting to and reflecting an individual's unique writing style~\cite{gabriel2015inkwell}, may enhance the user's writing experience.

%% file: tables/user-designspace.tex
\begin{table*}
    \resizebox{1.\linewidth}{!}{
        \renewcommand{\arraystretch}{1.45}
        \setlength{\tabcolsep}{4pt}
        \begin{tabular}{p{0.055\textwidth}p{0.2\textwidth}p{0.8\textwidth}}
            \toprule
            &
             \textbf{Code} &
             \textbf{Definition} \\
            \midrule
           \multirow{8}{*}{\rotatebox{90}
           {\textbf{\userdimensiont{Demographic Profile}}}} &
             \multicolumn{2}{l}{\emph{What are the demographic details of the users considered?}} \\
             \dashedline{2-3}
             &
             \usercode{Gender} &
             User's gender \\
            &
             \usercode{Race} &
             User's race or ethnicity \\
            &
             \usercode{Socioeconomic status} &
             User's socioeconomic status \\
            &
             \usercode{Language \& Culture} &
             User's primary language and cultural background \\
            &
             \usercode{Age} &
             User's age \\
            &
            \usercode{Education} &
             User's educational background \\
            &
             \usercode{Profession} &
             User's profession \\
              \midrule
           \multirow{10}{*}{\rotatebox{90}{\textbf{\userdimensiont{User Capabilities}}}} & 
            \multicolumn{2}{l}{\emph{What user attributes associated with the writing process can be
influenced by writing assistants?}} \\
             \dashedline{2-3}
           &
             \usercode{Writing expertise} &
             User's writing expertise in terms of writing quality or genre specializations \\
            &
             \usercode{Efficiency} &
             User's writing efficiency, often measured as number of words written, time spent, or effort expended  \\ 
            &
             \usercode{Technical proficiency} &
             User's understanding of and comfort level with the underlying technology \\
            &
             \usercode{Confidence} &
             User's confidence in the writing process and final product, such as self-efficacy or perceived skill level \\
            &
             \usercode{Creativity} &
             User's engagement in creative exploration, including fostering curiosity and innovative thinking \\
            &
             \usercode{Emotion} &
             User's emotional state before, during, or after using the writing assistant \\
            &
             \usercode{Empathy} &
             User's ability to emotionally and cognitively empathize within the context of writing \\
            &
             \usercode{Cognition} &
             User's cognitive aspects, such as focus, sense of immersion, cognitive load, and writer's block \\ 
             &
             \usercode{Neurodiversity} &
             User's neurological profiles (both neurotypical and neurodivergent) \\
             \midrule
           \multirow{8}{*}[-.1em]{\rotatebox{90}{\textbf{\userdimensiont{Relationship to System}}}} &
            \multicolumn{2}{l}{\emph{What aspects characterize users' perceptions and their relationship with the system?}} \\
             \dashedline{2-3}
           &
             \usercode{Agency} &
             User's sense of control or autonomy in their interactions with the writing assistant \\
            &
             \usercode{Ownership} &
             User's sense of ownership or authenticity over the written artifact when using the writing assistant \\
            &
             \usercode{Integrity} &
             User's concerns about plagiarism and sense of integrity when using the writing assistant \\
            &
             \usercode{Trust} &
             User's sense of trust in the writing assistant's ability or perception of its suitability for a task \\
            &
             \usercode{Availability} &
             User's expectation of the writing assistant being at hand when one needs or wants to use it \\
            &
             \usercode{Privacy} &
             User's concerns about how their data is handled by the writing assistant \\
            &
             \usercode{Transparency} &
             User’s understanding of the writing assistant’s mechanism, capabilities, and limitations \\
             \midrule
            \multirow{6}{*}[-.6em]{\rotatebox{90}
            {\renewcommand{\arraystretch}{1}\begin{tabular}[c]{@{}c@{}}\userdimensiont{\textbf{System Output}}\\ \userdimensiont{\textbf{Preferences}} 
                    \end{tabular}}} & 
            \multicolumn{2}{l}{\emph{What influences users' perceptions of system outputs?}} \\
             \dashedline{2-3}
           &
             \usercode{Textual coherence} &
             Outputs that are coherent in terms of grammar, content, and tone\\
            &
             \usercode{Textual diversity} &
             Outputs that are novel and diverse, providing inspiration or surprises to the user \\ 
            &
             \usercode{Explainability} &
             Additional information for outputs that explains the rationale behind system outputs \\
            &
             \usercode{Bias} &
             Outputs that exhibit various forms of bias, such as skewed perspectives on topics and societal stereotypes \\
            &
             \usercode{Personalization} &
             Outputs that are personalized based on user preferences \\
        \bottomrule
        \end{tabular}
    }
\caption[]{User dimensions, codes, and definitions.}
\label{tab:user-designspace-1}
\end{table*}

%% file: sections/technology.tex
\input{tables/technology-designspace}

The technology aspect of writing assistants considers the advancements that underpin their intelligence and capabilities. 
In this section, we describe core building blocks of the end-to-end development process of models that power writing assistants.
In particular, we consider data, models, learning problems and techniques, and evaluation, all of which play a crucial role in determining the quality and degree of intelligence in the writing assistants.

\subsubsection{Dimensions and Codes}
Figure~\ref{fig:designspace} (``Technology'') shows technology dimensions in a broad context, while Tables~\ref{tab:technology-designspace-1} and \ref{tab:technology-designspace-2} list all dimensions, codes, and definitions.

\paragraph{\technologydimension{\change{Data - Source}}}
\change{
\question{Who is the creator of the data used to train or adapt a model?}
The source of the data used to develop a system or train a model can have a direct effect on the system's overall performance and reliability.
A dataset can be authored by \technologycode{experts} who have domain knowledge of the specific downstream task \cite{afrin2023predicting, weber2023structured, zhu2023visualize, karolus2023your}, or
\technologycode{users} of the system, during their interaction with the writing assistant \cite{Buschek2021emails, august2020writing, nagata2019toward, wang2018mirroru}. 
However, due to the difficulty of recruiting real experts and users, many researchers resort to
\technologycode{crowdworkers} to create data or annotate data entries \cite{zhong2023fiction, wang2023smart, chakrabarty2022help, kim2020lexichrome}. 
Sometimes, \technologycode{authors} themselves participate in the preparation and annotation of the dataset \cite{poddar2023aiwriting, jit2023semi, peng2020exploring, zhang2016using}.
Recently, we see more datasets that are generated by a \technologycode{machine} \cite{skitalinskaya2023revise, qi2022quoter, jiang2022arxivedits, hoque2022dramatvis}, which has the advantage of being relatively cheap and fast to generate at scale compared to human-generated datasets.
Finally, there are \technologycode{other} types of creators such as non-experts, unspecified individuals, or a broad set of creators (\eg in the case of web crawled data)~\cite{bhat2023interacting, yuan2022wordcraft, singh-etal-2021-drag, zomer2021beyond}.
}

\paragraph{\technologydimension{Data - Size}} \question{What is the size of the dataset\footnote{\googlechange{Note that we defined the size of a dataset with respect to the number of examples, not the total number of words in the dataset. 
As examples can vary in length, ranging from a single word to the length of a book, the actual size of the dataset doesn't necessarily correlate with the number of examples.}} used to train or adapt a model?}
Depending on the size of a dataset required to train or adapt (\eg fine-tuning or prompting) a model, there can be a huge overhead in terms of data collection.
While some models can be developed using very \technologycode{small} data (between 1 to 100 examples)~\cite{babaian2002awriters, Gero2022sparks,liu2022will}, the others require much larger data.
If the training needs more data (around 100 to few thousands of examples)
which is often the case for fine-tuned models, we categorize them as \technologycode{medium} \cite{halpin2004automatic,chang2015writeahead2,zhang2016argrewrite,wang-etal-2016-non,wang2018mirroru}. 
For larger datasets (around tens of 1000s of examples)  we denote this as \technologycode{large} \cite{watanabe2017lyrisys,rei2016compositional,dai2014wings}. 
For models that undergo extensive large-scale pre-training, we categorized data used in this process as \technologycode{extremely large} to indicate a dataset of millions of examples \cite{chollampatt2016adapting,singh-etal-2021-drag,zhu2023visualize,shi-etal-2023-effidit} or more. 
We also included an \technologycode{unknown} if the paper did not explicitly mention the dataset used for training \cite{park2008is,natalie2023supporting,sun2023songrewriter}.

\paragraph{\technologydimension{Model - \change{Type}}}
\question{What is the type of the underlying model?}
Advancements in AI accelerators and the availability of large amounts of data have led to an evolution in model architectures,\footnote{\change{\citet[][§1]{bommasani2021opportunities} provides an overview on AI research over the last 30 years.}} which we capture as the following four types.
First, \technologycode{rule-based models} rely on pre-defined logic, lookup tables, regular expressions, or other similar heuristic approaches that are deterministic in nature \cite{schneider1998recognizing, babaian2002awriters, gonzales2010motivating, cahill2021supporting}.
For \technologycode{statistical machine learning (ML) models}, we consider models that are trained from scratch on historical data, are not necessarily ``deep'' (as in deep neural networks), and are used to make future predictions (\eg support vector machines and logistic regression) \cite{huang2012transahead, zhang2016argrewrite, jakesch2019aimediated, peng2020exploring}.
Over the past decade, \technologycode{deep neural networks} have been the popular models of choice for writing assistants, including recurrent neural networks (RNNs) and long short-term memory networks (LSTMs) \cite{clark2018creative, roemmele2018automated, wu2019design, arnold2020predictive}.
Finally, 
recent works have increasingly utilized \technologycode{foundation models}, such as BERT \cite{devlin2019bert}, RoBERTa \cite{liu2019roberta}, GPT \cite{radford2018improving, radford2019language, brown2020gpt3}, and T5 \cite{raffel2019exploring}, to name a few.
A foundation model is ``any model that is trained on broad data that can be adapted
(e.g., fine-tuned) to a wide range of downstream tasks'' \cite{bommasani2021opportunities}.
These models can perform a wide range of tasks out of the box, learn from a few examples to provide tailored support to users, and be further fine-tuned for specific downstream task(s)~\cite{lee2022coauthor, weber2023structured, wang2023smart, shi-etal-2023-effidit}.

\paragraph{\technologydimension{\change{Model - External Resource Access}}}
\change{
\question{What additional access does a model have at inference time?}
Recently, models have been developed with access to additional tools or data at inference time to make them capable of providing assistance beyond the knowledge encoded in their parameters.
In the case of \technologycode{tool}, a model may access external software or third-party APIs to perform tasks like search, translation, or calculator, or even setting calendar events on behalf of users~\cite{natalie2023supporting, yimam2018demonstrating, chollampatt2016adapting, zhang2016using}.  \technologycode{Data} refers to external datasets or resources, such as information stored in a database, external knowledge repositories, or any other structured/unstructured data sources that the models might leverage to provide writing assistance \cite{shi-etal-2023-effidit, kim2023metaphorian, skitalinskaya2023revise, zhu2023visualize}.
}

\paragraph{\technologydimension{Learning - Problem}}
\question{How is the writing assistance task being formulated as a learning problem?}
How exactly writing assistants support their users usually varies based on the learning problem their underlying models are designed to solve.
\technologycode{Classification} refers to the class of problems that require categorizing data into predefined classes based on their attributes. 
It is one of the most widely formulated classes of problems in writing assistants, applicable to tasks such as detecting errors in writing \cite{tsai2020lingglewrite, francois2020amesure} 
and detecting the purpose of writing revisions \cite{jiang2022arxivedits}, among other tasks. 
In contrast, \technologycode{regression} problems involve the prediction of a continuous numerical value or quantity as the output instead of categorical labels or classes. 
This includes problems such as the prediction of the writer's sentiments \cite{wambsganss2022adaptive}, the readability \cite{karolus2023your}, or the emotional intensity \cite{suzuki2022emotional} of written text as numerical ratings or scores.
\technologycode{Structured prediction} refers to a class of learning problems that involve predicting structured outputs or sequences (\eg sequences, trees, and graphs) rather than single, isolated labels or values. 
Numerous works have focused on developing these approaches to make edits to improve the quality of written text during the revision stage
\cite{malmi-etal-2019-encode, stahlberg-kumar-2020-seq2edits, mallinson-etal-2020-felix, kim-etal-2022-improving, mallinson-etal-2022-edit5}.
\technologycode{Rewriting} problems involve sequence transduction tasks, where texts from one form are transformed to another while improving the quality by making them fluent, clear, readable, and coherent. 
These tasks are essential in various writing assistance applications, such as grammatical error correction \cite{chollampatt2016adapting, zomer2021beyond, chang2015writeahead2}, paraphrasing \cite{yimam2018demonstrating}, or general-purpose text editing \cite{faltings-etal-2021-text, du-etal-2022-understanding-iterative, raheja2023coedit, shu2023rewritelm} to name a few. 
\technologycode{Generation} refers primarily to problems that involve the creation of new, contextually relevant, coherent, and readable text from relatively limited inputs, such as autocomplete, paraphrasing, and story generation \cite{schmitt2021characterchat, arnold2020predictive, ippolito2022wordcraft, Chung2022talebrush}.
\change{\technologycode{Retrieval} problems take the input from a user as a query (\eg keywords) and search in a knowledge base or dataset for relevant information. 
Such problems may involve ranking the available data based on its relevance and similarity to the input but do not necessarily include the generation of new text beyond what is available in the knowledge base \cite{soyer2015crovewa, chang2015writeahead2, shi-etal-2023-effidit}.}

\paragraph{\technologydimension{Learning - Algorithm}}
\question{How is the underlying model trained?}
The models used as back-bones of writing assistants incorporate different training mechanisms based on the type of the available data, as well as the specific downstream tasks.
In \technologycode{supervised learning}, models are trained on a labeled dataset where each input is associated with the correct output. Some of the commonly used methods include Logistic Regression, Random Forests, and Naive Bayes \cite{halpin2004automatic, burstein2003toward, wang-etal-2016-non, bixler2013detecting}. Supervised learning also includes approaches such as Transfer Learning, which involves training a model on a large dataset and then fine-tuning it for a specific task or domain using a smaller dataset \cite{duval2021breaking, zomer2021beyond}.
In \technologycode{unsupervised learning}, models are trained on unlabeled data to learn patterns and structures within the data. This approach includes techniques such as representation learning and clustering methods, to name a few \cite{watanabe2017lyrisys, osone2021buncho}.
\technologycode{Self-supervised learning}
approaches train models on unlabeled data with a supervisory signal \cite{gamon2010using}. These approaches leverage the benefits of both supervised and unsupervised learning, especially in scenarios where obtaining a large amount of labeled data is challenging. This includes pre-training objectives for large language models such as Causal Language Modeling \cite{radford2019language} and Masked Language Modeling \cite{devlin2019bert}.
In \technologycode{reinforcement learning} (RL), models learn by interacting with an environment and receiving feedback in the form of rewards. This approach is useful for tasks requiring action sequences, such as language generation and dialogue systems \cite{singh-etal-2021-drag}.

\paragraph{\technologydimension{\change{Learning - Training and Adaptation}}}
\change{
\question{How is the underlying model being trained or adapted for a specific task at hand?}
The training and adaptation process is integral part of developing an intelligent model that can perform tasks at hand and support user needs.
Before foundation models, many models used to be \technologycode{trained from scratch} \cite{sun2023songrewriter, hanawa2021exploring, gero2019how, liu2019neuralbased}.
On the other hand, with the advance of foundation models, the common learning paradigm has been shifted to ``pre-training'' a large model on broad data and then ``adapting'' the model to a wide range of downstream tasks.
One way to adapt a model is \technologycode{fine-tuning}, where the pre-trained model is further trained on a specific dataset~\cite{singh2022where, poddar2023aiwriting, zhong2023fiction, bhat2023interacting}. 
Note that there are numerous variants of fine-tuning, such as \textit{transfer learning}, \textit{instruction tuning}, \textit{alignment tuning}, \textit{prompt tuning}, \textit{prefix tuning}, and \textit{adapter tuning}, among others.
Another way to adapt a model is \technologycode{prompt engineering} (or ``prompting''), where one can simply provide a natural language description of the task (or ``prompt'')~\cite{brown2020language} to guide model outputs~\cite{dang2023choice, mirowski2023cowriting, jakesch2023cowriting, lee2022coauthor}. 
A prompt may include a few examples for a model to learn from (``few-shot learning'' or ``in-context learning''). 
Lastly, it is possible to \technologycode{tune decoding parameters} of a model to influence model outputs (\eg changing temperature to make outputs more or less predictable, manipulating logit bias to prevent some words from being generated)~\cite{goldfarb2019plan, schneider1998recognizing,lee2022coauthor}.
}

\paragraph{\technologydimension{Evaluation - \change{Evaluator}}} 
\question{Who evaluates the quality of model outputs?}
A core aspect of model development is its evaluation.
We consider four common types of evaluators who can review and evaluate various qualities of model outputs (as opposed to writing assistants or user interactions).
\technologycode{Automatic evaluation} 
compares machine-generated outputs with human-generated labels or texts using aggregate statistics or syntactic and semantic measures. 
These include metrics like precision, recall, F-measure, and accuracy, as well as ones commonly used in generation tasks, such as BLEU~\cite{papineni02bleu}, METEOR~\cite{lavie2009meteor}, and ROUGE~\cite{lin2004rouge} to name a few \cite{afrin2023predicting, sun2023songrewriter, chang2015writeahead2}.
\technologycode{Machine-learned evaluation} uses automated metrics, which are themselves produced by a machine-learned system. 
These are typically classification or regression models that are trained to evaluate the quality of model outputs 
\cite{zhang2019bertscore, sellam2020bleurt, wambsganss2022adaptive, jiang2022arxivedits, qi2022quoter}. 
In contrast, \technologycode{human evaluation} corresponds to evaluating the system with human annotators either directly interacting with, or evaluating the output of a writing assistant. 
Some evaluations may require the judging of task-specific criteria (\eg evaluating that certain entity names appear correctly in the text \cite{mishra2014text}), while others can be generalized for most text generation tasks (\eg evaluating the fluency or grammar of the generated text \cite{mackenzie2016empirical, yuan2022wordcraft, liu2022will, wambsganss2022modeling}). 
\technologycode{Human-machine evaluation} captures cases where both machine-learned metrics or models and human judges are involved in the evaluation of the outputs. 
This hybrid evaluation is particularly relevant in co-creative, mixed-initiative writing assistance settings. 
Such studies often involve expert users and participatory methodologies \cite{lin2023beyond, mirowski2023cowriting, lee2022coauthor, Chung2022talebrush}.

\paragraph{\technologydimension{\change{Evaluation - Focus}}}
\change{
\question{What is the focus of evaluation when evaluating individual model outputs?}
Evaluating (or benchmarking) models has been a long standing challenge in NLP~\cite{liang2023holistic}.
In particular, as we increasingly use foundation models (\eg GPT-4)
for a wide range of downstream tasks, it is difficult to evaluate the quality of model outputs across all tasks, let alone the difficulty of evaluating open-ended generation.
Here, we highlight four common evaluation foci relevant to writing assistants in the literature.
\technologycode{Linguistic quality} focuses on the grammatical correctness, readability, clarity, and accuracy of the model's outputs. This aspect ensures that the outputs are not only correct in terms of language use but also easily understandable and precise in conveying the intended message \cite{dang2023choice, wang2023smart, zhu2023visualize, karolus2023your}. 
\technologycode{Controllability} assesses how well the model's outputs reflect constraints (or control inputs) specified by users or designers (\eg how effectively the model adheres to any specific level of formality or writing style) \cite{poddar2023aiwriting, jakesch2023cowriting, schmitt2021characterchat, swanson2021story}.
Furthermore, it is crucial that the model's responses not only make sense in isolation, but also fit seamlessly within the broader context of the text.
\technologycode{Style and adequacy} pertain to the alignment between the model's outputs and their surrounding texts or contexts. 
This includes evaluating the stylistic and semantic coherence, relevance, and consistency of the outputs with the given context \cite{natalie2023supporting, yuan2022wordcraft, Gero2019metaphoria, lukasik2018content}. 
Finally, \technologycode{ethics} encompasses a broad range of crucial considerations such as bias, toxicity, factuality, and transparency. 
Ethics focuses on the adherence of the model's outputs to social norms and ethical standards, and seeks to avoid generating outputs containing harmful biases, misinformation, and other unethical elements \cite{biermann2022fromtool, hoque2022dramatvis, peng2020exploring, skitalinskaya2023revise}. 
This aspect of evaluation is particularly critical in maintaining the trustworthiness and societal acceptance of the model.
}

\paragraph{\technologydimension{Scalability}}
\question{What are the economic and computational considerations for developing and deploying a model?}
Recent models, especially \lms, have demonstrated exceptional performance across various tasks~\cite{chen2021codex,openai2023gpt4,bubeck2023sparks}. 
However, the significantly large size of these models has substantially increased the \technologycode{cost} of their development~\cite{kaplan2020scaling}.
In this regard, directly utilizing pre-trained \lms via prompting~\cite{wei2022cot, brown2020gpt3}, 
employing efficient fine-tuning methods like low-rank adaptation \cite{hu2022lora}, 
or prefix-tuning \cite{li-liang-2021-prefix} can help avoid the cost of full fine-tuning.
During deployment, this affects not only the inference costs but also system \technologycode{latency}, which often degrades user experience~\cite{cai2022context,lee2022evaluating}. 
Techniques such as quantization \cite{gholami2022survey} and knowledge distillation~\cite {hinton2015distilling} have shown promising results in addressing these issues.

%% file: tables/technology-designspace.tex
\begin{table*}
    \resizebox{1.\linewidth}{!}{
        \renewcommand{\arraystretch}{1.43}
        \setlength{\tabcolsep}{4pt}
        \begin{tabular}{p{0.05\textwidth}p{0.25\textwidth}p{0.8\textwidth}}
            \toprule
            &
             \textbf{Code} &
             \textbf{Definition} \\
             \midrule

             \multirow{9}{*}[2.5em]{\rotatebox{90}{\renewcommand{\arraystretch}{1}\begin{tabular}[c]{@{}c@{}}\technologydimensiont{\textbf{Data}}\\ \technologydimensiont{{\textbf{Source}}}\end{tabular}}} &
             \multicolumn{2}{l}{\emph{Who is the creator of the data used to train or adapt a model?}} \\
             \dashedline{2-3}
            &
             \technologycode{Experts} &
             Experts of the task or domain of interest \\
            &
             \technologycode{Users} &
             Current or target users of a downstream application \\
            &
             \technologycode{Crowdworkers} &
             Crowdworkers from various platforms, such as Amazon Mechanical Turk and Prolific \\
            &
             \technologycode{Authors} &
             Authors of the research paper \\
            &
             \technologycode{Machine} &
             Another model that generates synthetic data \\
            & 
             \technologycode{Other} & 
             Other creators, such as non-experts and unspecified individuals for web crawled data \\
             \midrule

            \multirow{7}{*}{\rotatebox{90}{\renewcommand{\arraystretch}{1}\begin{tabular}[c]{@{}c@{}}\technologydimensiont{\textbf{Data}}\\ \technologydimensiont{{\textbf{Size}}}\end{tabular}}}
            &
             \multicolumn{2}{l}{\emph{What is the size of the dataset that used to train or adapt a model?}} \\
             \dashedline{2-3}
            &
             \technologycode{Small (<100)} &
             Uses one to a hundred data points \\
            &
             \technologycode{Medium (<10k)} &
             Uses a hundred to a few thousand data points \\
            &
             \technologycode{Large (<1M)} &
             Uses tens to hundreds of thousands of data points \\
            &
             \technologycode{Extremely large (>1M)} &
             Uses millions of data points or more\\
             &
             \technologycode{Unknown} &
             Unknown data or dataset size \\
             \midrule

             \multirow{6}{*}{\rotatebox{90}{\renewcommand{\arraystretch}{1}\begin{tabular}[c]{@{}c@{}}\technologydimensiont{\textbf{Model}}\\ \technologydimensiont{{\textbf{Type}}}\end{tabular}}} &
             \multicolumn{2}{l}{\emph{What is the type of the underlying model?}} \\
             \dashedline{2-3}
            &
             \technologycode{Rule-based model} &
             Model relying on heuristics or deterministic approaches (\eg lookup tables and regular expressions) \\
            &
             \technologycode{Statistical ML model} &
             Machine learning (ML) model typically trained for a specific task (\eg logistic regression) \\
            &
             \technologycode{Deep neural network} &
             ML model that uses multiple layers between the input and output layers (\eg RNNs and LSTMs) \\
            &
             \technologycode{Foundation model} &
             Pre-trained deep neural network that can be adapted for downstream tasks (\eg BERT, GPT-4) \\
            \midrule

             \multirow{3}{*}[.05em]{\rotatebox{90}{\renewcommand{\arraystretch}{1}\begin{tabular}[c]{@{}c@{}}\technologydimensiont{\textbf{Model}}\\ \technologydimensiont{{\textbf{\small External Res.}}}\end{tabular}}} &
             \multicolumn{2}{l}{\emph{What additional access does a model have at inference time?}} \\
             \dashedline{2-3}
            &
             \technologycode{Tool} &
             External software or third party APIs the model might rely on to perform its task\\
            &
             \technologycode{Data} &
             External datasets or resources, such as external knowledge repositories\\
            \midrule
            
             \multirow{7}{*}[0em]{\rotatebox{90}{\renewcommand{\arraystretch}{1}\begin{tabular}[c]{@{}c@{}}\technologydimensiont{\textbf{Learning}}\\ \technologydimensiont{{\textbf{Problem}}}\end{tabular}}} &
             \multicolumn{2}{l}{\emph{How is the writing assistance task being formulated as a learning problem?}} \\
             \dashedline{2-3}
            &
             \technologycode{Classification} &
             Assigns inputs to predefined categories or classes \\
            &
             \technologycode{Regression} &
             Predicts continuous numeric values for a given input \\
            &
             \technologycode{Structured prediction} &
             Focuses on capturing dependencies, relationships, and patterns in language data \\
            &
             \technologycode{Rewriting} &
             Translates text from one form to another, while preserving its meaning and information content \\
            &
             \technologycode{Generation} &
             Creates new, coherent, and contextually relevant text \\
            & 
             \technologycode{Retrieval} &
             Finds and optionally ranks relevant instances for a given input \\
             \midrule

             \multirow{6}{*}[0.7em]{\rotatebox{90}{\renewcommand{\arraystretch}{1}\begin{tabular}[c]{@{}c@{}}\technologydimensiont{\textbf{Learning}}\\ \technologydimensiont{{\textbf{Algorithm}}}\end{tabular}}} &
             \multicolumn{2}{l}{\emph{How is the underlying model being trained?}} \\
             \dashedline{2-3}
            &
             \technologycode{Supervised learning} &
             Model is trained on labeled data where each input is associated with the correct output \\
            &
             \technologycode{Unsupervised learning} &
             Model is trained on unlabeled data to discover patterns and structures within the data \\
            &
             \technologycode{Self-supervised learning} &
             Model creates a supervisory signal from the data itself, without human-annotated labels \\
            &
             \technologycode{Reinforcement learning} &
             Model learns by interacting with an environment and receiving feedback in the form of rewards \\
             \bottomrule
            \end{tabular}
        }
    \caption{Technology dimensions, codes, and definitions (continued in the next table).}
    \label{tab:technology-designspace-1}
\end{table*}

\begin{table*}
    \resizebox{1.\linewidth}{!}{
        \renewcommand{\arraystretch}{1.45}
        \setlength{\tabcolsep}{4pt}
        \begin{tabular}{p{0.05\textwidth}p{0.3\textwidth}p{0.8\textwidth}}
            \toprule
            &
             \textbf{Code} &
             \textbf{Definition} \\
             \midrule

             \multirow{6}{*}[0.60em]{\rotatebox{90}{\renewcommand{\arraystretch}{1}\begin{tabular}[c]{@{}c@{}}\technologydimensiont{\textbf{Learning}}\\ \technologydimensiont{{\textbf{\small Training \& adaptation}}}\end{tabular}}} &
             \multicolumn{2}{l}{\emph{How is the underlying model being trained or adapted for a specific task at hand?}} \\
             \dashedline{2-3}
            &
             \technologycode{Training from scratch} &
             A new model is trained from scratch, or a foundation model is used without  adaptation \\
            &
             \technologycode{Fine-tuning} &
             A foundation model is further trained on a specific dataset \\
            &
             \technologycode{Prompt engineering} &
             A foundation model is further adapted via prompts (a.k.a. in-context learning) \\
            &
             \technologycode{Tuning decoding parameters} &
             A model's decoding parameters are adjusted to stir model outputs (\eg temperature and logit bias) \\
            \midrule
                        
            \multirow{9}{*}[2.5em]{\rotatebox{90}{\renewcommand{\arraystretch}{1}\begin{tabular}[c]{@{}c@{}}\technologydimensiont{\textbf{Evaluation}}\\ \technologydimensiont{{\textbf{Evaluator}}}\end{tabular}}} &
             \multicolumn{2}{l}{\emph{Who evaluates the quality of model outputs?}} \\
             \dashedline{2-3}
            &
             \technologycode{Automatic evaluation} &
             Quality is evaluated based on simple aggregate statistics or on similarity metrics\\
            &
             \technologycode{Machine-learned evaluation} &
             Quality is evaluated by another model to produce ratings or scores (e.g., BERTScore) \\
            &
             \technologycode{Human evaluation} &
             Quality is evaluated by human annotators \\
            &
             \technologycode{Human-machine evaluation} &
             Quality is evaluated by both  human annotators and another model, often by having the model filter a subset of outputs to be annotated by humans \\
            \midrule

            \multirow{9}{*}[3.5em]{\rotatebox{90}{\renewcommand{\arraystretch}{1}\begin{tabular}[c]{@{}c@{}}\technologydimensiont{\textbf{Evaluation}}\\ \technologydimensiont{{\textbf{Focus}}}\end{tabular}}} &
             \multicolumn{2}{l}{\emph{What is the focus of evaluation when evaluating individual model outputs?}} \\
             \dashedline{2-3}
            &
             \technologycode{Linguistic quality} &
             Evaluation focuses on qualities such as grammaticality, readability, clarity, and accuracy \\
            &
             \technologycode{Controllability} &
             Evaluation focuses on the reflectiveness to controls  or constraints specified by users or designers\\
            &
             \technologycode{Style \& Adequacy} &
             Evaluation focuses on the alignment between model ouputs and their surrounding texts\\
            &
             \technologycode{Ethics} &
             Evaluation focuses on social norms and ethics, such as bias, toxicity, factuality, and transparency \\
            \midrule            
            
            \multirow{3}{*}[-0.1em]{\rotatebox{90}{\textbf{\technologydimensiont{Scalability}}}} &
             \multicolumn{2}{l}{\textit{What are the economic and computational considerations for developing and deploying a model?}} \\
             \dashedline{2-3}
            &
             \technologycode{Cost} &
             Cost of deploying the model \\
            &
             \technologycode{Latency} &
             Delay between when the model receives an input and generates a corresponding output \\
        \bottomrule
        \end{tabular}
    }
\caption{Technology dimensions, codes, and definitions (continued from the previous table).}
\label{tab:technology-designspace-2}
\end{table*}

%% file: sections/interaction.tex
\input{figures/interaction}

\input{tables/interaction-designspace}

The interaction between a user and a writing assistant primarily involves three key components: User, user interface (UI; frontend), and system (backend).
The UI acts as a mediator, facilitating interaction between the user and the system, as illustrated in Figure~\ref{fig:InteractionFramework}.

\subsubsection{Dimensions and codes}
Figure~\ref{fig:designspace} (``Interaction'') shows interaction dimensions in a broad context, while Figure \ref{fig:InteractionFramework} shows a detailed visualization. Tables~\ref{tab:interaction-designspace-1} and \ref{tab:interaction-designspace-2} list all dimensions, codes, and definitions.

\paragraph{\interactiondimension{\change{UI} - Interaction Metaphor}}
\label{sec:ui_metaphor}
\question{What is the interaction metaphor for the system?} Interaction metaphors shape how the user relates to the system.
We identify three primary metaphors. 
First, systems designed as \interactioncode{agents} employ designs meant to evoke human-like interaction, including roles like ``collaborator'' or ``co-writer'' \cite{chakrabarty2022copoet,lee2022coauthor}, ``dialog partner'' \cite{park2021iwrote, wambsganss21arguetutor}, ``assistant'' \cite{gabriel2015inkwell}, and ``companion'' \cite{Gero2019metaphoria}. 
Techniques for evoking the agent metaphor include character-by-character text rendering to simulate typing \cite{jakesch2023cowriting}, avatars \cite{cahill2021supporting}, implicit and explicit conversational interaction \cite{park2021iwrote}, and first-person conversational styles \cite{peng2020exploring}. 
In contrast, other systems present as \interactioncode{tools}, where there is no sense of interacting with another agent.
These systems tend to avoid conversational styles and rather present feedback in imperative or factual style \cite{weber2023structured,wambsganss2022modeling,hui2023lettersmith}. They provide traditional GUI elements (\eg checkboxes, buttons) that spread out system capabilities rather than centralizing them into one ``agent.'' 
\interactioncode{Hybrid} systems blend ``agent'' and ``tool'' metaphors in their design or their authors' descriptions \cite{Gero2019metaphoria,lee2022coauthor}.

\paragraph{\interactiondimension{\change{UI} - Layout}}
\label{sec:ui_layout}
\question{Where are the system interactions situated in the UI?} 
Often, interaction with the underlying system takes place in the \interactioncode{writing area} where users create text. 
This supports the selection of the existing writing \citep{Dang2022beyond, shi-etal-2023-effidit} and/or seamless integration of output from the system \citep{dang2023choice,lee2022coauthor}. 
Alternatively, a design might choose to isolate the interaction with AI as a \interactioncode{separated} UI element, such as through a sidebar.
A separated design puts clear boundaries between the users' writing and the system output.  
For example, this design style is used to display information related to text diagnosis \citep{tsai2020lingglewrite}, provide inspiration \citep{kim2023metaphorian,Chung2022talebrush}, or support language learning \citep{chang2015writeahead2}. 
Interaction with the system can also be embedded with the user's text \interactioncode{input UI}, such as text suggestions on touchscreen keyboards \citep{arnold2020predictive}.
Finally, there are \interactioncode{custom} designs: for example, a tangible UI that triggers the system by lifting a coffee cup \cite{belakova2021sonami}, or an exploratory visualization for referencing information which has no user text input \cite{kim2020lexichrome}.

\paragraph{\interactiondimension{\change{UI} - Interface Paradigm}}
\label{sec:ui_paradigm}
\question{What is the general platform of the interface?}
The \interactioncode{text editor} is the prevalent interface for AI-assisted writing \citep{dang2023choice, shi-etal-2023-effidit}, providing writers with a traditional blank canvas while incorporating a variety of system-driven functionalities such as feedback \citep{hui2023lettersmith}, automated checklists \citep{gabriel2015inkwell}, or completion suggestions \citep{jakesch2023cowriting}. 
In contrast, \change{\interactioncode{chatbot}} interfaces use turn-taking interactions for motivation \citep{park2021iwrote} or suggestions \citep{chakrabarty2022help}. Conversational exchanges serve to progressively and iteratively achieve a goal, like fiction writing \citep{clark2021choose,schmitt2021characterchat,biermann2022fromtool}.
Finally, \interactioncode{other} interfaces cater to specific needs, introducing novel \cite{kim2020lexichrome, sadauskas2015mining}, sometimes multimodal \citep{Chung2022talebrush, singh2022where}, interactions. For example, such custom interfaces take lyrical structure into account for music generation~\cite{watanabe2017lyrisys}, and emphasize the conceptual and figurative expressions to generate metaphors for scientific writing ~\cite{kim2023metaphorian}.

\paragraph{\interactiondimension{\change{UI} - Visual Differentiation}}
\label{sec:ui_visdiff}
\question{How is system output differentiated from user output?}
This dimension identifies different visual designs for separating user-written text and system outputs. 
A system must present its outputs to the user in some way, or the interaction loop between the user and the system is incomplete. 
The most common mechanism is to use text \interactioncode{formatting} such as colors and underlines~\cite{yimam2018demonstrating, yuan2022wordcraft, dang2023choice}. 
Another common mechanism is to keep system outputs in a separate \interactioncode{location} like a wizard or panel~\cite{peng2020exploring, Dang2022beyond, wambsganss2022adaptive}. 
A system may not use any formatting when the system output has a different \interactioncode{media} type (\eg meta-analysis~\citep{hui2023lettersmith, weber2023structured} or audio~\cite{belakova2021sonami}) than the text written by the user. 
Alternatively, the differentiation is ``\interactioncode{none}'': System output is presented identically to user output after it is added into the text~\cite{arnold2020predictive, lee2022coauthor}.

\paragraph{\interactiondimension{\change{UI - Initiation}}}
\label{sec:ai_trigger}
\change{\question{How is system output triggered?}
System outputs can be triggered in two main ways: 
\interactioncode{User-initiated} triggers give the user control over when system output is created or presented in the UI. For example, phrase suggestions might be displayed whenever the user presses the ``tab'' key~\cite{lee2022coauthor} or clicks on a ``generate'' button~\cite{Gero2022sparks}. 
In contrast, with \interactioncode{system-initiated} triggers, the system has control over when it brings in its output.} This might be a rule, heuristic, or a dedicated trigger decision model. For example, the user and system might take turns such that the system output is triggered when the user submits a chat message~\cite{park2021iwrote} or sentence~\cite{clark2021choose}, or pauses writing for a certain amount of time \cite{jakesch2023cowriting, bhat2023interacting, Buschek2021emails}. For some systems, this initiative is almost real-time, or ``live.'' For example, several systems update suggestions and feedback in a dedicated panel while the user is typing \cite{arnold2020predictive,chang2015writeahead2, tsai2020lingglewrite}.

\paragraph{\interactiondimension{\change{User} - Steering the System}}
\label{sec:user_control}
\question{How can the user steer the system?} Users need to communicate their intentions and goals to a system to steer its behavior. In the \interactioncode{implicit} paradigm, users directly compose the artifact (typically, text in a workspace) as a way to communicate their intentions to the system which in turn provides support based on this information. 
Such systems often provide support in the form of text suggestions~\cite{lee2022coauthor,yuan2022wordcraft}, reflection~\cite{park2021iwrote,Dang2022beyond}, or inspiration and ideation~\cite{Gero2019metaphoria}. 
Alternatively, users have \interactioncode{explicit} controls over the system behavior. 
For example, users give thumbs up or down on shown text suggestions to steer future output towards user preferences~\cite{clarksmith2021choose}, control the diversity of generated text through numeric parameters \cite{goldfarb2019plan}, or steer story arcs via sketching \cite{Chung2022talebrush}. 
The two paradigms have implications for the users' workflow \cite{dang2023choice}: implicit input can be less disruptive to the writing process, while explicit input offers more ways for users to express intentions.
\change{Finally, some systems offer 
\interactioncode{no control} to the user, for example, if the \lm is prompted in the background to respond in a particular way regardless of 
the user's prior text~\cite{jakesch2023cowriting}.}

\paragraph{\interactiondimension{\change{User} - Integrating System Output}}
\label{sec:user_output}
\question{How does the user integrate system output?} 
After the system creates an output, the user can choose how to integrate it into the overall goal. For material intended to be included in the output, the user may take a \interactioncode{selection} action, such as accepting a suggestion using a button or key press~\cite{lee2022coauthor,kim2023metaphorian,arnold2020predictive}. 
When the system does not provide material for direct inclusion, the user may engage with the output as \interactioncode{inspiration} 
and choose to type their own text~\cite{wang2018mirroru,park2021iwrote}. 
If the system text is inserted into the final text with no explicit user interaction, the user must then choose whether to keep, edit, or remove the text, through \interactioncode{editing}~\cite{Chung2022talebrush,mirowski2023cowriting}.
Finally, the system may provide outputs that require \interactioncode{no integration}, if they are not meant to be added to an output or to inform direct contributions, as in analysis for future improvements~\cite{hui2023lettersmith,weber2023structured}.

\paragraph{\interactiondimension{System - \change{User} Data Access}}
\label{sec:ai_data}
\question{What user data can the system access through the UI?}
Systems can access different types of user data to produce the desired outputs. 
\change{Typically, the user's \interactioncode{input text} is the primary source of data; the current writing progress can be used to generate completions~\cite{lee2022coauthor,arnold2020predictive} or provide feedback~\cite{francois2020amesure, cahill2021supporting}.
It is also important to allow users to have additional controls over the text generations, which we characterize as \interactioncode{additional data}. 
For example, some systems allow users to specify the task for generation via explicit instruction~\cite{yuan2022wordcraft, chakrabarty2022help} or via sketching~\cite{Chung2022talebrush}.}

\paragraph{\interactiondimension{System - Output Type}}
\label{sec:ai_output}
\textbf{What type of output does the system create?} 
\change{Writing assistants can generate different types of outputs to support users' writing.}
System output might serve as an \interactioncode{analysis} of the user's writing, e.g., how clear text is in an administrative context \citep{francois2020amesure}, how empathetic a peer review is \citep{wambsganss2022adaptive}, or how structured and formal writing is in a professional context \citep{hui2023lettersmith}. 
Systems can provide this analysis in the form of annotations or highlights on the text \citep{hui2023lettersmith, wambsganss2022adaptive}, analytics in a separate UI \citep{kim2023metaphorian, karolus2023your}, or high-level statements about the text \citep{cahill2021supporting, rapp2015thesis}. 
In contrast, \interactioncode{generation} is usually realized as completions \citep{arnold2020predictive, watanabe2017lyrisys, bhat2023interacting}, although it can also encompass longer sections (\eg creating a paragraph of text \citep{Chung2022talebrush, duval2021breaking}). 
Finally, some systems' outputs only provide \interactioncode{proposals}: Content meant to be used as reference but not directly incorporated into the document. 
This content can be provided in the form of planning and outlining support \cite{rapp2015thesis,sadauskas2015mining}, or questions about the user's writing to encourage reflection \citep{cahill2021supporting} or summaries of the text so far \citep{Dang2022beyond}.

\paragraph{\interactiondimension{System - Curation Type}} 
\label{sec:ai_generation}
\question{How are system outputs curated?}
The predominant way that intelligent writing support systems curate outputs is to generate a custom response from an NLP \interactioncode{model}, especially with the recent dramatic increase in fluency and capability of \lms~\cite{lee2022coauthor, watanabe2017lyrisys, jakesch2023cowriting, roemmele2018automated}.
However, providing the natural language outputs of such models requires the designer to give up a great deal of control over the possible outputs.  
In sensitive contexts such as topics like mental health~\cite{peng2020exploring, park2021iwrote, wang2018mirroru} or to leverage the benefit of high-quality source material, designers may choose to trade off the flexibility of an NLP model for the control of a pre-made, \interactioncode{curated} list of responses~\cite{soyer2015crovewa}, from which an output is selected by the system~\cite{chang2015writeahead2, cahill2021supporting, schmitt2021characterchat}. 
To increase personalization, these pre-curated responses may also be dynamically \interactioncode{customized}~\cite{park2021iwrote}.  
Alternatively, the output may be generated from user input in a \interactioncode{deterministic} way, in writing systems that do not use AI and/or prefer rule-based automation~\cite{belakova2021sonami, kinnunen2012swan, sadauskas2015mining}.

%% file: figures/interaction.tex
\begin{figure*}[t]
    \centering
    \includegraphics[width=0.8\textwidth]{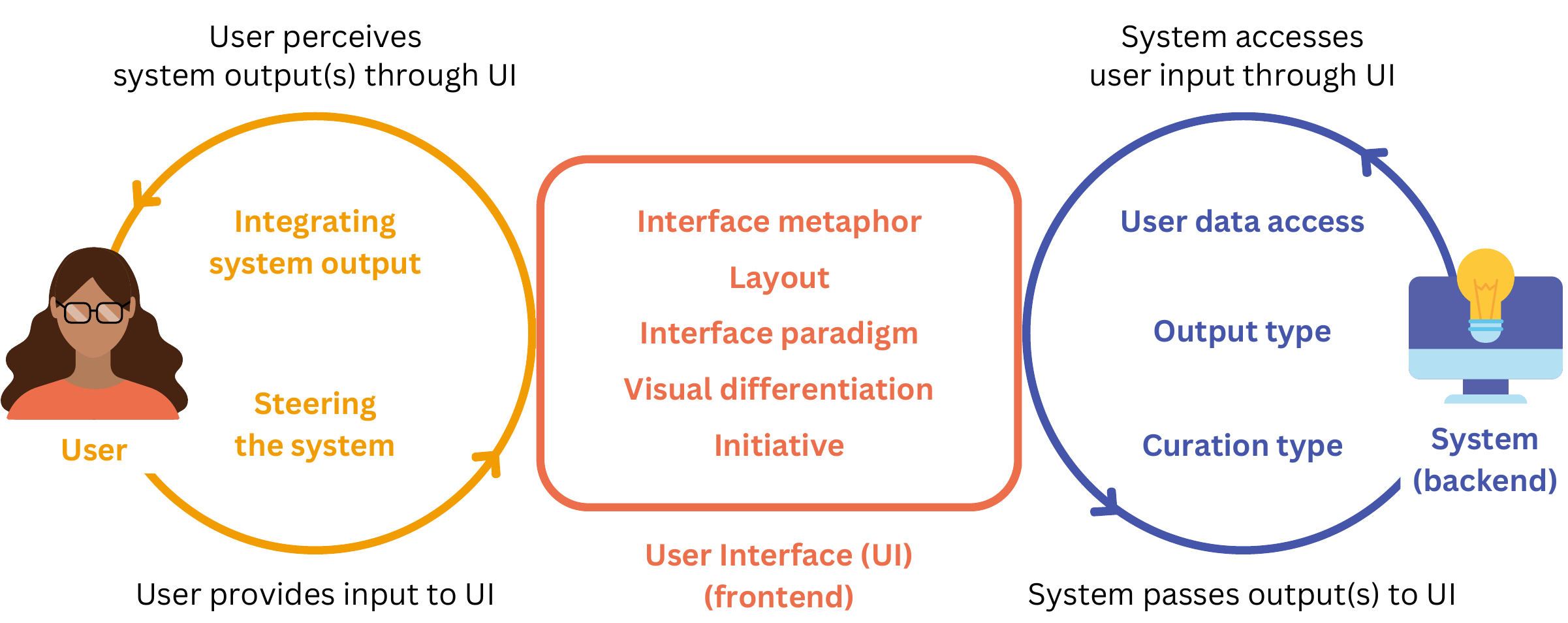}
    \caption[]{
    A visualization of the three main components for the interaction aspect in our design space. The visualization outlines the relationships between a user, user interface (frontend), and system (backend). Users interact with the system through the interface by perceiving system outputs and providing input to the system.
    The system interacts with the user through the interface by accessing user input and data and generating outputs that get rendered in the interface.
    }
    \Description{The figure shows how interaction happens between a user, an interface (frontend), and a system (backend). A connecting arrow from the user to the interface is labeled "User provides input to UI" and an arrow is connecting from the interface to the user with the label "User perceives system output(s) through UI." Between the user and the user interface, there are two boxes, "Integrating system output" and "Steering the system". The interface box has the header "Design of User Interface (UI)" and has smaller boxes "Interaction metaphor," "Layout," "Interface paradigm," "Visual differentiation," and "Initiative". The user interface is connected to the system with the label "System accesses user input through UI" and the system is connected to the interface with the label "System generates output(s) and passes it to UI." Between them, there are boxes of "User data access," "Output type," and "Curation type".}
    \label{fig:InteractionFramework}
\end{figure*}

%% file: tables/interaction-designspace.tex
\begin{table*}[t]
  \vspace{0.5cm}
  \resizebox{1.\linewidth}{!}{
      \small
      \renewcommand{\arraystretch}{1.95}
      \setlength{\tabcolsep}{4pt}
      \begin{tabular}{cp{0.2\textwidth}p{0.725\textwidth}}
          \toprule
          &
            \textbf{Code} &
            \textbf{Definition} \\
            \midrule
            \multirow{4}{*}[-0.3em]{\rotatebox{90}{\renewcommand{\arraystretch}{1}\begin{tabular}[c]{@{}c@{}}\interactiondimensiont{\textbf{UI}}\\ \interactiondimensiont{{\small	\textbf{Interaction metaphor}}}\end{tabular}}} &
          \multicolumn{2}{p{0.8\textwidth}}{\emph{What is the interaction metaphor for the system?}} \\
        \dashedline{2-3}
        &
          \interactioncode{Agent} &
          System is meant to evoke a sense of another agent acting on the interface or data \\
        &
          \interactioncode{Tool} &
          System is presented as tools, where there is not a sense of interacting with another agent \\
        &
          \interactioncode{Hybrid} &
          System draws from both “agent” and “tool” metaphors \\
        \midrule
        \multirow{5}{*}[-0.1em]{\rotatebox{90}{\renewcommand{\arraystretch}{1}\begin{tabular}[c]{@{}c@{}}\interactiondimensiont{\textbf{UI}} \\ \interactiondimensiont{{\small \textbf{Layout}}}\end{tabular}}} &
          \multicolumn{2}{p{0.8\textwidth}}{\emph{Where are the system interactions situated in the UI?}} \\
        \dashedline{2-3}
        &
          \interactioncode{Writing area} &
          In-situ interactions take place in the same place as user output text \\
        &
          \interactioncode{Separated} &
          Interface isolates the system outputs and controls in a separate layout panel \\
        &
          \interactioncode{Input UI} &
          Interactions are situated in the input device  \\
        &
          \interactioncode{Custom} &
          System has a dedicated custom interface \\
        \midrule
        \multirow{4}{*}[-0.2em]{\rotatebox{90}{\renewcommand{\arraystretch}{1}\begin{tabular}[c]{@{}c@{}}\interactiondimensiont{\textbf{UI}} \\ \interactiondimensiont{{\small \textbf{Interface paradigm}}}\end{tabular}}} &
          \multicolumn{2}{p{0.8\textwidth}}{\emph{What is the general platform of the interface?}} \\
        \dashedline{2-3}
        &
          \interactioncode{Text editor} &
          The main interface is a text editor \\
        &
          \interactioncode{Chatbot} &
          The main interface is a chat client \\
        &
          \interactioncode{Other} &
          Other UI paradigms, including those developed custom for the application at hand \\
        \midrule
        \multirow{5}{*}[-0.3em]{\rotatebox{90}{\renewcommand{\arraystretch}{1}\begin{tabular}[c]{@{}c@{}}\interactiondimensiont{\textbf{UI}}\\ \interactiondimensiont{{\small \textbf{Visual differentiation}}} 
        \end{tabular}}} &
          \multicolumn{2}{p{0.8\textwidth}}{\emph{How is system output differentiated from user output?}} \\
        \dashedline{2-3}
        &
          \interactioncode{None} &
          System output is not differentiated from user output after it is accepted into the text \\
        &
          \interactioncode{Formatting} &
          System output is included with user output, but is differentiated by color or other formatting \\
        &
          \interactioncode{Location} &
          System output is in a separate location or UI panel from user output \\
        &
          \interactioncode{Media} &
          System creates a different type of output than the user, such as images
          \\
  \midrule
  \multirow{3}{*}[0.0em]{\rotatebox{90}{\renewcommand{\arraystretch}{1}\begin{tabular}[c]{@{}c@{}}\interactiondimensiont{\textbf{UI}} \\ \interactiondimensiont{{\small \textbf{Initiation}}}\end{tabular}}} &
    \multicolumn{2}{p{0.8\textwidth}}{\emph{How is system output triggered?}} \\
   \dashedline{2-3}
   &
    \interactioncode{User-initiated} &
    Users initiate a request or prompt for system to generate an output  \\
   &
    \interactioncode{System-initiated} &
    System provides output based on internal rules \\
  \midrule

  \multirow{4}{*}[-0.1em]{\rotatebox{90}{\renewcommand{\arraystretch}{1}\begin{tabular}[c]{@{}c@{}}\interactiondimensiont{\textbf{User}}\\ \interactiondimensiont{{\small \textbf{Steering the system}}}\end{tabular}}} &
    \multicolumn{2}{p{0.8\textwidth}}{\emph{How can the user steer the system?}} \\
   \dashedline{2-3}
   &
    \interactioncode{Explicit} &
    User can control system behavior by selecting buttons, checkmarks, etc. \\
   &
    \interactioncode{Implicit} &
    User updates user text and the system takes it as input to generate output \\
   &
    \interactioncode{No control} &
    User cannot control system; the only controls were pre-decided by the designer \\

          \bottomrule
      \end{tabular}
  }
\caption[]{Interaction dimensions, codes, and definitions (continued in the next table).}
\vspace{0.5cm}
\label{tab:interaction-designspace-1}
\end{table*}

\begin{table*}[t]
\resizebox{1.\linewidth}{!}{
    \small
    \renewcommand{\arraystretch}{1.8}
    \setlength{\tabcolsep}{4pt}
    \begin{tabular}{cp{0.2\textwidth}p{0.725\textwidth}}
        \toprule
        &
          \textbf{Code} &
          \textbf{Definition} \\
          \midrule

  \multirow{5}{*}[-0.4em]{\rotatebox{90}{\renewcommand{\arraystretch}{1}\begin{tabular}[c]{@{}c@{}}\interactiondimensiont{\textbf{User}} \\ \interactiondimensiont{{\small \textbf{Integrating system output}}}\end{tabular}}} &
    \multicolumn{2}{p{0.8\textwidth}}{\emph{How does the user integrate system output?}} \\
   \dashedline{2-3}
   &
    \interactioncode{Selection} &
    Users select a system output, such as accepting a suggestion from multiple suggestions using a button or key press \\
   &
    \interactioncode{Inspiration} &
    Users do not keep system output, but may be inspired to create new text on their own \\
   &
    \interactioncode{Editing} &
    Users keep, edit, or remove system output \\
   &
    \interactioncode{No integration} &
    System may provide outputs that are not meant to be added or inform direct contributions \vspace{.6em} \\

          \midrule
          
  \multirow{3}{*}[-0.0em]{\rotatebox{90}{\renewcommand{\arraystretch}{1}\begin{tabular}[c]{@{}c@{}}\interactiondimensiont{\textbf{System}}\\ \interactiondimensiont{{\small \textbf{User Data Access}}}\end{tabular}}} &
    \multicolumn{2}{p{0.8\textwidth}}{\emph{What user data can the system access through the UI?}} \\
   \dashedline{2-3}
   &
    \interactioncode{Input text}&
    The text the user is working on \\
   &
    \interactioncode{Additional data}&
    Extra data that are not intended to be part of the input text, such as random seeds, control labels, or prompts\vspace{.6em}\\
  \midrule
  \multirow{4}{*}[-0.4em]{\rotatebox{90}{\renewcommand{\arraystretch}{1}\begin{tabular}[c]{@{}c@{}}\interactiondimensiont{\textbf{System}}\\ \interactiondimensiont{{\small \textbf{Output Type}}}\end{tabular}}} &
    \multicolumn{2}{p{0.8\textwidth}}{\emph{What type of output does the system create?}} \\
   \dashedline{2-3}
   &
    \interactioncode{Analysis} &
    Feedback, analytics, or context based on automatic analysis of the user's text \\
   &
    \interactioncode{Generation} &
    New content that is intended to be incorporated into the final product \\
   &
    \interactioncode{Proposal} &
    New content that is meant to be referenced but not directly incorporated into the final product \\
  \midrule
  
  \multirow{5}{*}[-0.6em]{\rotatebox{90}{\renewcommand{\arraystretch}{1}\begin{tabular}[c]{@{}c@{}} \interactiondimensiont{\textbf{System}} \\ \interactiondimensiont{{\small \textbf{Curation Type}}}\end{tabular}}} &
    \multicolumn{2}{p{0.8\textwidth}}{\emph{How are system outputs curated?}} \\
   \dashedline{2-3}
   &
    \interactioncode{Model} &
    Model generates outputs which are directly used by the system \\
   &
    \interactioncode{Curated} &
    System designers curates a list of outputs in advance, and the system picks one for the user \\
   &
    \interactioncode{Customized} &
    A response from a curated list is selected and then further customized by the system for the current context \\
   &
    \interactioncode{Deterministic} &
    User input automatically determines outputs \\
  \bottomrule
    \end{tabular}
}
\caption[]{Interaction dimensions, codes, and definitions (continued from the previous table).}
\label{tab:interaction-designspace-2}
\end{table*}

%% file: sections/ecosystem.tex
\input{tables/ecosystem-designspace}

As writing assistants become embedded in authentic contexts, we must consider the embodied, material, sociotechnical environment (see Appendix~\ref{app:background} for discussion on micro- vs. macro-HCI).
Following \citet{guggenberger2020ecosystem}, we consider the ecosystem as \textit{``the overarching sociotechnical context in which the writer and the tool are situated, encompassing a range of complex, interdependent actors that frequently play a role in the functioning of the writing assistant.''}
This aligns with the writing model proposed by \citet{hayes1996model}, updating the cognitive processes of writing proposed in 1981 \cite{flower1981cognitive} to add the \textit{social} and \textit{physical} environments. 
Although much of the ecosystem is beyond the immediate control of writing assistant designers, this aspect draws attention to how one might design in anticipation of its influences.

\subsubsection{Dimensions and codes}
Figure~\ref{fig:designspace} (``Ecosystem'') shows ecosystem dimensions in a broad context, while Table~\ref{tab:ecosystem-designspace} lists all dimensions, codes, and definitions.

\paragraph{\ecosystemdimension{Digital \change{Infrastructure}}} 
\question{What compatibility issues are considered?} 
A key compatibility issue to consider is \ecosystemcode{usability consistency}, \ie the extent to which the writing assistant's user experience intentionally aligns with other systems in the writer's ecosystem. 
Examples in the corpus include integrating AI language technologies into everyday writing apps \cite{hagiwara2019teaspn}, extending Google Docs \cite{turkay2018itero}, and using familiar-looking BibTeX style codes to enable writers to invoke remote bibliographic searches \cite{babaian2002awriters}. 
This example also illustrates \ecosystemcode{technical interoperability} with external services, which designers may conceive in many ways. 
For instance, another project integrated tangible media with their writing assistant, such that lifting a mug on a digital coaster triggered text-to-speech replay of recent sentences to assist reflection \cite{belakova2021sonami}.

\paragraph{\ecosystemdimension{Access Model}} 
\question{How does the openness of data, models, and products influence design decisions?}
This dimension taps into how the openness and licensing of data, models, and products may influence design decisions and dissemination of artifacts. 
First, when a new writing assistant is developed to be embedded within existing \ecosystemcode{commercial software}, products, or platforms,
the aesthetic and display space of these products exerts a strong influence over the UX and UI design of the writing assistant, as illustrated by the examples of Airbnb \cite{jakesch2019aimediated} and Facebook \cite{park2021iwrote}.
This influence could further affect the choice of data (\eg proprietary data) and models (\eg closed models).
On the other hand, for many researchers, it is a common and often desired practice to build and work with \ecosystemcode{free and open-source software} (as well as open-source data and models~\cite{touvron2023llama,alpaca,workshop2023bloom}) given their transparency and free availability.
Lastly, researchers in turn can open source their writing assistants~\cite{neate2019empowering,lee2022coauthor} as well as other associated artifacts, such as interaction traces for human-\lm collaborative writing~\cite{lee2022coauthor} and programming languages for generating text~\cite{howe2009rita}, to promote collaboration and reproducibility.


\paragraph{\ecosystemdimension{Social \change{Factors}}} 
\question{Who affects the design and use of writing assistants?}
This dimension draws attention to how \textit{designers} engage with stakeholders, and the social support around \textit{authors}. 
First, \change{\ecosystemcode{designing with stakeholders}} concerns human-centered and participatory design methods to meaningfully involve target authors, and other groups such as educators, coaches, or subject-matter experts. 
This was by far the most prevalent ecosystem dimension in the corpus, attributable to design concept interviews, co-design sessions, and usability studies with target authors \cite{singh2022where,padmakumar2022machine,mirowski2023cowriting}. 
While not represented in the corpus, other research focuses on co-designing tools with educators to build trust in their design \cite{roscoe2014writing,shibani2020educator}.  
Second, \ecosystemcode{designing for social writing} covers the formal and informal support network which writers may call on, including co-authors, peers, mentors, friends and family.
For instance, children writing at home may involve informal social support from their parents~\cite{sulzby1989emergent,strickland1989emerging}.

\paragraph{\ecosystemdimension{\change{Locale}}} 
\question{Does the writing assistant's design take into account features of a physical locale?}
Digital systems are used in physical environments, whose affordances could aid writing (\eg whiteboard used to plan a document, sticky notes on screens, and opportunistic conversations with passing colleagues) or could impede writing (\eg a noisy environment may permit notetaking but disrupt the attention needed for detailed writing; fieldwork with limited Internet access may prevent certain writing until back in the office). 
\change{This dimension draws attention to whether the writer is engaged in \ecosystemcode{local writing} (\ie at what is considered to be ``home base'' with full social and technical networking), or \ecosystemcode{remote writing} (\ie away from home base, with different affordances and constraints) such as fieldwork or commuting}. 
Much of the research to date did not attend to this explicitly, but examples include designing an intentionally calm interface to help reduce classroom distractions for vulnerable young people \cite{goncalves2015you}, \change{and using a messaging platform to encourage more reflective writing for well-being \cite{wang2018mirroru,park2021iwrote}.} 

\paragraph{\ecosystemdimension{Norms \& Rules}} 
\question{What norms and rules affect the design and use of a writing assistant?}
As writing is embedded in countless societal processes, it will inevitably be governed by various norms and rules, both formal and informal. 
First, there can be alignments or misalignments with \ecosystemcode{laws}. 
\change{While none of the corpus papers addressed legal issues, this is of course an active field of scholarship \cite{hacker2023regulatinggpt}. 
Furthermore, U.S. and E.U. legislation is changing \cite{eu2023aiact,whitehouse2023execai} with intellectual property legal cases under way \cite{reuters2023openai,forbes2023openai}. 
Consequently, this code draws designers' attention to legal changes, which could conceivably shift market preference to \lms trained on ethically sourced data, or passing a particular algorithmic impact assessment. 
Second, writing assistants could account for societal \ecosystemcode{conventions}. 
User-centered design builds on concepts and practices familiar to users,
such as writing conventions in job application letters~\citep{hui2023lettersmith} and 
a system based on an established typology of expository phrases for science writing that readers and writers would recognize~\cite{Gero2022sparks}.
\change{Other types of conventions guiding design decisions include features of good metaphors \cite{Gero2019metaphoria}, the social acceptability of automating emails, and clinical principles to support writers with aphasia \cite{neate2019empowering} or their mental health \cite{park2021iwrote}. 
Emerging evidence \cite{cambon2023early,dellacqua2023jagged}} may raise expectations around employee productivity, with the Writers Guild of America strike in the U.S. demonstrating the conflict that the proliferation of AI writing is now provoking \cite{wgastrike2023whatwewon}. 
While there were no corpus papers documenting the embedding of writing assistants into established work practices and conventions, we see this beginning to happen in K-12 \cite{roscoe2014writing} and higher education \cite{cotos2020understanding,knight2020acawriter}.}

\paragraph{\ecosystemdimension{Change Over Time}} 
\question{What are the key temporal considerations when designing a writing assistant?} 
\change{This last dimension recognizes that people, technology, regulation, and the broader information environment are in motion, not static. 
Like the 5,000 year history of writing systems, we anticipate writing assistants and the ecosystem influencing each other over different timescales, from instantaneous to longer-term change. 
As detailed in the user and interaction dimensions, writing assistants can be shown to have demonstrable effects on \ecosystemcode{authors} and their \ecosystemcode{writing} outcomes, with empirical studies documenting immediate effects on product reviews~\cite{arnold2020predictive,jakesch2023cowriting}, stories~\cite{singh2022where}, screenplays~\cite{mirowski2023cowriting} and business pitches~\cite{wambsganss2022modeling}, to name a few. 
However, while we found no empirical studies beyond a single writing session in our corpus, some researchers anticipate the longer-term risks of homogenization in creative writing \cite{Gero2019metaphoria,padmakumar2024does,anderson2024homogenization}, professional deskilling in written communication \cite{Buschek2021emails}, and loss of author agency simply through fatigue in reviewing AI suggestions \cite{bhat2023interacting}. 
Designers should consider the risk that AI-generated text is ingested as training data by other AI projects (``model collapse'' \cite{shumailov2023curse,taori2023data}), an example of change in \ecosystemcode{information environment} \cite{arnold2020predictive}. 

We can also anticipate changes in \ecosystemcode{readers}. 
As \lms can (co-)author complex writing that is hard to distinguish from human-authored text~\cite{clark-etal-2021-thats,jakesch2023human}, designers should consider different readers' criteria for trustworthiness \cite{jakesch2019aimediated}. 
However, as with authors, there were no longitudinal studies of readers or reading practices with writing assistants. 
Given the current pace of technological development, frequent changes in \ecosystemcode{technology} will be the norm, much more so than what we observed in the past~\cite{macarthur2006effects,haas2013writing}.
An HCI perspective might ask how systems can be designed to gracefully adapt to the evolution of technology, for instance, by assisting authors who dislike an updated \lm to roll back to their older, more personally-tuned version (cf. user protests about companion chatbot updates \cite{washpost2023botupdate} resulting in version management \cite{kuyda2023replikaversions}).} 
Finally, as \ecosystemcode{regulation} catches up with technological advances,
it could affect model performance (\eg models trained on copyright material could be banned), procurement (\eg corporate AI governance restricts products meeting new ethical standards), or subscription models (\eg more ``ethical'' \lms might cost more).

%% file: tables/ecosystem-designspace.tex
\begin{table*}
    \resizebox{1.\linewidth}{!}{
        \renewcommand{\arraystretch}{1.6}
        \setlength{\tabcolsep}{4pt}
        \begin{tabular}{p{0.05\textwidth}p{0.275\textwidth}p{0.8\textwidth}}
            \toprule
            &
              \textbf{Code} &
              \textbf{Definition} \\
              \midrule

            \multirow{3}{*}[-0.2em]{\textbf{\rotatebox{90}{\renewcommand{\arraystretch}{1}\begin{tabular}[c]{@{}c@{}}\ecosystemdimensiont{Digital} \\ \ecosystemdimensiont{{\small Infrastructure}}\end{tabular}}}} &
              \multicolumn{2}{l}{\emph{What compatibility issues are considered?}} \\
              \dashedline{2-3}
             &
              \ecosystemcode{Usability consistency} &
              Alignment of user experience with other systems that the user is familiar with \\
             &
              \ecosystemcode{Technical interoperability} &
              Ability to communicate and work together with other systems, applications, or devices seamlessly \\
              \midrule

             \multirow{3}{*}[-0.6em]{\textbf{\rotatebox{90}{\renewcommand{\arraystretch}{1}\begin{tabular}[c]{@{}c@{}} \ecosystemdimensiont{Access}\\\ecosystemdimensiont{Model }\end{tabular}}}} &
              \multicolumn{2}{l}{\emph{How does the openness of data, models, and products influence design decisions?}} \\
              \dashedline{2-3}
             &
              \ecosystemcode{Free and open-source software}  &
              Factors related to free access, derivation of work, and redistribution \\
             &
              \ecosystemcode{Commercial software} &
              Factors related to the use of or integration as part of a commercial product \\
              \midrule

            \multirow{3}{*}[-.5em]{\textbf{\rotatebox{90}{\renewcommand{\arraystretch}{1}\begin{tabular}[c]{@{}c@{}}\ecosystemdimensiont{Social} \\ \ecosystemdimensiont{Factors}\end{tabular}}}} &
              \multicolumn{2}{l}{\emph{Who affects the design and use of writing assistants?}} \\
              \dashedline{2-3}
             &
              \ecosystemcode{Design with stakeholders} &
              Accounting for stakeholders' perspectives and behaviors \\
            &
              \ecosystemcode{Design for social writing} &
              Accounting for writers' social context of writing, such as co-authors \\ 
 
              \midrule

            \multirow{6}{*}[+2.5em]{\textbf{\rotatebox{90}{\renewcommand{\arraystretch}{1}\begin{tabular}[c]{@{}c@{}}\ecosystemdimensiont{Locale}\end{tabular}}}} &
              \multicolumn{2}{l}{\emph{Does the writing assistant’s design take into account features of a physical locale?}} \\
              \dashedline{2-3}
             &
              \ecosystemcode{Local writing} &
              Design for writing at ``home base'' with full sociotechnical networking  \\
             &
              \ecosystemcode{Remote writing} &
              Design for writing remotely from ``home base'' \\
              
              \midrule
              
            \multirow{4}{*}[0.5em]{\textbf{\rotatebox{90}{\renewcommand{\arraystretch}{1}\begin{tabular}[c]{@{}c@{}}\ecosystemdimensiont{Norms and}\\\ecosystemdimensiont{Rules} \end{tabular}}}} &
              \multicolumn{2}{l}{\emph{What norms and rules affect the design and use of a writing assistant?}} \\
              \dashedline{2-3}
             &
              \ecosystemcode{Laws} &
              Legal requirements, such as privacy, copyright, and age-appropriate content \\ 
              
             &
              \ecosystemcode{Conventions} &
              Cultural, professional, or organizational norms \& standards \\ 
              \midrule

            \multirow{4}{*}[-3em]{\textbf{\rotatebox{90}{\renewcommand{\arraystretch}{1}\begin{tabular}[c]{@{}c@{}}\ecosystemdimensiont{Change} \\ \ecosystemdimensiont{Over Time}\end{tabular}}}} &
              \multicolumn{2}{l}{\emph{What are the key temporal considerations when designing a writing assistant?}} \\
              \dashedline{2-3}
             &
              \ecosystemcode{Authors} &
              Changes in authors' perception, knowledge, and skills regarding writing assistants \\
             &
              \ecosystemcode{Readers} &
              Changes in readers' perception, knowledge, and skills regarding writing assistants \\
             &
              \ecosystemcode{Writing} &
              Changes in written artifacts due to the use of writing assistants \\
             &
              \ecosystemcode{Information environment} &
              Changes in the existing text and knowledge landscape due to the use of writing assistants \\
             &
              \ecosystemcode{Technologies} &
              Changes in the technologies powering writing assistants \\
             &
              \ecosystemcode{Regulation} &
              Changes in the laws and conventions that govern the use of writing assistants in the ecosystem \\
        \bottomrule
        \end{tabular}
    }
\caption[]{Ecosystem dimensions, codes, and definitions.}
\label{tab:ecosystem-designspace}
\end{table*}

%% file: sections/discussion.tex
In this section, we share use cases for our design space, our reflections on the current literature and ethical implications, the challenges and limitations we encountered in creating the design space, and our plans for future work.

\input{figures/papers}

\subsection{\change{Use Case Scenarios for the Design Space}}
\label{sec:usecases}
\change{
In this section, we present two use case scenarios for our design space. 
These scenarios illustrate how the design space can be utilized (\eg generative and analytical) and demonstrate its value for a range of stakeholders (\eg researchers and policymakers). 
They also underscore the interdependencies and trade-offs between different dimensions and codes.

\paragraph{Generative scenario.} 
Suppose a group of researchers aims to create a writing assistant, specifically designed to aid non-native English writers in choosing the best paraphrase among multiple paraphrases generated by AI.\footnote{Note that this is a hypothetical scenario based on \citet{kim2023explainable}.}
Motivated by previous work, the researchers plan to build a prototype of the writing assistant to improve the writing quality of the target user group by offering support features.
Upon examining the design space, the researchers find they have already factored in numerous dimensions, such as the \taskdimensiont{writing process} [\taskcode{revision}], \taskdimensiont{writing context} [\taskcode{academic}], \userdimensiont{demographic profile} [\usercode{language \& culture}], \userdimensiont{system output preferences} [\usercode{explainability}], and \ecosystemdimensiont{digital infrastructure} [\ecosystemcode{usability consistency}]. 
However, they realize that they overlooked certain key dimensions like \technologydimensiont{data - source} and \technologydimensiont{evaluation - focus} of the foundation model they intend to use, as well as dimensions like \userdimensiont{relationship to system} [\usercode{trust, transparency}] and \interactiondimensiont{interaction metaphor} that could influence how the user perceives the system. 
These insights prompt them to do investigations into their options for foundation models that they may not have previously considered, take into account user concerns of trust and transparency, and think about various ways to frame and present the system to users. 
Overall, the researchers find that the design space ensures they do not overlook important design decisions, resulting in a richer and more thoughtful design.

\paragraph{Analytic scenario.} 
Suppose a group of policymakers is concerned about the unintended consequences of AI-powered writing assistants swaying public opinions.
This worry stems from a research report suggesting that co-writing with opinionated language models (\lms) can influence writers' views.\footnote{Note that this is a hypothetical scenario based on \citet{jakesch2023cowriting}.}
To gain a comprehensive understanding of the context from which these findings originate, the policymakers refer to the research paper as well as our design space.
By mapping the writing assistant used in the paper to the design space, they gain a nuanced understanding of the experimental context (\eg \taskdimensiont{writing stage} [\taskcode{drafting}], \taskdimensiont{specificity} [\taskcode{general direction}], and \technologydimensiont{model - type} [\technologycode{foundation model}]). 
More importantly, they identify several factors that could potentially alter the findings.
For example, the writing assistant in the study automatically provided suggestions to users (\interactiondimensiont{UI - initiative} [\interactioncode{system-initiated}]), rather than allowing users to request suggestions when needed. 
Furthermore, users had no way to \emph{explicitly} control or guide the system's output, and had limited \emph{implicit} control;
even though the underlying model took user text as input to account for the user's writing style and existing contents,
it was consistently prompted by the system to output text in favor of a pre-determined stance on the topic that the user was asked to write about (\interactiondimensiont{user - steering the system} [\interactioncode{implicit}]). 
Recognizing these factors, the policymakers realize they could potentially introduce regulations to make user-initiated interactions mandatory and to allow users to explicitly steer the system's output. 
They could also recommend that designers  visually differentiate between user-generated and AI-generated text (\interactiondimensiont{UI - visual differentiation}). 
Finally, the design space draws attention to a state regulatory proposal to categorize as ``high risk'' any AI system that could subtly bias voting behavior through nudging (\ecosystemdimensiont{norms \& rules} [\ecosystemcode{laws}]).
}

\input{figures/dimensions}

\subsection{Trends and Gaps in the Literature}
\label{sec:analysis}

\change{Based on an analysis of our corpus of papers, we see that there is a sharp increase in papers about writing assistants starting in the mid-2010s (Figure~\ref{fig:numpapers}). This increase is roughly equal among HCI and NLP venues, though there are slightly more HCI than NLP papers in our corpus. It is this increase that spurred our interest in creating a design space to support the increasing number of researchers and designers working in this space.}

\change{Based on our final coding of all papers, we observe that certain dimensions in the design space are over- or under-represented. Figure~\ref{fig:papersperdimension} shows the number of papers per dimension which were coded as relevant. 
To highlight a few notable trends, 
we see \taskdimensiont{audience} is under-represented compared to the other task dimensions, suggesting future work may want to more explicitly consider who is the audience of a piece of writing. \technologydimensiont{Scalability} is quite under-represented overall, as well as relative to other technology dimensions, suggesting that there may currently not be enough consideration of the economic and computational costs of training and using recent large models in the context of writing assistants.
Finally, most ecosystem dimensions are, as previously noted, under-represented, representing a rich area of future study as writing assistants become more widely adopted and the circumstance and context of their adoption becomes increasingly important. Longitudinal studies should illuminate if, and in what ways, writers' \userdimensiont{relationship to system} changes through extended use, and how it will affect not only the writers, but also readers and information environment (\ecosystemdimensiont{change over time}).}

\change{We also note that technological advances are driving changes in writing assistants. The use of \technologycode{foundation models} has rapidly increased in just the past few years; we see 13 papers with this code in 2023 versus 1 in 2020 in our corpus. 
We expect this number to grow substantially in the coming years.
However, we have not yet seen a corresponding increase in codes that seem relevant to their increased usage, such as user concerns of \usercode{trust} and \usercode{transparency}, or technological evaluations of \technologycode{controllability} or issues of \technologycode{ethics}. 
We hope that the provision of our design space can help researchers and designers think about these issues as they become increasingly important with rapid technological advances.}

\subsection{Ethical Implications of Writing Assistants}
\googlechange{Writing assistants, while beneficial, hold a potential for serious risks, particularly when intentionally designed or misused by individuals or organizations to plagiarize content~\cite{lee2023language,perkins2023academic}, generate deceptive content~\cite{hutchens2023language}, or systematically sway opinions~\cite{jakesch2023cowriting}, thus requiring careful scrutiny and ethical considerations.}
Additionally, uses of writing assistants have begun to affect labor markets \citep{eloundou2023gpts,brynjolfsson2023generative,noy2023productivity,wgastrike2023whatwewon}, signifying substantial societal consequences and evidencing the need for monitoring such effects. 
As these systems integrate into various industries, a comprehensive, multi-faceted approach to ethical governance is essential to tackle sector-specific concerns.

Another ethical consideration is accommodating the needs and preferences of diverse users, including those from differently-abled, under-represented, and marginalized communities~\googlechange{\cite{bi2022accessibility,abid2021persistent,Hill2017GenderInclusivenessPV}}. 
Beyond traits inherent to users, such as primary language and culture, such accommodations should consider user preferences or capabilities like \usercode{writing expertise} and \usercode{technical proficiency}, which may widely vary across educational, socioeconomic and neurocognitive backgrounds. 
Failure to address these factors can lead to misalignment of expectations or biased outputs that further perpetuate inequalities and stereotypes. 
Future work could consider, for instance, \emph{value-sensitive design}, which acknowledges the importance of understanding the needs, preferences, and concerns of different user groups. 
This is especially important in sensitive contexts, such as education, counseling, or healthcare.

\subsection{Challenges in Developing a Design Space}
We underscore that the five aspects within the design space have blurry boundaries, as some dimensions may straddle multiple aspects. 
When defining dimensions, we sought to increase coverage and make them as mutually exclusive as possible.
However, in some cases (\eg the relationship between \taskdimensiont{writing context} and \taskdimensiont{purpose}), this cleaving was simply impossible. 
Defining dimensions and codes that were not frequently mentioned or implicitly addressed in research papers posed additional challenges. 
For instance, many dimensions for ecosystem (\eg \ecosystemdimensiont{digital infrastructure}, \ecosystemdimensiont{locale}, and \ecosystemdimensiont{access model}) were sometimes possible to infer from papers but were often not explicitly mentioned.

\change{In addition, we noticed that the dimensions of some aspects have inherently different natures.
For instance, the user and ecosystem aspects focus on the very \emph{existence} of codes in work (\eg whether a paper takes \userdimensiont{demographic profile} [\usercode{age}] or \ecosystemdimensiont{digital infrastructure} [\ecosystemcode{usability consistency}] into account when designing a writing assistant).
On the other hand, other aspects (\eg interaction) presume the existence, and focus on the \emph{classification} (\eg whether \interactiondimensiont{UI - interface metaphor} is close to \interactioncode{agent}, \interactioncode{tool}, or \interactioncode{hybrid}).
Furthermore, we find that it is possible for user dimensions to be not only design choices, but reported properties from user studies.
For example, researchers might use a general-purpose writing assistant in their user studies, but focus their evaluation and analysis on the users' \userdimensiont{relationship to system} [\usercode{agency}].
To handle this contingency, we duplicated user dimensions and coded for both design choices and reported properties, while keeping the official set of user dimensions without duplicates.
}

Some codes intrinsically have continuous values, and converting these into discrete codes remained challenging (\eg \taskdimensiont{specificity} and \technologydimensiont{data - size}).
Even when codes are discrete, the space of possible codes can be vast; in this case, we focus on the elements that are explicitly mentioned and frequently observed in the coding process (\eg \userdimensiont{system output preferences} and \technologydimensiont{evaluation - focus}), leaving room for future extension.
When applicable, we abstracted codes to increase their coverage and generalizability, while trading off their specificity (\eg \technologycode{generation} instead of ``dialogue,'' ``story generation,'' and ``question answering'').
\change{During the coding process, we were able to select multiple codes for a dimension, to account for a writing assistant's various functionalities and purposes.}

\subsection{Limitations \change{and Future Work}}
One limitation of our work is the coverage of sampled and filtered papers, restricted by the search criteria. 
Using ``write'' in titles or keywords may not be particularly suitable for NLP papers. 
While expanding the search to include abstracts would increase the corpus, this approach could also lead to a diminishing return rate. 
Consequently, the focus remained on developing useful dimensions and codes, rather than striving for an exhaustive collection of papers. 
Another limitation was not explicitly including commercial writing assistants (without research papers) in our search.
In our coding process, it is possible that we may have misinterpreted authors' intentions, which may have resulted in errors in our annotation.
Despite these limitations, we argue that this design space serves as a reference for examining existing assistants and developing new ones, while preventing implicit assumptions or overlooked considerations, thereby facilitating a holistic understanding of the factors that drive design considerations and choices. 
For future work, we hope to continue to refine our dimensions and codes and code commercial writing assistants to further understand the gaps and opportunities in the current research landscape and suggest possible directions for future research and development.

\subsection{\change{Lessons for Future Writing Assistants}}
\change{
To promote creativity and exploration in this emerging area,
we intentionally avoid imposing subjective views, prescribing how to design writing assistants, or when and where writing assistants should be used.
Instead, we share lessons learned along the way as helpful for our fellow researchers and designers.
We believe it is important to recognize the interconnection of dimensions and codes across the five aspects and their trade-offs, and utilize them as a reference when designing writing assistants. 
As technology continues to evolve, we anticipate the emergence of new capabilities, interaction designs, and insights about user preferences and behaviors. 
Therefore, it is essential to remain adaptive to these changes and heed them when designing and analyzing writing assistants.
Lastly, there are many under-represented or under-explored dimensions in the design space (Section~\ref{sec:analysis}).
We encourage researchers and designers to venture into these overlooked dimensions and offer their innovative ideas and unique insights, contributing to the holistic development of better writing assistants.}

%% file: figures/papers.tex
\begin{figure*}
    \centering
    \includegraphics[width=0.6\textwidth]{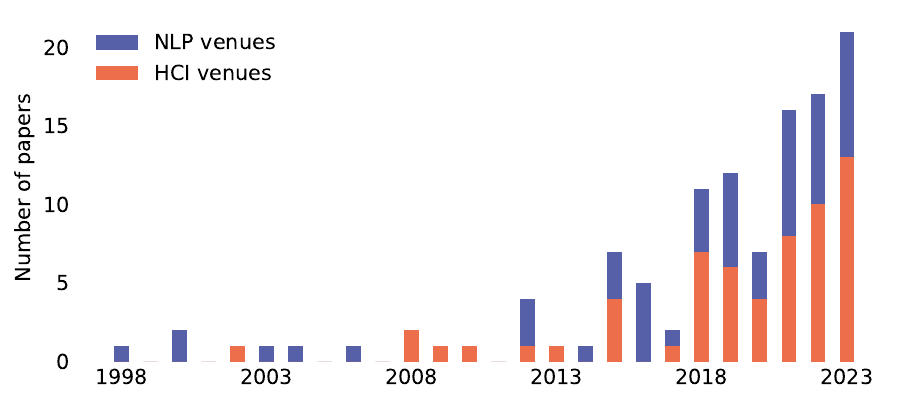}
    \caption[]{\change{The number of papers in our corpus over the years from NLP and HCI venues (retrieved in August 2023). We see a sharp increase in papers on writing assistants starting in the mid-2010s, with roughly equal increase from NLP and HCI venues.
    }}
    \label{fig:numpapers}
    \Description{
    The figure describes the number of papers from NLP and HCI venues included in our literature review over the years. Between 1998 and 2010, the number of paper is low per year, ranging between 1 and 2. From the mid-2010s, there is a noticeable surge in the number of paper. In 2012, the number of paper increases to four. For 2013 and 2014, the number of paper is one for each, but for 2015, the number increases to seven. 2016 and 2015 see a bit of decrease to five and two, but in 2018, it spikes to 11, then in 2019, to 12. 2020 sees a bit of decrease to seven, but it increases again to 16 in 2021, 17 in 2022, and 21 in 2023. This increase is roughly equal among HCI and NLP venues, though there are slightly more HCI than NLP papers in our corpus.
    }
\end{figure*}

%% file: figures/dimensions.tex
\begin{figure*}
    \centering
    \includegraphics[width=\textwidth]{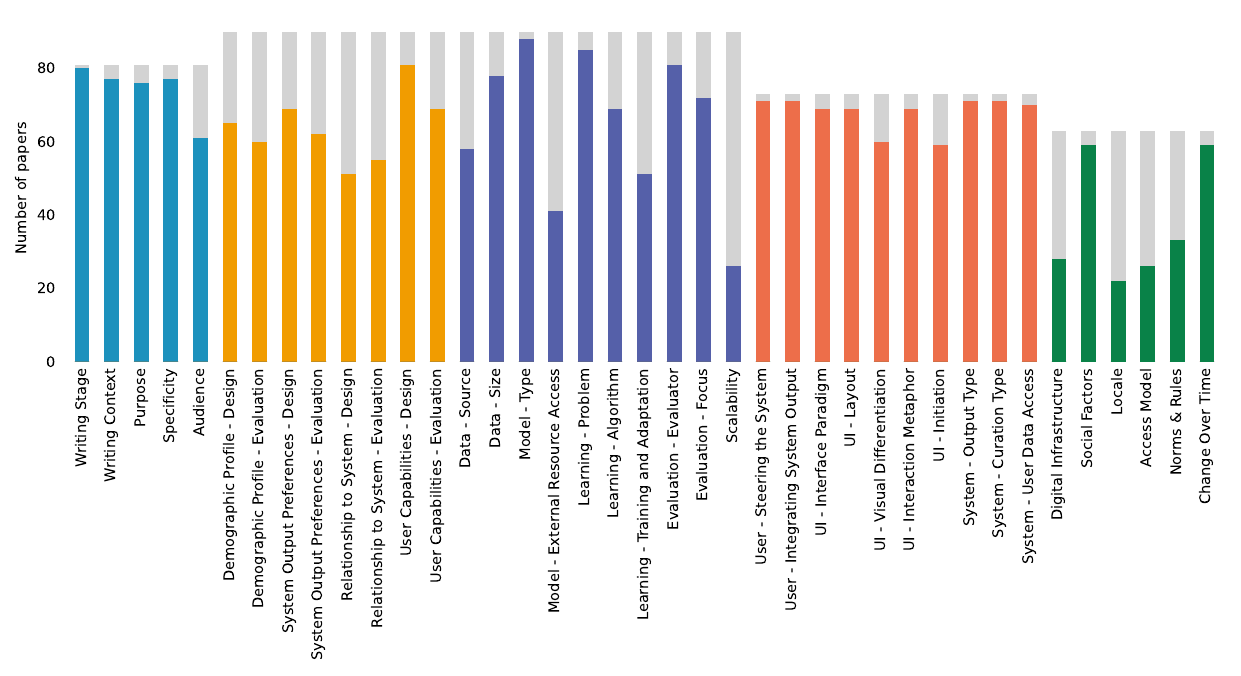}
    \caption[]{\change{The number of papers for each dimension in the design space. Out of the papers that are relevant to each aspect (\ie gray bars), we show the number of relevant papers for each dimension (\ie colored bars). We observe that certain dimensions are over- or under-represented in the current literature and highlight them in Section~\ref{sec:discussion}.
    }}
    \label{fig:papersperdimension}
    \Description{
    The figure shows the number of papers relevant to each dimension in the design space. It first shows the dimensions for Task, with all dimensions hitting around 75 except for Audience, which only hits around 60. Second, it shows User dimensions, with each dimension hitting between 50 and 70, except for User Capabilities - Design, which hits about 80. Technology dimensions show a wider range of numbers ranging from 50 to 80. Scalability and Model - External Resource Access has low counts, each hitting around 25 and 40. Model - Type is highest, showing the number above 80. Interaction dimensions show quite flat counts, around 70, but UI - Visual Differentiation and UI - Initiative have low counts, around 60. Ecosystem dimensions have the lowest counts compared to other aspects, with Digital Infrastructure, Locale, Access Model, and Norms and Rules hitting around 25. Social factors and Change Over Time have higher counts, around 60. 
    }
\end{figure*}

%% file: sections/conclusion.tex
In this work, we present a design space as a structured way to examine and explore the multidimensional space of intelligent and interactive writing assistants.
Through community collaboration and systematic literature review across multiple disciplines, we define \numdim dimensions and \numcode codes exploring five key aspects of writing assistants. 
We hope that this design space provides researchers and designers with a practical tool to navigate, comprehend, and compare the various possibilities of writing assistants, and aids them in the design of new writing assistants.

%% file: sections/appendix.tex

\section{Author contributions}
\label{app:author}

This project was a large collaboration with $36$ researchers across $27$ institutions. This team effort was built on countless contributions from everyone involved.
To acknowledge individual authors' contributions and enable future inquiries to be directed appropriately, we listed authors in three different ways in our paper and artifact.

\subsection{Overall Author List}

In the beginning of the project, each author signed up for one of the four roles in the project.
Project leads oversaw the entire project, supporting team leads and members. 
Team leads kept team members on track, provided feedback on literature review and writing, and maintained alignment with the project's direction. 
Team members contributed to decision making, conducted extensive literature review, and wrote the paper. Advisors, although sometimes not directly involved in literature review and writing, provided additional guidance and feedback. 
Some authors took on two roles, occasionally blurring the role distinctions.
The following list contains each author's name, affiliation, and contributions, grouped by their main self-assigned role.

\vspace{0.2cm}
\noindent \textbf{Project leads}
\begin{itemize}
    \item \textbf{Mina Lee} (University of Chicago \& Microsoft Research): Led and managed the overall project, prepared weekly project meetings, filtered papers, designed dimensions and codes (initial, revision), coded papers (initial, revision), participated in writing (initial, revision), designed figures, and open-sourced the artifact
    
    \item \textbf{Katy Ilonka Gero} (Harvard University): Led and managed the overall project, led the user team, created and managed resources for coding papers, filtered papers, designed dimensions and codes (initial, revision), coded papers (initial, revision), and participated in writing (initial, revision)
    
    \item \textbf{John Joon Young Chung} (Midjourney): Led the systematic literature review process, sampled papers, filtered papers, designed dimensions and codes (initial), coded papers (initial), participated in writing (initial), and analyzed annotations
\end{itemize}

\noindent \textbf{Team leads (alphabetical)}
\begin{itemize}
    \item \textbf{Simon Buckingham Shum} (University of Technology Sydney): Led the ecosystem team, prepared weekly team meetings, filtered papers, designed dimensions and codes (initial, revision), coded papers (initial, revision), and participated in writing (initial, revision)
    
    \item \textbf{Vipul Raheja} (Grammarly): Led the technology team, prepared weekly team meetings, sampled extra papers, filtered papers, designed dimensions and codes (initial), coded papers (initial), and participated in writing (initial, revision)
    
    \item \textbf{Hua Shen} (University of Michigan): Led the interaction team, prepared weekly team meetings, filtered papers, designed dimensions and codes (initial, revision), coded papers (initial, revision), and participated in writing (initial, revision)
    
    \item \textbf{Subhashini Venugopalan} (Google): Led the technology team, sampled extra papers, filtered papers, designed dimensions and codes (initial), coded papers (initial), and participated in writing (initial)
    
    \item \textbf{Thiemo Wambsganss} (Bern University of Applied Sciences): Led the interaction team, prepared weekly team meetings, filtered papers, designed dimensions and codes (initial), coded papers (initial), and participated in writing (initial)
    
    \item \textbf{David Zhou} (University of Illinois, Urbana-Champaign): Led the task team, prepared weekly team meetings, filtered papers, designed dimensions and codes (initial, revision), coded papers (initial, revision), and participated in writing (initial, revision)
\end{itemize}

\noindent \textbf{Team members  (alphabetical)}
\begin{itemize}
    \item \textbf{Emad A. Alghamdi} (King Abdulaziz University): Filtered papers, designed dimensions and codes (initial, revision), coded papers (initial, revision), and participated in writing (initial)
    
    \item \textbf{Tal August} (Allen Institute for AI): Designed dimensions and codes (initial), coded papers (initial), and participated in writing (initial)
    
    \item \textbf{Avinash Bhat} (McGill University): Filtered papers, designed dimensions and codes (initial, revision), coded papers (initial, revision), identified issues in papers and coding results, and participated in writing (initial)
    
    \item \textbf{Madiha Zahrah Choksi} (Cornell Tech): Filtered papers, designed dimensions and codes (initial, revision), coded papers (initial, revision), and participated in writing (initial)
    
    \item \textbf{Senjuti Dutta} (University of Tennessee, Knoxville): Filtered papers, designed dimensions and codes (initial), coded papers (initial), and participated in writing (initial)
    
    \item \textbf{Jin L.C. Guo} (McGill University): Designed dimensions and codes (initial, revision), and coded papers (initial, revision), and participated in writing (initial)
    
    \item \textbf{Md Naimul Hoque} (University of Maryland, College Park): Filtered papers, designed dimensions and codes (initial, revision), and coded papers (initial, revision)
    
    \item \textbf{Yewon Kim} (KAIST): Filtered papers, designed dimensions and codes (initial, revision), coded papers (initial, revision), participated in writing (initial, revision), and identified issues in papers and coding results
    
    \item \textbf{Simon Knight} (University of Technology Sydney): Designed dimensions and codes (initial, revision)
    
    \item \textbf{Seyed Parsa Neshaei} (EPFL): Filtered papers, designed dimensions and codes (initial, revision), coded papers (initial, revision), and participated in writing (initial, revision)

    \item \textbf{Agnia Sergeyuk} (JetBrains Research): Filtered papers, designed dimensions and codes (initial, revision), coded papers (initial, revision), participated in writing (initial, revision), and identified issues in papers and coding results
    
    \item \textbf{Antonette Shibani} (University of Technology Sydney): Designed dimensions and codes (initial, revision) and coded papers (revision)
    
    \item \textbf{Disha Shrivastava} (Google DeepMind): Designed dimensions and codes (initial) and coded papers (initial)
    
    \item \textbf{Lila Shroff} (Stanford University): Filtered papers, designed dimensions and codes (initial), coded papers (initial), and participated in writing (initial)
    
    \item \textbf{Jessi Stark} (University of Toronto): Filtered papers, designed dimensions and codes (initial), coded papers (initial), and participated in writing (initial)
    
    \item \textbf{Sarah Sterman} (University of Illinois, Urbana-Champaign): Filtered papers, designed dimensions and codes (initial, revision), coded papers (initial, revision), participated in writing (initial, revision), designed interaction framework, and analyzed annotations
    
    \item \textbf{Sitong Wang} (Columbia University): Filtered papers, designed dimensions and codes (initial), coded papers (initial), and participated in writing (initial)
    
\end{itemize}

\noindent \textbf{Advisors  (alphabetical)}
\begin{itemize}
    \item \textbf{Antoine Bosselut} (EPFL): Filtered papers, coded papers (initial), participated in writing (initial)
    
    \item \textbf{Daniel Buschek} (University of Bayreuth): Filtered papers, designed dimensions and codes (initial, revision), coded papers (initial, revision), and participated in writing (initial, revision)
    
    \item \textbf{Joseph Chee Chang} (Allen Institute for AI): Filtered papers, designed dimensions and codes (initial), coded papers (initial), and participated in writing (initial)
    
    \item \textbf{Sherol Chen} (Google): Designed dimensions and codes (initial)
    
    \item \textbf{Max Kreminski} (Midjourney): Filtered papers and designed dimensions and codes (initial)
    
    \item \textbf{Joonsuk Park} (University of Richmond): Sampled extra papers, designed dimensions and codes (initial), and participated in writing (initial)
    
    \item \textbf{Roy Pea} (Stanford University): Designed dimensions and codes (initial) and participated in writing (initial, revision)
    
    \item \textbf{Eugenia H. Rho} (Virginia Tech): Designed dimensions and codes (initial) and participated in writing (initial, revision)

    \item \textbf{Shannon Zejiang Shen} (Massachusetts Institute of Technology):
    Designed dimensions and codes (initial), participated in writing (initial), designed tables, and ideated and open-sourced the artifact
    
    \item \textbf{Pao Siangliulue} (B12):
    Filtered papers, designed dimensions and codes (initial), and participated in writing (initial)
\end{itemize}

\subsection{Team-Specific Author Lists}

As described in Section~\ref{sec:coding}, the authors were split into five teams to develop team-specific dimensions and codes based on the five aspects.
Each team operated as its own project group in that most teams had separate weekly meetings.
Below, we specify a team-specific author list for each team.
The ordering follows the convention in Computer Science (\eg the first person in the team member group has the most contribution and the last person in the advisor group has the most contribution within each group).

\begin{itemize}
    \item \textbf{\textcolor{mina-skyblue}{Task}}: David Zhou (team lead), Agnia Sergeyuk (team member), Jessi Stark (team member), Emad A. Alghamdi (team member), Sitong Wang (team member), Roy Pea (advisor)
    \item \textbf{\textcolor{mina-yellow}{User}}: Katy Ilonka Gero (team lead), Yewon Kim (team member), John Joon Young Chung (team member), Senjuti Dutta (team member), Lila Shroff (team member), Disha Shrivastava (team member), Eugenia H. Rho (advisor)
    \item \textbf{\textcolor{mina-blue}{Technology}}: Vipul Raheja (team lead), Subhashini Venugopalan (team lead), Seyed Parsa Neshaei (team member), Disha Shrivastava (team member), Antoine Bosselut (advisor), Sherol Chen (advisor), Joonsuk Park (advisor)
    \item \textbf{\textcolor{mina-orange}{Interaction}}: Hua Shen (team lead), Thiemo Wambsganss (team lead), Sarah Sterman (team member), Tal August (team member), Avinash Bhat (team member), Md Naimul Hoque (team member), Jin L.C. Guo (team member), Pao Siangliulue (advisor), Joseph Chee Chang (advisor), Max Kreminski (advisor), Shannon Zejiang Shen (advisor), Daniel Buschek (advisor)
    \item \textbf{\textcolor{mina-green}{Ecosystem}}: Simon Buckingham Shum (team lead), Madiha Zahrah Choksi (team member), Antonette Shibani (team member), Simon Knight (team member)
\end{itemize}

\subsection{Core Group of Annotators}

Most authors coded papers as part of designing and refining dimensions and codes.
During this process, they focused on the team-specific dimensions and codes and looked at a subset of the papers that were relevant to their teams.
After several iterations within each team, we created an initial version of the design space. 
Then, a subset of the authors volunteered to form the core group of annotators and coded all papers for all dimensions and codes (beyond their own teams).
This relatively small group allowed us to be more efficient and reduced communication overhead.
Here, we list the authors who spearheaded and annotated the papers as part of this core group in the order of their contributions (\ie the first person annotated the highest number of papers) as well as the authors who helped with creating the living artifact.

\begin{itemize}
    \item \textbf{Annotators}: Avinash Bhat, Simon Buckingham Shum, Agnia Sergeyuk, Yewon Kim, David Zhou, Emad A. Alghamdi, Jin L.C. Guo, Seyed Parsa Neshaei, Hua Shen, Md Naimul Hoque, Madiha Zahrah Choksi, Katy Ilonka Gero, Sarah Sterman, Antonette Shibani, Mina Lee
    \item \textbf{Artifact designers}: Shannon Zejiang Shen, Mina Lee
\end{itemize}

\section{Terminology}
\label{app:terminology}

\change{Throughout the paper, we use ``intelligent and interactive writing assistants'' and ``writing assistants'' interchangeably.
Here, we describe the distinctions we make between ``writing assistants'' and other related terms: ``models,'' ``systems,'' and ``technology''.
Firstly, we use ``writing assistants'' to refer to computational systems that assist users with their writing.
These writing assistants must have the frontend that can interface with users, whereas the other three terms do not.
We use ``model'' to refer to a specific model (\eg a specific instance of GPT-3.5, such as gpt-3.5-turbo).
We use ``system'' quite broadly in the paper.
Beyond using it as a concise way to refer to writing assistants,
it can refer to 1) a model (but in this case, we prefer to say ``model'' to be more specific), 2) a model + alpha (e.g., ChatGPT which is GPT-3.5 with extra safety filters on top, a tool-augmented \lm where a model has access to external resources), and 3) anything that is not a model-based system (\eg rule-based system).
The use of ``system'' in Section~\ref{sec:interaction} is an example of the second case.
We use ``technology'' as a much broader concept than ``model'' or ``system'' in that it incorporates data, model, learning, evaluation, and beyond.}

\section{Additional Background}
\label{app:background}

\subsection{Technological Evolution in Writing Assistants}
Since the model architectures used to learn patterns have evolved hand-in-hand to consume and capture patterns in increasing amounts of data, we discuss these two aspects together here.
A number of works on writing assistants in the early 2000s \cite{schneider1998recognizing,liu2000pens,burstein2003toward,park2008is} used purely human-labeled data to develop rule-based methods \cite{burstein2003toward} or train statistical models \cite{babaian2002awriters,park2008is}, which were often used to detect errors \cite{burstein2003toward,park2008is} or suggest corrections. 
These models were trained on much smaller datasets consisting of hundreds or a few thousands of examples. 
With developments in model architecture such as statistical ML models and deep neural networks \cite{mikolov2013efficient,collobert2008unified}, models behind writing tools could take advantage of larger sources of unlabeled data. 
For instance, some work started using learned word embeddings or trained embeddings specifically for a task \cite{dai2014wings, yimam2018demonstrating, Gero2019metaphoria}. 
They were then able to bootstrap off this to use smaller human-labeled datasets for further modeling and evaluation. 
With deep sequence to sequence models~\cite{cho2014gru,sutskever2014sequence} writing assistants were able to use a combination of human-labeled and machine-labeled data with several thousand sentences or examples ~\cite{lukasik2018content,liu2019neuralbased}. 
Several recent works~\cite{yuan2022wordcraft,Chung2022talebrush,du-etal-2022-read,padmakumar2022machine,chakrabarty2022help,mirowski2023cowriting,dang2023choice} take advantage of the large language models that are already pre-trained on corpora of millions of sentences. 
They are then able to use small amounts of human-labeled data to tune the model (\eg instruction tune \cite{raheja2023coedit, shu2023rewritelm, schick2022peer}) and in many cases simply prompt the model in a zero-shot manner to help with the writing task.

\subsection{User Interaction \& Interface Evolution in Writing Assistants}

The origin of generating text to support writers lies in augmentative and alternative communication (AAC) research. 
The goal of these text entry methods is to reduce manual typing, in particular, for people with motor impairments. This was typically realized by predicting next words, and showing them in the user interface for people to select directly, saving letter-by-letter input efforts~\cite{higginbotham1992evaluation, Fowler2015effects}.
These ideas were later applied more generally to improve input efficiency~\cite{Kristensson2014inviscid}. Here we see two interlinked developments: On the technical side, systems evolved from using simple n-gram models to deep learning with today's \lms. This enabled coherent, longer generation as well as the recent prompting paradigm. This in turn shaped the interaction and UI design.

Generated text shown in the UI can now be longer, evolving from single words~\cite{Dunlop2012multidimensional, Fowler2015effects, Gordon2016WatchWriter, Quinn2016costbenefit}, to phrases~\cite{Arnold2016phrases_vs_words, Buschek2021emails, Chen2019smartcompose, lee2022coauthor, Vertanen2015velocitap}, to whole drafts (\eg based on keywords or an incoming email~\cite{Kannan2016smartreply}), depending on use cases.
Relatedly, interactions for controlling text generation have become much more varied. 
As one key distinction~\cite{dang2023choice}, users can implicitly steer text completion with their preceding draft text, or write explicit instructions to the system. 
This instruction style is currently often combined with a user interface design---a chat history with the system (\eg ChatGPT \cite{openai2022chatgpt})---that differs from the traditional page view. 
Other emerging UI patterns in this context include sidebars for suggestions~\cite{yuan2022wordcraft} or annotations~\cite{Dang2022beyond}, as well as other separate views and tools for engaging with AI text~\cite{Gero2019metaphoria, Gero2022sparks}. 
In other designs, AI-generated text is directly integrated into the user's writing area (\eg previewed in a light grey font color~\cite{Chen2019smartcompose}) or can be selected from a pop-up list at the cursor~\cite{Buschek2021emails, dang2023choice, lee2022coauthor}. Beyond linear text, further UI concepts include sketching~\cite{Chung2022talebrush} and views that arrange (generated) text on a 2D canvas, for example, as a graph~\cite{Jiang2023graphologue} or post-it notes.\footnote{\url{https://fermat.app/}}

\subsection{Ecosystem: Going from Micro-HCI to Macro-HCI}
Historical accounts of HCI \citep{rogers2012hcitheory,shneiderman2012charting} document how the definition of ``the system'' was first enlarged from the computer to include the human user, starting with attempts to apply psychological models of the individual human using a computer. 
This frame then expanded to include the ways in which the wider sociotechnical system impinged on (and was shaped by) how interacting groups of people appropriated technology, requiring theory and methods from many more disciplines such as design, ecological psychology, sociology, anthropology, linguistics, critical theory and more, to create today's HCI landscape. 
Borrowing an art history metaphor, \citet{rogers2012hcitheory} characterizes this as the evolution of HCI theory from classical, to modern to contemporary, while \citet{shneiderman2012charting} uses the language of micro-HCI and macro-HCI:

\textit{``Micro-HCI researchers and developers design and build innovative interfaces and deliver validated guidelines for use across the range of desktop, Web, mobile, and ubiquitous devices. The challenges for micro-HCI are to deal with rapidly changing technologies while accommodating the wide range of users.'' [...] ``Macro-HCI researchers and developers design and build interfaces in expanding areas, such as affective experience, aesthetics, motivation, social participation, trust, empathy, responsibility, and privacy. [...] Macro-HCI researchers have to face the challenge of more open tasks, unanticipated user goals, new measures of system efficacy, and even conflicts among users in large communities.''}

In these terms, the design space relating to the task, user, technology, and interaction aspects describes the \textit{micro-HCI} level, but as writing assistants become embedded in broader sociotechnical contexts, this must extend to \textit{macro-HCI} concerns beyond the individual writer and software, to what we term the \textit{ecosystem}.

\section{Systematic Literature Review}
\label{app:litreview}

\subsection{Venues}
\label{sec:venues}

\change{To keep the number of papers reasonable, we decided to focus on papers from the following venues.
We included all their paper tracks (e.g., CHI Late-Breaking Work and ACL Findings), but excluded workshops.}

\begin{itemize}
    \item \textbf{HCI}: CHI, CSCW, UIST, IUI, C\&C, DIS, and ToCHI
    \item \textbf{NLP}: ACL, NAACL, EMNLP, EACL, and TACL
\end{itemize}

\subsection{Keywords}

\change{When retrieving candidate papers from ACM DL, we simply used ``write'' as our keyword, since ACM DL supports automatic matching of variations.
Any paper that has the keyword in its title or its keywords was retrieved.
When retrieving candidate papers from ACL Anthology,
we used ``writ'' and ``wrote'' as our keywords to manually account for the verb ``write'''s variations (``write,'' ``writes,'' ``writing,'' ``wrote,'' and ``written'').
Because papers on ACL Anthology do not have associated keywords, we retrieved any paper that has at least one of the keywords in its title.
Despite the differences, we retrieved the similar number of papers: \numhcipaper papers in HCI and \numnlppaper papers in NLP.}

\subsection{Collected Papers}

\begin{itemize}
    \item \textbf{HCI}: \cite{babaian2002awriters, kim2008common, park2008is, howe2009rita, gonzales2010motivating, dong2012towards, bixler2013detecting, sadauskas2015mining, rapp2015thesis, goncalves2015you, gabriel2015inkwell, watanabe2017lyrisys, hui2018introassist, turkay2018itero, wang2018mirroru, roemmele2018automated, clark2018creative, wu2019design, fan2019character, jakesch2019aimediated, hsu2019on, Gero2019metaphoria, neate2019empowering, gero2019how, peng2020exploring, arnold2020predictive, kim2020lexichrome, afrin2021effective, osone2021buncho, buschek2021impact, robertson2021icant, park2021iwrote, booten2021poetry, belakova2021sonami, schmitt2021characterchat, lee2022coauthor, liu2022will, wambsganss2022adaptive, Chung2022talebrush, yuan2022wordcraft, Dang2022beyond, Gero2022sparks, hoque2022dramatvis, biermann2022fromtool, carrera2022watch, dang2023choice, gero2023social, jakesch2023cowriting, poddar2023aiwriting, mirowski2023cowriting, hui2023lettersmith, lehmann2023mixed, jit2023semi, natalie2023supporting, bhat2023interacting, kim2023metaphorian, karolus2023your, singh2022where}
    \item \textbf{NLP}: \cite{schneider1998recognizing, miltsakaki2000therole, liu2000pens, burstein2003toward, halpin2004automatic, ishioka2006automated, gurevich2007document,  
    huang2012transahead-computer, huang2012transahead, kinnunen2012swan, louis2013makes, dai2014wings, chang2015linguistic, soyer2015crovewa, chang2015writeahead2, rei2016compositional, zhang2016argrewrite, zhang2016using, chollampatt2016adapting, dacunha2017artext, yimam2018demonstrating, lukasik2018content, somasundaran2018towards, liu2019neuralbased, lee2019icomposer, goldfarb2019plan, xu2019alter, hagiwara2019teaspn, nagata2019toward, august2020writing, beigman2020automated, tsai2020lingglewrite, francois2020amesure, clark2021choose, cahill2021supporting, zomer2021beyond, hanawa2021exploring, swanson2021story, duval2021breaking, wambsganss2022supporting, padmakumar2022machine, chakrabarty2022help, jiang2022arxivedits, qi2022quoter, wambsganss2022modeling, zhu2023visualize, afrin2023predicting, weber2023structured, wang2023smart, sun2023songrewriter, zhong2023fiction, skitalinskaya2023revise, shi-etal-2023-effidit, hong2023visualwriting, singh-etal-2021-drag, francois2020amesure}
\end{itemize}

\subsection{\change{Additional References for Technology}}
\label{app:external}

As described in Section~\ref{sec:coding}, the technology team selected 25 additional papers to ensure a broader, deeper, and more relevant coverage of recent technologies.
The paper selection process was identical to the one in Section~\ref{sec:litreview}, but with an expanded set of search keywords, such as ``text revision'' and ``text editing,'' that are more likely to appear in NLP papers.
Furthermore, the team retrieved papers from an expanded set of venues. 
The initial set was 80 papers, which were then deduplicated based on the common pool, and then filtered and adjudicated by two authors based on their relevance to the technology aspect. 
Finally, the team referenced the papers chosen by both the authors as relevant to the aspect, leading to a set of 25 papers.
Note that some of these papers were considered as out of scope based on the criteria in Section~\ref{sec:scope}, but were still relevant as they are concerned with AI models built for writing-related tasks (\eg \lms fine-tuned for specific writing tasks, such as text composition or revision).
The team used these papers to develop their codes, but did not include them as part of their literature review.

\begin{itemize}
    \item \textbf{Technology}: \cite{malmi-etal-2019-encode, stahlberg-kumar-2020-seq2edits, mallinson-etal-2020-felix, faltings-etal-2021-text, sun-etal-2021-iga, wu-etal-2021-automatic, du-etal-2022-understanding-iterative, shu2023rewritelm, iv-etal-2022-fruit, li-etal-2022-text, mallinson-etal-2022-edit5, kim-etal-2022-improving, reid-neubig-2022-learning, schick2022peer, raheja2023coedit, ito-etal-2020-langsmith, mori-etal-2022-plug, su-etal-2023-reviewriter, darcy2023aries, mita2022automated, yang-etal-2017-identifying-semantic, faruqui-etal-2018-wikiatomicedits, zhang-etal-2019-modeling, anthonio-etal-2020-wikihowtoimprove, akoury-etal-2020-storium, rajagopal-etal-2022-one}
\end{itemize}


%% file: main.bbl

\begin{thebibliography}{277}


\ifx \showCODEN    \undefined \def \showCODEN     #1{\unskip}     \fi
\ifx \showDOI      \undefined \def \showDOI       #1{#1}\fi
\ifx \showISBNx    \undefined \def \showISBNx     #1{\unskip}     \fi
\ifx \showISBNxiii \undefined \def \showISBNxiii  #1{\unskip}     \fi
\ifx \showISSN     \undefined \def \showISSN      #1{\unskip}     \fi
\ifx \showLCCN     \undefined \def \showLCCN      #1{\unskip}     \fi
\ifx \shownote     \undefined \def \shownote      #1{#1}          \fi
\ifx \showarticletitle \undefined \def \showarticletitle #1{#1}   \fi
\ifx \showURL      \undefined \def \showURL       {\relax}        \fi
\providecommand\bibfield[2]{#2}
\providecommand\bibinfo[2]{#2}
\providecommand\natexlab[1]{#1}
\providecommand\showeprint[2][]{arXiv:#2}

\bibitem[Abid et~al\mbox{.}(2021)]%
        {abid2021persistent}
\bibfield{author}{\bibinfo{person}{Abubakar Abid}, \bibinfo{person}{Maheen
  Farooqi}, {and} \bibinfo{person}{James Zou}.}
  \bibinfo{year}{2021}\natexlab{}.
\newblock \showarticletitle{Persistent anti-muslim bias in large language
  models}.
\newblock \bibinfo{journal}{\emph{arXiv preprint arXiv:2101.05783}}
  (\bibinfo{year}{2021}).
\newblock


\bibitem[Afrin et~al\mbox{.}(2021)]%
        {afrin2021effective}
\bibfield{author}{\bibinfo{person}{Tazin Afrin}, \bibinfo{person}{Omid
  Kashefi}, \bibinfo{person}{Christopher Olshefski}, \bibinfo{person}{Diane
  Litman}, \bibinfo{person}{Rebecca Hwa}, {and} \bibinfo{person}{Amanda
  Godley}.} \bibinfo{year}{2021}\natexlab{}.
\newblock \showarticletitle{Effective Interfaces for Student-Driven Revision
  Sessions for Argumentative Writing}. In \bibinfo{booktitle}{\emph{Proceedings
  of the 2021 CHI Conference on Human Factors in Computing Systems}} (Yokohama,
  Japan) \emph{(\bibinfo{series}{CHI '21})}. \bibinfo{publisher}{Association
  for Computing Machinery}, \bibinfo{address}{New York, NY, USA}, Article
  \bibinfo{articleno}{58}, \bibinfo{numpages}{13}~pages.
\newblock
\showISBNx{9781450380966}


\bibitem[Afrin and Litman(2023)]%
        {afrin2023predicting}
\bibfield{author}{\bibinfo{person}{Tazin Afrin} {and} \bibinfo{person}{Diane
  Litman}.} \bibinfo{year}{2023}\natexlab{}.
\newblock \showarticletitle{Predicting Desirable Revisions of Evidence and
  Reasoning in Argumentative Writing}. In \bibinfo{booktitle}{\emph{Findings of
  the Association for Computational Linguistics: EACL 2023}}.
  \bibinfo{publisher}{Association for Computational Linguistics},
  \bibinfo{address}{Dubrovnik, Croatia}, \bibinfo{pages}{2550--2561}.
\newblock


\bibitem[Akoury et~al\mbox{.}(2020)]%
        {akoury-etal-2020-storium}
\bibfield{author}{\bibinfo{person}{Nader Akoury}, \bibinfo{person}{Shufan
  Wang}, \bibinfo{person}{Josh Whiting}, \bibinfo{person}{Stephen Hood},
  \bibinfo{person}{Nanyun Peng}, {and} \bibinfo{person}{Mohit Iyyer}.}
  \bibinfo{year}{2020}\natexlab{}.
\newblock \showarticletitle{{STORIUM}: {A} {D}ataset and {E}valuation
  {P}latform for {M}achine-in-the-{L}oop {S}tory {G}eneration}. In
  \bibinfo{booktitle}{\emph{Proceedings of the 2020 Conference on Empirical
  Methods in Natural Language Processing (EMNLP)}},
  \bibfield{editor}{\bibinfo{person}{Bonnie Webber}, \bibinfo{person}{Trevor
  Cohn}, \bibinfo{person}{Yulan He}, {and} \bibinfo{person}{Yang Liu}} (Eds.).
  \bibinfo{publisher}{Association for Computational Linguistics},
  \bibinfo{address}{Online}, \bibinfo{pages}{6470--6484}.
\newblock


\bibitem[Anderson et~al\mbox{.}(2024)]%
        {anderson2024homogenization}
\bibfield{author}{\bibinfo{person}{Barrett~R. Anderson},
  \bibinfo{person}{Jash~Hemant Shah}, {and} \bibinfo{person}{Max Kreminski}.}
  \bibinfo{year}{2024}\natexlab{}.
\newblock \bibinfo{title}{Homogenization Effects of Large Language Models on
  Human Creative Ideation}.
\newblock
\newblock
\showeprint[arxiv]{2402.01536}~[cs.HC]


\bibitem[Anthonio et~al\mbox{.}(2020)]%
        {anthonio-etal-2020-wikihowtoimprove}
\bibfield{author}{\bibinfo{person}{Talita Anthonio}, \bibinfo{person}{Irshad
  Bhat}, {and} \bibinfo{person}{Michael Roth}.}
  \bibinfo{year}{2020}\natexlab{}.
\newblock \showarticletitle{wiki{H}ow{T}o{I}mprove: A Resource and Analyses on
  Edits in Instructional Texts}. In \bibinfo{booktitle}{\emph{Proceedings of
  the Twelfth Language Resources and Evaluation Conference}},
  \bibfield{editor}{\bibinfo{person}{Nicoletta Calzolari},
  \bibinfo{person}{Fr{\'e}d{\'e}ric B{\'e}chet}, \bibinfo{person}{Philippe
  Blache}, \bibinfo{person}{Khalid Choukri}, \bibinfo{person}{Christopher
  Cieri}, \bibinfo{person}{Thierry Declerck}, \bibinfo{person}{Sara Goggi},
  \bibinfo{person}{Hitoshi Isahara}, \bibinfo{person}{Bente Maegaard},
  \bibinfo{person}{Joseph Mariani}, \bibinfo{person}{H{\'e}l{\`e}ne Mazo},
  \bibinfo{person}{Asuncion Moreno}, \bibinfo{person}{Jan Odijk}, {and}
  \bibinfo{person}{Stelios Piperidis}} (Eds.). \bibinfo{publisher}{European
  Language Resources Association}, \bibinfo{address}{Marseille, France},
  \bibinfo{pages}{5721--5729}.
\newblock
\showISBNx{979-10-95546-34-4}


\bibitem[Arnold et~al\mbox{.}(2020)]%
        {arnold2020predictive}
\bibfield{author}{\bibinfo{person}{Kenneth~C. Arnold}, \bibinfo{person}{Krysta
  Chauncey}, {and} \bibinfo{person}{Krzysztof~Z. Gajos}.}
  \bibinfo{year}{2020}\natexlab{}.
\newblock \showarticletitle{Predictive Text Encourages Predictable Writing}. In
  \bibinfo{booktitle}{\emph{Proceedings of the 25th International Conference on
  Intelligent User Interfaces}} (Cagliari, Italy) \emph{(\bibinfo{series}{IUI
  '20})}. \bibinfo{publisher}{Association for Computing Machinery},
  \bibinfo{address}{New York, NY, USA}, \bibinfo{pages}{128–138}.
\newblock
\showISBNx{9781450371186}


\bibitem[Arnold et~al\mbox{.}(2016)]%
        {Arnold2016phrases_vs_words}
\bibfield{author}{\bibinfo{person}{Kenneth~C. Arnold},
  \bibinfo{person}{Krzysztof~Z. Gajos}, {and} \bibinfo{person}{Adam~T. Kalai}.}
  \bibinfo{year}{2016}\natexlab{}.
\newblock \showarticletitle{On {Suggesting} {Phrases} vs. {Predicting} {Words}
  for {Mobile} {Text} {Composition}}. In \bibinfo{booktitle}{\emph{Proceedings
  of the 29th {Annual} {Symposium} on {User} {Interface} {Software} and
  {Technology}}}. \bibinfo{publisher}{ACM}, \bibinfo{address}{Tokyo Japan},
  \bibinfo{pages}{603--608}.
\newblock
\showISBNx{978-1-4503-4189-9}


\bibitem[August et~al\mbox{.}(2020)]%
        {august2020writing}
\bibfield{author}{\bibinfo{person}{Tal August}, \bibinfo{person}{Lauren Kim},
  \bibinfo{person}{Katharina Reinecke}, {and} \bibinfo{person}{Noah~A. Smith}.}
  \bibinfo{year}{2020}\natexlab{}.
\newblock \showarticletitle{Writing Strategies for Science Communication: Data
  and Computational Analysis}. In \bibinfo{booktitle}{\emph{Proceedings of the
  2020 Conference on Empirical Methods in Natural Language Processing
  (EMNLP)}}. \bibinfo{publisher}{Association for Computational Linguistics},
  \bibinfo{address}{Online}, \bibinfo{pages}{5327--5344}.
\newblock


\bibitem[Babaian et~al\mbox{.}(2002)]%
        {babaian2002awriters}
\bibfield{author}{\bibinfo{person}{Tamara Babaian}, \bibinfo{person}{Barbara~J.
  Grosz}, {and} \bibinfo{person}{Stuart~M. Shieber}.}
  \bibinfo{year}{2002}\natexlab{}.
\newblock \showarticletitle{A Writer's Collaborative Assistant}. In
  \bibinfo{booktitle}{\emph{Proceedings of the 7th International Conference on
  Intelligent User Interfaces}} (San Francisco, California, USA)
  \emph{(\bibinfo{series}{IUI '02})}. \bibinfo{publisher}{Association for
  Computing Machinery}, \bibinfo{address}{New York, NY, USA},
  \bibinfo{pages}{7–14}.
\newblock
\showISBNx{1581134592}


\bibitem[Beigman~Klebanov and Madnani(2020)]%
        {beigman2020automated}
\bibfield{author}{\bibinfo{person}{Beata Beigman~Klebanov} {and}
  \bibinfo{person}{Nitin Madnani}.} \bibinfo{year}{2020}\natexlab{}.
\newblock \showarticletitle{Automated Evaluation of Writing {--} 50 Years and
  Counting}. In \bibinfo{booktitle}{\emph{Proceedings of the 58th Annual
  Meeting of the Association for Computational Linguistics}}.
  \bibinfo{publisher}{Association for Computational Linguistics},
  \bibinfo{address}{Online}, \bibinfo{pages}{7796--7810}.
\newblock


\bibitem[Belakova and Mackay(2021)]%
        {belakova2021sonami}
\bibfield{author}{\bibinfo{person}{Jekaterina Belakova} {and}
  \bibinfo{person}{Wendy~E. Mackay}.} \bibinfo{year}{2021}\natexlab{}.
\newblock \showarticletitle{SonAmi: A Tangible Creativity Support Tool for
  Productive Procrastination}. In \bibinfo{booktitle}{\emph{Proceedings of the
  13th Conference on Creativity and Cognition}} (Virtual Event, Italy)
  \emph{(\bibinfo{series}{C\&C '21})}. \bibinfo{publisher}{Association for
  Computing Machinery}, \bibinfo{address}{New York, NY, USA}, Article
  \bibinfo{articleno}{7}, \bibinfo{numpages}{10}~pages.
\newblock
\showISBNx{9781450383769}


\bibitem[Bhat et~al\mbox{.}(2023)]%
        {bhat2023interacting}
\bibfield{author}{\bibinfo{person}{Advait Bhat}, \bibinfo{person}{Saaket
  Agashe}, \bibinfo{person}{Parth Oberoi}, \bibinfo{person}{Niharika Mohile},
  \bibinfo{person}{Ravi Jangir}, {and} \bibinfo{person}{Anirudha Joshi}.}
  \bibinfo{year}{2023}\natexlab{}.
\newblock \showarticletitle{Interacting with Next-Phrase Suggestions: How
  Suggestion Systems Aid and Influence the Cognitive Processes of Writing}. In
  \bibinfo{booktitle}{\emph{Proceedings of the 28th International Conference on
  Intelligent User Interfaces}} (Sydney, NSW, Australia)
  \emph{(\bibinfo{series}{IUI '23})}. \bibinfo{publisher}{Association for
  Computing Machinery}, \bibinfo{address}{New York, NY, USA},
  \bibinfo{pages}{436–452}.
\newblock
\showISBNx{9798400701061}


\bibitem[Bi et~al\mbox{.}(2022)]%
        {bi2022accessibility}
\bibfield{author}{\bibinfo{person}{Tingting Bi}, \bibinfo{person}{Xin Xia},
  \bibinfo{person}{David Lo}, \bibinfo{person}{John Grundy},
  \bibinfo{person}{Thomas Zimmermann}, {and} \bibinfo{person}{Denae Ford}.}
  \bibinfo{year}{2022}\natexlab{}.
\newblock \showarticletitle{Accessibility in software practice: A
  practitioner’s perspective}.
\newblock \bibinfo{journal}{\emph{ACM Transactions on Software Engineering and
  Methodology (TOSEM)}} \bibinfo{volume}{31}, \bibinfo{number}{4}
  (\bibinfo{year}{2022}), \bibinfo{pages}{1--26}.
\newblock


\bibitem[Biermann et~al\mbox{.}(2022)]%
        {biermann2022fromtool}
\bibfield{author}{\bibinfo{person}{Oloff~C. Biermann}, \bibinfo{person}{Ning~F.
  Ma}, {and} \bibinfo{person}{Dongwook Yoon}.} \bibinfo{year}{2022}\natexlab{}.
\newblock \showarticletitle{From Tool to Companion: Storywriters Want AI
  Writers to Respect Their Personal Values and Writing Strategies}. In
  \bibinfo{booktitle}{\emph{Proceedings of the 2022 ACM Designing Interactive
  Systems Conference}} (Virtual Event, Australia) \emph{(\bibinfo{series}{DIS
  '22})}. \bibinfo{publisher}{Association for Computing Machinery},
  \bibinfo{address}{New York, NY, USA}, \bibinfo{pages}{1209–1227}.
\newblock
\showISBNx{9781450393584}


\bibitem[Bixler and D'Mello(2013)]%
        {bixler2013detecting}
\bibfield{author}{\bibinfo{person}{Robert Bixler} {and} \bibinfo{person}{Sidney
  D'Mello}.} \bibinfo{year}{2013}\natexlab{}.
\newblock \showarticletitle{Detecting Boredom and Engagement during Writing
  with Keystroke Analysis, Task Appraisals, and Stable Traits}. In
  \bibinfo{booktitle}{\emph{Proceedings of the 2013 International Conference on
  Intelligent User Interfaces}} (Santa Monica, California, USA)
  \emph{(\bibinfo{series}{IUI '13})}. \bibinfo{publisher}{Association for
  Computing Machinery}, \bibinfo{address}{New York, NY, USA},
  \bibinfo{pages}{225–234}.
\newblock
\showISBNx{9781450319652}


\bibitem[Bommasani et~al\mbox{.}(2021)]%
        {bommasani2021opportunities}
\bibfield{author}{\bibinfo{person}{Rishi Bommasani}, \bibinfo{person}{Drew~A.
  Hudson}, \bibinfo{person}{Ehsan Adeli}, \bibinfo{person}{Russ Altman},
  \bibinfo{person}{Simran Arora}, \bibinfo{person}{Sydney von Arx},
  \bibinfo{person}{Michael~S. Bernstein}, \bibinfo{person}{Jeannette Bohg},
  \bibinfo{person}{Antoine Bosselut}, \bibinfo{person}{Emma Brunskill},
  \bibinfo{person}{Erik Brynjolfsson}, \bibinfo{person}{Shyamal Buch},
  \bibinfo{person}{Dallas Card}, \bibinfo{person}{Rodrigo Castellon},
  \bibinfo{person}{Niladri Chatterji}, \bibinfo{person}{Annie Chen},
  \bibinfo{person}{Kathleen Creel}, \bibinfo{person}{Jared~Quincy Davis},
  \bibinfo{person}{Dorottya Demszky}, \bibinfo{person}{Chris Donahue},
  \bibinfo{person}{Moussa Doumbouya}, \bibinfo{person}{Esin Durmus},
  \bibinfo{person}{Stefano Ermon}, \bibinfo{person}{John Etchemendy},
  \bibinfo{person}{Kawin Ethayarajh}, \bibinfo{person}{Li Fei-Fei},
  \bibinfo{person}{Chelsea Finn}, \bibinfo{person}{Trevor Gale},
  \bibinfo{person}{Lauren Gillespie}, \bibinfo{person}{Karan Goel},
  \bibinfo{person}{Noah Goodman}, \bibinfo{person}{Shelby Grossman},
  \bibinfo{person}{Neel Guha}, \bibinfo{person}{Tatsunori Hashimoto},
  \bibinfo{person}{Peter Henderson}, \bibinfo{person}{John Hewitt},
  \bibinfo{person}{Daniel~E. Ho}, \bibinfo{person}{Jenny Hong},
  \bibinfo{person}{Kyle Hsu}, \bibinfo{person}{Jing Huang},
  \bibinfo{person}{Thomas Icard}, \bibinfo{person}{Saahil Jain},
  \bibinfo{person}{Dan Jurafsky}, \bibinfo{person}{Pratyusha Kalluri},
  \bibinfo{person}{Siddharth Karamcheti}, \bibinfo{person}{Geoff Keeling},
  \bibinfo{person}{Fereshte Khani}, \bibinfo{person}{Omar Khattab},
  \bibinfo{person}{Pang~Wei Koh}, \bibinfo{person}{Mark Krass},
  \bibinfo{person}{Ranjay Krishna}, \bibinfo{person}{Rohith Kuditipudi},
  \bibinfo{person}{Ananya Kumar}, \bibinfo{person}{Faisal Ladhak},
  \bibinfo{person}{Mina Lee}, \bibinfo{person}{Tony Lee}, \bibinfo{person}{Jure
  Leskovec}, \bibinfo{person}{Isabelle Levent}, \bibinfo{person}{Xiang~Lisa
  Li}, \bibinfo{person}{Xuechen Li}, \bibinfo{person}{Tengyu Ma},
  \bibinfo{person}{Ali Malik}, \bibinfo{person}{Christopher~D. Manning},
  \bibinfo{person}{Suvir Mirchandani}, \bibinfo{person}{Eric Mitchell},
  \bibinfo{person}{Zanele Munyikwa}, \bibinfo{person}{Suraj Nair},
  \bibinfo{person}{Avanika Narayan}, \bibinfo{person}{Deepak Narayanan},
  \bibinfo{person}{Ben Newman}, \bibinfo{person}{Allen Nie},
  \bibinfo{person}{Juan~Carlos Niebles}, \bibinfo{person}{Hamed Nilforoshan},
  \bibinfo{person}{Julian Nyarko}, \bibinfo{person}{Giray Ogut},
  \bibinfo{person}{Laurel Orr}, \bibinfo{person}{Isabel Papadimitriou},
  \bibinfo{person}{Joon~Sung Park}, \bibinfo{person}{Chris Piech},
  \bibinfo{person}{Eva Portelance}, \bibinfo{person}{Christopher Potts},
  \bibinfo{person}{Aditi Raghunathan}, \bibinfo{person}{Rob Reich},
  \bibinfo{person}{Hongyu Ren}, \bibinfo{person}{Frieda Rong},
  \bibinfo{person}{Yusuf Roohani}, \bibinfo{person}{Camilo Ruiz},
  \bibinfo{person}{Jack Ryan}, \bibinfo{person}{Christopher Ré},
  \bibinfo{person}{Dorsa Sadigh}, \bibinfo{person}{Shiori Sagawa},
  \bibinfo{person}{Keshav Santhanam}, \bibinfo{person}{Andy Shih},
  \bibinfo{person}{Krishnan Srinivasan}, \bibinfo{person}{Alex Tamkin},
  \bibinfo{person}{Rohan Taori}, \bibinfo{person}{Armin~W. Thomas},
  \bibinfo{person}{Florian Tramèr}, \bibinfo{person}{Rose~E. Wang},
  \bibinfo{person}{William Wang}, \bibinfo{person}{Bohan Wu},
  \bibinfo{person}{Jiajun Wu}, \bibinfo{person}{Yuhuai Wu},
  \bibinfo{person}{Sang~Michael Xie}, \bibinfo{person}{Michihiro Yasunaga},
  \bibinfo{person}{Jiaxuan You}, \bibinfo{person}{Matei Zaharia},
  \bibinfo{person}{Michael Zhang}, \bibinfo{person}{Tianyi Zhang},
  \bibinfo{person}{Xikun Zhang}, \bibinfo{person}{Yuhui Zhang},
  \bibinfo{person}{Lucia Zheng}, \bibinfo{person}{Kaitlyn Zhou}, {and}
  \bibinfo{person}{Percy Liang}.} \bibinfo{year}{2021}\natexlab{}.
\newblock \showarticletitle{On the Opportunities and Risks of Foundation
  Models}.
\newblock \bibinfo{journal}{\emph{arXiv preprint arXiv:2108.07258}}
  (\bibinfo{year}{2021}).
\newblock


\bibitem[Booten and Gero(2021)]%
        {booten2021poetry}
\bibfield{author}{\bibinfo{person}{Kyle Booten} {and}
  \bibinfo{person}{Katy~Ilonka Gero}.} \bibinfo{year}{2021}\natexlab{}.
\newblock \showarticletitle{Poetry Machines: Eliciting Designs for Interactive
  Writing Tools from Poets}. In \bibinfo{booktitle}{\emph{Proceedings of the
  13th Conference on Creativity and Cognition}} (Virtual Event, Italy)
  \emph{(\bibinfo{series}{C\&C '21})}. \bibinfo{publisher}{Association for
  Computing Machinery}, \bibinfo{address}{New York, NY, USA}, Article
  \bibinfo{articleno}{51}, \bibinfo{numpages}{5}~pages.
\newblock
\showISBNx{9781450383769}


\bibitem[Bostrom and Heinen(1977)]%
        {bostrom1977mis}
\bibfield{author}{\bibinfo{person}{Robert~P Bostrom} {and}
  \bibinfo{person}{J~Stephen Heinen}.} \bibinfo{year}{1977}\natexlab{}.
\newblock \showarticletitle{MIS problems and failures: A socio-technical
  perspective. Part I: The causes}.
\newblock \bibinfo{journal}{\emph{MIS quarterly}} (\bibinfo{year}{1977}),
  \bibinfo{pages}{17--32}.
\newblock


\bibitem[Bowman(2022)]%
        {bowman2022chatbot}
\bibfield{author}{\bibinfo{person}{Emma Bowman}.}
  \bibinfo{year}{2022}\natexlab{}.
\newblock \bibinfo{title}{A new {AI} chatbot might do your homework for you.
  But it's still not an {A+} student}.
\newblock
\newblock
\newblock
\shownote{Accessed: Jan 26, 2024}.


\bibitem[Britton et~al\mbox{.}(1975)]%
        {britton1975development}
\bibfield{author}{\bibinfo{person}{James Britton} {et~al\mbox{.}}}
  \bibinfo{year}{1975}\natexlab{}.
\newblock \showarticletitle{The Development of Writing Abilities.}
\newblock  (\bibinfo{year}{1975}).
\newblock


\bibitem[Brown et~al\mbox{.}(2020a)]%
        {brown2020language}
\bibfield{author}{\bibinfo{person}{Tom~B. Brown}, \bibinfo{person}{Benjamin
  Mann}, \bibinfo{person}{Nick Ryder}, \bibinfo{person}{Melanie Subbiah},
  \bibinfo{person}{Jared Kaplan}, \bibinfo{person}{Prafulla Dhariwal},
  \bibinfo{person}{Arvind Neelakantan}, \bibinfo{person}{Pranav Shyam},
  \bibinfo{person}{Girish Sastry}, \bibinfo{person}{Amanda Askell},
  \bibinfo{person}{Sandhini Agarwal}, \bibinfo{person}{Ariel Herbert-Voss},
  \bibinfo{person}{Gretchen Krueger}, \bibinfo{person}{Tom Henighan},
  \bibinfo{person}{Rewon Child}, \bibinfo{person}{Aditya Ramesh},
  \bibinfo{person}{Daniel~M. Ziegler}, \bibinfo{person}{Jeffrey Wu},
  \bibinfo{person}{Clemens Winter}, \bibinfo{person}{Christopher Hesse},
  \bibinfo{person}{Mark Chen}, \bibinfo{person}{Eric Sigler},
  \bibinfo{person}{Mateusz Litwin}, \bibinfo{person}{Scott Gray},
  \bibinfo{person}{Benjamin Chess}, \bibinfo{person}{Jack Clark},
  \bibinfo{person}{Christopher Berner}, \bibinfo{person}{Sam McCandlish},
  \bibinfo{person}{Alec Radford}, \bibinfo{person}{Ilya Sutskever}, {and}
  \bibinfo{person}{Dario Amodei}.} \bibinfo{year}{2020}\natexlab{a}.
\newblock \bibinfo{title}{Language Models are Few-Shot Learners}.
\newblock
\newblock
\showeprint[arxiv]{2005.14165}~[cs.CL]


\bibitem[Brown et~al\mbox{.}(2020b)]%
        {brown2020gpt3}
\bibfield{author}{\bibinfo{person}{Tom~B. Brown}, \bibinfo{person}{Benjamin
  Mann}, \bibinfo{person}{Nick Ryder}, \bibinfo{person}{Melanie Subbiah},
  \bibinfo{person}{Jared Kaplan}, \bibinfo{person}{Prafulla Dhariwal},
  \bibinfo{person}{Arvind Neelakantan}, \bibinfo{person}{Pranav Shyam},
  \bibinfo{person}{Girish Sastry}, \bibinfo{person}{Amanda Askell},
  \bibinfo{person}{Sandhini Agarwal}, \bibinfo{person}{Ariel Herbert-Voss},
  \bibinfo{person}{Gretchen Krueger}, \bibinfo{person}{Tom Henighan},
  \bibinfo{person}{Rewon Child}, \bibinfo{person}{Aditya Ramesh},
  \bibinfo{person}{Daniel~M. Ziegler}, \bibinfo{person}{Jeffrey Wu},
  \bibinfo{person}{Clemens Winter}, \bibinfo{person}{Christopher Hesse},
  \bibinfo{person}{Mark Chen}, \bibinfo{person}{Eric Sigler},
  \bibinfo{person}{Mateusz Litwin}, \bibinfo{person}{Scott Gray},
  \bibinfo{person}{Benjamin Chess}, \bibinfo{person}{Jack Clark},
  \bibinfo{person}{Christopher Berner}, \bibinfo{person}{Sam McCandlish},
  \bibinfo{person}{Alec Radford}, \bibinfo{person}{Ilya Sutskever}, {and}
  \bibinfo{person}{Dario Amodei}.} \bibinfo{year}{2020}\natexlab{b}.
\newblock \showarticletitle{Language Models are Few-Shot Learners}.
\newblock \bibinfo{journal}{\emph{arXiv preprint arXiv:2005.14165}}
  (\bibinfo{year}{2020}).
\newblock


\bibitem[Brynjolfsson et~al\mbox{.}(2023)]%
        {brynjolfsson2023generative}
\bibfield{author}{\bibinfo{person}{Erik Brynjolfsson},
  \bibinfo{person}{Danielle Li}, {and} \bibinfo{person}{Lindsey~R Raymond}.}
  \bibinfo{year}{2023}\natexlab{}.
\newblock \bibinfo{booktitle}{\emph{Generative AI at work}}.
\newblock \bibinfo{type}{{T}echnical {R}eport}. \bibinfo{institution}{National
  Bureau of Economic Research}.
\newblock


\bibitem[Bubeck et~al\mbox{.}(2023)]%
        {bubeck2023sparks}
\bibfield{author}{\bibinfo{person}{Sébastien Bubeck}, \bibinfo{person}{Varun
  Chandrasekaran}, \bibinfo{person}{Ronen Eldan}, \bibinfo{person}{Johannes
  Gehrke}, \bibinfo{person}{Eric Horvitz}, \bibinfo{person}{Ece Kamar},
  \bibinfo{person}{Peter Lee}, \bibinfo{person}{Yin~Tat Lee},
  \bibinfo{person}{Yuanzhi Li}, \bibinfo{person}{Scott Lundberg},
  \bibinfo{person}{Harsha Nori}, \bibinfo{person}{Hamid Palangi},
  \bibinfo{person}{Marco~Tulio Ribeiro}, {and} \bibinfo{person}{Yi Zhang}.}
  \bibinfo{year}{2023}\natexlab{}.
\newblock \bibinfo{title}{Sparks of Artificial General Intelligence: Early
  experiments with GPT-4}.
\newblock
\newblock
\showeprint[arxiv]{2303.12712}~[cs.CL]


\bibitem[Burstein and Wolska(2003)]%
        {burstein2003toward}
\bibfield{author}{\bibinfo{person}{Jill Burstein} {and}
  \bibinfo{person}{Magdalena Wolska}.} \bibinfo{year}{2003}\natexlab{}.
\newblock \showarticletitle{Toward Evaluation of Writing Style: Overly
  Repetitious Word Use}. In \bibinfo{booktitle}{\emph{10th Conference of the
  {E}uropean Chapter of the Association for Computational Linguistics}}.
  \bibinfo{publisher}{Association for Computational Linguistics},
  \bibinfo{address}{Budapest, Hungary}.
\newblock


\bibitem[Buschek et~al\mbox{.}(2021a)]%
        {Buschek2021emails}
\bibfield{author}{\bibinfo{person}{Daniel Buschek}, \bibinfo{person}{Martin
  Z\"{u}rn}, {and} \bibinfo{person}{Malin Eiband}.}
  \bibinfo{year}{2021}\natexlab{a}.
\newblock \showarticletitle{The Impact of Multiple Parallel Phrase Suggestions
  on Email Input and Composition Behaviour of Native and Non-Native English
  Writers}. In \bibinfo{booktitle}{\emph{Proceedings of the 2021 CHI Conference
  on Human Factors in Computing Systems}} (Yokohama, Japan)
  \emph{(\bibinfo{series}{CHI '21})}. \bibinfo{publisher}{Association for
  Computing Machinery}, \bibinfo{address}{New York, NY, USA}, Article
  \bibinfo{articleno}{732}, \bibinfo{numpages}{13}~pages.
\newblock
\showISBNx{9781450380966}


\bibitem[Buschek et~al\mbox{.}(2021b)]%
        {buschek2021impact}
\bibfield{author}{\bibinfo{person}{Daniel Buschek}, \bibinfo{person}{Martin
  Z\"{u}rn}, {and} \bibinfo{person}{Malin Eiband}.}
  \bibinfo{year}{2021}\natexlab{b}.
\newblock \showarticletitle{The Impact of Multiple Parallel Phrase Suggestions
  on Email Input and Composition Behaviour of Native and Non-Native {E}nglish
  Writers}. In \bibinfo{booktitle}{\emph{Conference on Human Factors in
  Computing Systems (CHI)}}.
\newblock


\bibitem[Cahill et~al\mbox{.}(2021)]%
        {cahill2021supporting}
\bibfield{author}{\bibinfo{person}{Aoife Cahill}, \bibinfo{person}{James
  Bruno}, \bibinfo{person}{James Ramey}, \bibinfo{person}{Gilmar
  Ayala~Meneses}, \bibinfo{person}{Ian Blood}, \bibinfo{person}{Florencia
  Tolentino}, \bibinfo{person}{Tamar Lavee}, {and} \bibinfo{person}{Slava
  Andreyev}.} \bibinfo{year}{2021}\natexlab{}.
\newblock \showarticletitle{Supporting {S}panish Writers using Automated
  Feedback}. In \bibinfo{booktitle}{\emph{Proceedings of the 2021 Conference of
  the North American Chapter of the Association for Computational Linguistics:
  Human Language Technologies: Demonstrations}}.
  \bibinfo{publisher}{Association for Computational Linguistics},
  \bibinfo{address}{Online}, \bibinfo{pages}{116--124}.
\newblock


\bibitem[Cai et~al\mbox{.}(2022)]%
        {cai2022context}
\bibfield{author}{\bibinfo{person}{Shanqing Cai}, \bibinfo{person}{Subhashini
  Venugopalan}, \bibinfo{person}{Katrin Tomanek}, \bibinfo{person}{Ajit
  Narayanan}, \bibinfo{person}{Meredith~Ringel Morris}, {and}
  \bibinfo{person}{Michael~P Brenner}.} \bibinfo{year}{2022}\natexlab{}.
\newblock \showarticletitle{Context-Aware Abbreviation Expansion Using Large
  Language Models}.
\newblock \bibinfo{journal}{\emph{arXiv preprint arXiv:2205.03767}}
  (\bibinfo{year}{2022}).
\newblock


\bibitem[Cambon et~al\mbox{.}(2023)]%
        {cambon2023early}
\bibfield{author}{\bibinfo{person}{Alexia Cambon}, \bibinfo{person}{Brent
  Hecht}, \bibinfo{person}{Benjamin Edelman}, \bibinfo{person}{Donald Ngwe},
  \bibinfo{person}{Sonia Jaffe}, \bibinfo{person}{Amy Heger},
  \bibinfo{person}{Mihaela Vorvoreanu}, \bibinfo{person}{Sida Peng},
  \bibinfo{person}{Jake Hofman}, \bibinfo{person}{Alex Farach},
  \bibinfo{person}{Margarita Bermejo-Cano}, \bibinfo{person}{Eric Knudsen},
  \bibinfo{person}{James Bono}, \bibinfo{person}{Hardik Sanghavi},
  \bibinfo{person}{Sofia Spatharioti}, \bibinfo{person}{David Rothschild},
  \bibinfo{person}{Daniel~G. Goldstein}, \bibinfo{person}{Eirini Kalliamvakou},
  \bibinfo{person}{Peter Cihon}, \bibinfo{person}{Mert Demirer},
  \bibinfo{person}{Michael Schwarz}, {and} \bibinfo{person}{Jaime Teevan}.}
  \bibinfo{year}{2023}\natexlab{}.
\newblock \bibinfo{booktitle}{\emph{Early LLM-based Tools for Enterprise
  Information Workers Likely Provide Meaningful Boosts to Productivity}}.
\newblock \bibinfo{type}{{T}echnical {R}eport} MSR-TR-2023-43.
  \bibinfo{institution}{Microsoft}.
\newblock


\bibitem[Card et~al\mbox{.}(1990)]%
        {card1990thedesignspace}
\bibfield{author}{\bibinfo{person}{Stuart~K. Card}, \bibinfo{person}{Jock~D.
  Mackinlay}, {and} \bibinfo{person}{George~G. Robertson}.}
  \bibinfo{year}{1990}\natexlab{}.
\newblock \showarticletitle{The Design Space of Input Devices}. In
  \bibinfo{booktitle}{\emph{Proceedings of the SIGCHI Conference on Human
  Factors in Computing Systems}} (Seattle, Washington, USA)
  \emph{(\bibinfo{series}{CHI '90})}. \bibinfo{publisher}{Association for
  Computing Machinery}, \bibinfo{address}{New York, NY, USA},
  \bibinfo{pages}{117–124}.
\newblock
\showISBNx{0201509326}


\bibitem[Card et~al\mbox{.}(1991)]%
        {card1991morphological}
\bibfield{author}{\bibinfo{person}{Stuart~K Card}, \bibinfo{person}{Jock~D
  Mackinlay}, {and} \bibinfo{person}{George~G Robertson}.}
  \bibinfo{year}{1991}\natexlab{}.
\newblock \showarticletitle{A morphological analysis of the design space of
  input devices}.
\newblock \bibinfo{journal}{\emph{ACM Transactions on Information Systems
  (TOIS)}} \bibinfo{volume}{9}, \bibinfo{number}{2} (\bibinfo{year}{1991}),
  \bibinfo{pages}{99--122}.
\newblock


\bibitem[Carrera and Lee(2022)]%
        {carrera2022watch}
\bibfield{author}{\bibinfo{person}{Dashiel Carrera} {and}
  \bibinfo{person}{Sang~Won Lee}.} \bibinfo{year}{2022}\natexlab{}.
\newblock \showarticletitle{Watch Me Write: Exploring the Effects of Revealing
  Creative Writing Process through Writing Replay}. In
  \bibinfo{booktitle}{\emph{Proceedings of the 14th Conference on Creativity
  and Cognition}} (Venice, Italy) \emph{(\bibinfo{series}{C\&C '22})}.
  \bibinfo{publisher}{Association for Computing Machinery},
  \bibinfo{address}{New York, NY, USA}, \bibinfo{pages}{146–160}.
\newblock
\showISBNx{9781450393270}


\bibitem[Chakrabarty et~al\mbox{.}(2022a)]%
        {chakrabarty2022help}
\bibfield{author}{\bibinfo{person}{Tuhin Chakrabarty}, \bibinfo{person}{Vishakh
  Padmakumar}, {and} \bibinfo{person}{He He}.}
  \bibinfo{year}{2022}\natexlab{a}.
\newblock \showarticletitle{Help me write a Poem: Instruction Tuning as a
  Vehicle for Collaborative Poetry Writing}. In
  \bibinfo{booktitle}{\emph{Proceedings of the 2022 Conference on Empirical
  Methods in Natural Language Processing}}. \bibinfo{publisher}{Association for
  Computational Linguistics}, \bibinfo{address}{Abu Dhabi, United Arab
  Emirates}, \bibinfo{pages}{6848--6863}.
\newblock


\bibitem[Chakrabarty et~al\mbox{.}(2022b)]%
        {chakrabarty2022copoet}
\bibfield{author}{\bibinfo{person}{Tuhin Chakrabarty}, \bibinfo{person}{Vishakh
  Padmakumar}, {and} \bibinfo{person}{He He}.}
  \bibinfo{year}{2022}\natexlab{b}.
\newblock \showarticletitle{Help me write a poem: Instruction Tuning as a
  Vehicle for Collaborative Poetry Writing}. In
  \bibinfo{booktitle}{\emph{Empirical Methods in Natural Language Processing
  (EMNLP)}}.
\newblock


\bibitem[Chang and Chang(2015)]%
        {chang2015writeahead2}
\bibfield{author}{\bibinfo{person}{Jim Chang} {and} \bibinfo{person}{Jason
  Chang}.} \bibinfo{year}{2015}\natexlab{}.
\newblock \showarticletitle{{W}rite{A}head2: Mining Lexical Grammar Patterns
  for Assisted Writing}. In \bibinfo{booktitle}{\emph{Proceedings of the 2015
  Conference of the North {A}merican Chapter of the Association for
  Computational Linguistics: Demonstrations}}. \bibinfo{publisher}{Association
  for Computational Linguistics}, \bibinfo{address}{Denver, Colorado},
  \bibinfo{pages}{106--110}.
\newblock


\bibitem[Chang et~al\mbox{.}(2015)]%
        {chang2015linguistic}
\bibfield{author}{\bibinfo{person}{Yung-Chun Chang}, \bibinfo{person}{Cen-Chieh
  Chen}, \bibinfo{person}{Yu-Lun Hsieh}, \bibinfo{person}{Chien~Chin Chen},
  {and} \bibinfo{person}{Wen-Lian Hsu}.} \bibinfo{year}{2015}\natexlab{}.
\newblock \showarticletitle{Linguistic Template Extraction for Recognizing
  Reader-Emotion and Emotional Resonance Writing Assistance}. In
  \bibinfo{booktitle}{\emph{Proceedings of the 53rd Annual Meeting of the
  Association for Computational Linguistics and the 7th International Joint
  Conference on Natural Language Processing (Volume 2: Short Papers)}}.
  \bibinfo{publisher}{Association for Computational Linguistics},
  \bibinfo{address}{Beijing, China}, \bibinfo{pages}{775--780}.
\newblock


\bibitem[Chen and Nath(2008)]%
        {chen2008aperspective}
\bibfield{author}{\bibinfo{person}{Leida Chen} {and} \bibinfo{person}{Ravi
  Nath}.} \bibinfo{year}{2008}\natexlab{}.
\newblock \showarticletitle{A Socio-Technical Perspective of Mobile Work}.
\newblock \bibinfo{journal}{\emph{Inf. Knowl. Syst. Manag.}}
  \bibinfo{volume}{7}, \bibinfo{number}{1,2} (\bibinfo{date}{apr}
  \bibinfo{year}{2008}), \bibinfo{pages}{41–60}.
\newblock
\showISSN{1389-1995}


\bibitem[Chen et~al\mbox{.}(2021)]%
        {chen2021codex}
\bibfield{author}{\bibinfo{person}{Mark Chen}, \bibinfo{person}{Jerry Tworek},
  \bibinfo{person}{Heewoo Jun}, \bibinfo{person}{Qiming Yuan},
  \bibinfo{person}{Henrique~Ponde de Oliveira~Pinto}, \bibinfo{person}{Jared
  Kaplan}, \bibinfo{person}{Harri Edwards}, \bibinfo{person}{Yuri Burda},
  \bibinfo{person}{Nicholas Joseph}, \bibinfo{person}{Greg Brockman},
  \bibinfo{person}{Alex Ray}, \bibinfo{person}{Raul Puri},
  \bibinfo{person}{Gretchen Krueger}, \bibinfo{person}{Michael Petrov},
  \bibinfo{person}{Heidy Khlaaf}, \bibinfo{person}{Girish Sastry},
  \bibinfo{person}{Pamela Mishkin}, \bibinfo{person}{Brooke Chan},
  \bibinfo{person}{Scott Gray}, \bibinfo{person}{Nick Ryder},
  \bibinfo{person}{Mikhail Pavlov}, \bibinfo{person}{Alethea Power},
  \bibinfo{person}{Lukasz Kaiser}, \bibinfo{person}{Mohammad Bavarian},
  \bibinfo{person}{Clemens Winter}, \bibinfo{person}{Philippe Tillet},
  \bibinfo{person}{Felipe~Petroski Such}, \bibinfo{person}{Dave Cummings},
  \bibinfo{person}{Matthias Plappert}, \bibinfo{person}{Fotios Chantzis},
  \bibinfo{person}{Elizabeth Barnes}, \bibinfo{person}{Ariel Herbert-Voss},
  \bibinfo{person}{William~Hebgen Guss}, \bibinfo{person}{Alex Nichol},
  \bibinfo{person}{Alex Paino}, \bibinfo{person}{Nikolas Tezak},
  \bibinfo{person}{Jie Tang}, \bibinfo{person}{Igor Babuschkin},
  \bibinfo{person}{Suchir Balaji}, \bibinfo{person}{Shantanu Jain},
  \bibinfo{person}{William Saunders}, \bibinfo{person}{Christopher Hesse},
  \bibinfo{person}{Andrew~N. Carr}, \bibinfo{person}{Jan Leike},
  \bibinfo{person}{Josh Achiam}, \bibinfo{person}{Vedant Misra},
  \bibinfo{person}{Evan Morikawa}, \bibinfo{person}{Alec Radford},
  \bibinfo{person}{Matthew Knight}, \bibinfo{person}{Miles Brundage},
  \bibinfo{person}{Mira Murati}, \bibinfo{person}{Katie Mayer},
  \bibinfo{person}{Peter Welinder}, \bibinfo{person}{Bob McGrew},
  \bibinfo{person}{Dario Amodei}, \bibinfo{person}{Sam McCandlish},
  \bibinfo{person}{Ilya Sutskever}, {and} \bibinfo{person}{Wojciech Zaremba}.}
  \bibinfo{year}{2021}\natexlab{}.
\newblock \showarticletitle{Evaluating Large Language Models Trained on Code}.
\newblock \bibinfo{journal}{\emph{arXiv preprint arXiv:2107.03374}}
  (\bibinfo{year}{2021}).
\newblock


\bibitem[Chen et~al\mbox{.}(2019)]%
        {Chen2019smartcompose}
\bibfield{author}{\bibinfo{person}{Mia~Xu Chen}, \bibinfo{person}{Benjamin~N.
  Lee}, \bibinfo{person}{Gagan Bansal}, \bibinfo{person}{Yuan Cao},
  \bibinfo{person}{Shuyuan Zhang}, \bibinfo{person}{Justin Lu},
  \bibinfo{person}{Jackie Tsay}, \bibinfo{person}{Yinan Wang},
  \bibinfo{person}{Andrew~M. Dai}, \bibinfo{person}{Zhifeng Chen},
  \bibinfo{person}{Timothy Sohn}, {and} \bibinfo{person}{Yonghui Wu}.}
  \bibinfo{year}{2019}\natexlab{}.
\newblock \showarticletitle{Gmail Smart Compose: Real-Time Assisted Writing}.
  In \bibinfo{booktitle}{\emph{Proceedings of the 25th ACM SIGKDD International
  Conference on Knowledge Discovery \& Data Mining}} (Anchorage, AK, USA)
  \emph{(\bibinfo{series}{KDD ’19})}. \bibinfo{publisher}{Association for
  Computing Machinery}, \bibinfo{address}{New York, NY, USA},
  \bibinfo{pages}{2287–2295}.
\newblock
\showISBNx{9781450362016}


\bibitem[Cho et~al\mbox{.}(2014)]%
        {cho2014gru}
\bibfield{author}{\bibinfo{person}{Kyunghyun Cho}, \bibinfo{person}{Bart van
  Merri{\"e}nboer}, \bibinfo{person}{Dzmitry Bahdanau}, {and}
  \bibinfo{person}{Yoshua Bengio}.} \bibinfo{year}{2014}\natexlab{}.
\newblock \showarticletitle{On the properties of neural machine translation:
  Encoder-decoder approaches}.
\newblock \bibinfo{journal}{\emph{arXiv preprint arXiv:1409.1259}}
  (\bibinfo{year}{2014}).
\newblock


\bibitem[Chollampatt et~al\mbox{.}(2016)]%
        {chollampatt2016adapting}
\bibfield{author}{\bibinfo{person}{Shamil Chollampatt},
  \bibinfo{person}{Duc~Tam Hoang}, {and} \bibinfo{person}{Hwee~Tou Ng}.}
  \bibinfo{year}{2016}\natexlab{}.
\newblock \showarticletitle{Adapting Grammatical Error Correction Based on the
  Native Language of Writers with Neural Network Joint Models}. In
  \bibinfo{booktitle}{\emph{Proceedings of the 2016 Conference on Empirical
  Methods in Natural Language Processing}}. \bibinfo{publisher}{Association for
  Computational Linguistics}, \bibinfo{address}{Austin, Texas},
  \bibinfo{pages}{1901--1911}.
\newblock


\bibitem[Chung et~al\mbox{.}(2021)]%
        {chung2021intersection}
\bibfield{author}{\bibinfo{person}{John Joon~Young Chung},
  \bibinfo{person}{Shiqing He}, {and} \bibinfo{person}{Eytan Adar}.}
  \bibinfo{year}{2021}\natexlab{}.
\newblock \showarticletitle{The intersection of users, roles, interactions, and
  technologies in creativity support tools}. In
  \bibinfo{booktitle}{\emph{Designing Interactive Systems Conference 2021}}.
  \bibinfo{pages}{1817--1833}.
\newblock


\bibitem[Chung et~al\mbox{.}(2022)]%
        {Chung2022talebrush}
\bibfield{author}{\bibinfo{person}{John Joon~Young Chung},
  \bibinfo{person}{Wooseok Kim}, \bibinfo{person}{Kang~Min Yoo},
  \bibinfo{person}{Hwaran Lee}, \bibinfo{person}{Eytan Adar}, {and}
  \bibinfo{person}{Minsuk Chang}.} \bibinfo{year}{2022}\natexlab{}.
\newblock \showarticletitle{TaleBrush: Sketching Stories with Generative
  Pretrained Language Models}. In \bibinfo{booktitle}{\emph{Proceedings of the
  2022 CHI Conference on Human Factors in Computing Systems}} (New Orleans, LA,
  USA) \emph{(\bibinfo{series}{CHI '22})}. \bibinfo{publisher}{Association for
  Computing Machinery}, \bibinfo{address}{New York, NY, USA}, Article
  \bibinfo{articleno}{209}, \bibinfo{numpages}{19}~pages.
\newblock
\showISBNx{9781450391573}


\bibitem[Clark et~al\mbox{.}(2021)]%
        {clark-etal-2021-thats}
\bibfield{author}{\bibinfo{person}{Elizabeth Clark}, \bibinfo{person}{Tal
  August}, \bibinfo{person}{Sofia Serrano}, \bibinfo{person}{Nikita Haduong},
  \bibinfo{person}{Suchin Gururangan}, {and} \bibinfo{person}{Noah~A. Smith}.}
  \bibinfo{year}{2021}\natexlab{}.
\newblock \showarticletitle{All That{'}s {`}Human{'} Is Not Gold: Evaluating
  Human Evaluation of Generated Text}. In \bibinfo{booktitle}{\emph{Proceedings
  of the 59th Annual Meeting of the Association for Computational Linguistics
  and the 11th International Joint Conference on Natural Language Processing
  (Volume 1: Long Papers)}}, \bibfield{editor}{\bibinfo{person}{Chengqing
  Zong}, \bibinfo{person}{Fei Xia}, \bibinfo{person}{Wenjie Li}, {and}
  \bibinfo{person}{Roberto Navigli}} (Eds.). \bibinfo{publisher}{Association
  for Computational Linguistics}, \bibinfo{address}{Online},
  \bibinfo{pages}{7282--7296}.
\newblock


\bibitem[Clark et~al\mbox{.}(2018)]%
        {clark2018creative}
\bibfield{author}{\bibinfo{person}{Elizabeth Clark},
  \bibinfo{person}{Anne~Spencer Ross}, \bibinfo{person}{Chenhao Tan},
  \bibinfo{person}{Yangfeng Ji}, {and} \bibinfo{person}{Noah~A. Smith}.}
  \bibinfo{year}{2018}\natexlab{}.
\newblock \showarticletitle{Creative Writing with a Machine in the Loop: Case
  Studies on Slogans and Stories}. In \bibinfo{booktitle}{\emph{23rd
  International Conference on Intelligent User Interfaces}} (Tokyo, Japan)
  \emph{(\bibinfo{series}{IUI '18})}. \bibinfo{publisher}{Association for
  Computing Machinery}, \bibinfo{address}{New York, NY, USA},
  \bibinfo{pages}{329–340}.
\newblock
\showISBNx{9781450349451}


\bibitem[Clark and Smith(2021a)]%
        {clark2021choose}
\bibfield{author}{\bibinfo{person}{Elizabeth Clark} {and}
  \bibinfo{person}{Noah~A. Smith}.} \bibinfo{year}{2021}\natexlab{a}.
\newblock \showarticletitle{Choose Your Own Adventure: Paired Suggestions in
  Collaborative Writing for Evaluating Story Generation Models}. In
  \bibinfo{booktitle}{\emph{Proceedings of the 2021 Conference of the North
  American Chapter of the Association for Computational Linguistics: Human
  Language Technologies}}. \bibinfo{publisher}{Association for Computational
  Linguistics}, \bibinfo{address}{Online}, \bibinfo{pages}{3566--3575}.
\newblock


\bibitem[Clark and Smith(2021b)]%
        {clarksmith2021choose}
\bibfield{author}{\bibinfo{person}{Elizabeth Clark} {and}
  \bibinfo{person}{Noah~A. Smith}.} \bibinfo{year}{2021}\natexlab{b}.
\newblock \showarticletitle{Choose Your Own Adventure: Paired Suggestions in
  Collaborative Writing for Evaluating Story Generation Models}. In
  \bibinfo{booktitle}{\emph{Association for Computational Linguistics (ACL)}}.
\newblock


\bibitem[Collins and Brown(1988)]%
        {collins1988thecomputer}
\bibfield{author}{\bibinfo{person}{Allan Collins} {and}
  \bibinfo{person}{John~Seely Brown}.} \bibinfo{year}{1988}\natexlab{}.
\newblock \bibinfo{booktitle}{\emph{The Computer as a Tool for Learning Through
  Reflection}}.
\newblock \bibinfo{publisher}{Springer US}, \bibinfo{address}{New York, NY},
  \bibinfo{pages}{1--18}.
\newblock
\showISBNx{978-1-4684-6350-7}


\bibitem[Collobert and Weston(2008)]%
        {collobert2008unified}
\bibfield{author}{\bibinfo{person}{Ronan Collobert} {and}
  \bibinfo{person}{Jason Weston}.} \bibinfo{year}{2008}\natexlab{}.
\newblock \showarticletitle{A unified architecture for natural language
  processing: Deep neural networks with multitask learning}. In
  \bibinfo{booktitle}{\emph{International Conference on Machine Learning
  (ICML)}}. \bibinfo{pages}{160--167}.
\newblock


\bibitem[Cotos et~al\mbox{.}(2020)]%
        {cotos2020understanding}
\bibfield{author}{\bibinfo{person}{Elena Cotos}, \bibinfo{person}{Sarah
  Huffman}, {and} \bibinfo{person}{Stephanie Link}.}
  \bibinfo{year}{2020}\natexlab{}.
\newblock \showarticletitle{Understanding graduate writers’ interaction with
  and impact of the Research Writing Tutor during revision}.
\newblock \bibinfo{journal}{\emph{Journal of Writing Research}}
  \bibinfo{volume}{12}, \bibinfo{number}{1} (\bibinfo{year}{2020}),
  \bibinfo{pages}{187--232}.
\newblock


\bibitem[Council(2023)]%
        {eu2023aiact}
\bibfield{author}{\bibinfo{person}{European Council}.}
  \bibinfo{year}{2023}\natexlab{}.
\newblock \bibinfo{title}{Artificial intelligence act: Council and Parliament
  strike a deal on the first rules for AI in the world}.
\newblock
\newblock
\newblock
\shownote{Accessed: Dec 12, 2023}.


\bibitem[da~Cunha et~al\mbox{.}(2017)]%
        {dacunha2017artext}
\bibfield{author}{\bibinfo{person}{Iria da Cunha}, \bibinfo{person}{M.~Amor
  Montan{\'e}}, {and} \bibinfo{person}{Luis Hysa}.}
  \bibinfo{year}{2017}\natexlab{}.
\newblock \showarticletitle{The ar{T}ext prototype: An automatic system for
  writing specialized texts}. In \bibinfo{booktitle}{\emph{Proceedings of the
  Software Demonstrations of the 15th Conference of the {E}uropean Chapter of
  the Association for Computational Linguistics}}.
  \bibinfo{publisher}{Association for Computational Linguistics},
  \bibinfo{address}{Valencia, Spain}, \bibinfo{pages}{57--60}.
\newblock


\bibitem[D'Agostino(2023)]%
        {dagostino2023chatgpt}
\bibfield{author}{\bibinfo{person}{Susan D'Agostino}.}
  \bibinfo{year}{2023}\natexlab{}.
\newblock \bibinfo{title}{{ChatGPT} Advice Academics Can Use Now}.
\newblock
\newblock
\newblock
\shownote{Accessed: Jan 26, 2024}.


\bibitem[Dai et~al\mbox{.}(2014)]%
        {dai2014wings}
\bibfield{author}{\bibinfo{person}{Xianjun Dai}, \bibinfo{person}{Yuanchao
  Liu}, \bibinfo{person}{Xiaolong Wang}, {and} \bibinfo{person}{Bingquan Liu}.}
  \bibinfo{year}{2014}\natexlab{}.
\newblock \showarticletitle{{WINGS}:Writing with Intelligent Guidance and
  Suggestions}. In \bibinfo{booktitle}{\emph{Proceedings of 52nd Annual Meeting
  of the Association for Computational Linguistics: System Demonstrations}}.
  \bibinfo{publisher}{Association for Computational Linguistics},
  \bibinfo{address}{Baltimore, Maryland}, \bibinfo{pages}{25--30}.
\newblock


\bibitem[Dang et~al\mbox{.}(2022)]%
        {Dang2022beyond}
\bibfield{author}{\bibinfo{person}{Hai Dang}, \bibinfo{person}{Karim
  Benharrak}, \bibinfo{person}{Florian Lehmann}, {and} \bibinfo{person}{Daniel
  Buschek}.} \bibinfo{year}{2022}\natexlab{}.
\newblock \showarticletitle{Beyond Text Generation: Supporting Writers with
  Continuous Automatic Text Summaries}. In
  \bibinfo{booktitle}{\emph{Proceedings of the 35th Annual ACM Symposium on
  User Interface Software and Technology}} (Bend, OR, USA)
  \emph{(\bibinfo{series}{UIST '22})}. \bibinfo{publisher}{Association for
  Computing Machinery}, \bibinfo{address}{New York, NY, USA}, Article
  \bibinfo{articleno}{98}, \bibinfo{numpages}{13}~pages.
\newblock
\showISBNx{9781450393201}


\bibitem[Dang et~al\mbox{.}(2023)]%
        {dang2023choice}
\bibfield{author}{\bibinfo{person}{Hai Dang}, \bibinfo{person}{Sven Goller},
  \bibinfo{person}{Florian Lehmann}, {and} \bibinfo{person}{Daniel Buschek}.}
  \bibinfo{year}{2023}\natexlab{}.
\newblock \showarticletitle{Choice Over Control: How Users Write with Large
  Language Models Using Diegetic and Non-Diegetic Prompting}. In
  \bibinfo{booktitle}{\emph{Proceedings of the 2023 CHI Conference on Human
  Factors in Computing Systems}} (Hamburg, Germany) \emph{(\bibinfo{series}{CHI
  '23})}. \bibinfo{publisher}{Association for Computing Machinery},
  \bibinfo{address}{New York, NY, USA}, Article \bibinfo{articleno}{408},
  \bibinfo{numpages}{17}~pages.
\newblock
\showISBNx{9781450394215}


\bibitem[D'Arcy et~al\mbox{.}(2023)]%
        {darcy2023aries}
\bibfield{author}{\bibinfo{person}{Mike D'Arcy}, \bibinfo{person}{Alexis Ross},
  \bibinfo{person}{Erin Bransom}, \bibinfo{person}{Bailey Kuehl},
  \bibinfo{person}{Jonathan Bragg}, \bibinfo{person}{Tom Hope}, {and}
  \bibinfo{person}{Doug Downey}.} \bibinfo{year}{2023}\natexlab{}.
\newblock \bibinfo{title}{ARIES: A Corpus of Scientific Paper Edits Made in
  Response to Peer Reviews}.
\newblock
\newblock
\showeprint[arxiv]{2306.12587}~[cs.CL]


\bibitem[Dell'Acqua et~al\mbox{.}(2023)]%
        {dellacqua2023jagged}
\bibfield{author}{\bibinfo{person}{Fabrizio Dell'Acqua},
  \bibinfo{person}{Edward McFowland}, \bibinfo{person}{Ethan~R. Mollick},
  \bibinfo{person}{Hila Lifshitz-Assaf}, \bibinfo{person}{Katherine Kellogg},
  \bibinfo{person}{Saran Rajendran}, \bibinfo{person}{Lisa Krayer},
  \bibinfo{person}{François Candelon}, {and} \bibinfo{person}{Karim~R.
  Lakhani}.} \bibinfo{year}{2023}\natexlab{}.
\newblock \bibinfo{booktitle}{\emph{Navigating the Jagged Technological
  Frontier: Field Experimental Evidence of the Effects of AI on Knowledge
  Worker Productivity and Quality}}.
\newblock \bibinfo{type}{Report}. \bibinfo{institution}{Harvard Business
  School}.
\newblock


\bibitem[Devlin et~al\mbox{.}(2019)]%
        {devlin2019bert}
\bibfield{author}{\bibinfo{person}{Jacob Devlin}, \bibinfo{person}{Ming-Wei
  Chang}, \bibinfo{person}{Kenton Lee}, {and} \bibinfo{person}{Kristina
  Toutanova}.} \bibinfo{year}{2019}\natexlab{}.
\newblock \showarticletitle{{BERT}: Pre-training of Deep Bidirectional
  Transformers for Language Understanding}. In
  \bibinfo{booktitle}{\emph{Association for Computational Linguistics (ACL)}}.
  \bibinfo{pages}{4171--4186}.
\newblock


\bibitem[Donahue et~al\mbox{.}(2020)]%
        {donahue2020infilling}
\bibfield{author}{\bibinfo{person}{Chris Donahue}, \bibinfo{person}{Mina Lee},
  {and} \bibinfo{person}{Percy Liang}.} \bibinfo{year}{2020}\natexlab{}.
\newblock \showarticletitle{Enabling Language Models to Fill in the Blanks}. In
  \bibinfo{booktitle}{\emph{Association for Computational Linguistics (ACL)}}.
\newblock


\bibitem[Dong et~al\mbox{.}(2012a)]%
        {dong2012first}
\bibfield{author}{\bibinfo{person}{Ruihai Dong}, \bibinfo{person}{Kevin
  McCarthy}, \bibinfo{person}{Michael O'Mahony}, \bibinfo{person}{Markus
  Schaal}, {and} \bibinfo{person}{Barry Smyth}.}
  \bibinfo{year}{2012}\natexlab{a}.
\newblock \showarticletitle{First Demonstration of the Intelligent Reviewer's
  Assistant}. In \bibinfo{booktitle}{\emph{Proceedings of the 2012 ACM
  International Conference on Intelligent User Interfaces}} (Lisbon, Portugal)
  \emph{(\bibinfo{series}{IUI '12})}. \bibinfo{publisher}{Association for
  Computing Machinery}, \bibinfo{address}{New York, NY, USA},
  \bibinfo{pages}{337–338}.
\newblock
\showISBNx{9781450310482}


\bibitem[Dong et~al\mbox{.}(2012b)]%
        {dong2012towards}
\bibfield{author}{\bibinfo{person}{Ruihai Dong}, \bibinfo{person}{Kevin
  McCarthy}, \bibinfo{person}{Michael O'Mahony}, \bibinfo{person}{Markus
  Schaal}, {and} \bibinfo{person}{Barry Smyth}.}
  \bibinfo{year}{2012}\natexlab{b}.
\newblock \showarticletitle{Towards an Intelligent Reviewer's Assistant:
  Recommending Topics to Help Users to Write Better Product Reviews}. In
  \bibinfo{booktitle}{\emph{Proceedings of the 2012 ACM International
  Conference on Intelligent User Interfaces}} (Lisbon, Portugal)
  \emph{(\bibinfo{series}{IUI '12})}. \bibinfo{publisher}{Association for
  Computing Machinery}, \bibinfo{address}{New York, NY, USA},
  \bibinfo{pages}{159–168}.
\newblock
\showISBNx{9781450310482}


\bibitem[Du et~al\mbox{.}(2022a)]%
        {du-etal-2022-read}
\bibfield{author}{\bibinfo{person}{Wanyu Du}, \bibinfo{person}{Zae~Myung Kim},
  \bibinfo{person}{Vipul Raheja}, \bibinfo{person}{Dhruv Kumar}, {and}
  \bibinfo{person}{Dongyeop Kang}.} \bibinfo{year}{2022}\natexlab{a}.
\newblock \showarticletitle{Read, Revise, Repeat: A System Demonstration for
  Human-in-the-loop Iterative Text Revision}. In
  \bibinfo{booktitle}{\emph{Proceedings of the First Workshop on Intelligent
  and Interactive Writing Assistants (In2Writing 2022)}}.
  \bibinfo{publisher}{Association for Computational Linguistics},
  \bibinfo{address}{Dublin, Ireland}, \bibinfo{pages}{96--108}.
\newblock


\bibitem[Du et~al\mbox{.}(2022b)]%
        {du-etal-2022-understanding-iterative}
\bibfield{author}{\bibinfo{person}{Wanyu Du}, \bibinfo{person}{Vipul Raheja},
  \bibinfo{person}{Dhruv Kumar}, \bibinfo{person}{Zae~Myung Kim},
  \bibinfo{person}{Melissa Lopez}, {and} \bibinfo{person}{Dongyeop Kang}.}
  \bibinfo{year}{2022}\natexlab{b}.
\newblock \showarticletitle{Understanding Iterative Revision from Human-Written
  Text}. In \bibinfo{booktitle}{\emph{Proceedings of the 60th Annual Meeting of
  the Association for Computational Linguistics (Volume 1: Long Papers)}}.
  \bibinfo{publisher}{Association for Computational Linguistics},
  \bibinfo{address}{Dublin, Ireland}, \bibinfo{pages}{3573--3590}.
\newblock


\bibitem[Dunlop and Levine(2012)]%
        {Dunlop2012multidimensional}
\bibfield{author}{\bibinfo{person}{Mark Dunlop} {and} \bibinfo{person}{John
  Levine}.} \bibinfo{year}{2012}\natexlab{}.
\newblock \showarticletitle{Multidimensional Pareto Optimization of Touchscreen
  Keyboards for Speed, Familiarity and Improved Spell Checking}. In
  \bibinfo{booktitle}{\emph{Proceedings of the SIGCHI Conference on Human
  Factors in Computing Systems}} (Austin, Texas, USA)
  \emph{(\bibinfo{series}{CHI ’12})}. \bibinfo{publisher}{Association for
  Computing Machinery}, \bibinfo{address}{New York, NY, USA},
  \bibinfo{pages}{2669–2678}.
\newblock
\showISBNx{9781450310154}


\bibitem[Duval et~al\mbox{.}(2021)]%
        {duval2021breaking}
\bibfield{author}{\bibinfo{person}{Alexandre Duval}, \bibinfo{person}{Thomas
  Lamson}, \bibinfo{person}{Ga{\"e}l de~L{\'e}s{\'e}leuc~de K{\'e}rouara},
  {and} \bibinfo{person}{Matthias Gall{\'e}}.} \bibinfo{year}{2021}\natexlab{}.
\newblock \showarticletitle{Breaking Writer{'}s Block: Low-cost Fine-tuning of
  Natural Language Generation Models}. In \bibinfo{booktitle}{\emph{Proceedings
  of the 16th Conference of the European Chapter of the Association for
  Computational Linguistics: System Demonstrations}}.
  \bibinfo{publisher}{Association for Computational Linguistics},
  \bibinfo{address}{Online}, \bibinfo{pages}{278--287}.
\newblock


\bibitem[Eloundou et~al\mbox{.}(2023)]%
        {eloundou2023gpts}
\bibfield{author}{\bibinfo{person}{Tyna Eloundou}, \bibinfo{person}{Sam
  Manning}, \bibinfo{person}{Pamela Mishkin}, {and} \bibinfo{person}{Daniel
  Rock}.} \bibinfo{year}{2023}\natexlab{}.
\newblock \bibinfo{title}{GPTs are GPTs: An Early Look at the Labor Market
  Impact Potential of Large Language Models}.
\newblock
\newblock
\showeprint[arxiv]{2303.10130}~[econ.GN]


\bibitem[Faltings et~al\mbox{.}(2021)]%
        {faltings-etal-2021-text}
\bibfield{author}{\bibinfo{person}{Felix Faltings}, \bibinfo{person}{Michel
  Galley}, \bibinfo{person}{Gerold Hintz}, \bibinfo{person}{Chris Brockett},
  \bibinfo{person}{Chris Quirk}, \bibinfo{person}{Jianfeng Gao}, {and}
  \bibinfo{person}{Bill Dolan}.} \bibinfo{year}{2021}\natexlab{}.
\newblock \showarticletitle{Text Editing by Command}. In
  \bibinfo{booktitle}{\emph{Proceedings of the 2021 Conference of the North
  American Chapter of the Association for Computational Linguistics: Human
  Language Technologies}}. \bibinfo{publisher}{Association for Computational
  Linguistics}, \bibinfo{address}{Online}, \bibinfo{pages}{5259--5274}.
\newblock


\bibitem[Fan et~al\mbox{.}(2019)]%
        {fan2019character}
\bibfield{author}{\bibinfo{person}{Min Fan}, \bibinfo{person}{Jianyu Fan},
  \bibinfo{person}{Alissa~N. Antle}, \bibinfo{person}{Sheng Jin},
  \bibinfo{person}{Dongxu Yin}, {and} \bibinfo{person}{Philippe Pasquier}.}
  \bibinfo{year}{2019}\natexlab{}.
\newblock \showarticletitle{Character Alive: A Tangible Reading and Writing
  System for Chinese Children At-Risk for Dyslexia}. In
  \bibinfo{booktitle}{\emph{Extended Abstracts of the 2019 CHI Conference on
  Human Factors in Computing Systems}} (Glasgow, Scotland Uk)
  \emph{(\bibinfo{series}{CHI EA '19})}. \bibinfo{publisher}{Association for
  Computing Machinery}, \bibinfo{address}{New York, NY, USA},
  \bibinfo{pages}{1–6}.
\newblock
\showISBNx{9781450359719}


\bibitem[Faruqui et~al\mbox{.}(2018)]%
        {faruqui-etal-2018-wikiatomicedits}
\bibfield{author}{\bibinfo{person}{Manaal Faruqui}, \bibinfo{person}{Ellie
  Pavlick}, \bibinfo{person}{Ian Tenney}, {and} \bibinfo{person}{Dipanjan
  Das}.} \bibinfo{year}{2018}\natexlab{}.
\newblock \showarticletitle{{W}iki{A}tomic{E}dits: A Multilingual Corpus of
  {W}ikipedia Edits for Modeling Language and Discourse}. In
  \bibinfo{booktitle}{\emph{Proceedings of the 2018 Conference on Empirical
  Methods in Natural Language Processing}},
  \bibfield{editor}{\bibinfo{person}{Ellen Riloff}, \bibinfo{person}{David
  Chiang}, \bibinfo{person}{Julia Hockenmaier}, {and}
  \bibinfo{person}{Jun{'}ichi Tsujii}} (Eds.). \bibinfo{publisher}{Association
  for Computational Linguistics}, \bibinfo{address}{Brussels, Belgium},
  \bibinfo{pages}{305--315}.
\newblock


\bibitem[Flower and Hayes(1981)]%
        {flower1981cognitive}
\bibfield{author}{\bibinfo{person}{Linda Flower} {and} \bibinfo{person}{John~R.
  Hayes}.} \bibinfo{year}{1981}\natexlab{}.
\newblock \showarticletitle{A Cognitive Process Theory of Writing}.
\newblock \bibinfo{journal}{\emph{College Composition and Communication}}
  \bibinfo{volume}{32}, \bibinfo{number}{4} (\bibinfo{year}{1981}),
  \bibinfo{pages}{365--387}.
\newblock
\showISSN{0010096X}


\bibitem[for Teaching Exellence at The University~of Kansas(2023)]%
        {ku2023ai}
\bibfield{author}{\bibinfo{person}{Center for Teaching Exellence at The
  University~of Kansas}.} \bibinfo{year}{2023}\natexlab{}.
\newblock \bibinfo{title}{Using AI ethically in writing assignments}.
\newblock
\newblock
\newblock
\shownote{Accessed: Jan 26, 2024}.


\bibitem[Forbes(2023)]%
        {forbes2023openai}
\bibfield{author}{\bibinfo{person}{Forbes}.} \bibinfo{year}{2023}\natexlab{}.
\newblock \bibinfo{title}{OpenAI, Microsoft hit with new US consumer privacy
  class action}.
\newblock
\newblock
\newblock
\shownote{Accessed: Dec 12, 2023}.


\bibitem[Fowler et~al\mbox{.}(2015)]%
        {Fowler2015effects}
\bibfield{author}{\bibinfo{person}{Andrew Fowler}, \bibinfo{person}{Kurt
  Partridge}, \bibinfo{person}{Ciprian Chelba}, \bibinfo{person}{Xiaojun Bi},
  \bibinfo{person}{Tom Ouyang}, {and} \bibinfo{person}{Shumin Zhai}.}
  \bibinfo{year}{2015}\natexlab{}.
\newblock \showarticletitle{Effects of Language Modeling and Its
  Personalization on Touchscreen Typing Performance}. In
  \bibinfo{booktitle}{\emph{Proceedings of the 33rd Annual ACM Conference on
  Human Factors in Computing Systems}} (Seoul, Republic of Korea)
  \emph{(\bibinfo{series}{CHI '15})}. \bibinfo{publisher}{Association for
  Computing Machinery}, \bibinfo{address}{New York, NY, USA},
  \bibinfo{pages}{649–658}.
\newblock
\showISBNx{9781450331456}


\bibitem[Fran{\c{c}}ois et~al\mbox{.}(2020)]%
        {francois2020amesure}
\bibfield{author}{\bibinfo{person}{Thomas Fran{\c{c}}ois},
  \bibinfo{person}{Adeline M{\"u}ller}, \bibinfo{person}{Eva Rolin}, {and}
  \bibinfo{person}{Magali Norr{\'e}}.} \bibinfo{year}{2020}\natexlab{}.
\newblock \showarticletitle{{AM}esure: A Web Platform to Assist the Clear
  Writing of Administrative Texts}. In \bibinfo{booktitle}{\emph{Proceedings of
  the 1st Conference of the Asia-Pacific Chapter of the Association for
  Computational Linguistics and the 10th International Joint Conference on
  Natural Language Processing: System Demonstrations}}.
  \bibinfo{publisher}{Association for Computational Linguistics},
  \bibinfo{address}{Suzhou, China}, \bibinfo{pages}{1--7}.
\newblock


\bibitem[Frase(1983)]%
        {frase1983unix}
\bibfield{author}{\bibinfo{person}{L.~T. Frase}.}
  \bibinfo{year}{1983}\natexlab{}.
\newblock \showarticletitle{Human factors and behavioral science: The UNIX™
  Writer'S workbench software: Philosophy}.
\newblock \bibinfo{journal}{\emph{The Bell System Technical Journal}}
  \bibinfo{volume}{62}, \bibinfo{number}{6} (\bibinfo{year}{1983}),
  \bibinfo{pages}{1883--1890}.
\newblock


\bibitem[Frich et~al\mbox{.}(2019)]%
        {frich2019mapping}
\bibfield{author}{\bibinfo{person}{Jonas Frich}, \bibinfo{person}{Lindsay
  MacDonald~Vermeulen}, \bibinfo{person}{Christian Remy},
  \bibinfo{person}{Michael~Mose Biskjaer}, {and} \bibinfo{person}{Peter
  Dalsgaard}.} \bibinfo{year}{2019}\natexlab{}.
\newblock \showarticletitle{Mapping the Landscape of Creativity Support Tools
  in HCI}. In \bibinfo{booktitle}{\emph{Proceedings of the 2019 CHI Conference
  on Human Factors in Computing Systems}} (Glasgow, Scotland Uk)
  \emph{(\bibinfo{series}{CHI '19})}. \bibinfo{publisher}{Association for
  Computing Machinery}, \bibinfo{address}{New York, NY, USA},
  \bibinfo{pages}{1–18}.
\newblock
\showISBNx{9781450359702}


\bibitem[Gabriel et~al\mbox{.}(2015)]%
        {gabriel2015inkwell}
\bibfield{author}{\bibinfo{person}{Richard~P. Gabriel}, \bibinfo{person}{Jilin
  Chen}, {and} \bibinfo{person}{Jeffrey Nichols}.}
  \bibinfo{year}{2015}\natexlab{}.
\newblock \showarticletitle{InkWell: A Creative Writer's Creative Assistant}.
  In \bibinfo{booktitle}{\emph{Proceedings of the 2015 ACM SIGCHI Conference on
  Creativity and Cognition}} (Glasgow, United Kingdom)
  \emph{(\bibinfo{series}{C\&C '15})}. \bibinfo{publisher}{Association for
  Computing Machinery}, \bibinfo{address}{New York, NY, USA},
  \bibinfo{pages}{93–102}.
\newblock
\showISBNx{9781450335980}


\bibitem[Gamon(2010)]%
        {gamon2010using}
\bibfield{author}{\bibinfo{person}{Michael Gamon}.}
  \bibinfo{year}{2010}\natexlab{}.
\newblock \showarticletitle{Using Mostly Native Data to Correct Errors in
  Learners{'} Writing}. In \bibinfo{booktitle}{\emph{Human Language
  Technologies: The 2010 Annual Conference of the North {A}merican Chapter of
  the Association for Computational Linguistics}}.
  \bibinfo{publisher}{Association for Computational Linguistics},
  \bibinfo{address}{Los Angeles, California}, \bibinfo{pages}{163--171}.
\newblock


\bibitem[Gero et~al\mbox{.}(2022a)]%
        {gero2022design}
\bibfield{author}{\bibinfo{person}{Katy Gero}, \bibinfo{person}{Alex
  Calderwood}, \bibinfo{person}{Charlotte Li}, {and} \bibinfo{person}{Lydia
  Chilton}.} \bibinfo{year}{2022}\natexlab{a}.
\newblock \showarticletitle{A Design Space for Writing Support Tools Using a
  Cognitive Process Model of Writing}. In \bibinfo{booktitle}{\emph{Proceedings
  of the First Workshop on Intelligent and Interactive Writing Assistants
  (In2Writing 2022)}}. \bibinfo{publisher}{Association for Computational
  Linguistics}, \bibinfo{address}{Dublin, Ireland}, \bibinfo{pages}{11--24}.
\newblock


\bibitem[Gero and Chilton(2019a)]%
        {gero2019how}
\bibfield{author}{\bibinfo{person}{Katy~Ilonka Gero} {and}
  \bibinfo{person}{Lydia~B. Chilton}.} \bibinfo{year}{2019}\natexlab{a}.
\newblock \showarticletitle{How a Stylistic, Machine-Generated Thesaurus
  Impacts a Writer's Process}. In \bibinfo{booktitle}{\emph{Proceedings of the
  2019 Conference on Creativity and Cognition}} (San Diego, CA, USA)
  \emph{(\bibinfo{series}{C\&C '19})}. \bibinfo{publisher}{Association for
  Computing Machinery}, \bibinfo{address}{New York, NY, USA},
  \bibinfo{pages}{597–603}.
\newblock
\showISBNx{9781450359177}


\bibitem[Gero and Chilton(2019b)]%
        {Gero2019metaphoria}
\bibfield{author}{\bibinfo{person}{Katy~Ilonka Gero} {and}
  \bibinfo{person}{Lydia~B. Chilton}.} \bibinfo{year}{2019}\natexlab{b}.
\newblock \showarticletitle{Metaphoria: An Algorithmic Companion for Metaphor
  Creation}. In \bibinfo{booktitle}{\emph{Proceedings of the 2019 CHI
  Conference on Human Factors in Computing Systems}} (Glasgow, Scotland Uk)
  \emph{(\bibinfo{series}{CHI '19})}. \bibinfo{publisher}{Association for
  Computing Machinery}, \bibinfo{address}{New York, NY, USA},
  \bibinfo{pages}{1–12}.
\newblock
\showISBNx{9781450359702}


\bibitem[Gero et~al\mbox{.}(2022b)]%
        {Gero2022sparks}
\bibfield{author}{\bibinfo{person}{Katy~Ilonka Gero}, \bibinfo{person}{Vivian
  Liu}, {and} \bibinfo{person}{Lydia Chilton}.}
  \bibinfo{year}{2022}\natexlab{b}.
\newblock \showarticletitle{Sparks: Inspiration for Science Writing Using
  Language Models}. In \bibinfo{booktitle}{\emph{Proceedings of the 2022 ACM
  Designing Interactive Systems Conference}} (Virtual Event, Australia)
  \emph{(\bibinfo{series}{DIS '22})}. \bibinfo{publisher}{Association for
  Computing Machinery}, \bibinfo{address}{New York, NY, USA},
  \bibinfo{pages}{1002–1019}.
\newblock
\showISBNx{9781450393584}


\bibitem[Gero et~al\mbox{.}(2023)]%
        {gero2023social}
\bibfield{author}{\bibinfo{person}{Katy~Ilonka Gero}, \bibinfo{person}{Tao
  Long}, {and} \bibinfo{person}{Lydia~B Chilton}.}
  \bibinfo{year}{2023}\natexlab{}.
\newblock \showarticletitle{Social Dynamics of AI Support in Creative Writing}.
  In \bibinfo{booktitle}{\emph{Proceedings of the 2023 CHI Conference on Human
  Factors in Computing Systems}} (Hamburg, Germany) \emph{(\bibinfo{series}{CHI
  '23})}. \bibinfo{publisher}{Association for Computing Machinery},
  \bibinfo{address}{New York, NY, USA}, Article \bibinfo{articleno}{245},
  \bibinfo{numpages}{15}~pages.
\newblock
\showISBNx{9781450394215}


\bibitem[Gholami et~al\mbox{.}(2022)]%
        {gholami2022survey}
\bibfield{author}{\bibinfo{person}{Amir Gholami}, \bibinfo{person}{Sehoon Kim},
  \bibinfo{person}{Zhen Dong}, \bibinfo{person}{Zhewei Yao},
  \bibinfo{person}{Michael~W Mahoney}, {and} \bibinfo{person}{Kurt Keutzer}.}
  \bibinfo{year}{2022}\natexlab{}.
\newblock \showarticletitle{A Survey of Quantization Methods for Efficient
  Neural Network Inference}.
\newblock In \bibinfo{booktitle}{\emph{Low-Power Computer Vision}}.
  \bibinfo{publisher}{Chapman and Hall/CRC}, \bibinfo{pages}{291--326}.
\newblock


\bibitem[Goldfarb-Tarrant et~al\mbox{.}(2019)]%
        {goldfarb2019plan}
\bibfield{author}{\bibinfo{person}{Seraphina Goldfarb-Tarrant},
  \bibinfo{person}{Haining Feng}, {and} \bibinfo{person}{Nanyun Peng}.}
  \bibinfo{year}{2019}\natexlab{}.
\newblock \showarticletitle{Plan, Write, and Revise: an Interactive System for
  Open-Domain Story Generation}. In \bibinfo{booktitle}{\emph{Proceedings of
  the 2019 Conference of the North {A}merican Chapter of the Association for
  Computational Linguistics (Demonstrations)}}. \bibinfo{publisher}{Association
  for Computational Linguistics}, \bibinfo{address}{Minneapolis, Minnesota},
  \bibinfo{pages}{89--97}.
\newblock


\bibitem[Gon\c{c}alves et~al\mbox{.}(2015)]%
        {goncalves2015you}
\bibfield{author}{\bibinfo{person}{Frederica Gon\c{c}alves},
  \bibinfo{person}{Pedro Campos}, \bibinfo{person}{Julian Hanna}, {and}
  \bibinfo{person}{Simone Ashby}.} \bibinfo{year}{2015}\natexlab{}.
\newblock \showarticletitle{You're the Voice: Evaluating User Interfaces for
  Encouraging Underserved Youths to Express Themselves through Creative
  Writing}. In \bibinfo{booktitle}{\emph{Proceedings of the 2015 ACM SIGCHI
  Conference on Creativity and Cognition}} (Glasgow, United Kingdom)
  \emph{(\bibinfo{series}{C\&C '15})}. \bibinfo{publisher}{Association for
  Computing Machinery}, \bibinfo{address}{New York, NY, USA},
  \bibinfo{pages}{63–72}.
\newblock
\showISBNx{9781450335980}


\bibitem[Gonzales et~al\mbox{.}(2010)]%
        {gonzales2010motivating}
\bibfield{author}{\bibinfo{person}{Amy~L. Gonzales},
  \bibinfo{person}{Tiffany~Y. Ng}, \bibinfo{person}{OJ Zhao}, {and}
  \bibinfo{person}{Geri Gay}.} \bibinfo{year}{2010}\natexlab{}.
\newblock \showarticletitle{Motivating Expressive Writing with a Text-to-Sound
  Application}. In \bibinfo{booktitle}{\emph{Proceedings of the SIGCHI
  Conference on Human Factors in Computing Systems}} (Atlanta, Georgia, USA)
  \emph{(\bibinfo{series}{CHI '10})}. \bibinfo{publisher}{Association for
  Computing Machinery}, \bibinfo{address}{New York, NY, USA},
  \bibinfo{pages}{1937–1940}.
\newblock
\showISBNx{9781605589299}


\bibitem[Gordon et~al\mbox{.}(2016)]%
        {Gordon2016WatchWriter}
\bibfield{author}{\bibinfo{person}{Mitchell Gordon}, \bibinfo{person}{Tom
  Ouyang}, {and} \bibinfo{person}{Shumin Zhai}.}
  \bibinfo{year}{2016}\natexlab{}.
\newblock \showarticletitle{WatchWriter: Tap and Gesture Typing on a Smartwatch
  Miniature Keyboard with Statistical Decoding}. In
  \bibinfo{booktitle}{\emph{Proceedings of the 2016 CHI Conference on Human
  Factors in Computing Systems}} (San Jose, California, USA)
  \emph{(\bibinfo{series}{CHI ’16})}. \bibinfo{publisher}{Association for
  Computing Machinery}, \bibinfo{address}{New York, NY, USA},
  \bibinfo{pages}{3817–3821}.
\newblock
\showISBNx{9781450333627}


\bibitem[Guggenberger et~al\mbox{.}(2020)]%
        {guggenberger2020ecosystem}
\bibfield{author}{\bibinfo{person}{Tobias~Moritz Guggenberger},
  \bibinfo{person}{Frederik M{\"o}ller}, \bibinfo{person}{Tim Haarhaus},
  \bibinfo{person}{Inan G{\"u}r}, {and} \bibinfo{person}{Boris Otto}.}
  \bibinfo{year}{2020}\natexlab{}.
\newblock \showarticletitle{Ecosystem Types in Information Systems}. In
  \bibinfo{booktitle}{\emph{Twenty-Eighth European Conference on Information
  Systems (ECIS2020)}}.
\newblock


\bibitem[Gurevich and Deane(2007)]%
        {gurevich2007document}
\bibfield{author}{\bibinfo{person}{Olga Gurevich} {and} \bibinfo{person}{Paul
  Deane}.} \bibinfo{year}{2007}\natexlab{}.
\newblock \showarticletitle{Document Similarity Measures to Distinguish Native
  vs. Non-Native Essay Writers}. In \bibinfo{booktitle}{\emph{Human Language
  Technologies 2007: The Conference of the North {A}merican Chapter of the
  Association for Computational Linguistics; Companion Volume, Short Papers}}.
  \bibinfo{publisher}{Association for Computational Linguistics},
  \bibinfo{address}{Rochester, New York}, \bibinfo{pages}{49--52}.
\newblock


\bibitem[Haas(2013)]%
        {haas2013writing}
\bibfield{author}{\bibinfo{person}{Christina Haas}.}
  \bibinfo{year}{2013}\natexlab{}.
\newblock \bibinfo{booktitle}{\emph{Writing technology: Studies on the
  materiality of literacy}}.
\newblock \bibinfo{publisher}{Routledge}.
\newblock


\bibitem[Hacker et~al\mbox{.}(2023)]%
        {hacker2023regulatinggpt}
\bibfield{author}{\bibinfo{person}{Philipp Hacker}, \bibinfo{person}{Andreas
  Engel}, {and} \bibinfo{person}{Marco Mauer}.}
  \bibinfo{year}{2023}\natexlab{}.
\newblock \showarticletitle{Regulating ChatGPT and Other Large Generative AI
  Models}.
\newblock  (\bibinfo{year}{2023}), \bibinfo{pages}{1112–1123}.
\newblock
\showISBNx{9798400701924}


\bibitem[Hagiwara et~al\mbox{.}(2019)]%
        {hagiwara2019teaspn}
\bibfield{author}{\bibinfo{person}{Masato Hagiwara}, \bibinfo{person}{Takumi
  Ito}, \bibinfo{person}{Tatsuki Kuribayashi}, \bibinfo{person}{Jun Suzuki},
  {and} \bibinfo{person}{Kentaro Inui}.} \bibinfo{year}{2019}\natexlab{}.
\newblock \showarticletitle{{TEASPN}: Framework and Protocol for Integrated
  Writing Assistance Environments}. In \bibinfo{booktitle}{\emph{Proceedings of
  the 2019 Conference on Empirical Methods in Natural Language Processing and
  the 9th International Joint Conference on Natural Language Processing
  (EMNLP-IJCNLP): System Demonstrations}}. \bibinfo{publisher}{Association for
  Computational Linguistics}, \bibinfo{address}{Hong Kong, China},
  \bibinfo{pages}{229--234}.
\newblock


\bibitem[Halpin et~al\mbox{.}(2004)]%
        {halpin2004automatic}
\bibfield{author}{\bibinfo{person}{Harry Halpin}, \bibinfo{person}{Johanna~D.
  Moore}, {and} \bibinfo{person}{Judy Robertson}.}
  \bibinfo{year}{2004}\natexlab{}.
\newblock \showarticletitle{Automatic Analysis of Plot for Story Rewriting}. In
  \bibinfo{booktitle}{\emph{Proceedings of the 2004 Conference on Empirical
  Methods in Natural Language Processing}}. \bibinfo{publisher}{Association for
  Computational Linguistics}, \bibinfo{address}{Barcelona, Spain},
  \bibinfo{pages}{127--133}.
\newblock


\bibitem[Hanawa et~al\mbox{.}(2021)]%
        {hanawa2021exploring}
\bibfield{author}{\bibinfo{person}{Kazuaki Hanawa}, \bibinfo{person}{Ryo
  Nagata}, {and} \bibinfo{person}{Kentaro Inui}.}
  \bibinfo{year}{2021}\natexlab{}.
\newblock \showarticletitle{Exploring Methods for Generating Feedback Comments
  for Writing Learning}. In \bibinfo{booktitle}{\emph{Proceedings of the 2021
  Conference on Empirical Methods in Natural Language Processing}}.
  \bibinfo{publisher}{Association for Computational Linguistics},
  \bibinfo{address}{Online and Punta Cana, Dominican Republic},
  \bibinfo{pages}{9719--9730}.
\newblock


\bibitem[Hayes(1996)]%
        {hayes1996model}
\bibfield{author}{\bibinfo{person}{J.R. Hayes}.}
  \bibinfo{year}{1996}\natexlab{}.
\newblock \showarticletitle{A new framework for understanding cognition and
  affect in writing}.
\newblock In \bibinfo{booktitle}{\emph{The Science of Writing: Theories,
  Methods, Individual Differences, and Applications}},
  \bibfield{editor}{\bibinfo{person}{C.M. Levy} {and}
  \bibinfo{person}{S.~Randall}} (Eds.). \bibinfo{publisher}{Lawrence Erlbaum
  Associates}, \bibinfo{address}{Mahwah, NJ}, \bibinfo{pages}{6--44}.
\newblock


\bibitem[Hayes(2012)]%
        {hayes2012modeling}
\bibfield{author}{\bibinfo{person}{John~R. Hayes}.}
  \bibinfo{year}{2012}\natexlab{}.
\newblock \showarticletitle{Modeling and Remodeling Writing}.
\newblock \bibinfo{journal}{\emph{Written Communication}} \bibinfo{volume}{29},
  \bibinfo{number}{3} (\bibinfo{year}{2012}), \bibinfo{pages}{369--388}.
\newblock


\bibitem[Hayes and Flower(1986)]%
        {hayes1986writing}
\bibfield{author}{\bibinfo{person}{John~R Hayes} {and} \bibinfo{person}{Linda~S
  Flower}.} \bibinfo{year}{1986}\natexlab{}.
\newblock \showarticletitle{Writing research and the writer}.
\newblock \bibinfo{journal}{\emph{American psychologist}} \bibinfo{volume}{41},
  \bibinfo{number}{10} (\bibinfo{year}{1986}), \bibinfo{pages}{1106--1113}.
\newblock


\bibitem[Higginbotham(1992)]%
        {higginbotham1992evaluation}
\bibfield{author}{\bibinfo{person}{D.~Jeffery Higginbotham}.}
  \bibinfo{year}{1992}\natexlab{}.
\newblock \showarticletitle{Evaluation of keystroke savings across five
  assistive communication technologies}.
\newblock \bibinfo{journal}{\emph{Augmentative and Alternative Communication}}
  \bibinfo{volume}{8}, \bibinfo{number}{4} (\bibinfo{date}{Jan.}
  \bibinfo{year}{1992}), \bibinfo{pages}{258--272}.
\newblock
\showISSN{0743-4618, 1477-3848}


\bibitem[Hill et~al\mbox{.}(2017)]%
        {Hill2017GenderInclusivenessPV}
\bibfield{author}{\bibinfo{person}{Charles Hill}, \bibinfo{person}{Maren Haag},
  \bibinfo{person}{Alannah Oleson}, \bibinfo{person}{Christopher~J. Mendez},
  \bibinfo{person}{Nicola Marsden}, \bibinfo{person}{Anita Sarma}, {and}
  \bibinfo{person}{Margaret~M. Burnett}.} \bibinfo{year}{2017}\natexlab{}.
\newblock \showarticletitle{Gender-Inclusiveness Personas vs. Stereotyping: Can
  We Have it Both Ways?}
\newblock \bibinfo{journal}{\emph{Proceedings of the 2017 CHI Conference on
  Human Factors in Computing Systems}} (\bibinfo{year}{2017}).
\newblock


\bibitem[Hinton et~al\mbox{.}(2015)]%
        {hinton2015distilling}
\bibfield{author}{\bibinfo{person}{Geoffrey Hinton}, \bibinfo{person}{Oriol
  Vinyals}, {and} \bibinfo{person}{Jeffrey Dean}.}
  \bibinfo{year}{2015}\natexlab{}.
\newblock \showarticletitle{Distilling the Knowledge in a Neural Network}. In
  \bibinfo{booktitle}{\emph{NIPS Deep Learning and Representation Learning
  Workshop}}.
\newblock


\bibitem[Hong et~al\mbox{.}(2023)]%
        {hong2023visualwriting}
\bibfield{author}{\bibinfo{person}{Xudong Hong}, \bibinfo{person}{Asad Sayeed},
  \bibinfo{person}{Khushboo Mehra}, \bibinfo{person}{Vera Demberg}, {and}
  \bibinfo{person}{Bernt Schiele}.} \bibinfo{year}{2023}\natexlab{}.
\newblock \showarticletitle{Visual Writing Prompts: Character-Grounded Story
  Generation with Curated Image Sequences}.
\newblock \bibinfo{journal}{\emph{Transactions of the Association for
  Computational Linguistics}}  \bibinfo{volume}{11} (\bibinfo{year}{2023}),
  \bibinfo{pages}{565--581}.
\newblock


\bibitem[Hoque et~al\mbox{.}(2022)]%
        {hoque2022dramatvis}
\bibfield{author}{\bibinfo{person}{Md~Naimul Hoque}, \bibinfo{person}{Bhavya
  Ghai}, {and} \bibinfo{person}{Niklas Elmqvist}.}
  \bibinfo{year}{2022}\natexlab{}.
\newblock \showarticletitle{DramatVis Personae: Visual Text Analytics for
  Identifying Social Biases in Creative Writing}. In
  \bibinfo{booktitle}{\emph{Proceedings of the 2022 ACM Designing Interactive
  Systems Conference}} (Virtual Event, Australia) \emph{(\bibinfo{series}{DIS
  '22})}. \bibinfo{publisher}{Association for Computing Machinery},
  \bibinfo{address}{New York, NY, USA}, \bibinfo{pages}{1260–1276}.
\newblock
\showISBNx{9781450393584}


\bibitem[House(2023)]%
        {whitehouse2023execai}
\bibfield{author}{\bibinfo{person}{The~White House}.}
  \bibinfo{year}{2023}\natexlab{}.
\newblock \bibinfo{title}{Executive Order on the Safe, Secure, and Trustworthy
  Development and Use of Artificial Intelligence}.
\newblock
\newblock
\newblock
\shownote{Accessed: Dec 12, 2023}.


\bibitem[Howe(2009)]%
        {howe2009rita}
\bibfield{author}{\bibinfo{person}{Daniel~C. Howe}.}
  \bibinfo{year}{2009}\natexlab{}.
\newblock \showarticletitle{RiTa: Creativity Support for Computational
  Literature}. In \bibinfo{booktitle}{\emph{Proceedings of the Seventh ACM
  Conference on Creativity and Cognition}} (Berkeley, California, USA)
  \emph{(\bibinfo{series}{C\&C '09})}. \bibinfo{publisher}{Association for
  Computing Machinery}, \bibinfo{address}{New York, NY, USA},
  \bibinfo{pages}{205–210}.
\newblock
\showISBNx{9781605588650}


\bibitem[Hsu et~al\mbox{.}(2019)]%
        {hsu2019on}
\bibfield{author}{\bibinfo{person}{Ting-Yao Hsu}, \bibinfo{person}{Yen-Chia
  Hsu}, {and} \bibinfo{person}{Ting-Hao~(Kenneth) Huang}.}
  \bibinfo{year}{2019}\natexlab{}.
\newblock \showarticletitle{On How Users Edit Computer-Generated Visual
  Stories}. In \bibinfo{booktitle}{\emph{Extended Abstracts of the 2019 CHI
  Conference on Human Factors in Computing Systems}} (Glasgow, Scotland Uk)
  \emph{(\bibinfo{series}{CHI EA '19})}. \bibinfo{publisher}{Association for
  Computing Machinery}, \bibinfo{address}{New York, NY, USA},
  \bibinfo{pages}{1–6}.
\newblock
\showISBNx{9781450359719}


\bibitem[Hu et~al\mbox{.}(2022)]%
        {hu2022lora}
\bibfield{author}{\bibinfo{person}{Edward~J. Hu}, \bibinfo{person}{Yelong
  Shen}, \bibinfo{person}{Phillip Wallis}, \bibinfo{person}{Zeyuan Allen-Zhu},
  \bibinfo{person}{Yuanzhi Li}, \bibinfo{person}{Shean Wang},
  \bibinfo{person}{Lu Wang}, {and} \bibinfo{person}{Weizhu Chen}.}
  \bibinfo{year}{2022}\natexlab{}.
\newblock \showarticletitle{LoRA: Low-Rank Adaptation of Large Language
  Models}. In \bibinfo{booktitle}{\emph{International Conference on Learning
  Representations (ICLR)}}.
\newblock


\bibitem[Huang et~al\mbox{.}(2012a)]%
        {huang2012transahead-computer}
\bibfield{author}{\bibinfo{person}{Chung-chi Huang}, \bibinfo{person}{Ping-che
  Yang}, \bibinfo{person}{Keh-jiann Chen}, {and} \bibinfo{person}{Jason~S.
  Chang}.} \bibinfo{year}{2012}\natexlab{a}.
\newblock \showarticletitle{{T}rans{A}head: A Computer-Assisted Translation and
  Writing Tool}. In \bibinfo{booktitle}{\emph{Proceedings of the 2012
  Conference of the North {A}merican Chapter of the Association for
  Computational Linguistics: Human Language Technologies}}.
  \bibinfo{publisher}{Association for Computational Linguistics},
  \bibinfo{address}{Montr{\'e}al, Canada}, \bibinfo{pages}{352--356}.
\newblock


\bibitem[Huang et~al\mbox{.}(2012b)]%
        {huang2012transahead}
\bibfield{author}{\bibinfo{person}{Chung-chi Huang}, \bibinfo{person}{Ping-che
  Yang}, \bibinfo{person}{Mei-hua Chen}, \bibinfo{person}{Hung-ting Hsieh},
  \bibinfo{person}{Ting-hui Kao}, {and} \bibinfo{person}{Jason~S. Chang}.}
  \bibinfo{year}{2012}\natexlab{b}.
\newblock \showarticletitle{{T}rans{A}head: A Writing Assistant for {CAT} and
  {CALL}}. In \bibinfo{booktitle}{\emph{Proceedings of the Demonstrations at
  the 13th Conference of the {E}uropean Chapter of the Association for
  Computational Linguistics}}. \bibinfo{publisher}{Association for
  Computational Linguistics}, \bibinfo{address}{Avignon, France},
  \bibinfo{pages}{16--19}.
\newblock


\bibitem[Huang et~al\mbox{.}(2018)]%
        {huang2018feedback}
\bibfield{author}{\bibinfo{person}{Yi-Ching Huang}, \bibinfo{person}{Hao-Chuan
  Wang}, {and} \bibinfo{person}{Jane Yung-jen Hsu}.}
  \bibinfo{year}{2018}\natexlab{}.
\newblock \showarticletitle{Feedback Orchestration: Structuring Feedback for
  Facilitating Reflection and Revision in Writing}. In
  \bibinfo{booktitle}{\emph{Companion of the 2018 ACM Conference on Computer
  Supported Cooperative Work and Social Computing}} (<conf-loc>, <city>Jersey
  City</city>, <state>NJ</state>, <country>USA</country>, </conf-loc>)
  \emph{(\bibinfo{series}{CSCW '18 Companion})}.
  \bibinfo{publisher}{Association for Computing Machinery},
  \bibinfo{address}{New York, NY, USA}, \bibinfo{pages}{257–260}.
\newblock
\showISBNx{9781450360180}


\bibitem[Hui and Sprouse(2023)]%
        {hui2023lettersmith}
\bibfield{author}{\bibinfo{person}{Julie Hui} {and}
  \bibinfo{person}{Michelle~L. Sprouse}.} \bibinfo{year}{2023}\natexlab{}.
\newblock \showarticletitle{Lettersmith: Scaffolding Written Professional
  Communication Among College Students}. In
  \bibinfo{booktitle}{\emph{Proceedings of the 2023 CHI Conference on Human
  Factors in Computing Systems}} (Hamburg, Germany) \emph{(\bibinfo{series}{CHI
  '23})}. \bibinfo{publisher}{Association for Computing Machinery},
  \bibinfo{address}{New York, NY, USA}, Article \bibinfo{articleno}{703},
  \bibinfo{numpages}{17}~pages.
\newblock
\showISBNx{9781450394215}


\bibitem[Hui et~al\mbox{.}(2018)]%
        {hui2018introassist}
\bibfield{author}{\bibinfo{person}{Julie~S. Hui}, \bibinfo{person}{Darren
  Gergle}, {and} \bibinfo{person}{Elizabeth~M. Gerber}.}
  \bibinfo{year}{2018}\natexlab{}.
\newblock \showarticletitle{IntroAssist: A Tool to Support Writing Introductory
  Help Requests}. In \bibinfo{booktitle}{\emph{Proceedings of the 2018 CHI
  Conference on Human Factors in Computing Systems}} (Montreal QC, Canada)
  \emph{(\bibinfo{series}{CHI '18})}. \bibinfo{publisher}{Association for
  Computing Machinery}, \bibinfo{address}{New York, NY, USA},
  \bibinfo{pages}{1–13}.
\newblock
\showISBNx{9781450356206}


\bibitem[Hutchens(2023)]%
        {hutchens2023language}
\bibfield{author}{\bibinfo{person}{Justin Hutchens}.}
  \bibinfo{year}{2023}\natexlab{}.
\newblock \bibinfo{booktitle}{\emph{The Language of Deception: Weaponizing Next
  Generation AI}}.
\newblock \bibinfo{publisher}{John Wiley \& Sons}.
\newblock


\bibitem[Ippolito et~al\mbox{.}(2022)]%
        {ippolito2022wordcraft}
\bibfield{author}{\bibinfo{person}{Daphne Ippolito}, \bibinfo{person}{Ann
  Yuan}, \bibinfo{person}{Andy Coenen}, {and} \bibinfo{person}{Sehmon Burnam}.}
  \bibinfo{year}{2022}\natexlab{}.
\newblock \showarticletitle{Creative Writing with an {AI}-Powered Writing
  Assistant: Perspectives from Professional Writers}.
\newblock \bibinfo{journal}{\emph{arXiv preprint arXiv:2211.05030}}
  (\bibinfo{year}{2022}).
\newblock


\bibitem[Ishioka and Kameda(2006)]%
        {ishioka2006automated}
\bibfield{author}{\bibinfo{person}{Tsunenori Ishioka} {and}
  \bibinfo{person}{Masayuki Kameda}.} \bibinfo{year}{2006}\natexlab{}.
\newblock \showarticletitle{Automated {J}apanese Essay Scoring System based on
  Articles Written by Experts}. In \bibinfo{booktitle}{\emph{Proceedings of the
  21st International Conference on Computational Linguistics and 44th Annual
  Meeting of the Association for Computational Linguistics}}.
  \bibinfo{publisher}{Association for Computational Linguistics},
  \bibinfo{address}{Sydney, Australia}, \bibinfo{pages}{233--240}.
\newblock


\bibitem[Ito et~al\mbox{.}(2020)]%
        {ito-etal-2020-langsmith}
\bibfield{author}{\bibinfo{person}{Takumi Ito}, \bibinfo{person}{Tatsuki
  Kuribayashi}, \bibinfo{person}{Masatoshi Hidaka}, \bibinfo{person}{Jun
  Suzuki}, {and} \bibinfo{person}{Kentaro Inui}.}
  \bibinfo{year}{2020}\natexlab{}.
\newblock \showarticletitle{Langsmith: An Interactive Academic Text Revision
  System}. In \bibinfo{booktitle}{\emph{Proceedings of the 2020 Conference on
  Empirical Methods in Natural Language Processing: System Demonstrations}},
  \bibfield{editor}{\bibinfo{person}{Qun Liu} {and} \bibinfo{person}{David
  Schlangen}} (Eds.). \bibinfo{publisher}{Association for Computational
  Linguistics}, \bibinfo{address}{Online}, \bibinfo{pages}{216--226}.
\newblock


\bibitem[Iv et~al\mbox{.}(2022)]%
        {iv-etal-2022-fruit}
\bibfield{author}{\bibinfo{person}{Robert Iv}, \bibinfo{person}{Alexandre
  Passos}, \bibinfo{person}{Sameer Singh}, {and} \bibinfo{person}{Ming-Wei
  Chang}.} \bibinfo{year}{2022}\natexlab{}.
\newblock \showarticletitle{{FRUIT}: Faithfully Reflecting Updated Information
  in Text}. In \bibinfo{booktitle}{\emph{Proceedings of the 2022 Conference of
  the North American Chapter of the Association for Computational Linguistics:
  Human Language Technologies}}, \bibfield{editor}{\bibinfo{person}{Marine
  Carpuat}, \bibinfo{person}{Marie-Catherine de~Marneffe}, {and}
  \bibinfo{person}{Ivan~Vladimir Meza~Ruiz}} (Eds.).
  \bibinfo{publisher}{Association for Computational Linguistics},
  \bibinfo{address}{Seattle, United States}, \bibinfo{pages}{3670--3686}.
\newblock


\bibitem[Jakesch et~al\mbox{.}(2023a)]%
        {jakesch2023cowriting}
\bibfield{author}{\bibinfo{person}{Maurice Jakesch}, \bibinfo{person}{Advait
  Bhat}, \bibinfo{person}{Daniel Buschek}, \bibinfo{person}{Lior Zalmanson},
  {and} \bibinfo{person}{Mor Naaman}.} \bibinfo{year}{2023}\natexlab{a}.
\newblock \showarticletitle{Co-Writing with Opinionated Language Models Affects
  Users’ Views}. In \bibinfo{booktitle}{\emph{Proceedings of the 2023 CHI
  Conference on Human Factors in Computing Systems}} (Hamburg, Germany)
  \emph{(\bibinfo{series}{CHI '23})}. \bibinfo{publisher}{Association for
  Computing Machinery}, \bibinfo{address}{New York, NY, USA}, Article
  \bibinfo{articleno}{111}, \bibinfo{numpages}{15}~pages.
\newblock
\showISBNx{9781450394215}


\bibitem[Jakesch et~al\mbox{.}(2019)]%
        {jakesch2019aimediated}
\bibfield{author}{\bibinfo{person}{Maurice Jakesch}, \bibinfo{person}{Megan
  French}, \bibinfo{person}{Xiao Ma}, \bibinfo{person}{Jeffrey~T. Hancock},
  {and} \bibinfo{person}{Mor Naaman}.} \bibinfo{year}{2019}\natexlab{}.
\newblock \showarticletitle{AI-Mediated Communication: How the Perception That
  Profile Text Was Written by AI Affects Trustworthiness}. In
  \bibinfo{booktitle}{\emph{Proceedings of the 2019 CHI Conference on Human
  Factors in Computing Systems}} (Glasgow, Scotland Uk)
  \emph{(\bibinfo{series}{CHI '19})}. \bibinfo{publisher}{Association for
  Computing Machinery}, \bibinfo{address}{New York, NY, USA},
  \bibinfo{pages}{1–13}.
\newblock
\showISBNx{9781450359702}


\bibitem[Jakesch et~al\mbox{.}(2023b)]%
        {jakesch2023human}
\bibfield{author}{\bibinfo{person}{Maurice Jakesch},
  \bibinfo{person}{Jeffrey~T. Hancock}, {and} \bibinfo{person}{Mor Naaman}.}
  \bibinfo{year}{2023}\natexlab{b}.
\newblock \showarticletitle{Human heuristics for AI-generated language are
  flawed}.
\newblock \bibinfo{journal}{\emph{Proceedings of the National Academy of
  Sciences}} \bibinfo{volume}{120}, \bibinfo{number}{11}
  (\bibinfo{year}{2023}), \bibinfo{pages}{e2208839120}.
\newblock


\bibitem[Jiang et~al\mbox{.}(2022)]%
        {jiang2022arxivedits}
\bibfield{author}{\bibinfo{person}{Chao Jiang}, \bibinfo{person}{Wei Xu}, {and}
  \bibinfo{person}{Samuel Stevens}.} \bibinfo{year}{2022}\natexlab{}.
\newblock \showarticletitle{ar{X}iv{E}dits: Understanding the Human Revision
  Process in Scientific Writing}. In \bibinfo{booktitle}{\emph{Proceedings of
  the 2022 Conference on Empirical Methods in Natural Language Processing}}.
  \bibinfo{publisher}{Association for Computational Linguistics},
  \bibinfo{address}{Abu Dhabi, United Arab Emirates},
  \bibinfo{pages}{9420--9435}.
\newblock


\bibitem[Jiang et~al\mbox{.}(2023)]%
        {Jiang2023graphologue}
\bibfield{author}{\bibinfo{person}{Peiling Jiang}, \bibinfo{person}{Jude
  Rayan}, \bibinfo{person}{Steven~P Dow}, {and} \bibinfo{person}{Haijun Xia}.}
  \bibinfo{year}{2023}\natexlab{}.
\newblock \showarticletitle{Graphologue: Exploring Large Language Model
  Responses with Interactive Diagrams}.
\newblock \bibinfo{journal}{\emph{arXiv preprint arXiv:2305.11473}}
  (\bibinfo{year}{2023}).
\newblock


\bibitem[Jit et~al\mbox{.}(2023)]%
        {jit2023semi}
\bibfield{author}{\bibinfo{person}{Sophia Jit}, \bibinfo{person}{Jennifer
  Spinney}, \bibinfo{person}{Priyank Chandra}, {and} \bibinfo{person}{Robert
  Soden}.} \bibinfo{year}{2023}\natexlab{}.
\newblock \showarticletitle{Semi-Automated Approach for Evaluating Severe
  Weather Risk Communication}. In \bibinfo{booktitle}{\emph{Extended Abstracts
  of the 2023 CHI Conference on Human Factors in Computing Systems}} (Hamburg,
  Germany) \emph{(\bibinfo{series}{CHI EA '23})}.
  \bibinfo{publisher}{Association for Computing Machinery},
  \bibinfo{address}{New York, NY, USA}, Article \bibinfo{articleno}{249},
  \bibinfo{numpages}{8}~pages.
\newblock
\showISBNx{9781450394222}


\bibitem[Kannan et~al\mbox{.}(2016)]%
        {Kannan2016smartreply}
\bibfield{author}{\bibinfo{person}{Anjuli Kannan}, \bibinfo{person}{Karol
  Kurach}, \bibinfo{person}{Sujith Ravi}, \bibinfo{person}{Tobias Kaufmann},
  \bibinfo{person}{Andrew Tomkins}, \bibinfo{person}{Balint Miklos},
  \bibinfo{person}{Greg Corrado}, \bibinfo{person}{Laszlo Lukacs},
  \bibinfo{person}{Marina Ganea}, \bibinfo{person}{Peter Young}, {and}
  \bibinfo{person}{Vivek Ramavajjala}.} \bibinfo{year}{2016}\natexlab{}.
\newblock \showarticletitle{Smart Reply: Automated Response Suggestion for
  Email}. In \bibinfo{booktitle}{\emph{Proceedings of the 22nd ACM SIGKDD
  International Conference on Knowledge Discovery and Data Mining}} (San
  Francisco, California, USA) \emph{(\bibinfo{series}{KDD ’16})}.
  \bibinfo{publisher}{Association for Computing Machinery},
  \bibinfo{address}{New York, NY, USA}, \bibinfo{pages}{955–964}.
\newblock
\showISBNx{9781450342322}


\bibitem[Kaplan et~al\mbox{.}(2020)]%
        {kaplan2020scaling}
\bibfield{author}{\bibinfo{person}{Jared Kaplan}, \bibinfo{person}{Sam
  McCandlish}, \bibinfo{person}{Tom Henighan}, \bibinfo{person}{Tom~B. Brown},
  \bibinfo{person}{Benjamin Chess}, \bibinfo{person}{Rewon Child},
  \bibinfo{person}{Scott Gray}, \bibinfo{person}{Alec Radford},
  \bibinfo{person}{Jeffrey Wu}, {and} \bibinfo{person}{Dario Amodei}.}
  \bibinfo{year}{2020}\natexlab{}.
\newblock \showarticletitle{Scaling Laws for Neural Language Models}.
\newblock \bibinfo{journal}{\emph{arXiv}} (\bibinfo{year}{2020}).
\newblock


\bibitem[Karolus et~al\mbox{.}(2023)]%
        {karolus2023your}
\bibfield{author}{\bibinfo{person}{Jakob Karolus},
  \bibinfo{person}{Sebastian~S. Feger}, \bibinfo{person}{Albrecht Schmidt},
  {and} \bibinfo{person}{Pawe\l{}~W. Wo\'{z}niak}.}
  \bibinfo{year}{2023}\natexlab{}.
\newblock \showarticletitle{Your Text Is Hard to Read: Facilitating Readability
  Awareness to Support Writing Proficiency in Text Production}. In
  \bibinfo{booktitle}{\emph{Proceedings of the 2023 ACM Designing Interactive
  Systems Conference}} (Pittsburgh, PA, USA) \emph{(\bibinfo{series}{DIS
  '23})}. \bibinfo{publisher}{Association for Computing Machinery},
  \bibinfo{address}{New York, NY, USA}, \bibinfo{pages}{147–160}.
\newblock
\showISBNx{9781450398930}


\bibitem[Kaur et~al\mbox{.}(2018)]%
        {kaur2018using}
\bibfield{author}{\bibinfo{person}{Harmanpreet Kaur}, \bibinfo{person}{Alex~C.
  Williams}, \bibinfo{person}{Anne~Loomis Thompson}, \bibinfo{person}{Walter~S.
  Lasecki}, \bibinfo{person}{Shamsi Iqbal}, {and} \bibinfo{person}{Jaime
  Teevan}.} \bibinfo{year}{2018}\natexlab{}.
\newblock \showarticletitle{Using Vocabularies to Collaboratively Create Better
  Plans for Writing Tasks}. In \bibinfo{booktitle}{\emph{Extended Abstracts of
  the 2018 CHI Conference on Human Factors in Computing Systems}} (Montreal QC,
  Canada) \emph{(\bibinfo{series}{CHI EA '18})}.
  \bibinfo{publisher}{Association for Computing Machinery},
  \bibinfo{address}{New York, NY, USA}, \bibinfo{pages}{1–6}.
\newblock
\showISBNx{9781450356213}


\bibitem[Kim et~al\mbox{.}(2020)]%
        {kim2020lexichrome}
\bibfield{author}{\bibinfo{person}{Chris Kim}, \bibinfo{person}{Uta Hinrichs},
  \bibinfo{person}{Saif~M. Mohammad}, {and} \bibinfo{person}{Christopher
  Collins}.} \bibinfo{year}{2020}\natexlab{}.
\newblock \showarticletitle{Lexichrome: Text Construction and Lexical Discovery
  with Word-Color Associations Using Interactive Visualization}. In
  \bibinfo{booktitle}{\emph{Proceedings of the 2020 ACM Designing Interactive
  Systems Conference}} (Eindhoven, Netherlands) \emph{(\bibinfo{series}{DIS
  '20})}. \bibinfo{publisher}{Association for Computing Machinery},
  \bibinfo{address}{New York, NY, USA}, \bibinfo{pages}{477–488}.
\newblock
\showISBNx{9781450369749}


\bibitem[Kim(2022)]%
        {kim2022interaction}
\bibfield{author}{\bibinfo{person}{Juho Kim}.} \bibinfo{year}{2022}\natexlab{}.
\newblock \bibinfo{title}{Interaction-Centric {AI} ({NeurIPS} 2022 Keynote)}.
\newblock
\newblock
\newblock
\shownote{Accessed: Jan 26, 2024}.


\bibitem[Kim et~al\mbox{.}(2023b)]%
        {kim2023metaphorian}
\bibfield{author}{\bibinfo{person}{Jeongyeon Kim}, \bibinfo{person}{Sangho
  Suh}, \bibinfo{person}{Lydia~B Chilton}, {and} \bibinfo{person}{Haijun Xia}.}
  \bibinfo{year}{2023}\natexlab{b}.
\newblock \showarticletitle{Metaphorian: Leveraging Large Language Models to
  Support Extended Metaphor Creation for Science Writing}. In
  \bibinfo{booktitle}{\emph{Proceedings of the 2023 ACM Designing Interactive
  Systems Conference}} (Pittsburgh, PA, USA) \emph{(\bibinfo{series}{DIS
  '23})}. \bibinfo{publisher}{Association for Computing Machinery},
  \bibinfo{address}{New York, NY, USA}, \bibinfo{pages}{115–135}.
\newblock
\showISBNx{9781450398930}


\bibitem[Kim et~al\mbox{.}(2008)]%
        {kim2008common}
\bibfield{author}{\bibinfo{person}{Kyunghee Kim}, \bibinfo{person}{Rosalind~W.
  Picard}, {and} \bibinfo{person}{Henry Lieberman}.}
  \bibinfo{year}{2008}\natexlab{}.
\newblock \showarticletitle{Common Sense Assistant for Writing Stories That
  Teach Social Skills}. In \bibinfo{booktitle}{\emph{CHI '08 Extended Abstracts
  on Human Factors in Computing Systems}} (Florence, Italy)
  \emph{(\bibinfo{series}{CHI EA '08})}. \bibinfo{publisher}{Association for
  Computing Machinery}, \bibinfo{address}{New York, NY, USA},
  \bibinfo{pages}{2805–2810}.
\newblock
\showISBNx{9781605580128}


\bibitem[Kim et~al\mbox{.}(2019)]%
        {kim2019love}
\bibfield{author}{\bibinfo{person}{Taewook Kim}, \bibinfo{person}{Jung~Soo
  Lee}, \bibinfo{person}{Zhenhui Peng}, {and} \bibinfo{person}{Xiaojuan Ma}.}
  \bibinfo{year}{2019}\natexlab{}.
\newblock \showarticletitle{Love in Lyrics: An Exploration of Supporting
  Textual Manifestation of Affection in Social Messaging}.
\newblock \bibinfo{journal}{\emph{Proc. ACM Hum.-Comput. Interact.}}
  \bibinfo{volume}{3}, \bibinfo{number}{CSCW}, Article \bibinfo{articleno}{79}
  (\bibinfo{date}{nov} \bibinfo{year}{2019}), \bibinfo{numpages}{27}~pages.
\newblock


\bibitem[Kim et~al\mbox{.}(2023a)]%
        {kim2023explainable}
\bibfield{author}{\bibinfo{person}{Yewon Kim}, \bibinfo{person}{Mina Lee},
  \bibinfo{person}{Donghwi Kim}, {and} \bibinfo{person}{Sung-Ju Lee}.}
  \bibinfo{year}{2023}\natexlab{a}.
\newblock \showarticletitle{Towards Explainable AI Writing Assistants for
  Non-native English Speakers}.
\newblock
\showeprint[arxiv]{2304.02625}~[cs.CL]


\bibitem[Kim et~al\mbox{.}(2022)]%
        {kim-etal-2022-improving}
\bibfield{author}{\bibinfo{person}{Zae~Myung Kim}, \bibinfo{person}{Wanyu Du},
  \bibinfo{person}{Vipul Raheja}, \bibinfo{person}{Dhruv Kumar}, {and}
  \bibinfo{person}{Dongyeop Kang}.} \bibinfo{year}{2022}\natexlab{}.
\newblock \showarticletitle{Improving Iterative Text Revision by Learning Where
  to Edit from Other Revision Tasks}. In \bibinfo{booktitle}{\emph{Proceedings
  of the 2022 Conference on Empirical Methods in Natural Language Processing}}.
  \bibinfo{publisher}{Association for Computational Linguistics},
  \bibinfo{address}{Abu Dhabi, United Arab Emirates},
  \bibinfo{pages}{9986--9999}.
\newblock


\bibitem[Kinnunen et~al\mbox{.}(2012)]%
        {kinnunen2012swan}
\bibfield{author}{\bibinfo{person}{Tomi Kinnunen}, \bibinfo{person}{Henri
  Leisma}, \bibinfo{person}{Monika Machunik}, \bibinfo{person}{Tuomo Kakkonen},
  {and} \bibinfo{person}{Jean-Luc LeBrun}.} \bibinfo{year}{2012}\natexlab{}.
\newblock \showarticletitle{{SWAN} - Scientific Writing {A}ssista{N}t. A Tool
  for Helping Scholars to Write Reader-Friendly Manuscripts}. In
  \bibinfo{booktitle}{\emph{Proceedings of the Demonstrations at the 13th
  Conference of the {E}uropean Chapter of the Association for Computational
  Linguistics}}. \bibinfo{publisher}{Association for Computational
  Linguistics}, \bibinfo{address}{Avignon, France}, \bibinfo{pages}{20--24}.
\newblock


\bibitem[Knight et~al\mbox{.}(2020)]%
        {knight2020acawriter}
\bibfield{author}{\bibinfo{person}{Simon Knight}, \bibinfo{person}{Antonette
  Shibani}, \bibinfo{person}{Sophie Abel}, \bibinfo{person}{Andrew Gibson},
  \bibinfo{person}{Philippa Ryan}, \bibinfo{person}{Nicole Sutton},
  \bibinfo{person}{Raechel Wight}, \bibinfo{person}{Cherie Lucas},
  \bibinfo{person}{Agnes Sandor}, \bibinfo{person}{Kirsty Kitto},
  \bibinfo{person}{Ming Liu}, \bibinfo{person}{Radhika~Vijay Mogarkar}, {and}
  \bibinfo{person}{Simon Buckingham~Shum}.} \bibinfo{year}{2020}\natexlab{}.
\newblock \showarticletitle{AcaWriter: A learning analytics tool for formative
  feedback on academic writing}.
\newblock \bibinfo{journal}{\emph{Journal of Writing Research}}
  \bibinfo{volume}{12}, \bibinfo{number}{1} (\bibinfo{year}{2020}),
  \bibinfo{pages}{141--186}.
\newblock


\bibitem[Knittel et~al\mbox{.}(2019)]%
        {knittel2019bitcoin}
\bibfield{author}{\bibinfo{person}{Megan Knittel}, \bibinfo{person}{Shelby
  Pitts}, {and} \bibinfo{person}{Rick Wash}.} \bibinfo{year}{2019}\natexlab{}.
\newblock \showarticletitle{"The Most Trustworthy Coin": How Ideological
  Tensions Drive Trust in Bitcoin}.
\newblock \bibinfo{journal}{\emph{Proc. ACM Hum.-Comput. Interact.}}
  \bibinfo{volume}{3}, \bibinfo{number}{CSCW}, Article \bibinfo{articleno}{36}
  (\bibinfo{date}{nov} \bibinfo{year}{2019}), \bibinfo{numpages}{23}~pages.
\newblock


\bibitem[Krippendorff(2011)]%
        {krippendorff2011agreement}
\bibfield{author}{\bibinfo{person}{Klaus Krippendorff}.}
  \bibinfo{year}{2011}\natexlab{}.
\newblock \showarticletitle{Agreement and information in the reliability of
  coding}.
\newblock \bibinfo{journal}{\emph{Communication methods and measures}}
  \bibinfo{volume}{5}, \bibinfo{number}{2} (\bibinfo{year}{2011}),
  \bibinfo{pages}{93--112}.
\newblock


\bibitem[Kristensson and Vertanen(2014)]%
        {Kristensson2014inviscid}
\bibfield{author}{\bibinfo{person}{Per~Ola Kristensson} {and}
  \bibinfo{person}{Keith Vertanen}.} \bibinfo{year}{2014}\natexlab{}.
\newblock \showarticletitle{The inviscid text entry rate and its application as
  a grand goal for mobile text entry}. In \bibinfo{booktitle}{\emph{Proceedings
  of the 16th international conference on {Human}-computer interaction with
  mobile devices \& services}} \emph{(\bibinfo{series}{{MobileHCI} '14})}.
  \bibinfo{publisher}{Association for Computing Machinery},
  \bibinfo{address}{Toronto, ON, Canada}, \bibinfo{pages}{335--338}.
\newblock
\showISBNx{978-1-4503-3004-6}


\bibitem[Kuyda(2023)]%
        {kuyda2023replikaversions}
\bibfield{author}{\bibinfo{person}{Eugenia Kuyda}.}
  \bibinfo{year}{2023}\natexlab{}.
\newblock \bibinfo{title}{Replika Updates}.
\newblock
\newblock
\newblock
\shownote{Accessed: Feb. 24, 2024}.


\bibitem[Lavie and Denkowski(2009)]%
        {lavie2009meteor}
\bibfield{author}{\bibinfo{person}{Alon Lavie} {and} \bibinfo{person}{Michael
  Denkowski}.} \bibinfo{year}{2009}\natexlab{}.
\newblock \showarticletitle{The Meteor Metric for Automatic Evaluation of
  Machine Translation}.
\newblock \bibinfo{journal}{\emph{Machine Translation}}  \bibinfo{volume}{23}
  (\bibinfo{year}{2009}).
\newblock


\bibitem[Leavitt(1965)]%
        {leavitt1965applied}
\bibfield{author}{\bibinfo{person}{Harold~J Leavitt}.}
  \bibinfo{year}{1965}\natexlab{}.
\newblock \showarticletitle{Applied organizational change in industry:
  Structural, technological and humanistic approaches}.
\newblock In \bibinfo{booktitle}{\emph{Handbook of Organizations (RLE:
  Organizations)}}. \bibinfo{publisher}{Routledge},
  \bibinfo{pages}{1144--1170}.
\newblock


\bibitem[Lee et~al\mbox{.}(2019)]%
        {lee2019icomposer}
\bibfield{author}{\bibinfo{person}{Hsin-Pei Lee}, \bibinfo{person}{Jhih-Sheng
  Fang}, {and} \bibinfo{person}{Wei-Yun Ma}.} \bibinfo{year}{2019}\natexlab{}.
\newblock \showarticletitle{i{C}omposer: An Automatic Songwriting System for
  {C}hinese Popular Music}. In \bibinfo{booktitle}{\emph{Proceedings of the
  2019 Conference of the North {A}merican Chapter of the Association for
  Computational Linguistics (Demonstrations)}}. \bibinfo{publisher}{Association
  for Computational Linguistics}, \bibinfo{address}{Minneapolis, Minnesota},
  \bibinfo{pages}{84--88}.
\newblock


\bibitem[Lee et~al\mbox{.}(2023)]%
        {lee2023language}
\bibfield{author}{\bibinfo{person}{Jooyoung Lee}, \bibinfo{person}{Thai Le},
  \bibinfo{person}{Jinghui Chen}, {and} \bibinfo{person}{Dongwon Lee}.}
  \bibinfo{year}{2023}\natexlab{}.
\newblock \showarticletitle{Do language models plagiarize?}. In
  \bibinfo{booktitle}{\emph{Proceedings of the ACM Web Conference 2023}}.
  \bibinfo{pages}{3637--3647}.
\newblock


\bibitem[Lee et~al\mbox{.}(2022a)]%
        {lee2022coauthor}
\bibfield{author}{\bibinfo{person}{Mina Lee}, \bibinfo{person}{Percy Liang},
  {and} \bibinfo{person}{Qian Yang}.} \bibinfo{year}{2022}\natexlab{a}.
\newblock \showarticletitle{CoAuthor: Designing a Human-AI Collaborative
  Writing Dataset for Exploring Language Model Capabilities}. In
  \bibinfo{booktitle}{\emph{Proceedings of the 2022 CHI Conference on Human
  Factors in Computing Systems}} (New Orleans, LA, USA)
  \emph{(\bibinfo{series}{CHI '22})}. \bibinfo{publisher}{Association for
  Computing Machinery}, \bibinfo{address}{New York, NY, USA}, Article
  \bibinfo{articleno}{388}, \bibinfo{numpages}{19}~pages.
\newblock
\showISBNx{9781450391573}


\bibitem[Lee et~al\mbox{.}(2022b)]%
        {lee2022evaluating}
\bibfield{author}{\bibinfo{person}{Mina Lee}, \bibinfo{person}{Megha
  Srivastava}, \bibinfo{person}{Amelia Hardy}, \bibinfo{person}{John
  Thickstun}, \bibinfo{person}{Esin Durmus}, \bibinfo{person}{Ashwin
  Paranjape}, \bibinfo{person}{Ines Gerard-Ursin}, \bibinfo{person}{Xiang~Lisa
  Li}, \bibinfo{person}{Faisal Ladhak}, \bibinfo{person}{Frieda Rong},
  \bibinfo{person}{Rose~E. Wang}, \bibinfo{person}{Minae Kwon},
  \bibinfo{person}{Joon~Sung Park}, \bibinfo{person}{Hancheng Cao},
  \bibinfo{person}{Tony Lee}, \bibinfo{person}{Rishi Bommasani},
  \bibinfo{person}{Michael Bernstein}, {and} \bibinfo{person}{Percy Liang}.}
  \bibinfo{year}{2022}\natexlab{b}.
\newblock \showarticletitle{Evaluating Human-Language Model Interaction}.
\newblock \bibinfo{journal}{\emph{arXiv preprint arXiv:2212.09746}}
  (\bibinfo{year}{2022}).
\newblock


\bibitem[Lehmann(2023)]%
        {lehmann2023mixed}
\bibfield{author}{\bibinfo{person}{Florian Lehmann}.}
  \bibinfo{year}{2023}\natexlab{}.
\newblock \showarticletitle{Mixed-Initiative Interaction with Computational
  Generative Systems}. In \bibinfo{booktitle}{\emph{Extended Abstracts of the
  2023 CHI Conference on Human Factors in Computing Systems}} (Hamburg,
  Germany) \emph{(\bibinfo{series}{CHI EA '23})}.
  \bibinfo{publisher}{Association for Computing Machinery},
  \bibinfo{address}{New York, NY, USA}, Article \bibinfo{articleno}{501},
  \bibinfo{numpages}{6}~pages.
\newblock
\showISBNx{9781450394222}


\bibitem[Lewicki et~al\mbox{.}(2023)]%
        {lewicki2023service}
\bibfield{author}{\bibinfo{person}{Kornel Lewicki}, \bibinfo{person}{Michelle
  Seng~Ah Lee}, \bibinfo{person}{Jennifer Cobbe}, {and}
  \bibinfo{person}{Jatinder Singh}.} \bibinfo{year}{2023}\natexlab{}.
\newblock \showarticletitle{Out of Context: Investigating the Bias and Fairness
  Concerns of “Artificial Intelligence as a Service”}. In
  \bibinfo{booktitle}{\emph{Proceedings of the 2023 CHI Conference on Human
  Factors in Computing Systems}} (Hamburg, Germany) \emph{(\bibinfo{series}{CHI
  '23})}. \bibinfo{publisher}{Association for Computing Machinery},
  \bibinfo{address}{New York, NY, USA}, Article \bibinfo{articleno}{135},
  \bibinfo{numpages}{17}~pages.
\newblock
\showISBNx{9781450394215}


\bibitem[Li et~al\mbox{.}(2022)]%
        {li-etal-2022-text}
\bibfield{author}{\bibinfo{person}{Jingjing Li}, \bibinfo{person}{Zichao Li},
  \bibinfo{person}{Tao Ge}, \bibinfo{person}{Irwin King}, {and}
  \bibinfo{person}{Michael Lyu}.} \bibinfo{year}{2022}\natexlab{}.
\newblock \showarticletitle{Text Revision by On-the-Fly Representation
  Optimization}. In \bibinfo{booktitle}{\emph{Proceedings of the First Workshop
  on Intelligent and Interactive Writing Assistants (In2Writing 2022)}},
  \bibfield{editor}{\bibinfo{person}{Ting-Hao~'Kenneth' Huang},
  \bibinfo{person}{Vipul Raheja}, \bibinfo{person}{Dongyeop Kang},
  \bibinfo{person}{John Joon~Young Chung}, \bibinfo{person}{Daniel Gissin},
  \bibinfo{person}{Mina Lee}, {and} \bibinfo{person}{Katy~Ilonka Gero}} (Eds.).
  \bibinfo{publisher}{Association for Computational Linguistics},
  \bibinfo{address}{Dublin, Ireland}, \bibinfo{pages}{58--59}.
\newblock


\bibitem[Li and Liang(2021)]%
        {li-liang-2021-prefix}
\bibfield{author}{\bibinfo{person}{Xiang~Lisa Li} {and} \bibinfo{person}{Percy
  Liang}.} \bibinfo{year}{2021}\natexlab{}.
\newblock \showarticletitle{Prefix-Tuning: Optimizing Continuous Prompts for
  Generation}. In \bibinfo{booktitle}{\emph{Proceedings of the 59th Annual
  Meeting of the Association for Computational Linguistics and the 11th
  International Joint Conference on Natural Language Processing (Volume 1: Long
  Papers)}}. \bibinfo{publisher}{Association for Computational Linguistics},
  \bibinfo{address}{Online}, \bibinfo{pages}{4582--4597}.
\newblock


\bibitem[Liang et~al\mbox{.}(2023)]%
        {liang2023holistic}
\bibfield{author}{\bibinfo{person}{Percy Liang}, \bibinfo{person}{Rishi
  Bommasani}, \bibinfo{person}{Tony Lee}, \bibinfo{person}{Dimitris Tsipras},
  \bibinfo{person}{Dilara Soylu}, \bibinfo{person}{Michihiro Yasunaga},
  \bibinfo{person}{Yian Zhang}, \bibinfo{person}{Deepak Narayanan},
  \bibinfo{person}{Yuhuai Wu}, \bibinfo{person}{Ananya Kumar},
  \bibinfo{person}{Benjamin Newman}, \bibinfo{person}{Binhang Yuan},
  \bibinfo{person}{Bobby Yan}, \bibinfo{person}{Ce Zhang},
  \bibinfo{person}{Christian~Alexander Cosgrove},
  \bibinfo{person}{Christopher~D Manning}, \bibinfo{person}{Christopher Re},
  \bibinfo{person}{Diana Acosta-Navas}, \bibinfo{person}{Drew~Arad Hudson},
  \bibinfo{person}{Eric Zelikman}, \bibinfo{person}{Esin Durmus},
  \bibinfo{person}{Faisal Ladhak}, \bibinfo{person}{Frieda Rong},
  \bibinfo{person}{Hongyu Ren}, \bibinfo{person}{Huaxiu Yao},
  \bibinfo{person}{Jue WANG}, \bibinfo{person}{Keshav Santhanam},
  \bibinfo{person}{Laurel Orr}, \bibinfo{person}{Lucia Zheng},
  \bibinfo{person}{Mert Yuksekgonul}, \bibinfo{person}{Mirac Suzgun},
  \bibinfo{person}{Nathan Kim}, \bibinfo{person}{Neel Guha},
  \bibinfo{person}{Niladri~S. Chatterji}, \bibinfo{person}{Omar Khattab},
  \bibinfo{person}{Peter Henderson}, \bibinfo{person}{Qian Huang},
  \bibinfo{person}{Ryan~Andrew Chi}, \bibinfo{person}{Sang~Michael Xie},
  \bibinfo{person}{Shibani Santurkar}, \bibinfo{person}{Surya Ganguli},
  \bibinfo{person}{Tatsunori Hashimoto}, \bibinfo{person}{Thomas Icard},
  \bibinfo{person}{Tianyi Zhang}, \bibinfo{person}{Vishrav Chaudhary},
  \bibinfo{person}{William Wang}, \bibinfo{person}{Xuechen Li},
  \bibinfo{person}{Yifan Mai}, \bibinfo{person}{Yuhui Zhang}, {and}
  \bibinfo{person}{Yuta Koreeda}.} \bibinfo{year}{2023}\natexlab{}.
\newblock \showarticletitle{Holistic Evaluation of Language Models}.
\newblock \bibinfo{journal}{\emph{Transactions on Machine Learning Research}}
  (\bibinfo{year}{2023}).
\newblock
\showISSN{2835-8856}
\newblock
\shownote{Featured Certification, Expert Certification}.


\bibitem[Liao and Vaughan(2023)]%
        {liao2023ai}
\bibfield{author}{\bibinfo{person}{Q.~Vera Liao} {and}
  \bibinfo{person}{Jennifer~Wortman Vaughan}.} \bibinfo{year}{2023}\natexlab{}.
\newblock \bibinfo{title}{AI Transparency in the Age of LLMs: A Human-Centered
  Research Roadmap}.
\newblock
\newblock
\showeprint[arxiv]{2306.01941}~[cs.HC]


\bibitem[Lin et~al\mbox{.}(2023)]%
        {lin2023beyond}
\bibfield{author}{\bibinfo{person}{Zhiyu Lin}, \bibinfo{person}{Upol Ehsan},
  \bibinfo{person}{Rohan Agarwal}, \bibinfo{person}{Samihan Dani},
  \bibinfo{person}{Vidushi Vashishth}, {and} \bibinfo{person}{Mark Riedl}.}
  \bibinfo{year}{2023}\natexlab{}.
\newblock \showarticletitle{Beyond Prompts: Exploring the Design Space of
  Mixed-Initiative Co-Creativity Systems}.
\newblock \bibinfo{journal}{\emph{arXiv preprint arXiv:2305.07465}}
  (\bibinfo{year}{2023}).
\newblock


\bibitem[Liu et~al\mbox{.}(2000)]%
        {liu2000pens}
\bibfield{author}{\bibinfo{person}{Ting Liu}, \bibinfo{person}{Ming Zhou},
  \bibinfo{person}{Jianfeng Gao}, \bibinfo{person}{Endong Xun}, {and}
  \bibinfo{person}{Changning Huang}.} \bibinfo{year}{2000}\natexlab{}.
\newblock \showarticletitle{PENS: A Machine-Aided English Writing System for
  Chinese Users}. In \bibinfo{booktitle}{\emph{Proceedings of the 38th Annual
  Meeting on Association for Computational Linguistics}} (Hong Kong)
  \emph{(\bibinfo{series}{ACL '00})}. \bibinfo{publisher}{Association for
  Computational Linguistics}, \bibinfo{address}{USA},
  \bibinfo{pages}{529–536}.
\newblock


\bibitem[Liu et~al\mbox{.}(2022)]%
        {liu2022will}
\bibfield{author}{\bibinfo{person}{Yihe Liu}, \bibinfo{person}{Anushk Mittal},
  \bibinfo{person}{Diyi Yang}, {and} \bibinfo{person}{Amy Bruckman}.}
  \bibinfo{year}{2022}\natexlab{}.
\newblock \showarticletitle{Will AI Console Me When I Lose My Pet?
  Understanding Perceptions of AI-Mediated Email Writing}. In
  \bibinfo{booktitle}{\emph{Proceedings of the 2022 CHI Conference on Human
  Factors in Computing Systems}} (New Orleans, LA, USA)
  \emph{(\bibinfo{series}{CHI '22})}. \bibinfo{publisher}{Association for
  Computing Machinery}, \bibinfo{address}{New York, NY, USA}, Article
  \bibinfo{articleno}{474}, \bibinfo{numpages}{13}~pages.
\newblock
\showISBNx{9781450391573}


\bibitem[Liu et~al\mbox{.}(2019a)]%
        {liu2019roberta}
\bibfield{author}{\bibinfo{person}{Yinhan Liu}, \bibinfo{person}{Myle Ott},
  \bibinfo{person}{Naman Goyal}, \bibinfo{person}{Jingfei Du},
  \bibinfo{person}{Mandar Joshi}, \bibinfo{person}{Danqi Chen},
  \bibinfo{person}{Omer Levy}, \bibinfo{person}{Mike Lewis},
  \bibinfo{person}{Luke Zettlemoyer}, {and} \bibinfo{person}{Veselin
  Stoyanov}.} \bibinfo{year}{2019}\natexlab{a}.
\newblock \showarticletitle{{R}o{BERT}a: A Robustly Optimized {BERT}
  Pretraining Approach}.
\newblock \bibinfo{journal}{\emph{arXiv preprint arXiv:1907.11692}}
  (\bibinfo{year}{2019}).
\newblock


\bibitem[Liu et~al\mbox{.}(2019b)]%
        {liu2019neuralbased}
\bibfield{author}{\bibinfo{person}{Yuanchao Liu}, \bibinfo{person}{Bo Pang},
  {and} \bibinfo{person}{Bingquan Liu}.} \bibinfo{year}{2019}\natexlab{b}.
\newblock \showarticletitle{Neural-based {C}hinese Idiom Recommendation for
  Enhancing Elegance in Essay Writing}. In
  \bibinfo{booktitle}{\emph{Proceedings of the 57th Annual Meeting of the
  Association for Computational Linguistics}}. \bibinfo{publisher}{Association
  for Computational Linguistics}, \bibinfo{address}{Florence, Italy},
  \bibinfo{pages}{5522--5526}.
\newblock


\bibitem[Louis and Nenkova(2013)]%
        {louis2013makes}
\bibfield{author}{\bibinfo{person}{Annie Louis} {and} \bibinfo{person}{Ani
  Nenkova}.} \bibinfo{year}{2013}\natexlab{}.
\newblock \showarticletitle{What Makes Writing Great? First Experiments on
  Article Quality Prediction in the Science Journalism Domain}.
\newblock \bibinfo{journal}{\emph{Transactions of the Association for
  Computational Linguistics}}  \bibinfo{volume}{1} (\bibinfo{year}{2013}),
  \bibinfo{pages}{341--352}.
\newblock


\bibitem[Lukasik and Zens(2018)]%
        {lukasik2018content}
\bibfield{author}{\bibinfo{person}{Michal Lukasik} {and}
  \bibinfo{person}{Richard Zens}.} \bibinfo{year}{2018}\natexlab{}.
\newblock \showarticletitle{Content Explorer: Recommending Novel Entities for a
  Document Writer}. In \bibinfo{booktitle}{\emph{Proceedings of the 2018
  Conference on Empirical Methods in Natural Language Processing}}.
  \bibinfo{publisher}{Association for Computational Linguistics},
  \bibinfo{address}{Brussels, Belgium}, \bibinfo{pages}{3371--3380}.
\newblock


\bibitem[MacArthur(2006)]%
        {macarthur2006effects}
\bibfield{author}{\bibinfo{person}{Charles~A MacArthur}.}
  \bibinfo{year}{2006}\natexlab{}.
\newblock \showarticletitle{The effects of new technologies on writing and
  writing processes}.
\newblock \bibinfo{journal}{\emph{Handbook of writing research}}
  (\bibinfo{year}{2006}), \bibinfo{pages}{248--262}.
\newblock


\bibitem[Macdonald(1983)]%
        {macdonald1983unix}
\bibfield{author}{\bibinfo{person}{N.~H. Macdonald}.}
  \bibinfo{year}{1983}\natexlab{}.
\newblock \showarticletitle{Human factors and behavioral science: The UNIX™
  Writer's Workbench software: Rationale and design}.
\newblock \bibinfo{journal}{\emph{The Bell System Technical Journal}}
  \bibinfo{volume}{62}, \bibinfo{number}{6} (\bibinfo{year}{1983}),
  \bibinfo{pages}{1891--1908}.
\newblock


\bibitem[MacKenzie and Castellucci(2016)]%
        {mackenzie2016empirical}
\bibfield{author}{\bibinfo{person}{I~Scott MacKenzie} {and}
  \bibinfo{person}{Steven~J Castellucci}.} \bibinfo{year}{2016}\natexlab{}.
\newblock \showarticletitle{Empirical research methods for human-computer
  interaction}. In \bibinfo{booktitle}{\emph{Proceedings of the 2016 CHI
  Conference Extended Abstracts on Human Factors in Computing Systems}}.
  \bibinfo{pages}{996--999}.
\newblock


\bibitem[MacLean et~al\mbox{.}(1991)]%
        {maclean1991questions}
\bibfield{author}{\bibinfo{person}{Allan MacLean}, \bibinfo{person}{Richard
  Young}, \bibinfo{person}{Victoria Bellotti}, {and} \bibinfo{person}{Thomas
  Moran}.} \bibinfo{year}{1991}\natexlab{}.
\newblock \showarticletitle{Questions, Options, and Criteria: Elements of
  Design Space Analysis}.
\newblock \bibinfo{journal}{\emph{Human-Computer Interaction}}
  \bibinfo{volume}{6} (\bibinfo{date}{09} \bibinfo{year}{1991}),
  \bibinfo{pages}{201--250}.
\newblock


\bibitem[Mallinson et~al\mbox{.}(2022)]%
        {mallinson-etal-2022-edit5}
\bibfield{author}{\bibinfo{person}{Jonathan Mallinson}, \bibinfo{person}{Jakub
  Adamek}, \bibinfo{person}{Eric Malmi}, {and} \bibinfo{person}{Aliaksei
  Severyn}.} \bibinfo{year}{2022}\natexlab{}.
\newblock \showarticletitle{{E}di{T}5: Semi-Autoregressive Text Editing with T5
  Warm-Start}. In \bibinfo{booktitle}{\emph{Findings of the Association for
  Computational Linguistics: EMNLP 2022}}. \bibinfo{publisher}{Association for
  Computational Linguistics}, \bibinfo{address}{Abu Dhabi, United Arab
  Emirates}, \bibinfo{pages}{2126--2138}.
\newblock


\bibitem[Mallinson et~al\mbox{.}(2020)]%
        {mallinson-etal-2020-felix}
\bibfield{author}{\bibinfo{person}{Jonathan Mallinson},
  \bibinfo{person}{Aliaksei Severyn}, \bibinfo{person}{Eric Malmi}, {and}
  \bibinfo{person}{Guillermo Garrido}.} \bibinfo{year}{2020}\natexlab{}.
\newblock \showarticletitle{{FELIX}: Flexible Text Editing Through Tagging and
  Insertion}. In \bibinfo{booktitle}{\emph{Findings of the Association for
  Computational Linguistics: EMNLP 2020}}. \bibinfo{publisher}{Association for
  Computational Linguistics}, \bibinfo{address}{Online},
  \bibinfo{pages}{1244--1255}.
\newblock


\bibitem[Malmi et~al\mbox{.}(2019)]%
        {malmi-etal-2019-encode}
\bibfield{author}{\bibinfo{person}{Eric Malmi}, \bibinfo{person}{Sebastian
  Krause}, \bibinfo{person}{Sascha Rothe}, \bibinfo{person}{Daniil Mirylenka},
  {and} \bibinfo{person}{Aliaksei Severyn}.} \bibinfo{year}{2019}\natexlab{}.
\newblock \showarticletitle{Encode, Tag, Realize: High-Precision Text Editing}.
  In \bibinfo{booktitle}{\emph{Proceedings of the 2019 Conference on Empirical
  Methods in Natural Language Processing and the 9th International Joint
  Conference on Natural Language Processing (EMNLP-IJCNLP)}}.
  \bibinfo{publisher}{Association for Computational Linguistics},
  \bibinfo{address}{Hong Kong, China}, \bibinfo{pages}{5054--5065}.
\newblock


\bibitem[Meyer et~al\mbox{.}(2023)]%
        {meyer2023chatgpt}
\bibfield{author}{\bibinfo{person}{Jesse~G Meyer}, \bibinfo{person}{Ryan~J
  Urbanowicz}, \bibinfo{person}{Patrick~CN Martin}, \bibinfo{person}{Karen
  O’Connor}, \bibinfo{person}{Ruowang Li}, \bibinfo{person}{Pei-Chen Peng},
  \bibinfo{person}{Tiffani~J Bright}, \bibinfo{person}{Nicholas Tatonetti},
  \bibinfo{person}{Kyoung~Jae Won}, \bibinfo{person}{Graciela
  Gonzalez-Hernandez}, {et~al\mbox{.}}} \bibinfo{year}{2023}\natexlab{}.
\newblock \showarticletitle{{ChatGPT} and large language models in academia:
  opportunities and challenges}.
\newblock \bibinfo{journal}{\emph{BioData Mining}} \bibinfo{volume}{16},
  \bibinfo{number}{1} (\bibinfo{year}{2023}), \bibinfo{pages}{20}.
\newblock


\bibitem[Mieczkowski and Hancock(2023)]%
        {mieczkowski2023thesis}
\bibfield{author}{\bibinfo{person}{Hannah Mieczkowski} {and}
  \bibinfo{person}{Jeffrey Hancock}.} \bibinfo{year}{2023}\natexlab{}.
\newblock \showarticletitle{Examining Agency, Expertise, and Roles of {AI}
  Systems in {AI}-Mediated Communication}. In
  \bibinfo{booktitle}{\emph{Computer-Supported Cooperative Work And Social
  Computing (CSCW)}}.
\newblock


\bibitem[Mikolov et~al\mbox{.}(2013)]%
        {mikolov2013efficient}
\bibfield{author}{\bibinfo{person}{Tomas Mikolov}, \bibinfo{person}{Kai Chen},
  \bibinfo{person}{Greg Corrado}, {and} \bibinfo{person}{Jeffrey Dean}.}
  \bibinfo{year}{2013}\natexlab{}.
\newblock \showarticletitle{Efficient Estimation of Word Representations in
  Vector Space}.
\newblock \bibinfo{journal}{\emph{arXiv preprint arXiv:1301.3781}}
  (\bibinfo{year}{2013}).
\newblock


\bibitem[Miltsakaki and Kukich(2000)]%
        {miltsakaki2000therole}
\bibfield{author}{\bibinfo{person}{Eleni Miltsakaki} {and}
  \bibinfo{person}{Karen Kukich}.} \bibinfo{year}{2000}\natexlab{}.
\newblock \showarticletitle{The Role of Centering Theory's Rough-Shift in the
  Teaching and Evaluation of Writing Skills}. In
  \bibinfo{booktitle}{\emph{Proceedings of the 38th Annual Meeting on
  Association for Computational Linguistics}} (Hong Kong)
  \emph{(\bibinfo{series}{ACL '00})}. \bibinfo{publisher}{Association for
  Computational Linguistics}, \bibinfo{address}{USA},
  \bibinfo{pages}{408–415}.
\newblock


\bibitem[Mirowski et~al\mbox{.}(2023)]%
        {mirowski2023cowriting}
\bibfield{author}{\bibinfo{person}{Piotr Mirowski}, \bibinfo{person}{Kory~W.
  Mathewson}, \bibinfo{person}{Jaylen Pittman}, {and} \bibinfo{person}{Richard
  Evans}.} \bibinfo{year}{2023}\natexlab{}.
\newblock \showarticletitle{Co-Writing Screenplays and Theatre Scripts with
  Language Models: Evaluation by Industry Professionals}. In
  \bibinfo{booktitle}{\emph{Proceedings of the 2023 CHI Conference on Human
  Factors in Computing Systems}} (Hamburg, Germany) \emph{(\bibinfo{series}{CHI
  '23})}. \bibinfo{publisher}{Association for Computing Machinery},
  \bibinfo{address}{New York, NY, USA}, Article \bibinfo{articleno}{355},
  \bibinfo{numpages}{34}~pages.
\newblock
\showISBNx{9781450394215}


\bibitem[Mishra et~al\mbox{.}(2014)]%
        {mishra2014text}
\bibfield{author}{\bibinfo{person}{Rashmi Mishra}, \bibinfo{person}{Jiantao
  Bian}, \bibinfo{person}{Marcelo Fiszman}, \bibinfo{person}{Charlene~R Weir},
  \bibinfo{person}{Siddhartha Jonnalagadda}, \bibinfo{person}{Javed Mostafa},
  {and} \bibinfo{person}{Guilherme Del~Fiol}.} \bibinfo{year}{2014}\natexlab{}.
\newblock \showarticletitle{Text summarization in the biomedical domain: a
  systematic review of recent research}.
\newblock \bibinfo{journal}{\emph{Journal of biomedical informatics}}
  \bibinfo{volume}{52} (\bibinfo{year}{2014}), \bibinfo{pages}{457--467}.
\newblock


\bibitem[Mita et~al\mbox{.}(2022)]%
        {mita2022automated}
\bibfield{author}{\bibinfo{person}{Masato Mita}, \bibinfo{person}{Keisuke
  Sakaguchi}, \bibinfo{person}{Masato Hagiwara}, \bibinfo{person}{Tomoya
  Mizumoto}, \bibinfo{person}{Jun Suzuki}, {and} \bibinfo{person}{Kentaro
  Inui}.} \bibinfo{year}{2022}\natexlab{}.
\newblock \bibinfo{title}{Towards Automated Document Revision: Grammatical
  Error Correction, Fluency Edits, and Beyond}.
\newblock
\newblock
\showeprint[arxiv]{2205.11484}~[cs.CL]


\bibitem[Mori et~al\mbox{.}(2022)]%
        {mori-etal-2022-plug}
\bibfield{author}{\bibinfo{person}{Yusuke Mori}, \bibinfo{person}{Hiroaki
  Yamane}, \bibinfo{person}{Ryohei Shimizu}, {and} \bibinfo{person}{Tatsuya
  Harada}.} \bibinfo{year}{2022}\natexlab{}.
\newblock \showarticletitle{Plug-and-Play Controller for Story Completion: A
  Pilot Study toward Emotion-aware Story Writing Assistance}. In
  \bibinfo{booktitle}{\emph{Proceedings of the First Workshop on Intelligent
  and Interactive Writing Assistants (In2Writing 2022)}},
  \bibfield{editor}{\bibinfo{person}{Ting-Hao~'Kenneth' Huang},
  \bibinfo{person}{Vipul Raheja}, \bibinfo{person}{Dongyeop Kang},
  \bibinfo{person}{John Joon~Young Chung}, \bibinfo{person}{Daniel Gissin},
  \bibinfo{person}{Mina Lee}, {and} \bibinfo{person}{Katy~Ilonka Gero}} (Eds.).
  \bibinfo{publisher}{Association for Computational Linguistics},
  \bibinfo{address}{Dublin, Ireland}, \bibinfo{pages}{46--57}.
\newblock


\bibitem[Morris et~al\mbox{.}(2023)]%
        {morris2023design}
\bibfield{author}{\bibinfo{person}{Meredith~Ringel Morris},
  \bibinfo{person}{Carrie~J Cai}, \bibinfo{person}{Jess Holbrook},
  \bibinfo{person}{Chinmay Kulkarni}, {and} \bibinfo{person}{Michael Terry}.}
  \bibinfo{year}{2023}\natexlab{}.
\newblock \showarticletitle{The design space of generative models}.
\newblock \bibinfo{journal}{\emph{arXiv preprint arXiv:2304.10547}}
  (\bibinfo{year}{2023}).
\newblock


\bibitem[Nagata(2019)]%
        {nagata2019toward}
\bibfield{author}{\bibinfo{person}{Ryo Nagata}.}
  \bibinfo{year}{2019}\natexlab{}.
\newblock \showarticletitle{Toward a Task of Feedback Comment Generation for
  Writing Learning}. In \bibinfo{booktitle}{\emph{Proceedings of the 2019
  Conference on Empirical Methods in Natural Language Processing and the 9th
  International Joint Conference on Natural Language Processing
  (EMNLP-IJCNLP)}}. \bibinfo{publisher}{Association for Computational
  Linguistics}, \bibinfo{address}{Hong Kong, China},
  \bibinfo{pages}{3206--3215}.
\newblock


\bibitem[Natalie et~al\mbox{.}(2023)]%
        {natalie2023supporting}
\bibfield{author}{\bibinfo{person}{Rosiana Natalie}, \bibinfo{person}{Joshua
  Tseng}, \bibinfo{person}{Hernisa Kacorri}, {and} \bibinfo{person}{Kotaro
  Hara}.} \bibinfo{year}{2023}\natexlab{}.
\newblock \showarticletitle{Supporting Novices Author Audio Descriptions via
  Automatic Feedback}. In \bibinfo{booktitle}{\emph{Proceedings of the 2023 CHI
  Conference on Human Factors in Computing Systems}} (Hamburg, Germany)
  \emph{(\bibinfo{series}{CHI '23})}. \bibinfo{publisher}{Association for
  Computing Machinery}, \bibinfo{address}{New York, NY, USA}, Article
  \bibinfo{articleno}{77}, \bibinfo{numpages}{18}~pages.
\newblock
\showISBNx{9781450394215}


\bibitem[Neate et~al\mbox{.}(2019)]%
        {neate2019empowering}
\bibfield{author}{\bibinfo{person}{Timothy Neate}, \bibinfo{person}{Abi Roper},
  \bibinfo{person}{Stephanie Wilson}, {and} \bibinfo{person}{Jane Marshall}.}
  \bibinfo{year}{2019}\natexlab{}.
\newblock \showarticletitle{Empowering Expression for Users with Aphasia
  through Constrained Creativity}. In \bibinfo{booktitle}{\emph{Proceedings of
  the 2019 CHI Conference on Human Factors in Computing Systems}} (Glasgow,
  Scotland Uk) \emph{(\bibinfo{series}{CHI '19})}.
  \bibinfo{publisher}{Association for Computing Machinery},
  \bibinfo{address}{New York, NY, USA}, \bibinfo{pages}{1–12}.
\newblock
\showISBNx{9781450359702}


\bibitem[Noy and Zhang(2023)]%
        {noy2023productivity}
\bibfield{author}{\bibinfo{person}{Shakked Noy} {and} \bibinfo{person}{Whitney
  Zhang}.} \bibinfo{year}{2023}\natexlab{}.
\newblock \showarticletitle{Experimental evidence on the productivity effects
  of generative artificial intelligence}.
\newblock \bibinfo{journal}{\emph{Science}} \bibinfo{volume}{381},
  \bibinfo{number}{6654} (\bibinfo{year}{2023}), \bibinfo{pages}{187--192}.
\newblock


\bibitem[of~America(2023)]%
        {wgastrike2023whatwewon}
\bibfield{author}{\bibinfo{person}{Writers~Guild of America}.}
  \bibinfo{year}{2023}\natexlab{}.
\newblock \bibinfo{title}{What We Won}.
\newblock
\newblock
\newblock
\shownote{Accessed: Dec. 12, 2023}.


\bibitem[Olson(2002)]%
        {olson2002writing}
\bibfield{author}{\bibinfo{person}{David~R Olson}.}
  \bibinfo{year}{2002}\natexlab{}.
\newblock \showarticletitle{What writing does to the mind}.
\newblock \bibinfo{journal}{\emph{Language, literacy, and cognitive
  development: The development and consequences of symbolic communication}}
  (\bibinfo{year}{2002}), \bibinfo{pages}{153--166}.
\newblock


\bibitem[{OpenAI}(2022)]%
        {openai2022chatgpt}
\bibfield{author}{\bibinfo{person}{{OpenAI}}.} \bibinfo{year}{2022}\natexlab{}.
\newblock \bibinfo{title}{Introducing {ChatGPT}}.
\newblock \bibinfo{howpublished}{\url{https://openai.com/blog/chatgpt}}.
\newblock


\bibitem[{OpenAI}(2023)]%
        {openai2023gpt4}
\bibfield{author}{\bibinfo{person}{{OpenAI}}.} \bibinfo{year}{2023}\natexlab{}.
\newblock \showarticletitle{{GPT}-4 Technical Report}.
\newblock \bibinfo{journal}{\emph{arXiv preprint arXiv:2303.08774}}
  (\bibinfo{year}{2023}).
\newblock


\bibitem[Osone et~al\mbox{.}(2021)]%
        {osone2021buncho}
\bibfield{author}{\bibinfo{person}{Hiroyuki Osone}, \bibinfo{person}{Jun-Li
  Lu}, {and} \bibinfo{person}{Yoichi Ochiai}.} \bibinfo{year}{2021}\natexlab{}.
\newblock \showarticletitle{BunCho: AI Supported Story Co-Creation via
  Unsupervised Multitask Learning to Increase Writers’ Creativity in
  Japanese}. In \bibinfo{booktitle}{\emph{Extended Abstracts of the 2021 CHI
  Conference on Human Factors in Computing Systems}} (Yokohama, Japan)
  \emph{(\bibinfo{series}{CHI EA '21})}. \bibinfo{publisher}{Association for
  Computing Machinery}, \bibinfo{address}{New York, NY, USA}, Article
  \bibinfo{articleno}{19}, \bibinfo{numpages}{10}~pages.
\newblock
\showISBNx{9781450380959}


\bibitem[Padmakumar and He(2022)]%
        {padmakumar2022machine}
\bibfield{author}{\bibinfo{person}{Vishakh Padmakumar} {and}
  \bibinfo{person}{He He}.} \bibinfo{year}{2022}\natexlab{}.
\newblock \showarticletitle{Machine-in-the-Loop Rewriting for Creative Image
  Captioning}. In \bibinfo{booktitle}{\emph{Proceedings of the 2022 Conference
  of the North American Chapter of the Association for Computational
  Linguistics: Human Language Technologies}}. \bibinfo{publisher}{Association
  for Computational Linguistics}, \bibinfo{address}{Seattle, United States},
  \bibinfo{pages}{573--586}.
\newblock


\bibitem[Padmakumar and He(2024)]%
        {padmakumar2024does}
\bibfield{author}{\bibinfo{person}{Vishakh Padmakumar} {and}
  \bibinfo{person}{He He}.} \bibinfo{year}{2024}\natexlab{}.
\newblock \showarticletitle{Does Writing with Language Models Reduce Content
  Diversity?}. In \bibinfo{booktitle}{\emph{The Twelfth International
  Conference on Learning Representations}}.
\newblock


\bibitem[Papineni et~al\mbox{.}(2002)]%
        {papineni02bleu}
\bibfield{author}{\bibinfo{person}{Kishore Papineni}, \bibinfo{person}{Salim
  Roukos}, \bibinfo{person}{Todd Ward}, {and} \bibinfo{person}{Wei-Jing Zhu}.}
  \bibinfo{year}{2002}\natexlab{}.
\newblock \showarticletitle{{BLEU}: A Method for Automatic Evaluation of
  Machine Translation}. In \bibinfo{booktitle}{\emph{Association for
  Computational Linguistics (ACL)}}.
\newblock


\bibitem[Park et~al\mbox{.}(2021)]%
        {park2021iwrote}
\bibfield{author}{\bibinfo{person}{SoHyun Park}, \bibinfo{person}{Anja Thieme},
  \bibinfo{person}{Jeongyun Han}, \bibinfo{person}{Sungwoo Lee},
  \bibinfo{person}{Wonjong Rhee}, {and} \bibinfo{person}{Bongwon Suh}.}
  \bibinfo{year}{2021}\natexlab{}.
\newblock \showarticletitle{“I Wrote as If I Were Telling a Story to Someone
  I Knew.”: Designing Chatbot Interactions for Expressive Writing in Mental
  Health}. In \bibinfo{booktitle}{\emph{Proceedings of the 2021 ACM Designing
  Interactive Systems Conference}} (Virtual Event, USA)
  \emph{(\bibinfo{series}{DIS '21})}. \bibinfo{publisher}{Association for
  Computing Machinery}, \bibinfo{address}{New York, NY, USA},
  \bibinfo{pages}{926–941}.
\newblock
\showISBNx{9781450384766}


\bibitem[Park et~al\mbox{.}(2008)]%
        {park2008is}
\bibfield{author}{\bibinfo{person}{Taehyun Park}, \bibinfo{person}{Edward
  Lank}, \bibinfo{person}{Pascal Poupart}, {and} \bibinfo{person}{Michael
  Terry}.} \bibinfo{year}{2008}\natexlab{}.
\newblock \showarticletitle{Is the Sky Pure Today? AwkChecker: An Assistive
  Tool for Detecting and Correcting Collocation Errors}. In
  \bibinfo{booktitle}{\emph{Proceedings of the 21st Annual ACM Symposium on
  User Interface Software and Technology}} (Monterey, CA, USA)
  \emph{(\bibinfo{series}{UIST '08})}. \bibinfo{publisher}{Association for
  Computing Machinery}, \bibinfo{address}{New York, NY, USA},
  \bibinfo{pages}{121–130}.
\newblock
\showISBNx{9781595939753}


\bibitem[Pea(2018)]%
        {pea2018social}
\bibfield{author}{\bibinfo{person}{Roy~D Pea}.}
  \bibinfo{year}{2018}\natexlab{}.
\newblock \showarticletitle{The social and technological dimensions of
  scaffolding and related theoretical concepts for learning, education, and
  human activity}.
\newblock In \bibinfo{booktitle}{\emph{Scaffolding}}.
  \bibinfo{publisher}{Psychology Press}, \bibinfo{pages}{423--451}.
\newblock


\bibitem[Pea and Kurland(1987)]%
        {pea1987cognitive}
\bibfield{author}{\bibinfo{person}{Roy~D. Pea} {and} \bibinfo{person}{D.~Midian
  Kurland}.} \bibinfo{year}{1987}\natexlab{}.
\newblock \showarticletitle{Chapter 7: Cognitive Technologies for Writing}.
\newblock \bibinfo{journal}{\emph{Review of Research in Education}}
  \bibinfo{volume}{14}, \bibinfo{number}{1} (\bibinfo{year}{1987}),
  \bibinfo{pages}{277--326}.
\newblock


\bibitem[Peng et~al\mbox{.}(2020)]%
        {peng2020exploring}
\bibfield{author}{\bibinfo{person}{Zhenhui Peng}, \bibinfo{person}{Qingyu Guo},
  \bibinfo{person}{Ka~Wing Tsang}, {and} \bibinfo{person}{Xiaojuan Ma}.}
  \bibinfo{year}{2020}\natexlab{}.
\newblock \showarticletitle{Exploring the Effects of Technological Writing
  Assistance for Support Providers in Online Mental Health Community}. In
  \bibinfo{booktitle}{\emph{Proceedings of the 2020 CHI Conference on Human
  Factors in Computing Systems}} (Honolulu, HI, USA)
  \emph{(\bibinfo{series}{CHI '20})}. \bibinfo{publisher}{Association for
  Computing Machinery}, \bibinfo{address}{New York, NY, USA},
  \bibinfo{pages}{1–15}.
\newblock
\showISBNx{9781450367080}


\bibitem[Perkins(2023)]%
        {perkins2023academic}
\bibfield{author}{\bibinfo{person}{Mike Perkins}.}
  \bibinfo{year}{2023}\natexlab{}.
\newblock \showarticletitle{Academic Integrity considerations of AI Large
  Language Models in the post-pandemic era: ChatGPT and beyond}.
\newblock \bibinfo{journal}{\emph{Journal of University Teaching \& Learning
  Practice}} \bibinfo{volume}{20}, \bibinfo{number}{2} (\bibinfo{year}{2023}),
  \bibinfo{pages}{07}.
\newblock


\bibitem[Poddar et~al\mbox{.}(2023)]%
        {poddar2023aiwriting}
\bibfield{author}{\bibinfo{person}{Ritika Poddar}, \bibinfo{person}{Rashmi
  Sinha}, \bibinfo{person}{Mor Naaman}, {and} \bibinfo{person}{Maurice
  Jakesch}.} \bibinfo{year}{2023}\natexlab{}.
\newblock \showarticletitle{AI Writing Assistants Influence Topic Choice in
  Self-Presentation}. In \bibinfo{booktitle}{\emph{Extended Abstracts of the
  2023 CHI Conference on Human Factors in Computing Systems}} (Hamburg,
  Germany) \emph{(\bibinfo{series}{CHI EA '23})}.
  \bibinfo{publisher}{Association for Computing Machinery},
  \bibinfo{address}{New York, NY, USA}, Article \bibinfo{articleno}{29},
  \bibinfo{numpages}{6}~pages.
\newblock
\showISBNx{9781450394222}


\bibitem[Post(2023)]%
        {washpost2023botupdate}
\bibfield{author}{\bibinfo{person}{Washington Post}.}
  \bibinfo{year}{2023}\natexlab{}.
\newblock \bibinfo{title}{They fell in love with AI bots. A software update
  broke their hearts}.
\newblock
\newblock
\newblock
\shownote{Accessed: Feb. 24, 2024}.


\bibitem[Qi et~al\mbox{.}(2022)]%
        {qi2022quoter}
\bibfield{author}{\bibinfo{person}{Fanchao Qi}, \bibinfo{person}{Yanhui Yang},
  \bibinfo{person}{Jing Yi}, \bibinfo{person}{Zhili Cheng},
  \bibinfo{person}{Zhiyuan Liu}, {and} \bibinfo{person}{Maosong Sun}.}
  \bibinfo{year}{2022}\natexlab{}.
\newblock \showarticletitle{{Q}uote{R}: A Benchmark of Quote Recommendation for
  Writing}. In \bibinfo{booktitle}{\emph{Proceedings of the 60th Annual Meeting
  of the Association for Computational Linguistics (Volume 1: Long Papers)}}.
  \bibinfo{publisher}{Association for Computational Linguistics},
  \bibinfo{address}{Dublin, Ireland}, \bibinfo{pages}{336--348}.
\newblock


\bibitem[Quinn and Zhai(2016)]%
        {Quinn2016costbenefit}
\bibfield{author}{\bibinfo{person}{Philip Quinn} {and} \bibinfo{person}{Shumin
  Zhai}.} \bibinfo{year}{2016}\natexlab{}.
\newblock \showarticletitle{A Cost-Benefit Study of Text Entry Suggestion
  Interaction}. In \bibinfo{booktitle}{\emph{Proceedings of the 2016 CHI
  Conference on Human Factors in Computing Systems}} (San Jose, California,
  USA) \emph{(\bibinfo{series}{CHI ’16})}. \bibinfo{publisher}{Association
  for Computing Machinery}, \bibinfo{address}{New York, NY, USA},
  \bibinfo{pages}{83–88}.
\newblock
\showISBNx{9781450333627}


\bibitem[Radford et~al\mbox{.}(2018)]%
        {radford2018improving}
\bibfield{author}{\bibinfo{person}{Alec Radford}, \bibinfo{person}{Karthik
  Narasimhan}, \bibinfo{person}{Tim Salimans}, {and} \bibinfo{person}{Ilya
  Sutskever}.} \bibinfo{year}{2018}\natexlab{}.
\newblock \bibinfo{booktitle}{\emph{Improving language understanding by
  generative pre-training}}.
\newblock \bibinfo{type}{{T}echnical {R}eport}. \bibinfo{institution}{OpenAI}.
\newblock


\bibitem[Radford et~al\mbox{.}(2019)]%
        {radford2019language}
\bibfield{author}{\bibinfo{person}{Alec Radford}, \bibinfo{person}{Jeff Wu},
  \bibinfo{person}{Rewon Child}, \bibinfo{person}{David Luan},
  \bibinfo{person}{Dario Amodei}, {and} \bibinfo{person}{Ilya Sutskever}.}
  \bibinfo{year}{2019}\natexlab{}.
\newblock \showarticletitle{Language Models are Unsupervised Multitask
  Learners}.
\newblock  (\bibinfo{year}{2019}).
\newblock


\bibitem[Raffel et~al\mbox{.}(2019)]%
        {raffel2019exploring}
\bibfield{author}{\bibinfo{person}{Colin Raffel}, \bibinfo{person}{Noam
  Shazeer}, \bibinfo{person}{Adam Roberts}, \bibinfo{person}{Katherine Lee},
  \bibinfo{person}{Sharan Narang}, \bibinfo{person}{Michael Matena},
  \bibinfo{person}{Yanqi Zhou}, \bibinfo{person}{Wei Li}, {and}
  \bibinfo{person}{Peter~J. Liu}.} \bibinfo{year}{2019}\natexlab{}.
\newblock \showarticletitle{Exploring the limits of transfer learning with a
  unified text-to-text transformer}.
\newblock \bibinfo{journal}{\emph{arXiv preprint arXiv:1910.10683}}
  (\bibinfo{year}{2019}).
\newblock


\bibitem[Raheja et~al\mbox{.}(2023)]%
        {raheja2023coedit}
\bibfield{author}{\bibinfo{person}{Vipul Raheja}, \bibinfo{person}{Dhruv
  Kumar}, \bibinfo{person}{Ryan Koo}, {and} \bibinfo{person}{Dongyeop Kang}.}
  \bibinfo{year}{2023}\natexlab{}.
\newblock \showarticletitle{CoEdIT: Text Editing by Task-Specific Instruction
  Tuning}.
\newblock  (\bibinfo{year}{2023}).
\newblock
\showeprint[arxiv]{2305.09857}~[cs.CL]


\bibitem[Rajagopal et~al\mbox{.}(2022)]%
        {rajagopal-etal-2022-one}
\bibfield{author}{\bibinfo{person}{Dheeraj Rajagopal}, \bibinfo{person}{Xuchao
  Zhang}, \bibinfo{person}{Michael Gamon}, \bibinfo{person}{Sujay~Kumar
  Jauhar}, \bibinfo{person}{Diyi Yang}, {and} \bibinfo{person}{Eduard Hovy}.}
  \bibinfo{year}{2022}\natexlab{}.
\newblock \showarticletitle{One Document, Many Revisions: A Dataset for
  Classification and Description of Edit Intents}. In
  \bibinfo{booktitle}{\emph{Proceedings of the Thirteenth Language Resources
  and Evaluation Conference}}, \bibfield{editor}{\bibinfo{person}{Nicoletta
  Calzolari}, \bibinfo{person}{Fr{\'e}d{\'e}ric B{\'e}chet},
  \bibinfo{person}{Philippe Blache}, \bibinfo{person}{Khalid Choukri},
  \bibinfo{person}{Christopher Cieri}, \bibinfo{person}{Thierry Declerck},
  \bibinfo{person}{Sara Goggi}, \bibinfo{person}{Hitoshi Isahara},
  \bibinfo{person}{Bente Maegaard}, \bibinfo{person}{Joseph Mariani},
  \bibinfo{person}{H{\'e}l{\`e}ne Mazo}, \bibinfo{person}{Jan Odijk}, {and}
  \bibinfo{person}{Stelios Piperidis}} (Eds.). \bibinfo{publisher}{European
  Language Resources Association}, \bibinfo{address}{Marseille, France},
  \bibinfo{pages}{5517--5524}.
\newblock


\bibitem[Rapp et~al\mbox{.}(2015)]%
        {rapp2015thesis}
\bibfield{author}{\bibinfo{person}{Christian Rapp}, \bibinfo{person}{Otto
  Kruse}, \bibinfo{person}{Jennifer Erlemann}, {and} \bibinfo{person}{Jakob
  Ott}.} \bibinfo{year}{2015}\natexlab{}.
\newblock \showarticletitle{Thesis Writer: A System for Supporting Academic
  Writing}. In \bibinfo{booktitle}{\emph{Proceedings of the 18th ACM Conference
  Companion on Computer Supported Cooperative Work \& Social Computing}}
  (Vancouver, BC, Canada) \emph{(\bibinfo{series}{CSCW'15 Companion})}.
  \bibinfo{publisher}{Association for Computing Machinery},
  \bibinfo{address}{New York, NY, USA}, \bibinfo{pages}{57–60}.
\newblock
\showISBNx{9781450329460}


\bibitem[Rei and Yannakoudakis(2016)]%
        {rei2016compositional}
\bibfield{author}{\bibinfo{person}{Marek Rei} {and} \bibinfo{person}{Helen
  Yannakoudakis}.} \bibinfo{year}{2016}\natexlab{}.
\newblock \showarticletitle{Compositional Sequence Labeling Models for Error
  Detection in Learner Writing}. In \bibinfo{booktitle}{\emph{Proceedings of
  the 54th Annual Meeting of the Association for Computational Linguistics
  (Volume 1: Long Papers)}}. \bibinfo{publisher}{Association for Computational
  Linguistics}, \bibinfo{address}{Berlin, Germany},
  \bibinfo{pages}{1181--1191}.
\newblock


\bibitem[Reid and Neubig(2022)]%
        {reid-neubig-2022-learning}
\bibfield{author}{\bibinfo{person}{Machel Reid} {and} \bibinfo{person}{Graham
  Neubig}.} \bibinfo{year}{2022}\natexlab{}.
\newblock \showarticletitle{Learning to Model Editing Processes}. In
  \bibinfo{booktitle}{\emph{Findings of the Association for Computational
  Linguistics: EMNLP 2022}}, \bibfield{editor}{\bibinfo{person}{Yoav Goldberg},
  \bibinfo{person}{Zornitsa Kozareva}, {and} \bibinfo{person}{Yue Zhang}}
  (Eds.). \bibinfo{publisher}{Association for Computational Linguistics},
  \bibinfo{address}{Abu Dhabi, United Arab Emirates},
  \bibinfo{pages}{3822--3832}.
\newblock


\bibitem[Reuters(2023)]%
        {reuters2023openai}
\bibfield{author}{\bibinfo{person}{Reuters}.} \bibinfo{year}{2023}\natexlab{}.
\newblock \bibinfo{title}{OpenAI, Microsoft hit with new US consumer privacy
  class action}.
\newblock
\newblock
\newblock
\shownote{Accessed: Dec 12, 2023}.


\bibitem[Robertson et~al\mbox{.}(2021)]%
        {robertson2021icant}
\bibfield{author}{\bibinfo{person}{Ronald~E Robertson},
  \bibinfo{person}{Alexandra Olteanu}, \bibinfo{person}{Fernando Diaz},
  \bibinfo{person}{Milad Shokouhi}, {and} \bibinfo{person}{Peter Bailey}.}
  \bibinfo{year}{2021}\natexlab{}.
\newblock \showarticletitle{“I Can’t Reply with That”: Characterizing
  Problematic Email Reply Suggestions}. In
  \bibinfo{booktitle}{\emph{Proceedings of the 2021 CHI Conference on Human
  Factors in Computing Systems}} (Yokohama, Japan) \emph{(\bibinfo{series}{CHI
  '21})}. \bibinfo{publisher}{Association for Computing Machinery},
  \bibinfo{address}{New York, NY, USA}, Article \bibinfo{articleno}{724},
  \bibinfo{numpages}{18}~pages.
\newblock
\showISBNx{9781450380966}


\bibitem[Roemmele and Gordon(2018)]%
        {roemmele2018automated}
\bibfield{author}{\bibinfo{person}{Melissa Roemmele} {and}
  \bibinfo{person}{Andrew~S. Gordon}.} \bibinfo{year}{2018}\natexlab{}.
\newblock \showarticletitle{Automated Assistance for Creative Writing with an
  RNN Language Model}. In \bibinfo{booktitle}{\emph{Proceedings of the 23rd
  International Conference on Intelligent User Interfaces Companion}} (Tokyo,
  Japan) \emph{(\bibinfo{series}{IUI '18 Companion})}.
  \bibinfo{publisher}{Association for Computing Machinery},
  \bibinfo{address}{New York, NY, USA}, Article \bibinfo{articleno}{21},
  \bibinfo{numpages}{2}~pages.
\newblock
\showISBNx{9781450355711}


\bibitem[Rogers(2012)]%
        {rogers2012hcitheory}
\bibfield{author}{\bibinfo{person}{Yvonne Rogers}.}
  \bibinfo{year}{2012}\natexlab{}.
\newblock \bibinfo{booktitle}{\emph{HCI Theory: Classical, Modern, and
  Contemporary}}.
\newblock \bibinfo{publisher}{Morgan \& Claypool Publishers}.
\newblock
\showISBNx{1608459004}


\bibitem[Rohman(1965)]%
        {rohman1965pre}
\bibfield{author}{\bibinfo{person}{D~Gordon Rohman}.}
  \bibinfo{year}{1965}\natexlab{}.
\newblock \showarticletitle{Pre-writing the stage of discovery in the writing
  process}.
\newblock \bibinfo{journal}{\emph{College composition and communication}}
  \bibinfo{volume}{16}, \bibinfo{number}{2} (\bibinfo{year}{1965}),
  \bibinfo{pages}{106--112}.
\newblock


\bibitem[Romer and Mattern(2004)]%
        {romer2004thedesignspace}
\bibfield{author}{\bibinfo{person}{K. Romer} {and} \bibinfo{person}{F.
  Mattern}.} \bibinfo{year}{2004}\natexlab{}.
\newblock \showarticletitle{The design space of wireless sensor networks}.
\newblock \bibinfo{journal}{\emph{IEEE Wireless Communications}}
  \bibinfo{volume}{11}, \bibinfo{number}{6} (\bibinfo{year}{2004}),
  \bibinfo{pages}{54--61}.
\newblock


\bibitem[Roscoe et~al\mbox{.}(2014)]%
        {roscoe2014writing}
\bibfield{author}{\bibinfo{person}{Rod~D Roscoe}, \bibinfo{person}{Laura~K
  Allen}, \bibinfo{person}{Jennifer~L Weston}, \bibinfo{person}{Scott~A
  Crossley}, {and} \bibinfo{person}{Danielle~S McNamara}.}
  \bibinfo{year}{2014}\natexlab{}.
\newblock \showarticletitle{The Writing Pal intelligent tutoring system:
  Usability testing and development}.
\newblock \bibinfo{journal}{\emph{Computers and Composition}}
  \bibinfo{volume}{34} (\bibinfo{year}{2014}), \bibinfo{pages}{39--59}.
\newblock


\bibitem[Sadauskas et~al\mbox{.}(2015)]%
        {sadauskas2015mining}
\bibfield{author}{\bibinfo{person}{John Sadauskas}, \bibinfo{person}{Daragh
  Byrne}, {and} \bibinfo{person}{Robert~K. Atkinson}.}
  \bibinfo{year}{2015}\natexlab{}.
\newblock \showarticletitle{Mining Memories: Designing a Platform to Support
  Social Media Based Writing}. In \bibinfo{booktitle}{\emph{Proceedings of the
  33rd Annual ACM Conference on Human Factors in Computing Systems}} (Seoul,
  Republic of Korea) \emph{(\bibinfo{series}{CHI '15})}.
  \bibinfo{publisher}{Association for Computing Machinery},
  \bibinfo{address}{New York, NY, USA}, \bibinfo{pages}{3691–3700}.
\newblock
\showISBNx{9781450331456}


\bibitem[Sawyer and Jarrahi(2014)]%
        {sawyer2014sociotechnical}
\bibfield{author}{\bibinfo{person}{Steve Sawyer} {and}
  \bibinfo{person}{Mohammad~Hossein Jarrahi}.} \bibinfo{year}{2014}\natexlab{}.
\newblock \showarticletitle{Sociotechnical approaches to the study of
  information systems}.
\newblock In \bibinfo{booktitle}{\emph{Computing handbook, third edition:
  Information systems and information technology}}. \bibinfo{publisher}{CRC
  Press}, \bibinfo{pages}{5--1}.
\newblock


\bibitem[Schick et~al\mbox{.}(2022)]%
        {schick2022peer}
\bibfield{author}{\bibinfo{person}{Timo Schick}, \bibinfo{person}{Jane
  Dwivedi-Yu}, \bibinfo{person}{Zhengbao Jiang}, \bibinfo{person}{Fabio
  Petroni}, \bibinfo{person}{Patrick Lewis}, \bibinfo{person}{Gautier Izacard},
  \bibinfo{person}{Qingfei You}, \bibinfo{person}{Christoforos Nalmpantis},
  \bibinfo{person}{Edouard Grave}, {and} \bibinfo{person}{Sebastian Riedel}.}
  \bibinfo{year}{2022}\natexlab{}.
\newblock \bibinfo{title}{PEER: A Collaborative Language Model}.
\newblock
\newblock
\showeprint[arxiv]{2208.11663}~[cs.CL]


\bibitem[Schmandt-Besserat(1992)]%
        {schmandt1992before}
\bibfield{author}{\bibinfo{person}{Denise Schmandt-Besserat}.}
  \bibinfo{year}{1992}\natexlab{}.
\newblock \bibinfo{booktitle}{\emph{Before writing, vol. I: from counting to
  cuneiform}}. Vol.~\bibinfo{volume}{1}.
\newblock \bibinfo{publisher}{University of Texas press}.
\newblock


\bibitem[Schmitt and Buschek(2021)]%
        {schmitt2021characterchat}
\bibfield{author}{\bibinfo{person}{Oliver Schmitt} {and}
  \bibinfo{person}{Daniel Buschek}.} \bibinfo{year}{2021}\natexlab{}.
\newblock \showarticletitle{CharacterChat: Supporting the Creation of Fictional
  Characters through Conversation and Progressive Manifestation with a
  Chatbot}. In \bibinfo{booktitle}{\emph{Proceedings of the 13th Conference on
  Creativity and Cognition}} (Virtual Event, Italy)
  \emph{(\bibinfo{series}{C\&C '21})}. \bibinfo{publisher}{Association for
  Computing Machinery}, \bibinfo{address}{New York, NY, USA}, Article
  \bibinfo{articleno}{10}, \bibinfo{numpages}{10}~pages.
\newblock
\showISBNx{9781450383769}


\bibitem[Schneider and McCoy(1998)]%
        {schneider1998recognizing}
\bibfield{author}{\bibinfo{person}{David Schneider} {and}
  \bibinfo{person}{Kathleen~F. McCoy}.} \bibinfo{year}{1998}\natexlab{}.
\newblock \showarticletitle{Recognizing Syntactic Errors in the Writing of
  Second Language Learners}. In \bibinfo{booktitle}{\emph{Proceedings of the
  36th Annual Meeting of the Association for Computational Linguistics and 17th
  International Conference on Computational Linguistics - Volume 2}} (Montreal,
  Quebec, Canada) \emph{(\bibinfo{series}{ACL '98/COLING '98})}.
  \bibinfo{publisher}{Association for Computational Linguistics},
  \bibinfo{address}{USA}, \bibinfo{pages}{1198–1204}.
\newblock


\bibitem[Sellam et~al\mbox{.}(2020)]%
        {sellam2020bleurt}
\bibfield{author}{\bibinfo{person}{Thibault Sellam}, \bibinfo{person}{Dipanjan
  Das}, {and} \bibinfo{person}{Ankur~P Parikh}.}
  \bibinfo{year}{2020}\natexlab{}.
\newblock \showarticletitle{BLEURT: Learning robust metrics for text
  generation}.
\newblock \bibinfo{journal}{\emph{arXiv preprint arXiv:2004.04696}}
  (\bibinfo{year}{2020}).
\newblock


\bibitem[Shi et~al\mbox{.}(2023)]%
        {shi-etal-2023-effidit}
\bibfield{author}{\bibinfo{person}{Shuming Shi}, \bibinfo{person}{Enbo Zhao},
  \bibinfo{person}{Wei Bi}, \bibinfo{person}{Deng Cai}, \bibinfo{person}{Leyang
  Cui}, \bibinfo{person}{Xinting Huang}, \bibinfo{person}{Haiyun Jiang},
  \bibinfo{person}{Duyu Tang}, \bibinfo{person}{Kaiqiang Song},
  \bibinfo{person}{Longyue Wang}, \bibinfo{person}{Chenyan Huang},
  \bibinfo{person}{Guoping Huang}, \bibinfo{person}{Yan Wang}, {and}
  \bibinfo{person}{Piji Li}.} \bibinfo{year}{2023}\natexlab{}.
\newblock \showarticletitle{Effidit: An Assistant for Improving Writing
  Efficiency}. In \bibinfo{booktitle}{\emph{Proceedings of the 61st Annual
  Meeting of the Association for Computational Linguistics (Volume 3: System
  Demonstrations)}}. \bibinfo{publisher}{Association for Computational
  Linguistics}, \bibinfo{address}{Toronto, Canada}, \bibinfo{pages}{508--515}.
\newblock


\bibitem[Shibani et~al\mbox{.}(2020)]%
        {shibani2020educator}
\bibfield{author}{\bibinfo{person}{Antonette Shibani}, \bibinfo{person}{Simon
  Knight}, {and} \bibinfo{person}{Simon~Buckingham Shum}.}
  \bibinfo{year}{2020}\natexlab{}.
\newblock \showarticletitle{Educator perspectives on learning analytics in
  classroom practice}.
\newblock \bibinfo{journal}{\emph{The Internet and Higher Education}}
  \bibinfo{volume}{46} (\bibinfo{year}{2020}), \bibinfo{pages}{100730}.
\newblock


\bibitem[Shneiderman(2011)]%
        {shneiderman2012charting}
\bibfield{author}{\bibinfo{person}{Ben Shneiderman}.}
  \bibinfo{year}{2011}\natexlab{}.
\newblock \showarticletitle{Claiming success, charting the future}.
\newblock \bibinfo{journal}{\emph{Interactions}} \bibinfo{volume}{18},
  \bibinfo{number}{5} (\bibinfo{year}{2011}), \bibinfo{pages}{10--11}.
\newblock
\showISSN{1072-5520}


\bibitem[Shu et~al\mbox{.}(2023)]%
        {shu2023rewritelm}
\bibfield{author}{\bibinfo{person}{Lei Shu}, \bibinfo{person}{Liangchen Luo},
  \bibinfo{person}{Jayakumar Hoskere}, \bibinfo{person}{Yun Zhu},
  \bibinfo{person}{Canoee Liu}, \bibinfo{person}{Simon Tong},
  \bibinfo{person}{Jindong Chen}, {and} \bibinfo{person}{Lei Meng}.}
  \bibinfo{year}{2023}\natexlab{}.
\newblock \bibinfo{title}{RewriteLM: An Instruction-Tuned Large Language Model
  for Text Rewriting}.
\newblock
\newblock
\showeprint[arxiv]{2305.15685}~[cs.CL]


\bibitem[Shumailov et~al\mbox{.}(2023)]%
        {shumailov2023curse}
\bibfield{author}{\bibinfo{person}{Ilia Shumailov}, \bibinfo{person}{Zakhar
  Shumaylov}, \bibinfo{person}{Yiren Zhao}, \bibinfo{person}{Yarin Gal},
  \bibinfo{person}{Nicolas Papernot}, {and} \bibinfo{person}{Ross Anderson}.}
  \bibinfo{year}{2023}\natexlab{}.
\newblock \bibinfo{title}{The Curse of Recursion: Training on Generated Data
  Makes Models Forget}.
\newblock
\newblock
\showeprint[arxiv]{2305.17493}~[cs.LG]


\bibitem[Singh et~al\mbox{.}(2021)]%
        {singh-etal-2021-drag}
\bibfield{author}{\bibinfo{person}{Hrituraj Singh}, \bibinfo{person}{Gaurav
  Verma}, \bibinfo{person}{Aparna Garimella}, {and}
  \bibinfo{person}{Balaji~Vasan Srinivasan}.} \bibinfo{year}{2021}\natexlab{}.
\newblock \showarticletitle{{DRAG}: Director-Generator Language Modelling
  Framework for Non-Parallel Author Stylized Rewriting}. In
  \bibinfo{booktitle}{\emph{Proceedings of the 16th Conference of the European
  Chapter of the Association for Computational Linguistics: Main Volume}}.
  \bibinfo{publisher}{Association for Computational Linguistics},
  \bibinfo{address}{Online}, \bibinfo{pages}{863--873}.
\newblock


\bibitem[Singh et~al\mbox{.}(2022)]%
        {singh2022where}
\bibfield{author}{\bibinfo{person}{Nikhil Singh}, \bibinfo{person}{Guillermo
  Bernal}, \bibinfo{person}{Daria Savchenko}, {and} \bibinfo{person}{Elena~L.
  Glassman}.} \bibinfo{year}{2022}\natexlab{}.
\newblock \showarticletitle{Where to Hide a Stolen Elephant: Leaps in Creative
  Writing with Multimodal Machine Intelligence}.
\newblock \bibinfo{journal}{\emph{ACM Trans. Comput.-Hum. Interact.}}
  (\bibinfo{date}{feb} \bibinfo{year}{2022}).
\newblock
\showISSN{1073-0516}
\newblock
\shownote{Just Accepted}.


\bibitem[Skitalinskaya and Wachsmuth(2023)]%
        {skitalinskaya2023revise}
\bibfield{author}{\bibinfo{person}{Gabriella Skitalinskaya} {and}
  \bibinfo{person}{Henning Wachsmuth}.} \bibinfo{year}{2023}\natexlab{}.
\newblock \showarticletitle{To Revise or Not to Revise: Learning to Detect
  Improvable Claims for Argumentative Writing Support}. In
  \bibinfo{booktitle}{\emph{Proceedings of the 61st Annual Meeting of the
  Association for Computational Linguistics (Volume 1: Long Papers)}}.
  \bibinfo{publisher}{Association for Computational Linguistics},
  \bibinfo{address}{Toronto, Canada}, \bibinfo{pages}{15799--15816}.
\newblock


\bibitem[Somasundaran et~al\mbox{.}(2018)]%
        {somasundaran2018towards}
\bibfield{author}{\bibinfo{person}{Swapna Somasundaran},
  \bibinfo{person}{Michael Flor}, \bibinfo{person}{Martin Chodorow},
  \bibinfo{person}{Hillary Molloy}, \bibinfo{person}{Binod Gyawali}, {and}
  \bibinfo{person}{Laura McCulla}.} \bibinfo{year}{2018}\natexlab{}.
\newblock \showarticletitle{Towards Evaluating Narrative Quality In Student
  Writing}.
\newblock \bibinfo{journal}{\emph{Transactions of the Association for
  Computational Linguistics}}  \bibinfo{volume}{6} (\bibinfo{year}{2018}),
  \bibinfo{pages}{91--106}.
\newblock


\bibitem[Soyer et~al\mbox{.}(2015)]%
        {soyer2015crovewa}
\bibfield{author}{\bibinfo{person}{Hubert Soyer}, \bibinfo{person}{Goran
  Topi{\'c}}, \bibinfo{person}{Pontus Stenetorp}, {and} \bibinfo{person}{Akiko
  Aizawa}.} \bibinfo{year}{2015}\natexlab{}.
\newblock \showarticletitle{{C}ro{V}e{WA}: Crosslingual Vector-Based Writing
  Assistance}. In \bibinfo{booktitle}{\emph{Proceedings of the 2015 Conference
  of the North {A}merican Chapter of the Association for Computational
  Linguistics: Demonstrations}}. \bibinfo{publisher}{Association for
  Computational Linguistics}, \bibinfo{address}{Denver, Colorado},
  \bibinfo{pages}{91--95}.
\newblock


\bibitem[Stahlberg and Kumar(2020)]%
        {stahlberg-kumar-2020-seq2edits}
\bibfield{author}{\bibinfo{person}{Felix Stahlberg} {and}
  \bibinfo{person}{Shankar Kumar}.} \bibinfo{year}{2020}\natexlab{}.
\newblock \showarticletitle{{S}eq2{E}dits: Sequence Transduction Using
  Span-level Edit Operations}. In \bibinfo{booktitle}{\emph{Proceedings of the
  2020 Conference on Empirical Methods in Natural Language Processing
  (EMNLP)}}. \bibinfo{publisher}{Association for Computational Linguistics},
  \bibinfo{address}{Online}, \bibinfo{pages}{5147--5159}.
\newblock


\bibitem[Strickland and Morrow(1989)]%
        {strickland1989emerging}
\bibfield{author}{\bibinfo{person}{Dorothy~S Strickland} {and}
  \bibinfo{person}{Lesley~Mandel Morrow}.} \bibinfo{year}{1989}\natexlab{}.
\newblock \bibinfo{booktitle}{\emph{Emerging literacy: Young children learn to
  read and write.}}
\newblock \bibinfo{publisher}{ERIC}.
\newblock


\bibitem[Su et~al\mbox{.}(2023)]%
        {su-etal-2023-reviewriter}
\bibfield{author}{\bibinfo{person}{Xiaotian Su}, \bibinfo{person}{Thiemo
  Wambsganss}, \bibinfo{person}{Roman Rietsche}, \bibinfo{person}{Seyed~Parsa
  Neshaei}, {and} \bibinfo{person}{Tanja K{\"a}ser}.}
  \bibinfo{year}{2023}\natexlab{}.
\newblock \showarticletitle{Reviewriter: {AI}-Generated Instructions For Peer
  Review Writing}. In \bibinfo{booktitle}{\emph{Proceedings of the 18th
  Workshop on Innovative Use of NLP for Building Educational Applications (BEA
  2023)}}, \bibfield{editor}{\bibinfo{person}{Ekaterina Kochmar},
  \bibinfo{person}{Jill Burstein}, \bibinfo{person}{Andrea Horbach},
  \bibinfo{person}{Ronja Laarmann-Quante}, \bibinfo{person}{Nitin Madnani},
  \bibinfo{person}{Ana{\"\i}s Tack}, \bibinfo{person}{Victoria Yaneva},
  \bibinfo{person}{Zheng Yuan}, {and} \bibinfo{person}{Torsten Zesch}} (Eds.).
  \bibinfo{publisher}{Association for Computational Linguistics},
  \bibinfo{address}{Toronto, Canada}, \bibinfo{pages}{57--71}.
\newblock


\bibitem[Sulzby et~al\mbox{.}(1989)]%
        {sulzby1989emergent}
\bibfield{author}{\bibinfo{person}{Elizabeth Sulzby},
  \bibinfo{person}{William~H Teale}, {and} \bibinfo{person}{George
  Kamberelis}.} \bibinfo{year}{1989}\natexlab{}.
\newblock \showarticletitle{Emergent writing in the classroom: Home and school
  connections}.
\newblock \bibinfo{journal}{\emph{Emerging literacy: Young children learn to
  read and write}} (\bibinfo{year}{1989}), \bibinfo{pages}{63--79}.
\newblock


\bibitem[Sun et~al\mbox{.}(2021)]%
        {sun-etal-2021-iga}
\bibfield{author}{\bibinfo{person}{Simeng Sun}, \bibinfo{person}{Wenlong Zhao},
  \bibinfo{person}{Varun Manjunatha}, \bibinfo{person}{Rajiv Jain},
  \bibinfo{person}{Vlad Morariu}, \bibinfo{person}{Franck Dernoncourt},
  \bibinfo{person}{Balaji~Vasan Srinivasan}, {and} \bibinfo{person}{Mohit
  Iyyer}.} \bibinfo{year}{2021}\natexlab{}.
\newblock \showarticletitle{{IGA}: An Intent-Guided Authoring Assistant}. In
  \bibinfo{booktitle}{\emph{Proceedings of the 2021 Conference on Empirical
  Methods in Natural Language Processing}},
  \bibfield{editor}{\bibinfo{person}{Marie-Francine Moens},
  \bibinfo{person}{Xuanjing Huang}, \bibinfo{person}{Lucia Specia}, {and}
  \bibinfo{person}{Scott Wen-tau Yih}} (Eds.). \bibinfo{publisher}{Association
  for Computational Linguistics}, \bibinfo{address}{Online and Punta Cana,
  Dominican Republic}, \bibinfo{pages}{5972--5985}.
\newblock


\bibitem[Sun et~al\mbox{.}(2023)]%
        {sun2023songrewriter}
\bibfield{author}{\bibinfo{person}{Yusen Sun}, \bibinfo{person}{Liangyou Li},
  \bibinfo{person}{Qun Liu}, {and} \bibinfo{person}{Dit-Yan Yeung}.}
  \bibinfo{year}{2023}\natexlab{}.
\newblock \showarticletitle{{S}ong{R}ewriter: A {C}hinese Song Rewriting System
  with Controllable Content and Rhyme Scheme}. In
  \bibinfo{booktitle}{\emph{Findings of the Association for Computational
  Linguistics: ACL 2023}}. \bibinfo{publisher}{Association for Computational
  Linguistics}, \bibinfo{address}{Toronto, Canada},
  \bibinfo{pages}{12863--12880}.
\newblock


\bibitem[Sutskever et~al\mbox{.}(2014)]%
        {sutskever2014sequence}
\bibfield{author}{\bibinfo{person}{Ilya Sutskever}, \bibinfo{person}{Oriol
  Vinyals}, {and} \bibinfo{person}{Quoc~V. Le}.}
  \bibinfo{year}{2014}\natexlab{}.
\newblock \showarticletitle{Sequence to sequence learning with neural
  networks}. In \bibinfo{booktitle}{\emph{Advances in Neural Information
  Processing Systems (NeurIPS)}}. \bibinfo{pages}{3104--3112}.
\newblock


\bibitem[Suzuki et~al\mbox{.}(2022)]%
        {suzuki2022emotional}
\bibfield{author}{\bibinfo{person}{Haruya Suzuki}, \bibinfo{person}{Sora
  Tarumoto}, \bibinfo{person}{Tomoyuki Kajiwara}, \bibinfo{person}{Takashi
  Ninomiya}, \bibinfo{person}{Yuta Nakashima}, {and} \bibinfo{person}{Hajime
  Nagahara}.} \bibinfo{year}{2022}\natexlab{}.
\newblock \showarticletitle{Emotional Intensity Estimation based on Writer{'}s
  Personality}. In \bibinfo{booktitle}{\emph{Proceedings of the 2nd Conference
  of the Asia-Pacific Chapter of the Association for Computational Linguistics
  and the 12th International Joint Conference on Natural Language Processing:
  Student Research Workshop}}. \bibinfo{publisher}{Association for
  Computational Linguistics}, \bibinfo{address}{Online}, \bibinfo{pages}{1--7}.
\newblock


\bibitem[Swanson et~al\mbox{.}(2021)]%
        {swanson2021story}
\bibfield{author}{\bibinfo{person}{Ben Swanson}, \bibinfo{person}{Kory
  Mathewson}, \bibinfo{person}{Ben Pietrzak}, \bibinfo{person}{Sherol Chen},
  {and} \bibinfo{person}{Monica Dinalescu}.} \bibinfo{year}{2021}\natexlab{}.
\newblock \showarticletitle{Story Centaur: Large Language Model Few Shot
  Learning as a Creative Writing Tool}. In
  \bibinfo{booktitle}{\emph{Proceedings of the 16th Conference of the European
  Chapter of the Association for Computational Linguistics: System
  Demonstrations}}. \bibinfo{publisher}{Association for Computational
  Linguistics}, \bibinfo{address}{Online}, \bibinfo{pages}{244--256}.
\newblock


\bibitem[Taori et~al\mbox{.}(2023)]%
        {alpaca}
\bibfield{author}{\bibinfo{person}{Rohan Taori}, \bibinfo{person}{Ishaan
  Gulrajani}, \bibinfo{person}{Tianyi Zhang}, \bibinfo{person}{Yann Dubois},
  \bibinfo{person}{Xuechen Li}, \bibinfo{person}{Carlos Guestrin},
  \bibinfo{person}{Percy Liang}, {and} \bibinfo{person}{Tatsunori~B.
  Hashimoto}.} \bibinfo{year}{2023}\natexlab{}.
\newblock \bibinfo{title}{Stanford Alpaca: An Instruction-following LLaMA
  model}.
\newblock
  \bibinfo{howpublished}{\url{https://github.com/tatsu-lab/stanford_alpaca}}.
\newblock


\bibitem[Taori and Hashimoto(2023)]%
        {taori2023data}
\bibfield{author}{\bibinfo{person}{Rohan Taori} {and}
  \bibinfo{person}{Tatsunori Hashimoto}.} \bibinfo{year}{2023}\natexlab{}.
\newblock \showarticletitle{Data feedback loops: Model-driven amplification of
  dataset biases}. In \bibinfo{booktitle}{\emph{International Conference on
  Machine Learning}}. PMLR, \bibinfo{pages}{33883--33920}.
\newblock


\bibitem[Touvron et~al\mbox{.}(2023)]%
        {touvron2023llama}
\bibfield{author}{\bibinfo{person}{Hugo Touvron}, \bibinfo{person}{Thibaut
  Lavril}, \bibinfo{person}{Gautier Izacard}, \bibinfo{person}{Xavier
  Martinet}, \bibinfo{person}{Marie-Anne Lachaux}, \bibinfo{person}{Timothée
  Lacroix}, \bibinfo{person}{Baptiste Rozière}, \bibinfo{person}{Naman Goyal},
  \bibinfo{person}{Eric Hambro}, \bibinfo{person}{Faisal Azhar},
  \bibinfo{person}{Aurelien Rodriguez}, \bibinfo{person}{Armand Joulin},
  \bibinfo{person}{Edouard Grave}, {and} \bibinfo{person}{Guillaume Lample}.}
  \bibinfo{year}{2023}\natexlab{}.
\newblock \showarticletitle{LLaMA: Open and Efficient Foundation Language
  Models}.
\newblock \bibinfo{journal}{\emph{arXiv}} (\bibinfo{year}{2023}).
\newblock


\bibitem[Trist(1981)]%
        {trist1981evolution}
\bibfield{author}{\bibinfo{person}{Eric~L Trist}.}
  \bibinfo{year}{1981}\natexlab{}.
\newblock \bibinfo{booktitle}{\emph{The evolution of socio-technical systems}}.
  Vol.~\bibinfo{volume}{2}.
\newblock \bibinfo{publisher}{Ontario Quality of Working Life Centre Toronto}.
\newblock


\bibitem[Tsai et~al\mbox{.}(2020)]%
        {tsai2020lingglewrite}
\bibfield{author}{\bibinfo{person}{Chung-Ting Tsai}, \bibinfo{person}{Jhih-Jie
  Chen}, \bibinfo{person}{Ching-Yu Yang}, {and} \bibinfo{person}{Jason~S.
  Chang}.} \bibinfo{year}{2020}\natexlab{}.
\newblock \showarticletitle{{L}inggle{W}rite: a Coaching System for Essay
  Writing}. In \bibinfo{booktitle}{\emph{Proceedings of the 58th Annual Meeting
  of the Association for Computational Linguistics: System Demonstrations}}.
  \bibinfo{publisher}{Association for Computational Linguistics},
  \bibinfo{address}{Online}, \bibinfo{pages}{127--133}.
\newblock


\bibitem[T\"{u}rkay et~al\mbox{.}(2018)]%
        {turkay2018itero}
\bibfield{author}{\bibinfo{person}{Selen T\"{u}rkay}, \bibinfo{person}{Daniel
  Seaton}, {and} \bibinfo{person}{Andrew~M. Ang}.}
  \bibinfo{year}{2018}\natexlab{}.
\newblock \showarticletitle{Itero: A Revision History Analytics Tool for
  Exploring Writing Behavior and Reflection}. In
  \bibinfo{booktitle}{\emph{Extended Abstracts of the 2018 CHI Conference on
  Human Factors in Computing Systems}} (Montreal QC, Canada)
  \emph{(\bibinfo{series}{CHI EA '18})}. \bibinfo{publisher}{Association for
  Computing Machinery}, \bibinfo{address}{New York, NY, USA},
  \bibinfo{pages}{1–6}.
\newblock
\showISBNx{9781450356213}


\bibitem[Vertanen et~al\mbox{.}(2015)]%
        {Vertanen2015velocitap}
\bibfield{author}{\bibinfo{person}{Keith Vertanen}, \bibinfo{person}{Haythem
  Memmi}, \bibinfo{person}{Justin Emge}, \bibinfo{person}{Shyam Reyal}, {and}
  \bibinfo{person}{Per~Ola Kristensson}.} \bibinfo{year}{2015}\natexlab{}.
\newblock \showarticletitle{{VelociTap}: {Investigating} {Fast} {Mobile} {Text}
  {Entry} using {Sentence}-{Based} {Decoding} of {Touchscreen} {Keyboard}
  {Input}}. In \bibinfo{booktitle}{\emph{Proceedings of the 33rd {Annual} {ACM}
  {Conference} on {Human} {Factors} in {Computing} {Systems}}}
  \emph{(\bibinfo{series}{{CHI} '15})}. \bibinfo{publisher}{Association for
  Computing Machinery}, \bibinfo{address}{Seoul, Republic of Korea},
  \bibinfo{pages}{659--668}.
\newblock
\showISBNx{978-1-4503-3145-6}


\bibitem[Wambsganss et~al\mbox{.}(2021a)]%
        {wambsganss21arguetutor}
\bibfield{author}{\bibinfo{person}{Thiemo Wambsganss}, \bibinfo{person}{Tobias
  Kueng}, \bibinfo{person}{Matthias Soellner}, {and} \bibinfo{person}{Jan~Marco
  Leimeister}.} \bibinfo{year}{2021}\natexlab{a}.
\newblock \showarticletitle{ArgueTutor: An Adaptive Dialog-Based Learning
  System for Argumentation Skills}. In \bibinfo{booktitle}{\emph{Proceedings of
  the 2021 CHI Conference on Human Factors in Computing Systems}} (Yokohama,
  Japan) \emph{(\bibinfo{series}{CHI '21})}. \bibinfo{publisher}{Association
  for Computing Machinery}, \bibinfo{address}{New York, NY, USA}, Article
  \bibinfo{articleno}{683}, \bibinfo{numpages}{13}~pages.
\newblock
\showISBNx{9781450380966}


\bibitem[Wambsganss and Niklaus(2022)]%
        {wambsganss2022modeling}
\bibfield{author}{\bibinfo{person}{Thiemo Wambsganss} {and}
  \bibinfo{person}{Christina Niklaus}.} \bibinfo{year}{2022}\natexlab{}.
\newblock \showarticletitle{Modeling Persuasive Discourse to Adaptively Support
  Students{'} Argumentative Writing}. In \bibinfo{booktitle}{\emph{Proceedings
  of the 60th Annual Meeting of the Association for Computational Linguistics
  (Volume 1: Long Papers)}}. \bibinfo{publisher}{Association for Computational
  Linguistics}, \bibinfo{address}{Dublin, Ireland},
  \bibinfo{pages}{8748--8760}.
\newblock


\bibitem[Wambsganss et~al\mbox{.}(2021b)]%
        {wambsganss2022supporting}
\bibfield{author}{\bibinfo{person}{Thiemo Wambsganss},
  \bibinfo{person}{Christina Niklaus}, \bibinfo{person}{Matthias S{\"o}llner},
  \bibinfo{person}{Siegfried Handschuh}, {and} \bibinfo{person}{Jan~Marco
  Leimeister}.} \bibinfo{year}{2021}\natexlab{b}.
\newblock \showarticletitle{Supporting Cognitive and Emotional Empathic Writing
  of Students}. In \bibinfo{booktitle}{\emph{Proceedings of the 59th Annual
  Meeting of the Association for Computational Linguistics and the 11th
  International Joint Conference on Natural Language Processing (Volume 1: Long
  Papers)}}. \bibinfo{publisher}{Association for Computational Linguistics},
  \bibinfo{address}{Online}, \bibinfo{pages}{4063--4077}.
\newblock


\bibitem[Wambsganss et~al\mbox{.}(2022)]%
        {wambsganss2022adaptive}
\bibfield{author}{\bibinfo{person}{Thiemo Wambsganss},
  \bibinfo{person}{Matthias Soellner}, \bibinfo{person}{Kenneth~R Koedinger},
  {and} \bibinfo{person}{Jan~Marco Leimeister}.}
  \bibinfo{year}{2022}\natexlab{}.
\newblock \showarticletitle{Adaptive Empathy Learning Support in Peer Review
  Scenarios}. In \bibinfo{booktitle}{\emph{Proceedings of the 2022 CHI
  Conference on Human Factors in Computing Systems}} (New Orleans, LA, USA)
  \emph{(\bibinfo{series}{CHI '22})}. \bibinfo{publisher}{Association for
  Computing Machinery}, \bibinfo{address}{New York, NY, USA}, Article
  \bibinfo{articleno}{227}, \bibinfo{numpages}{17}~pages.
\newblock
\showISBNx{9781450391573}


\bibitem[Wan et~al\mbox{.}(2022)]%
        {wan2022user}
\bibfield{author}{\bibinfo{person}{Ruyuan Wan}, \bibinfo{person}{Naome Etori},
  \bibinfo{person}{Karla Badillo-urquiola}, {and} \bibinfo{person}{Dongyeop
  Kang}.} \bibinfo{year}{2022}\natexlab{}.
\newblock \showarticletitle{User or Labor: An Interaction Framework for
  Human-Machine Relationships in {NLP}}. In
  \bibinfo{booktitle}{\emph{Proceedings of the Fourth Workshop on Data Science
  with Human-in-the-Loop (Language Advances)}}. \bibinfo{publisher}{Association
  for Computational Linguistics}, \bibinfo{address}{Abu Dhabi, United Arab
  Emirates (Hybrid)}, \bibinfo{pages}{112--121}.
\newblock


\bibitem[Wang et~al\mbox{.}(2023)]%
        {wang2023smart}
\bibfield{author}{\bibinfo{person}{Chenshuo Wang}, \bibinfo{person}{Shaoguang
  Mao}, \bibinfo{person}{Tao Ge}, \bibinfo{person}{Wenshan Wu},
  \bibinfo{person}{Xun Wang}, \bibinfo{person}{Yan Xia},
  \bibinfo{person}{Jonathan Tien}, {and} \bibinfo{person}{Dongyan Zhao}.}
  \bibinfo{year}{2023}\natexlab{}.
\newblock \showarticletitle{Smart Word Suggestions for Writing Assistance}. In
  \bibinfo{booktitle}{\emph{Findings of the Association for Computational
  Linguistics: ACL 2023}}. \bibinfo{publisher}{Association for Computational
  Linguistics}, \bibinfo{address}{Toronto, Canada},
  \bibinfo{pages}{11212--11225}.
\newblock


\bibitem[Wang et~al\mbox{.}(2018)]%
        {wang2018mirroru}
\bibfield{author}{\bibinfo{person}{Liuping Wang}, \bibinfo{person}{Xiangmin
  Fan}, \bibinfo{person}{Feng Tian}, \bibinfo{person}{Lingjia Deng},
  \bibinfo{person}{Shuai Ma}, \bibinfo{person}{Jin Huang}, {and}
  \bibinfo{person}{Hongan Wang}.} \bibinfo{year}{2018}\natexlab{}.
\newblock \showarticletitle{MirrorU: Scaffolding Emotional Reflection via
  In-Situ Assessment and Interactive Feedback}. In
  \bibinfo{booktitle}{\emph{Extended Abstracts of the 2018 CHI Conference on
  Human Factors in Computing Systems}} (Montreal QC, Canada)
  \emph{(\bibinfo{series}{CHI EA '18})}. \bibinfo{publisher}{Association for
  Computing Machinery}, \bibinfo{address}{New York, NY, USA},
  \bibinfo{pages}{1–6}.
\newblock
\showISBNx{9781450356213}


\bibitem[Wang et~al\mbox{.}(2016)]%
        {wang-etal-2016-non}
\bibfield{author}{\bibinfo{person}{Weibo Wang}, \bibinfo{person}{Abidalrahman
  Moh{'}d}, \bibinfo{person}{Aminul Islam}, \bibinfo{person}{Axel Soto}, {and}
  \bibinfo{person}{Evangelos Milios}.} \bibinfo{year}{2016}\natexlab{}.
\newblock \showarticletitle{Non-uniform Language Detection in Technical
  Writing}. In \bibinfo{booktitle}{\emph{Proceedings of the 2016 Conference on
  Empirical Methods in Natural Language Processing}}.
  \bibinfo{publisher}{Association for Computational Linguistics},
  \bibinfo{address}{Austin, Texas}, \bibinfo{pages}{1892--1900}.
\newblock


\bibitem[Watanabe et~al\mbox{.}(2017)]%
        {watanabe2017lyrisys}
\bibfield{author}{\bibinfo{person}{Kento Watanabe}, \bibinfo{person}{Yuichiroh
  Matsubayashi}, \bibinfo{person}{Kentaro Inui}, \bibinfo{person}{Tomoyasu
  Nakano}, \bibinfo{person}{Satoru Fukayama}, {and} \bibinfo{person}{Masataka
  Goto}.} \bibinfo{year}{2017}\natexlab{}.
\newblock \showarticletitle{LyriSys: An Interactive Support System for Writing
  Lyrics Based on Topic Transition}. In \bibinfo{booktitle}{\emph{Proceedings
  of the 22nd International Conference on Intelligent User Interfaces}}
  (Limassol, Cyprus) \emph{(\bibinfo{series}{IUI '17})}.
  \bibinfo{publisher}{Association for Computing Machinery},
  \bibinfo{address}{New York, NY, USA}, \bibinfo{pages}{559–563}.
\newblock
\showISBNx{9781450343480}


\bibitem[Weber et~al\mbox{.}(2023)]%
        {weber2023structured}
\bibfield{author}{\bibinfo{person}{Florian Weber}, \bibinfo{person}{Thiemo
  Wambsganss}, \bibinfo{person}{Seyed~Parsa Neshaei}, {and}
  \bibinfo{person}{Matthias Soellner}.} \bibinfo{year}{2023}\natexlab{}.
\newblock \showarticletitle{Structured Persuasive Writing Support in Legal
  Education: A Model and Tool for {G}erman Legal Case Solutions}. In
  \bibinfo{booktitle}{\emph{Findings of the Association for Computational
  Linguistics: ACL 2023}}. \bibinfo{publisher}{Association for Computational
  Linguistics}, \bibinfo{address}{Toronto, Canada},
  \bibinfo{pages}{2296--2313}.
\newblock


\bibitem[Weber et~al\mbox{.}(2021)]%
        {weber2021pedagogical}
\bibfield{author}{\bibinfo{person}{Florian Weber}, \bibinfo{person}{Thiemo
  Wambsganss}, \bibinfo{person}{Dominic R{\"u}ttimann}, {and}
  \bibinfo{person}{Matthias S{\"o}llner}.} \bibinfo{year}{2021}\natexlab{}.
\newblock \showarticletitle{Pedagogical agents for interactive learning: A
  taxonomy of conversational agents in education}. In
  \bibinfo{booktitle}{\emph{Forty-Second International Conference on
  Information Systems}}. \bibinfo{address}{Austin, Texas}.
\newblock


\bibitem[Wei et~al\mbox{.}(2022)]%
        {wei2022cot}
\bibfield{author}{\bibinfo{person}{Jason Wei}, \bibinfo{person}{Xuezhi Wang},
  \bibinfo{person}{Dale Schuurmans}, \bibinfo{person}{Maarten Bosma},
  \bibinfo{person}{Brian Ichter}, \bibinfo{person}{Fei Xia},
  \bibinfo{person}{Ed Chi}, \bibinfo{person}{Quoc Le}, {and}
  \bibinfo{person}{Denny Zhou}.} \bibinfo{year}{2022}\natexlab{}.
\newblock \showarticletitle{Chain-of-Thought Prompting Elicits Reasoning in
  Large Language Models}.
\newblock \bibinfo{journal}{\emph{arXiv preprint arXiv:2201.11903}}
  (\bibinfo{year}{2022}).
\newblock


\bibitem[Workshop et~al\mbox{.}(2023)]%
        {workshop2023bloom}
\bibfield{author}{\bibinfo{person}{BigScience Workshop}, \bibinfo{person}{:},
  \bibinfo{person}{Teven~Le Scao}, \bibinfo{person}{Angela Fan},
  \bibinfo{person}{Christopher Akiki}, \bibinfo{person}{Ellie Pavlick},
  \bibinfo{person}{Suzana Ilić}, \bibinfo{person}{Daniel Hesslow},
  \bibinfo{person}{Roman Castagné}, \bibinfo{person}{Alexandra~Sasha
  Luccioni}, \bibinfo{person}{François Yvon}, \bibinfo{person}{Matthias
  Gallé}, \bibinfo{person}{Jonathan Tow}, \bibinfo{person}{Alexander~M. Rush},
  \bibinfo{person}{Stella Biderman}, \bibinfo{person}{Albert Webson},
  \bibinfo{person}{Pawan~Sasanka Ammanamanchi}, \bibinfo{person}{Thomas Wang},
  \bibinfo{person}{Benoît Sagot}, \bibinfo{person}{Niklas Muennighoff},
  \bibinfo{person}{Albert~Villanova del Moral}, \bibinfo{person}{Olatunji
  Ruwase}, \bibinfo{person}{Rachel Bawden}, \bibinfo{person}{Stas Bekman},
  \bibinfo{person}{Angelina McMillan-Major}, \bibinfo{person}{Iz Beltagy},
  \bibinfo{person}{Huu Nguyen}, \bibinfo{person}{Lucile Saulnier},
  \bibinfo{person}{Samson Tan}, \bibinfo{person}{Pedro~Ortiz Suarez},
  \bibinfo{person}{Victor Sanh}, \bibinfo{person}{Hugo Laurençon},
  \bibinfo{person}{Yacine Jernite}, \bibinfo{person}{Julien Launay},
  \bibinfo{person}{Margaret Mitchell}, \bibinfo{person}{Colin Raffel},
  \bibinfo{person}{Aaron Gokaslan}, \bibinfo{person}{Adi Simhi},
  \bibinfo{person}{Aitor Soroa}, \bibinfo{person}{Alham~Fikri Aji},
  \bibinfo{person}{Amit Alfassy}, \bibinfo{person}{Anna Rogers},
  \bibinfo{person}{Ariel~Kreisberg Nitzav}, \bibinfo{person}{Canwen Xu},
  \bibinfo{person}{Chenghao Mou}, \bibinfo{person}{Chris Emezue},
  \bibinfo{person}{Christopher Klamm}, \bibinfo{person}{Colin Leong},
  \bibinfo{person}{Daniel van Strien}, \bibinfo{person}{David~Ifeoluwa
  Adelani}, \bibinfo{person}{Dragomir Radev},
  \bibinfo{person}{Eduardo~González Ponferrada}, \bibinfo{person}{Efrat
  Levkovizh}, \bibinfo{person}{Ethan Kim}, \bibinfo{person}{Eyal~Bar Natan},
  \bibinfo{person}{Francesco~De Toni}, \bibinfo{person}{Gérard Dupont},
  \bibinfo{person}{Germán Kruszewski}, \bibinfo{person}{Giada Pistilli},
  \bibinfo{person}{Hady Elsahar}, \bibinfo{person}{Hamza Benyamina},
  \bibinfo{person}{Hieu Tran}, \bibinfo{person}{Ian Yu}, \bibinfo{person}{Idris
  Abdulmumin}, \bibinfo{person}{Isaac Johnson}, \bibinfo{person}{Itziar
  Gonzalez-Dios}, \bibinfo{person}{Javier de~la Rosa}, \bibinfo{person}{Jenny
  Chim}, \bibinfo{person}{Jesse Dodge}, \bibinfo{person}{Jian Zhu},
  \bibinfo{person}{Jonathan Chang}, \bibinfo{person}{Jörg Frohberg},
  \bibinfo{person}{Joseph Tobing}, \bibinfo{person}{Joydeep Bhattacharjee},
  \bibinfo{person}{Khalid Almubarak}, \bibinfo{person}{Kimbo Chen},
  \bibinfo{person}{Kyle Lo}, \bibinfo{person}{Leandro~Von Werra},
  \bibinfo{person}{Leon Weber}, \bibinfo{person}{Long Phan},
  \bibinfo{person}{Loubna~Ben allal}, \bibinfo{person}{Ludovic Tanguy},
  \bibinfo{person}{Manan Dey}, \bibinfo{person}{Manuel~Romero Muñoz},
  \bibinfo{person}{Maraim Masoud}, \bibinfo{person}{María Grandury},
  \bibinfo{person}{Mario Šaško}, \bibinfo{person}{Max Huang},
  \bibinfo{person}{Maximin Coavoux}, \bibinfo{person}{Mayank Singh},
  \bibinfo{person}{Mike Tian-Jian Jiang}, \bibinfo{person}{Minh~Chien Vu},
  \bibinfo{person}{Mohammad~A. Jauhar}, \bibinfo{person}{Mustafa Ghaleb},
  \bibinfo{person}{Nishant Subramani}, \bibinfo{person}{Nora Kassner},
  \bibinfo{person}{Nurulaqilla Khamis}, \bibinfo{person}{Olivier Nguyen},
  \bibinfo{person}{Omar Espejel}, \bibinfo{person}{Ona de Gibert},
  \bibinfo{person}{Paulo Villegas}, \bibinfo{person}{Peter Henderson},
  \bibinfo{person}{Pierre Colombo}, \bibinfo{person}{Priscilla Amuok},
  \bibinfo{person}{Quentin Lhoest}, \bibinfo{person}{Rheza Harliman},
  \bibinfo{person}{Rishi Bommasani}, \bibinfo{person}{Roberto~Luis López},
  \bibinfo{person}{Rui Ribeiro}, \bibinfo{person}{Salomey Osei},
  \bibinfo{person}{Sampo Pyysalo}, \bibinfo{person}{Sebastian Nagel},
  \bibinfo{person}{Shamik Bose}, \bibinfo{person}{Shamsuddeen~Hassan Muhammad},
  \bibinfo{person}{Shanya Sharma}, \bibinfo{person}{Shayne Longpre},
  \bibinfo{person}{Somaieh Nikpoor}, \bibinfo{person}{Stanislav Silberberg},
  \bibinfo{person}{Suhas Pai}, \bibinfo{person}{Sydney Zink},
  \bibinfo{person}{Tiago~Timponi Torrent}, \bibinfo{person}{Timo Schick},
  \bibinfo{person}{Tristan Thrush}, \bibinfo{person}{Valentin Danchev},
  \bibinfo{person}{Vassilina Nikoulina}, \bibinfo{person}{Veronika Laippala},
  \bibinfo{person}{Violette Lepercq}, \bibinfo{person}{Vrinda Prabhu},
  \bibinfo{person}{Zaid Alyafeai}, \bibinfo{person}{Zeerak Talat},
  \bibinfo{person}{Arun Raja}, \bibinfo{person}{Benjamin Heinzerling},
  \bibinfo{person}{Chenglei Si}, \bibinfo{person}{Davut~Emre Taşar},
  \bibinfo{person}{Elizabeth Salesky}, \bibinfo{person}{Sabrina~J. Mielke},
  \bibinfo{person}{Wilson~Y. Lee}, \bibinfo{person}{Abheesht Sharma},
  \bibinfo{person}{Andrea Santilli}, \bibinfo{person}{Antoine Chaffin},
  \bibinfo{person}{Arnaud Stiegler}, \bibinfo{person}{Debajyoti Datta},
  \bibinfo{person}{Eliza Szczechla}, \bibinfo{person}{Gunjan Chhablani},
  \bibinfo{person}{Han Wang}, \bibinfo{person}{Harshit Pandey},
  \bibinfo{person}{Hendrik Strobelt}, \bibinfo{person}{Jason~Alan Fries},
  \bibinfo{person}{Jos Rozen}, \bibinfo{person}{Leo Gao},
  \bibinfo{person}{Lintang Sutawika}, \bibinfo{person}{M~Saiful Bari},
  \bibinfo{person}{Maged~S. Al-shaibani}, \bibinfo{person}{Matteo Manica},
  \bibinfo{person}{Nihal Nayak}, \bibinfo{person}{Ryan Teehan},
  \bibinfo{person}{Samuel Albanie}, \bibinfo{person}{Sheng Shen},
  \bibinfo{person}{Srulik Ben-David}, \bibinfo{person}{Stephen~H. Bach},
  \bibinfo{person}{Taewoon Kim}, \bibinfo{person}{Tali Bers},
  \bibinfo{person}{Thibault Fevry}, \bibinfo{person}{Trishala Neeraj},
  \bibinfo{person}{Urmish Thakker}, \bibinfo{person}{Vikas Raunak},
  \bibinfo{person}{Xiangru Tang}, \bibinfo{person}{Zheng-Xin Yong},
  \bibinfo{person}{Zhiqing Sun}, \bibinfo{person}{Shaked Brody},
  \bibinfo{person}{Yallow Uri}, \bibinfo{person}{Hadar Tojarieh},
  \bibinfo{person}{Adam Roberts}, \bibinfo{person}{Hyung~Won Chung},
  \bibinfo{person}{Jaesung Tae}, \bibinfo{person}{Jason Phang},
  \bibinfo{person}{Ofir Press}, \bibinfo{person}{Conglong Li},
  \bibinfo{person}{Deepak Narayanan}, \bibinfo{person}{Hatim Bourfoune},
  \bibinfo{person}{Jared Casper}, \bibinfo{person}{Jeff Rasley},
  \bibinfo{person}{Max Ryabinin}, \bibinfo{person}{Mayank Mishra},
  \bibinfo{person}{Minjia Zhang}, \bibinfo{person}{Mohammad Shoeybi},
  \bibinfo{person}{Myriam Peyrounette}, \bibinfo{person}{Nicolas Patry},
  \bibinfo{person}{Nouamane Tazi}, \bibinfo{person}{Omar Sanseviero},
  \bibinfo{person}{Patrick von Platen}, \bibinfo{person}{Pierre Cornette},
  \bibinfo{person}{Pierre~François Lavallée}, \bibinfo{person}{Rémi
  Lacroix}, \bibinfo{person}{Samyam Rajbhandari}, \bibinfo{person}{Sanchit
  Gandhi}, \bibinfo{person}{Shaden Smith}, \bibinfo{person}{Stéphane Requena},
  \bibinfo{person}{Suraj Patil}, \bibinfo{person}{Tim Dettmers},
  \bibinfo{person}{Ahmed Baruwa}, \bibinfo{person}{Amanpreet Singh},
  \bibinfo{person}{Anastasia Cheveleva}, \bibinfo{person}{Anne-Laure Ligozat},
  \bibinfo{person}{Arjun Subramonian}, \bibinfo{person}{Aurélie Névéol},
  \bibinfo{person}{Charles Lovering}, \bibinfo{person}{Dan Garrette},
  \bibinfo{person}{Deepak Tunuguntla}, \bibinfo{person}{Ehud Reiter},
  \bibinfo{person}{Ekaterina Taktasheva}, \bibinfo{person}{Ekaterina
  Voloshina}, \bibinfo{person}{Eli Bogdanov}, \bibinfo{person}{Genta~Indra
  Winata}, \bibinfo{person}{Hailey Schoelkopf}, \bibinfo{person}{Jan-Christoph
  Kalo}, \bibinfo{person}{Jekaterina Novikova}, \bibinfo{person}{Jessica~Zosa
  Forde}, \bibinfo{person}{Jordan Clive}, \bibinfo{person}{Jungo Kasai},
  \bibinfo{person}{Ken Kawamura}, \bibinfo{person}{Liam Hazan},
  \bibinfo{person}{Marine Carpuat}, \bibinfo{person}{Miruna Clinciu},
  \bibinfo{person}{Najoung Kim}, \bibinfo{person}{Newton Cheng},
  \bibinfo{person}{Oleg Serikov}, \bibinfo{person}{Omer Antverg},
  \bibinfo{person}{Oskar van~der Wal}, \bibinfo{person}{Rui Zhang},
  \bibinfo{person}{Ruochen Zhang}, \bibinfo{person}{Sebastian Gehrmann},
  \bibinfo{person}{Shachar Mirkin}, \bibinfo{person}{Shani Pais},
  \bibinfo{person}{Tatiana Shavrina}, \bibinfo{person}{Thomas Scialom},
  \bibinfo{person}{Tian Yun}, \bibinfo{person}{Tomasz Limisiewicz},
  \bibinfo{person}{Verena Rieser}, \bibinfo{person}{Vitaly Protasov},
  \bibinfo{person}{Vladislav Mikhailov}, \bibinfo{person}{Yada Pruksachatkun},
  \bibinfo{person}{Yonatan Belinkov}, \bibinfo{person}{Zachary Bamberger},
  \bibinfo{person}{Zdeněk Kasner}, \bibinfo{person}{Alice Rueda},
  \bibinfo{person}{Amanda Pestana}, \bibinfo{person}{Amir Feizpour},
  \bibinfo{person}{Ammar Khan}, \bibinfo{person}{Amy Faranak},
  \bibinfo{person}{Ana Santos}, \bibinfo{person}{Anthony Hevia},
  \bibinfo{person}{Antigona Unldreaj}, \bibinfo{person}{Arash Aghagol},
  \bibinfo{person}{Arezoo Abdollahi}, \bibinfo{person}{Aycha Tammour},
  \bibinfo{person}{Azadeh HajiHosseini}, \bibinfo{person}{Bahareh Behroozi},
  \bibinfo{person}{Benjamin Ajibade}, \bibinfo{person}{Bharat Saxena},
  \bibinfo{person}{Carlos~Muñoz Ferrandis}, \bibinfo{person}{Daniel McDuff},
  \bibinfo{person}{Danish Contractor}, \bibinfo{person}{David Lansky},
  \bibinfo{person}{Davis David}, \bibinfo{person}{Douwe Kiela},
  \bibinfo{person}{Duong~A. Nguyen}, \bibinfo{person}{Edward Tan},
  \bibinfo{person}{Emi Baylor}, \bibinfo{person}{Ezinwanne Ozoani},
  \bibinfo{person}{Fatima Mirza}, \bibinfo{person}{Frankline Ononiwu},
  \bibinfo{person}{Habib Rezanejad}, \bibinfo{person}{Hessie Jones},
  \bibinfo{person}{Indrani Bhattacharya}, \bibinfo{person}{Irene Solaiman},
  \bibinfo{person}{Irina Sedenko}, \bibinfo{person}{Isar Nejadgholi},
  \bibinfo{person}{Jesse Passmore}, \bibinfo{person}{Josh Seltzer},
  \bibinfo{person}{Julio~Bonis Sanz}, \bibinfo{person}{Livia Dutra},
  \bibinfo{person}{Mairon Samagaio}, \bibinfo{person}{Maraim Elbadri},
  \bibinfo{person}{Margot Mieskes}, \bibinfo{person}{Marissa Gerchick},
  \bibinfo{person}{Martha Akinlolu}, \bibinfo{person}{Michael McKenna},
  \bibinfo{person}{Mike Qiu}, \bibinfo{person}{Muhammed Ghauri},
  \bibinfo{person}{Mykola Burynok}, \bibinfo{person}{Nafis Abrar},
  \bibinfo{person}{Nazneen Rajani}, \bibinfo{person}{Nour Elkott},
  \bibinfo{person}{Nour Fahmy}, \bibinfo{person}{Olanrewaju Samuel},
  \bibinfo{person}{Ran An}, \bibinfo{person}{Rasmus Kromann},
  \bibinfo{person}{Ryan Hao}, \bibinfo{person}{Samira Alizadeh},
  \bibinfo{person}{Sarmad Shubber}, \bibinfo{person}{Silas Wang},
  \bibinfo{person}{Sourav Roy}, \bibinfo{person}{Sylvain Viguier},
  \bibinfo{person}{Thanh Le}, \bibinfo{person}{Tobi Oyebade},
  \bibinfo{person}{Trieu Le}, \bibinfo{person}{Yoyo Yang},
  \bibinfo{person}{Zach Nguyen}, \bibinfo{person}{Abhinav~Ramesh Kashyap},
  \bibinfo{person}{Alfredo Palasciano}, \bibinfo{person}{Alison Callahan},
  \bibinfo{person}{Anima Shukla}, \bibinfo{person}{Antonio Miranda-Escalada},
  \bibinfo{person}{Ayush Singh}, \bibinfo{person}{Benjamin Beilharz},
  \bibinfo{person}{Bo Wang}, \bibinfo{person}{Caio Brito},
  \bibinfo{person}{Chenxi Zhou}, \bibinfo{person}{Chirag Jain},
  \bibinfo{person}{Chuxin Xu}, \bibinfo{person}{Clémentine Fourrier},
  \bibinfo{person}{Daniel~León Periñán}, \bibinfo{person}{Daniel Molano},
  \bibinfo{person}{Dian Yu}, \bibinfo{person}{Enrique Manjavacas},
  \bibinfo{person}{Fabio Barth}, \bibinfo{person}{Florian Fuhrimann},
  \bibinfo{person}{Gabriel Altay}, \bibinfo{person}{Giyaseddin Bayrak},
  \bibinfo{person}{Gully Burns}, \bibinfo{person}{Helena~U. Vrabec},
  \bibinfo{person}{Imane Bello}, \bibinfo{person}{Ishani Dash},
  \bibinfo{person}{Jihyun Kang}, \bibinfo{person}{John Giorgi},
  \bibinfo{person}{Jonas Golde}, \bibinfo{person}{Jose~David Posada},
  \bibinfo{person}{Karthik~Rangasai Sivaraman}, \bibinfo{person}{Lokesh
  Bulchandani}, \bibinfo{person}{Lu Liu}, \bibinfo{person}{Luisa Shinzato},
  \bibinfo{person}{Madeleine~Hahn de Bykhovetz}, \bibinfo{person}{Maiko
  Takeuchi}, \bibinfo{person}{Marc Pàmies}, \bibinfo{person}{Maria~A
  Castillo}, \bibinfo{person}{Marianna Nezhurina}, \bibinfo{person}{Mario
  Sänger}, \bibinfo{person}{Matthias Samwald}, \bibinfo{person}{Michael
  Cullan}, \bibinfo{person}{Michael Weinberg}, \bibinfo{person}{Michiel~De
  Wolf}, \bibinfo{person}{Mina Mihaljcic}, \bibinfo{person}{Minna Liu},
  \bibinfo{person}{Moritz Freidank}, \bibinfo{person}{Myungsun Kang},
  \bibinfo{person}{Natasha Seelam}, \bibinfo{person}{Nathan Dahlberg},
  \bibinfo{person}{Nicholas~Michio Broad}, \bibinfo{person}{Nikolaus Muellner},
  \bibinfo{person}{Pascale Fung}, \bibinfo{person}{Patrick Haller},
  \bibinfo{person}{Ramya Chandrasekhar}, \bibinfo{person}{Renata Eisenberg},
  \bibinfo{person}{Robert Martin}, \bibinfo{person}{Rodrigo Canalli},
  \bibinfo{person}{Rosaline Su}, \bibinfo{person}{Ruisi Su},
  \bibinfo{person}{Samuel Cahyawijaya}, \bibinfo{person}{Samuele Garda},
  \bibinfo{person}{Shlok~S Deshmukh}, \bibinfo{person}{Shubhanshu Mishra},
  \bibinfo{person}{Sid Kiblawi}, \bibinfo{person}{Simon Ott},
  \bibinfo{person}{Sinee Sang-aroonsiri}, \bibinfo{person}{Srishti Kumar},
  \bibinfo{person}{Stefan Schweter}, \bibinfo{person}{Sushil Bharati},
  \bibinfo{person}{Tanmay Laud}, \bibinfo{person}{Théo Gigant},
  \bibinfo{person}{Tomoya Kainuma}, \bibinfo{person}{Wojciech Kusa},
  \bibinfo{person}{Yanis Labrak}, \bibinfo{person}{Yash~Shailesh Bajaj},
  \bibinfo{person}{Yash Venkatraman}, \bibinfo{person}{Yifan Xu},
  \bibinfo{person}{Yingxin Xu}, \bibinfo{person}{Yu Xu}, \bibinfo{person}{Zhe
  Tan}, \bibinfo{person}{Zhongli Xie}, \bibinfo{person}{Zifan Ye},
  \bibinfo{person}{Mathilde Bras}, \bibinfo{person}{Younes Belkada}, {and}
  \bibinfo{person}{Thomas Wolf}.} \bibinfo{year}{2023}\natexlab{}.
\newblock \bibinfo{title}{BLOOM: A 176B-Parameter Open-Access Multilingual
  Language Model}.
\newblock
\newblock
\showeprint[arxiv]{2211.05100}~[cs.CL]


\bibitem[Wu et~al\mbox{.}(2019)]%
        {wu2019design}
\bibfield{author}{\bibinfo{person}{Shaomei Wu}, \bibinfo{person}{Lindsay
  Reynolds}, \bibinfo{person}{Xian Li}, {and} \bibinfo{person}{Francisco
  Guzm\'{a}n}.} \bibinfo{year}{2019}\natexlab{}.
\newblock \showarticletitle{Design and Evaluation of a Social Media Writing
  Support Tool for People with Dyslexia}. In
  \bibinfo{booktitle}{\emph{Proceedings of the 2019 CHI Conference on Human
  Factors in Computing Systems}} (Glasgow, Scotland Uk)
  \emph{(\bibinfo{series}{CHI '19})}. \bibinfo{publisher}{Association for
  Computing Machinery}, \bibinfo{address}{New York, NY, USA},
  \bibinfo{pages}{1–14}.
\newblock
\showISBNx{9781450359702}


\bibitem[Wu et~al\mbox{.}(2021)]%
        {wu-etal-2021-automatic}
\bibfield{author}{\bibinfo{person}{Zeqiu Wu}, \bibinfo{person}{Michel Galley},
  \bibinfo{person}{Chris Brockett}, \bibinfo{person}{Yizhe Zhang}, {and}
  \bibinfo{person}{Bill Dolan}.} \bibinfo{year}{2021}\natexlab{}.
\newblock \showarticletitle{Automatic Document Sketching: Generating Drafts
  from Analogous Texts}. In \bibinfo{booktitle}{\emph{Findings of the
  Association for Computational Linguistics: ACL-IJCNLP 2021}},
  \bibfield{editor}{\bibinfo{person}{Chengqing Zong}, \bibinfo{person}{Fei
  Xia}, \bibinfo{person}{Wenjie Li}, {and} \bibinfo{person}{Roberto Navigli}}
  (Eds.). \bibinfo{publisher}{Association for Computational Linguistics},
  \bibinfo{address}{Online}, \bibinfo{pages}{2102--2113}.
\newblock


\bibitem[Xu et~al\mbox{.}(2019)]%
        {xu2019alter}
\bibfield{author}{\bibinfo{person}{Qiongkai Xu}, \bibinfo{person}{Chenchen Xu},
  {and} \bibinfo{person}{Lizhen Qu}.} \bibinfo{year}{2019}\natexlab{}.
\newblock \showarticletitle{{ALTER}: Auxiliary Text Rewriting Tool for Natural
  Language Generation}. In \bibinfo{booktitle}{\emph{Proceedings of the 2019
  Conference on Empirical Methods in Natural Language Processing and the 9th
  International Joint Conference on Natural Language Processing (EMNLP-IJCNLP):
  System Demonstrations}}. \bibinfo{publisher}{Association for Computational
  Linguistics}, \bibinfo{address}{Hong Kong, China}, \bibinfo{pages}{13--18}.
\newblock


\bibitem[Yang et~al\mbox{.}(2017)]%
        {yang-etal-2017-identifying-semantic}
\bibfield{author}{\bibinfo{person}{Diyi Yang}, \bibinfo{person}{Aaron
  Halfaker}, \bibinfo{person}{Robert Kraut}, {and} \bibinfo{person}{Eduard
  Hovy}.} \bibinfo{year}{2017}\natexlab{}.
\newblock \showarticletitle{Identifying Semantic Edit Intentions from Revisions
  in {W}ikipedia}. In \bibinfo{booktitle}{\emph{Proceedings of the 2017
  Conference on Empirical Methods in Natural Language Processing}},
  \bibfield{editor}{\bibinfo{person}{Martha Palmer}, \bibinfo{person}{Rebecca
  Hwa}, {and} \bibinfo{person}{Sebastian Riedel}} (Eds.).
  \bibinfo{publisher}{Association for Computational Linguistics},
  \bibinfo{address}{Copenhagen, Denmark}, \bibinfo{pages}{2000--2010}.
\newblock


\bibitem[yew Lin and Rey(2004)]%
        {lin2004rouge}
\bibfield{author}{\bibinfo{person}{Chin yew Lin} {and} \bibinfo{person}{Marina
  Rey}.} \bibinfo{year}{2004}\natexlab{}.
\newblock \showarticletitle{Looking for a Few Good Metrics: {ROUGE} and its
  Evaluation}. In \bibinfo{booktitle}{\emph{NTCIR Workshop}}.
\newblock


\bibitem[Yimam and Biemann(2018)]%
        {yimam2018demonstrating}
\bibfield{author}{\bibinfo{person}{Seid~Muhie Yimam} {and}
  \bibinfo{person}{Chris Biemann}.} \bibinfo{year}{2018}\natexlab{}.
\newblock \showarticletitle{Demonstrating {P}ar4{S}em - A Semantic Writing Aid
  with Adaptive Paraphrasing}. In \bibinfo{booktitle}{\emph{Proceedings of the
  2018 Conference on Empirical Methods in Natural Language Processing: System
  Demonstrations}}. \bibinfo{publisher}{Association for Computational
  Linguistics}, \bibinfo{address}{Brussels, Belgium}, \bibinfo{pages}{48--53}.
\newblock


\bibitem[Yuan et~al\mbox{.}(2022)]%
        {yuan2022wordcraft}
\bibfield{author}{\bibinfo{person}{Ann Yuan}, \bibinfo{person}{Andy Coenen},
  \bibinfo{person}{Emily Reif}, {and} \bibinfo{person}{Daphne Ippolito}.}
  \bibinfo{year}{2022}\natexlab{}.
\newblock \showarticletitle{Wordcraft: Story Writing With Large Language
  Models}. In \bibinfo{booktitle}{\emph{27th International Conference on
  Intelligent User Interfaces}} (Helsinki, Finland) \emph{(\bibinfo{series}{IUI
  '22})}. \bibinfo{publisher}{Association for Computing Machinery},
  \bibinfo{address}{New York, NY, USA}, \bibinfo{pages}{841–852}.
\newblock
\showISBNx{9781450391443}


\bibitem[Zamfirescu-Pereira et~al\mbox{.}(2023)]%
        {zamfirescu2023johnny}
\bibfield{author}{\bibinfo{person}{J.D. Zamfirescu-Pereira},
  \bibinfo{person}{Richmond~Y. Wong}, \bibinfo{person}{Bjoern Hartmann}, {and}
  \bibinfo{person}{Qian Yang}.} \bibinfo{year}{2023}\natexlab{}.
\newblock \showarticletitle{Why Johnny Can’t Prompt: How Non-AI Experts Try
  (and Fail) to Design LLM Prompts}. In \bibinfo{booktitle}{\emph{Proceedings
  of the 2023 CHI Conference on Human Factors in Computing Systems}}
  (<conf-loc>, <city>Hamburg</city>, <country>Germany</country>, </conf-loc>)
  \emph{(\bibinfo{series}{CHI '23})}. \bibinfo{publisher}{Association for
  Computing Machinery}, \bibinfo{address}{New York, NY, USA}, Article
  \bibinfo{articleno}{437}, \bibinfo{numpages}{21}~pages.
\newblock
\showISBNx{9781450394215}


\bibitem[Zarei et~al\mbox{.}(2020)]%
        {zarei2020investigating}
\bibfield{author}{\bibinfo{person}{Niloofar Zarei},
  \bibinfo{person}{Sharon~Lynn Chu}, \bibinfo{person}{Francis Quek},
  \bibinfo{person}{Nanjie'Jimmy' Rao}, {and} \bibinfo{person}{Sarah~Anne
  Brown}.} \bibinfo{year}{2020}\natexlab{}.
\newblock \showarticletitle{Investigating the Effects of Self-Avatars and
  Story-Relevant Avatars on Children's Creative Storytelling}. In
  \bibinfo{booktitle}{\emph{Proceedings of the 2020 CHI Conference on Human
  Factors in Computing Systems}}. \bibinfo{pages}{1--11}.
\newblock


\bibitem[Zhang et~al\mbox{.}(2016)]%
        {zhang2016argrewrite}
\bibfield{author}{\bibinfo{person}{Fan Zhang}, \bibinfo{person}{Rebecca Hwa},
  \bibinfo{person}{Diane Litman}, {and} \bibinfo{person}{Homa~B. Hashemi}.}
  \bibinfo{year}{2016}\natexlab{}.
\newblock \showarticletitle{{A}rg{R}ewrite: A Web-based Revision Assistant for
  Argumentative Writings}. In \bibinfo{booktitle}{\emph{Proceedings of the 2016
  Conference of the North {A}merican Chapter of the Association for
  Computational Linguistics: Demonstrations}}. \bibinfo{publisher}{Association
  for Computational Linguistics}, \bibinfo{address}{San Diego, California},
  \bibinfo{pages}{37--41}.
\newblock


\bibitem[Zhang and Litman(2016)]%
        {zhang2016using}
\bibfield{author}{\bibinfo{person}{Fan Zhang} {and} \bibinfo{person}{Diane
  Litman}.} \bibinfo{year}{2016}\natexlab{}.
\newblock \showarticletitle{Using Context to Predict the Purpose of
  Argumentative Writing Revisions}. In \bibinfo{booktitle}{\emph{Proceedings of
  the 2016 Conference of the North {A}merican Chapter of the Association for
  Computational Linguistics: Human Language Technologies}}.
  \bibinfo{publisher}{Association for Computational Linguistics},
  \bibinfo{address}{San Diego, California}, \bibinfo{pages}{1424--1430}.
\newblock


\bibitem[Zhang et~al\mbox{.}(2019a)]%
        {zhang2019bertscore}
\bibfield{author}{\bibinfo{person}{Tianyi Zhang}, \bibinfo{person}{Varsha
  Kishore}, \bibinfo{person}{Felix Wu}, \bibinfo{person}{Kilian~Q Weinberger},
  {and} \bibinfo{person}{Yoav Artzi}.} \bibinfo{year}{2019}\natexlab{a}.
\newblock \showarticletitle{Bertscore: Evaluating text generation with bert}.
\newblock \bibinfo{journal}{\emph{arXiv preprint arXiv:1904.09675}}
  (\bibinfo{year}{2019}).
\newblock


\bibitem[Zhang et~al\mbox{.}(2019b)]%
        {zhang-etal-2019-modeling}
\bibfield{author}{\bibinfo{person}{Xuchao Zhang}, \bibinfo{person}{Dheeraj
  Rajagopal}, \bibinfo{person}{Michael Gamon}, \bibinfo{person}{Sujay~Kumar
  Jauhar}, {and} \bibinfo{person}{ChangTien Lu}.}
  \bibinfo{year}{2019}\natexlab{b}.
\newblock \showarticletitle{Modeling the Relationship between User Comments and
  Edits in Document Revision}. In \bibinfo{booktitle}{\emph{Proceedings of the
  2019 Conference on Empirical Methods in Natural Language Processing and the
  9th International Joint Conference on Natural Language Processing
  (EMNLP-IJCNLP)}}, \bibfield{editor}{\bibinfo{person}{Kentaro Inui},
  \bibinfo{person}{Jing Jiang}, \bibinfo{person}{Vincent Ng}, {and}
  \bibinfo{person}{Xiaojun Wan}} (Eds.). \bibinfo{publisher}{Association for
  Computational Linguistics}, \bibinfo{address}{Hong Kong, China},
  \bibinfo{pages}{5002--5011}.
\newblock


\bibitem[Zhong et~al\mbox{.}(2023)]%
        {zhong2023fiction}
\bibfield{author}{\bibinfo{person}{Wenjie Zhong}, \bibinfo{person}{Jason
  Naradowsky}, \bibinfo{person}{Hiroya Takamura}, \bibinfo{person}{Ichiro
  Kobayashi}, {and} \bibinfo{person}{Yusuke Miyao}.}
  \bibinfo{year}{2023}\natexlab{}.
\newblock \showarticletitle{Fiction-Writing Mode: An Effective Control for
  Human-Machine Collaborative Writing}. In
  \bibinfo{booktitle}{\emph{Proceedings of the 17th Conference of the European
  Chapter of the Association for Computational Linguistics}}.
  \bibinfo{publisher}{Association for Computational Linguistics},
  \bibinfo{address}{Dubrovnik, Croatia}, \bibinfo{pages}{1752--1765}.
\newblock


\bibitem[Zhu et~al\mbox{.}(2023)]%
        {zhu2023visualize}
\bibfield{author}{\bibinfo{person}{Wanrong Zhu}, \bibinfo{person}{An Yan},
  \bibinfo{person}{Yujie Lu}, \bibinfo{person}{Wenda Xu}, \bibinfo{person}{Xin
  Wang}, \bibinfo{person}{Miguel Eckstein}, {and} \bibinfo{person}{William~Yang
  Wang}.} \bibinfo{year}{2023}\natexlab{}.
\newblock \showarticletitle{Visualize Before You Write: Imagination-Guided
  Open-Ended Text Generation}. In \bibinfo{booktitle}{\emph{Findings of the
  Association for Computational Linguistics: EACL 2023}}.
  \bibinfo{publisher}{Association for Computational Linguistics},
  \bibinfo{address}{Dubrovnik, Croatia}, \bibinfo{pages}{78--92}.
\newblock


\bibitem[Zomer and Frankenberg-Garcia(2021)]%
        {zomer2021beyond}
\bibfield{author}{\bibinfo{person}{Gustavo Zomer} {and} \bibinfo{person}{Ana
  Frankenberg-Garcia}.} \bibinfo{year}{2021}\natexlab{}.
\newblock \showarticletitle{Beyond Grammatical Error Correction: Improving
  {L}1-influenced research writing in {E}nglish using pre-trained
  encoder-decoder models}. In \bibinfo{booktitle}{\emph{Findings of the
  Association for Computational Linguistics: EMNLP 2021}}.
  \bibinfo{publisher}{Association for Computational Linguistics},
  \bibinfo{address}{Punta Cana, Dominican Republic},
  \bibinfo{pages}{2534--2540}.
\newblock


\end{thebibliography}
